%% file: edf.tex
\documentclass[twoside,12pt]{article}
\usepackage{epsfig,color,ulem,hyperref,float,amsmath,bm,amssymb}
\usepackage[T1]{fontenc}
 
\def\Journal#1#2#3#4{{#1} {#2} (#4) #3 }

\def\NPA{{\em Nucl. Phys.} A}

\def\PLB{{\em Phys. Lett.} B}

\def\PRL{\em Phys. Rev. Lett.}

\def\PREP{\em Phys. Rep.}

\def\PRD{{\em Phys. Rev.} D}
\def\PRC{{\em Phys. Rev.} C}

\def\r{\vec r}

\newcommand{\be}{\begin{equation}}
\newcommand{\ee}{\end{equation}}
\newcommand{\bea}{\begin{eqnarray}}
\newcommand{\eea}{\end{eqnarray}}

\begin{document}

\title{Nuclear Equation of State from ground and collective excited state properties of nuclei}

\author{X.\ Roca-Maza$^{1}$ and N.\ Paar$^2$\\ 
$^1$Dipartimento di Fisica, Universit\`a degli Studi di Milano\\ and INFN,  Sezione di Milano, 20133 Milano, Italy.\\
$^2$Department of Physics, Faculty of Science,\\ University of Zagreb, Bijeni$\check{\rm c}$ka c. 32, 10000 Zagreb, Croatia.\\
}

\maketitle
\begin{abstract}
  This contribution reviews the present status on the available constraints to the nuclear equation of state (EoS) around saturation density from nuclear structure calculations on ground and collective excited state properties of atomic nuclei. It concentrates on predictions based on self-consistent mean-field calculations, which can be considered as an approximate realization of an exact energy density functional (EDF). EDFs are derived from effective interactions commonly fitted to nuclear masses, charge radii and, in many cases, also to pseudo-data such as nuclear matter properties. Although in a model dependent way, EDFs constitute nowadays a unique tool to reliably and consistently access bulk ground state and collective excited state properties of atomic nuclei along the nuclear chart as well as the EoS. For comparison, some emphasis is also given to the results obtained with the so called {\it ab initio} approaches that aim at describing the nuclear EoS based on interactions fitted to few-body data only. Bridging the existent gap between these two frameworks will be essential since it may allow to improve our understanding on the diverse phenomenology observed in nuclei. Examples on observations from astrophysical objects and processes sensitive to the nuclear EoS are also briefly discussed. As the main conclusion, the isospin dependence of the nuclear EoS around saturation density and, to a lesser extent, the nuclear matter incompressibility remain to be accurately determined. Experimental and theoretical efforts in finding and measuring observables specially sensitive to the EoS properties are of paramount importance, not only for low-energy nuclear physics but also for nuclear astrophysics applications.

\end{abstract}
%\eject
\tableofcontents

\input{introduction.tex}

\input{phenomenology.tex}
\input{theory.tex}
\input{eos.tex}

\input{gs.tex}

\input{excitations.tex}

\input{conclusions.tex}

\section*{Acknowledgments}
This work was supported by the Croatian Science Foundation under the project
Structure and Dynamics of Exotic Femtosystems (IP-2014-09-9159) and
the QuantiXLie Centre of Excellence, a project
co-financed by the Croatian Government and European Union through the
European Regional Development Fund - the Competitiveness and Cohesion
Operational Programme (Grant KK.01.1.1.01.0004). Funding from the European Union's Horizon 2020 research and innovation programme under grant agreement No. 654002 is also acknowledged.
% bibliography
\bibliographystyle{unsrt}
\bibliography{bibliography}{}
\end{document}

%% file: introduction.tex
\section{Introduction}
The nuclear Equation of State (EoS) describes the energy per nucleon as a function of the neutron ($\rho_n$) and proton ($\rho_p$) densities of an uniform and infinite system at zero temperature that interact only via the residual strong interaction, or nuclear force \cite{baldo2011,lattimer2012,oertel2017}. Being the nuclear force of short-range (of the order of the fm), the energy per nucleon (of the order of the MeV) at a given nucleon density is finite and constant for a wide range of temperatures (1 MeV$\sim 10^{10}$ K). The EoS of an ideal system is, therefore, realized within a good approximation in the interior of nuclei --due to the short-range of the interaction as compared to the dimensions of most of the known nuclei-- and in the interior of cold neutron stars --due to the strong gravitational field able to host very high densities in its interior. The EoS has also been studied at non-zero temperatures relevant for the studies of some astrophysical processes \cite{lattimer2012,oertel2017}. In this review, we focus on laboratory data on some ground and collective excited state properties in nuclei that allow to probe the EoS at zero temperature and around nuclear saturation density ($\rho_0\approx0.16$ fm${}^{-3}$). The Coulomb interaction should be neglected for the study of the nuclear EoS. Being the Coulomb interaction of long-range, the energy per nucleon will diverge ($E_{\rm Coul.}\sim e^2Z^2/r$). Therefore, when studying finite systems, the Coulomb interaction should be accounted for and its effects subtracted before comparison with the EoS. In general, the separation between bulk and surface properties of nuclei is model dependent and, thus, to establish the relation between the properties of finite nuclei and the EoS will suffer from model intrinsic uncertainties difficult to evaluate.

Over the past years, different approaches to the nuclear EoS have been introduced. Simple macroscopic models allow one to assess some of the main properties such as the energy and density of symmetric nuclear matter at saturation, i.e., the EoS for the case $\rho_n=\rho_p$. Another relevant property is nuclear matter incompressibility, that is proportional to the curvature of the symmetric nuclear matter EoS. The knowledge on the EoS is insufficient for a quantitative description of finite nuclei.
% and, thus, limits our understanding of the EoS to a qualitative level.
From a microscopic perspective, the most successful methodology for an effective description of the bulk properties in nuclei along the periodic table corresponds to the self-consistent mean field approach. The respective models can be understood as an approximate realization of a nuclear energy density functional (EDF). The Density Functional Theory is a powerful and general approach already successfully implemented in physics, chemistry and material science \cite{martin2004}. The EDFs contain a relatively small number of parameters, of the order of ten, commonly adjusted to nuclear masses and charge radii. Within the EDF framework one can evaluate the EoS, and comparison of different models provides some insight into the model dependence of the EoS. By employing the experimental data on finite nuclei, constraints on the EoS are possible at and around the saturation density. EDFs, however, are not suitable for the study of single-particle dynamics in nuclei. Proof of that is the experimental evidence on the fragmentation of single-particle states or on the finite width (several MeV) in nuclear giant resonances. Therefore, one should calibrate these models to the proper observables before the nuclear EoS is analyzed in detail.

A different microscopic perspective, based on {\it ab initio} models with realistic nucleon-nucleon (NN) interaction in the vacuum, represents a more fundamental approach 
to the nuclear EoS.
There exist different types of NN potentials typically fitted to available nucleon-nucleon scattering data in the vacuum and few-body nuclear data. From these potentials, approximate many-body methods have been applied to study not only the EoS but also some properties in light and light-medium mass finite systems \cite{hofmann01,hirose07,soma13,ab1,ab2,ab3,ab4,ab5,ab6,ab7,ogawa11}. It has been shown that in addition to the NN interaction also three-body interaction is necessary for a reasonable description of the nuclear saturation point of the symmetric matter EoS, one of the basic properties needed to reproduce the experimental mass and size of finite nuclei. However, within the {\it ab initio} approaches medium effects are not well understood yet (see Fig.\ref{fig-eos-13} in Sec.\ref{eos}).

Hence, nuclear EDFs constitute nowadays a unique tool to reliably and consistently access bulk ground state and collective excited state properties of atomic nuclei along the nuclear chart as well as the EoS around saturation density. Accordingly, we present in this review the status on the available constraints to the nuclear EoS around saturation density from nuclear structure calculations based on EDFs.

The review is organized as follows. In Sec.\ref{pheno}, some basic phenomenology is presented and briefly discussed. The aim of this section is to provide a qualitative description of the nucleus, crucial for a reasonable understanding of the main properties of the nuclear EoS around the saturation density. In Sec.\ref{theo}, the basic methods employed to derive the EDFs of current use is given. This section is important to explain in some detail how this type of models access information on ground state and collective excited states in finite nuclei. The expressions for the nuclear EoS derived from EDFs is also given. Some basic ideas and examples on the relevance of assessing theoretical errors is discussed. In Sec.\ref{eos}, a summary on the main constraints to the EoS is given. This serves as an introduction to the more detailed discussions in the following sections. Purely theoretical predictions on the EoS are also briefly discussed. The latter is important for a better understanding of the current situation from other approaches and, thus, of the limitations that the community is facing in the study of the nuclear EoS. In Secs.\ref{gs} and \ref{excitations}, constraints on the EoS from masses and nuclear radii as well as collective excitations derived from the EDFs are presented. In some cases, for the sake of comparison, results from macroscopic-microscopic models and {\it ab initio} approaches are also provided. Finally, our conclusions lay in Sec.\ref{conclusions} together with a table summarizing the constraints on the parameters of the nuclear EoS around the saturation density discussed along this review.

%% file: phenomenology.tex
\section{Phenomenology}
\label{pheno}

In this section we will first briefly describe the phenomenology that allows for an overall understanding of the basic properties of the atomic nucleus and of the nuclear equation of state around the equilibrium densities present in the interior of stable nuclei --the so called saturation density $\rho_0=0.16$ fm$^{-3}$. We will then highlight different macroscopic models that have been used to get some deeper insights into the EoS by connecting properties of finite nuclei with parameters easily related to the EoS. 

\subsection{Ground state}

\begin{figure}[t!]
\includegraphics[width=0.5\linewidth,clip=true]{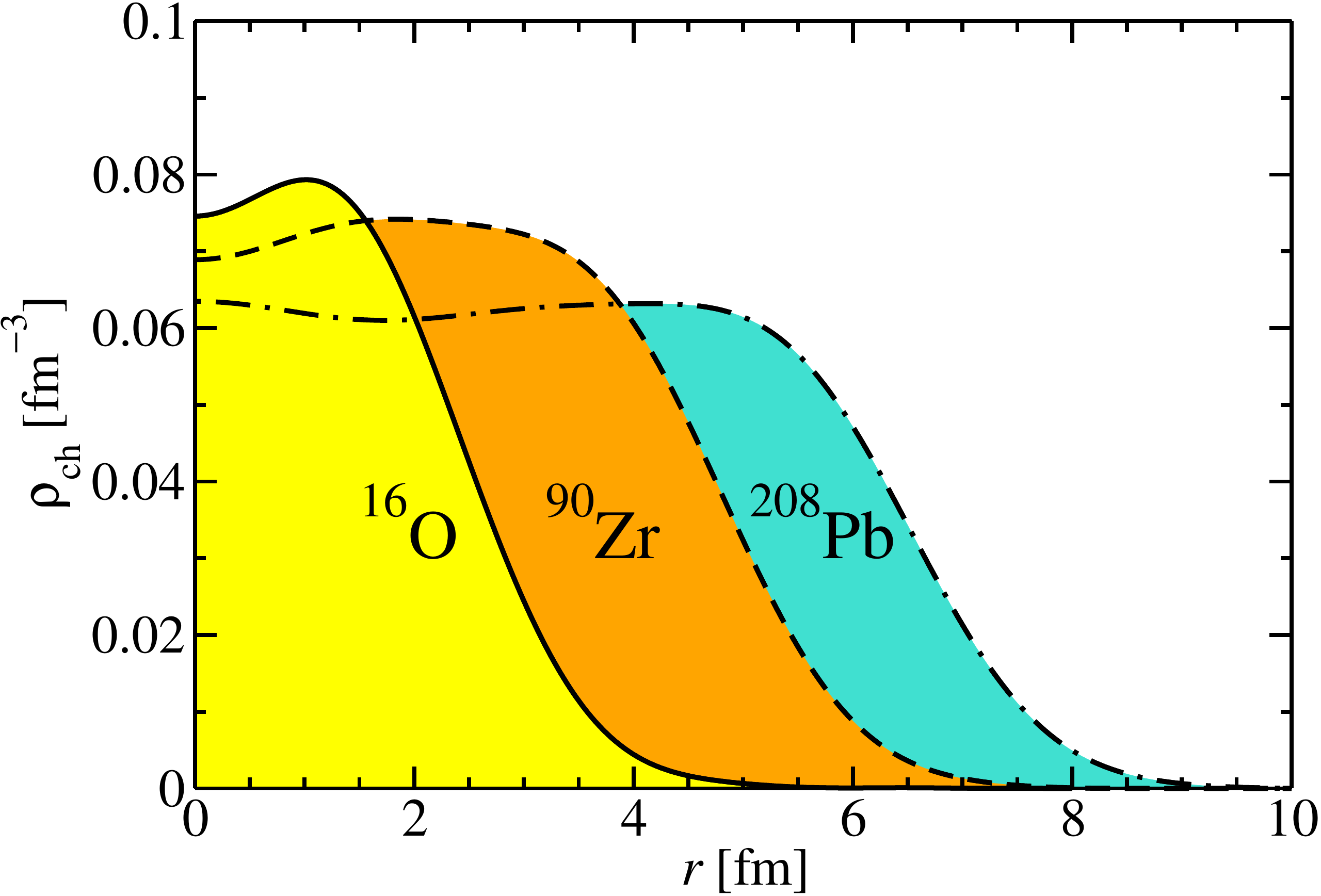}
\includegraphics[width=0.5\linewidth,clip=true]{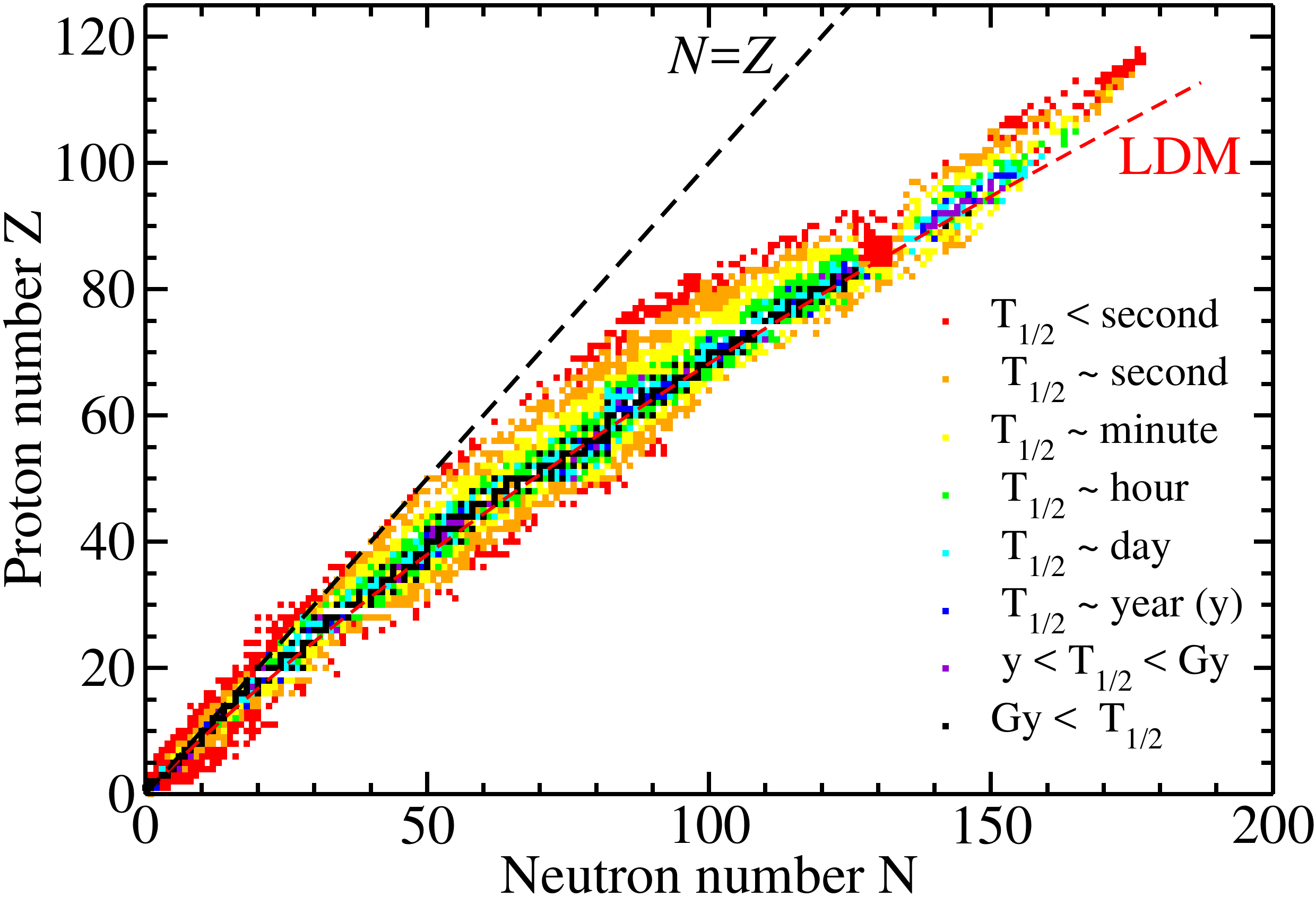}
\caption{\label{fig-pheno-1} Left panel: experimental charge density ($\rho_{\rm ch}$) as a function of the distance to the center of the nucleus ($r$, assuming spherical symmetry) derived from available data on elastic electron scattering for ${}^{16}$O, ${}^{90}$Zr and ${}^{208}$Pb \cite{devries1987}. Right panel: Chart of nuclides classified by half-life. Data taken from {\sc NUBASE16} \cite{nubase16}.}
\end{figure}

\begin{figure}[t!]
\includegraphics[width=0.5\linewidth,clip=true]{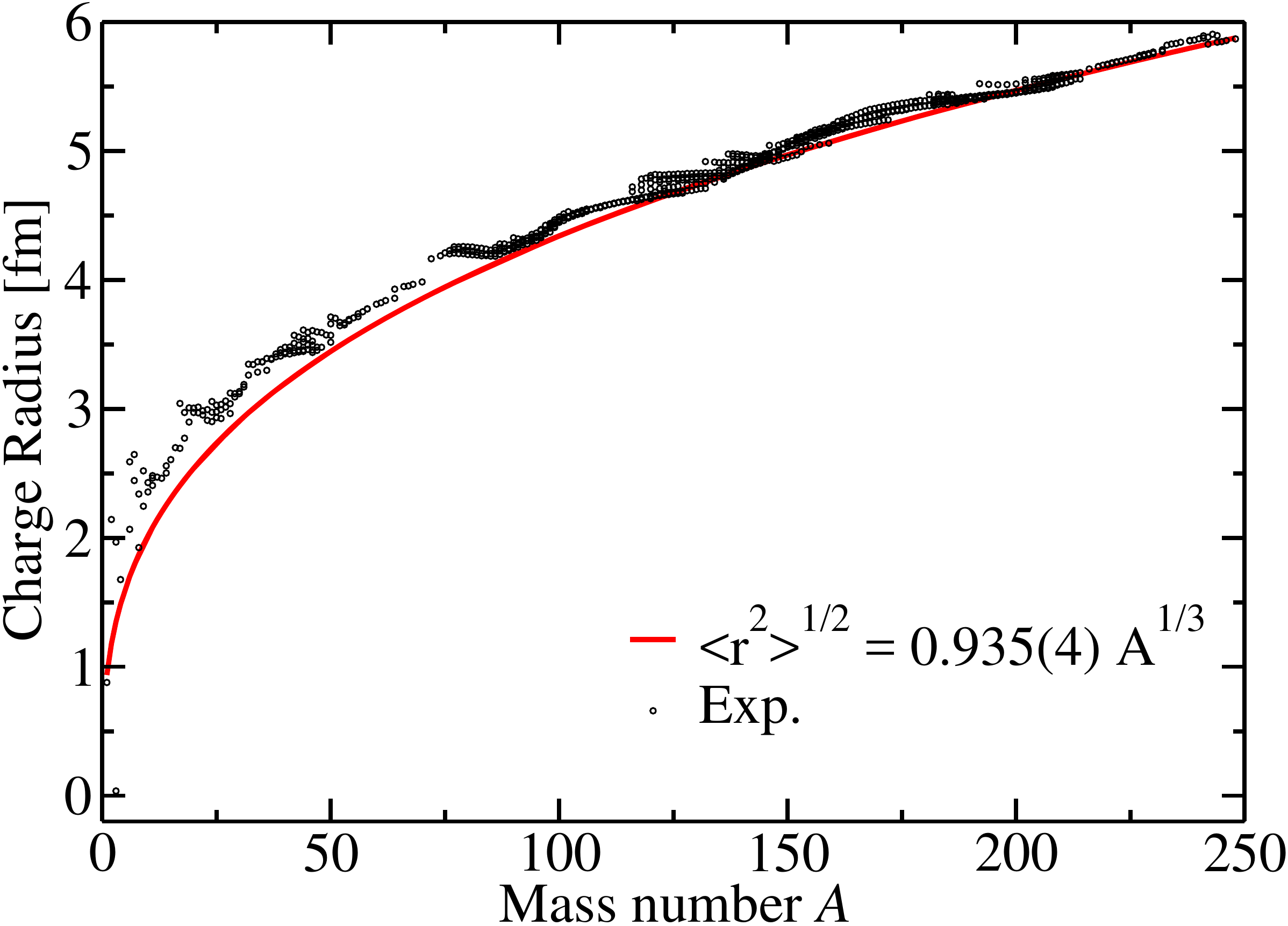}
\includegraphics[width=0.5\linewidth,clip=true]{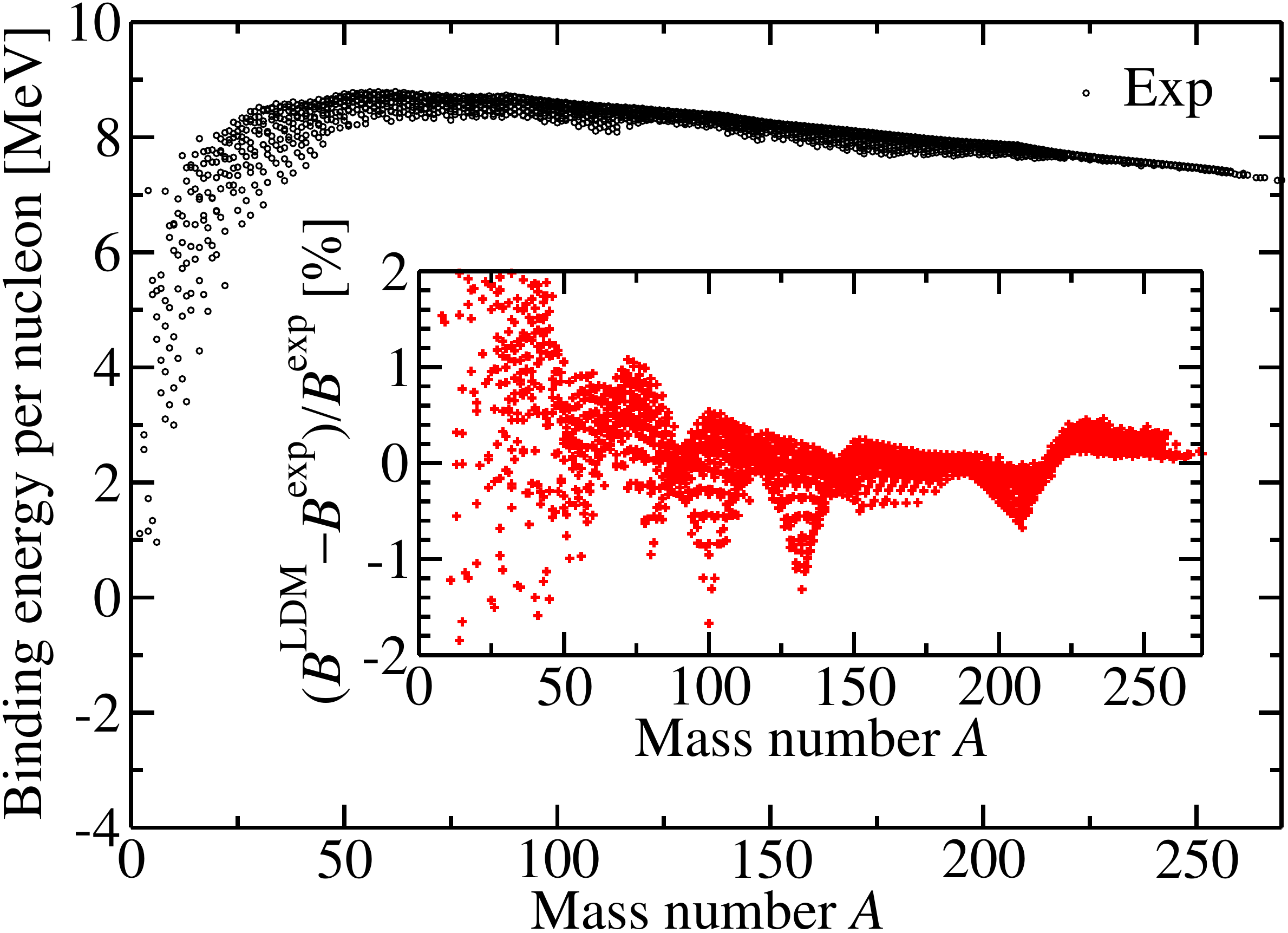}
\caption{\label{fig-pheno-2} Left panel: root mean square charge radii of all measured nuclei as a function of the mass number $A$ reported in Ref.\cite{angeli2013}. Right panel: binding energy per particle of all measured nuclei by 2016 that where compiled in {\sc AME16} \cite{ame16}. The inset shows the relative difference between the liquid drop model described in the text and the experimental data in \%.}
\end{figure}

In the left panel of Fig.\ref{fig-pheno-1}, ground state experimental charge density ($\rho_{\rm ch}$) as a function of the distance to the center of the nucleus ($r$, assuming spherical symmetry) derived from available data on elastic electron scattering for ${}^{16}$O, ${}^{90}$Zr and ${}^{208}$Pb is presented \cite{devries1987}. Within the plane wave Born approximation, the form factor derived from these experiments is the Fourier transform of the charge density. From this figure, it is interesting to note that the density in the interior of very different nuclei, is almost the same and, in good approximation, constant at around $\rho_{0p}=0.07$ fm${}^{-3}$. This means that there exists a saturation mechanism (equilibrium) that originates from the short-range nature of the nuclear force --much stronger than the Coulomb repulsion at the nuclear scale--, typically shorter than the size of most of the known nuclei. If this is true, the electromagnetic size of a nucleus should grow following a simple empirical law. Actually, considering nucleons as structureless --only protons ($Z$) would interact electromagnetically-- and distributed as a sharp sphere with constant density $\rho_{0p}=3Z/4\pi R_p^3$ of radius $R_p$ and, thus, a root mean square (rms) radius $\langle r_p^2\rangle^{1/2}=\sqrt{3/5} R_p$, one finds that $\langle r_p^2\rangle^{1/2}\approx 1.2 Z^{1/3}$. So far we have neglected neutrons ($N$). Knowing experimentally that stable nuclei are not far to be symmetric in the number of neutrons and protons (cf. right panel of Fig.\ref{fig-pheno-1} and Eq.(\ref{eq-pheno-2})), assuming the nuclear force is isospin invariant\footnote{The nucleus can be thought to be composed of nucleons with two possible isospin states, the neutron and the proton. Isospin invariance of the nuclear force is considered to be a good approximation and, thus, isospin a conserved quantum number as long as the Coulomb interaction is neglected in the description of nuclei.} and much stronger than the Coulomb force, one should find --in first approximation-- that the $\frac{\rho_{0p}}{Z}\approx\frac{\rho_{0n}}{N} \rightarrow \rho_{0p}\approx\rho_{0n}$ and $\langle r_p^2\rangle^{1/2}\approx\langle r_n^2\rangle^{1/2}$. Therefore, the neutron and proton rms radii in stable nuclei are equal $\langle r^2\rangle^{1/2}\approx 0.9 A^{1/3}$. All this can be easily tested from the hundreds of rms charge radii measured and compiled in Ref.\cite{angeli2013}. By performing a global least squares fit to this data as a function of $Z^{1/3}$ one finds $1.249(5) Z^{1/3}$ and by doing the same but using instead $A^{1/3}$ one obtains $0.935(4) A^{1/3}$ which is in almost perfect agreement with the very simple picture discussed here and that it is very well known in the literature since many decades \cite{bohr69, ringschuck}. In the left panel of Fig.\ref{fig-pheno-2}, the aforementioned data together with the latter fit is shown. Note that the functional form $A{}^{1/3}$ is too simple to accommodate a good description for light and heavy nuclei at the same time. 

The saturation mechanism inferred from the experimental data in elastic electron scattering allows to understand also nuclear masses. Historically, the most famous macroscopic model is based on the fact that the hard-sphere picture is a good first approximation $R\sim A^{1/3}$. This model known as the liquid drop model \cite{weizsacker1935} parameterizes, in its simpler version (neglecting pairing effects among others), the nuclear mass as $M(A,Z) = m_pZ+m_n(A-Z)-B(A,Z)$, where $B(A,Z)$ is the binding energy 
\begin{equation}
B(A,Z) = (a_V - a_SA^{-1/3})A - a_C\frac{Z(Z-1)}{A^{1/3}} - (a_A-a_{AS}A^{-1/3})\frac{(A-2Z)^2}{A} \ .
\label{eq-pheno-1}
\end{equation}  
The term in $a_V$ is called the volume term and should grow with $R^3$. The correction of this term ($a_S$) due to the presence of a surface should be of opposite sign and proportional to $R^{2}$. The correction $a_C$ due to the Coulomb repulsion between protons, should be also of opposite sign and follows the form $\int \rho_p Z/r dV\sim Z^2/R$. The so called asymmetry term $a_A$ should be taken into account due to the fact that neutrons and protons are distinguishable fermions that will occupy their own energy states in the nucleus without mixing. This introduces an extra term $a_A$ that should increase the mass of the nucleus with growing asymmetry $N-Z$ with respect to the symmetric configuration. This term, must be symmetric in the asymmetry $N-Z$ due to the isospin invariance of the nuclear force. Hence one can adopt the absolute value of $N-Z$ or, from the most used definition, just the square divided by the total number of nucleons. The latter form can be derived by assuming a two component ($N,Z$) free Fermi liquid\footnote{The accuracy of this approximation in nuclei relies on the fact that the many-body nuclear problem can be approximately solved by assuming an equivalent system of non-interacting fermions subject to an average potential that makes the nucleus bound, or analogously, by assuming a non-interacting system of particles with an effective mass} since its Fermi energy differs --in first approximation-- with a single component Fermi liquid with $A$ particles by a term that goes with $(N-Z)^2/A$. Finally $a_{AS}$ correspond to a surface correction to the asymmetry term. The expression (\ref{eq-pheno-1}) could be further extended by adding other terms such as a term that takes into account pairing, curvature ($A^{1/3}$), etc.    

In the right panel of Fig.\ref{fig-pheno-2} the binding energy per particle is shown for all measured nuclei by 2016 that where compiled in Ref.\cite{ame16}. Apart from the lightest nuclei, the binding energy can be considered as constant $E/A (A>20)\sim 8$ MeV for most of the measured nuclei. The decrease of the binding energy for heavy nuclei is mostly due to the Coulomb interaction between protons that grows with $Z^2/A^{1/3}$ and contributes to unbind the system. The main features of this data are easily described on average by the liquid drop model. The inset of the same figure shows the relative difference (in \%) of the model described in Eq.(\ref{eq-pheno-1}) with respect to the experimental data. The fitted parameters are $a_V=15.6(4)$ MeV, $a_S=18(1)$ MeV, $a_C=0.70(2)$ MeV $a_A=28(3)$ MeV and $a_{AS}=26(12)$ MeV. As it can be seen, the accuracy is remarkable for most of the measured nuclei, around or better than a 1\% (rms deviation is less than 3 MeV). It is however clear from the arch-like structure of these residuals that the liquid drop model lacks some physics. Essentially, the model is missing deformation effects, pairing effects and those that produce the arch-like structure: shell effects due to the quantum nature of the atomic nucleus.     

A nice feature of the liquid drop model is that it also reproduces very well the valley of stability (black color in Fig.\ref{fig-pheno-1}). This can be easily seen by imposing stability of $M(A,Z)$ with respect to $Z$, for example. The result is  
\begin{equation}
\frac{Z}{A}\approx\frac{1}{2}\frac{1}{1+\frac{a_C}{4(a_A-a_{AS}A^{-1/3})}A^{2/3}} \ ,
\label{eq-pheno-2}
\end{equation}
where we have considered that the neutron and proton masses are equal. The results of Eq.(\ref{eq-pheno-2}) are also plotted in Fig.\ref{fig-pheno-1} showing a nice overall agreement with experimental data. It is clear from this simple formula, that stability can be roughly understood as a competition between the Coulomb interaction that pushes the system to be as neutron-rich as possible and the asymmetry term that pushes the system to a symmetric configuration between neutrons and protons. 

Coming back to the size of the neutron and proton distributions in the atomic nucleus, in Ref.\cite{myers1974} (and references therein) a refined liquid drop model has been introduced with a diffuse surface rather than a sharp surface as assumed in the liquid drop model. This model is known as the droplet model (DM) and it is interesting to briefly present here the prediction for the so called neutron skin thickness ($\Delta r_{\rm np}\equiv \langle r_n^2\rangle^{1/2}-\langle r_p^2\rangle^{1/2}$), that is the difference between the neutron and proton rms radii. We have roughly analyzed the trends of the electric charge radius based on very simple considerations. Essentially, the measurements of the rms charge radii are model independent and very precise. They give information on the proton rms radii while the neutron rms radii remain very elusive to be directly measured in a very precise and accurate way. The reason is twofold. On the one side, neutrons can be probed by the nuclear force such as for example in proton elastic scattering (see \cite{zenihiro2010}) as well as in experiments using antiprotonic atoms --analyses of anti-proton annihilation residues and of strong-interaction effects on antiprotonic X-rays \cite{antip1,antip2}-- among others that we will discuss in Sec.\ref{gs}. The problem is that the different analyses are model dependent and subject in most of the cases to large systematic errors. On the other side, neutrons can also be probed via the weak interaction. The problem, in that case, is related to the technical difficulties on performing these experiments \cite{prex}. For these reasons it is very important that one has a guidance from a simple model in order to gain deeper insights into the formation of the difference between the neutron and proton rms radii.

\begin{figure}[t!]
\begin{center}  
\includegraphics[width=0.5\linewidth,clip=true]{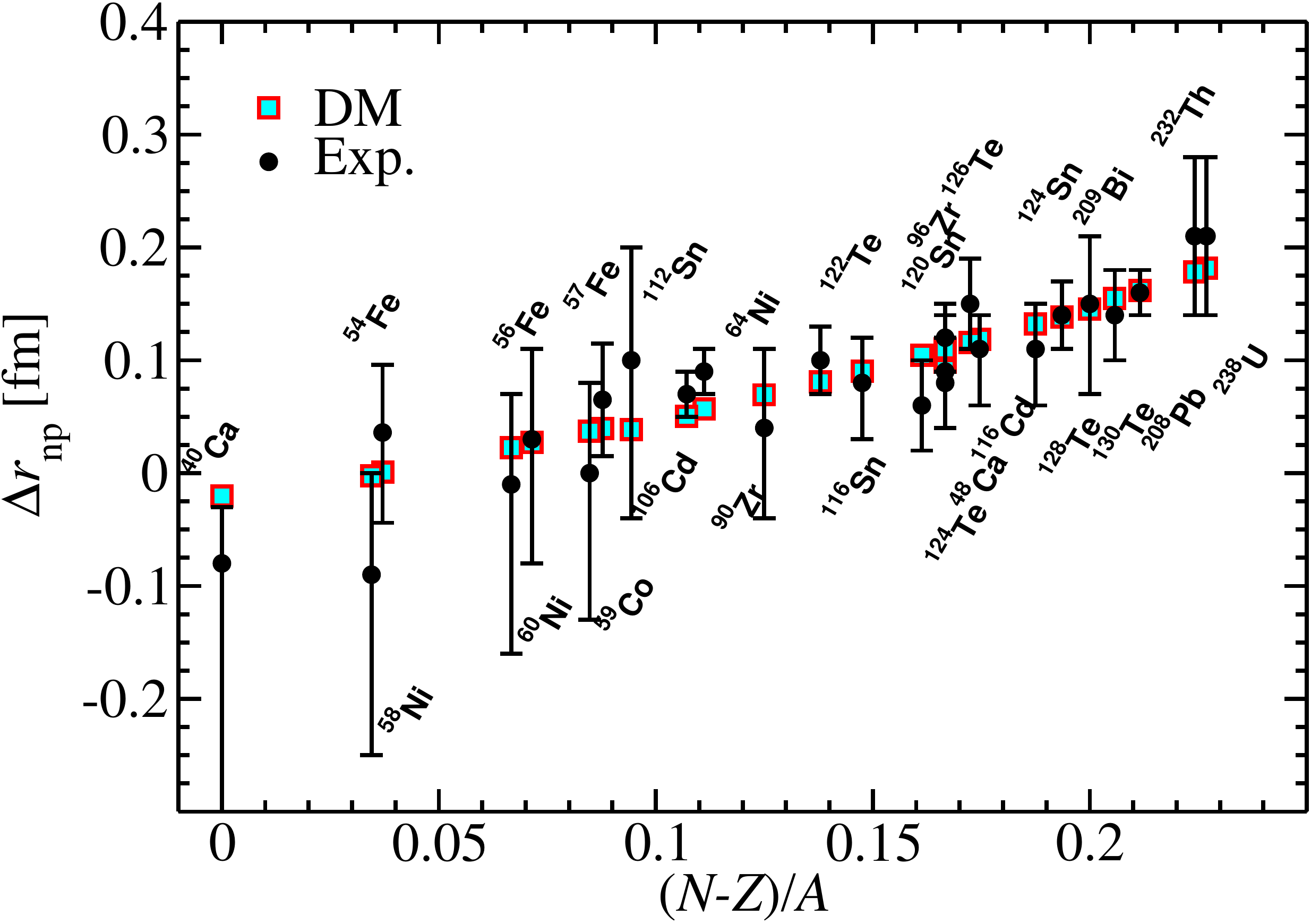}
\end{center}
\caption{\label{fig-pheno-3} Neutron skin thickness as a function of the neutron excess $(N-Z)/A$. Data from Ref.\cite{antip2}.}
\end{figure}

The neutron skin thickness in the DM can be written as follows
\begin{equation}
\Delta r_{\rm np}^{\rm DM} =\frac{2 \langle r^2\rangle^{1/2}}{3}\frac{a_{AS}}{a_A}(I-I_C) + \Delta r_{np}^C \ ,
\label{eq-pheno-3}
\end{equation}
where $I\equiv (N-Z)/A$ is the relative neutron excess; $\langle r^2\rangle^{1/2}=0.9A^{1/3}$ as determined above; $I_C\approx\sqrt{\frac{3}{5}}\frac{e^2Z}{20a_A\langle r^2\rangle^{1/2}} $ is a Coulomb correction to $I$; and $\Delta r_{np}^C\approx - \sqrt{\frac{3}{5}}\frac{e^2Z}{70 a_A}$ is a shift in the neutron skin due to the Coulomb interaction. For the sake of simplicity, in Eq.(\ref{eq-pheno-3}), the asymmetry term of the DM has been approximated by $a_A-a_{AS}A^{-1/3}$ and the bulk asymmetry term by $a_A$. It has also been assumed that the surface diffuseness of neutrons and protons are equal or close to be equal --this approximation is reasonable for stable nuclei. From this expression, one can learn that (macroscopically) the neutron skin thickness depends on the ratio between the parameters of the surface asymmetry and volume asymmetry terms $a_{AS}/a_{A}$: the larger $a_{AS}/a_A$, the larger the neutron skin thickness. Coulomb effects are non negligible but, within the DM, the larger uncertainty comes from the asymmetry terms. Noting that $a_A$ was much better determined by the fit to masses than $a_{AS}$, one can fix $a_A=28(3)$ MeV and fit the latter expression to experimental neutron skins. In Fig.\ref{fig-pheno-3} we show the neutron skins of different stable nuclei determined in antiprotonic atoms experiments \cite{antip1,antip2}. The data is subject to large error bars but allow us to test the presented variant of the DM.  The results of the fit agree qualitatively well with the experimental data, although $a_{AS}\approx 9(4)$ MeV is on the lower edge of the error in the same parameter when determined from measurements on nuclear masses. This simple exercise confirm that measured nuclear masses do not tightly constrain the surface asymmetry term and that more precise and/or systematic measurements of the neutron skin thickness are needed in order to better understand the surface properties of the asymmetry term \footnote{Note that the DM is able to be compatible with experimental data even though different approximations have been done and shell effects, pairing, deformation, etc. are completely neglected.}.

Up to now, it has been shown that measured nuclei in their ground state show a bulk or interior constant density of around $\rho_0=\rho_{0p}+\rho_{0n}\approx 0.15-0.16$ fm${}^{-3}$ (saturation density), a bulk binding energy per nucleon of about $a_V\approx 15-16$ MeV, a bulk asymmetry energy of $25-31$ MeV, and a surface asymmetry energy that is not well constrained by the explored data. This correspond to most of the information we know about the nuclear equation of state of symmetric matter ($\rho_n=\rho_p$) at saturation density from this simple perspective. Actually, it is customary and useful to generally define some parameters that characterize the nuclear EoS. This allows to provide estimates from different approaches and easily compare them. Since nuclear experiments probe the properties of atomic nuclei at or around the saturation density, it is customary to expand the binding energy per particle around saturation density. In addition, isospin asymmetries $\delta\equiv\frac{\rho_n-\rho_p}{\rho_n+\rho_p}$ are relatively small and, thus, an expansion for $\delta\rightarrow 0$ turns to be useful,     
\begin{equation}
e(\rho,\delta) = e(\rho,0) + S(\rho)\delta^2 + \mathcal{O}[\delta^4]
\label{eq-pheno-4}
\end{equation}
where the first term corresponds to the EoS of symmetric matter,
\begin{equation}
e(\rho,0) = e(\rho_0,0) + \frac{1}{2}K_0(\frac{\rho-\rho_0}{3\rho_0})^2 + \mathcal{O}[(\rho-\rho_0)^3]
\label{eq-pheno-5}
\end{equation}
and the second term to the so called symmetry energy $S(\rho)\equiv\frac{\partial e(\rho,\delta)}{\partial \delta}\vert_{\delta=0}$,  
\begin{equation}
S(\rho) = J + L\frac{\rho-\rho_0}{3\rho_0} + \frac{1}{2}K_{\rm sym}(\frac{\rho-\rho_0}{3\rho_0})^2+\mathcal{O}[(\rho-\rho_0)^3] \ ,  
\label{eq-pheno-6}
\end{equation}
where different parameters have been defined: the incompressibility of symmetric nuclear matter $K_0\equiv 9\rho_0^2\frac{\partial^2 e(\rho,0)}{\partial \rho^2}\vert_{\rho=\rho_0}$, symmetry energy at saturation $J\equiv S(\rho_0)$, the slope of the symmetry energy at saturation $L=3\rho_0\frac{\partial S(\rho)}{\partial \rho}\vert_{\rho=\rho_0}$ and the incompressibility (or curvature) of the symmetry energy at saturation $K_{\rm sym}\equiv 9\rho_0^2\frac{\partial^2 S(\rho)}{\partial \rho^2}\vert_{\rho=\rho_0}$. It has to be noted that the saturation energy of symmetric nuclear matter should be equal or close to the value of $a_V$. The parameter $J$ can be identified as the bulk asymmetry energy $a_A$. Again the reason why only even powers of the asymmetry parameter $\delta$ appear is due to the isospin invariance of the nuclear force. It is important to mention that consistent with available theoretical calculations, Eq.(\ref{eq-pheno-4}) is not only accurate for small values of $\delta$ at densities close to saturation but also for the most extreme value of $\delta =1$ which would describe an infinite system composed only by neutrons, the so called neutron matter EoS. This fact allows one to determine in good approximation $S(\rho)\approx e(\rho,1) - e(\rho,0)$, at least in the region of interest.     

\subsection{Nuclear response to small perturbations}

Parameters such as $K_0$, $L$ or $K_{\rm sym}$ probe the density dependence of the EoS around the saturation density and, thus, they might also be estimated in experiments that explore the behavior of nuclei when slightly perturbed from their ground state, i.e., by studying density oscillations\footnote{Keeping in mind an independent particle picture where nucleons are bound by an average one-body potential.}. 

\begin{figure}[t!]
\includegraphics[width=0.5\linewidth,height=0.39\linewidth,clip=true]{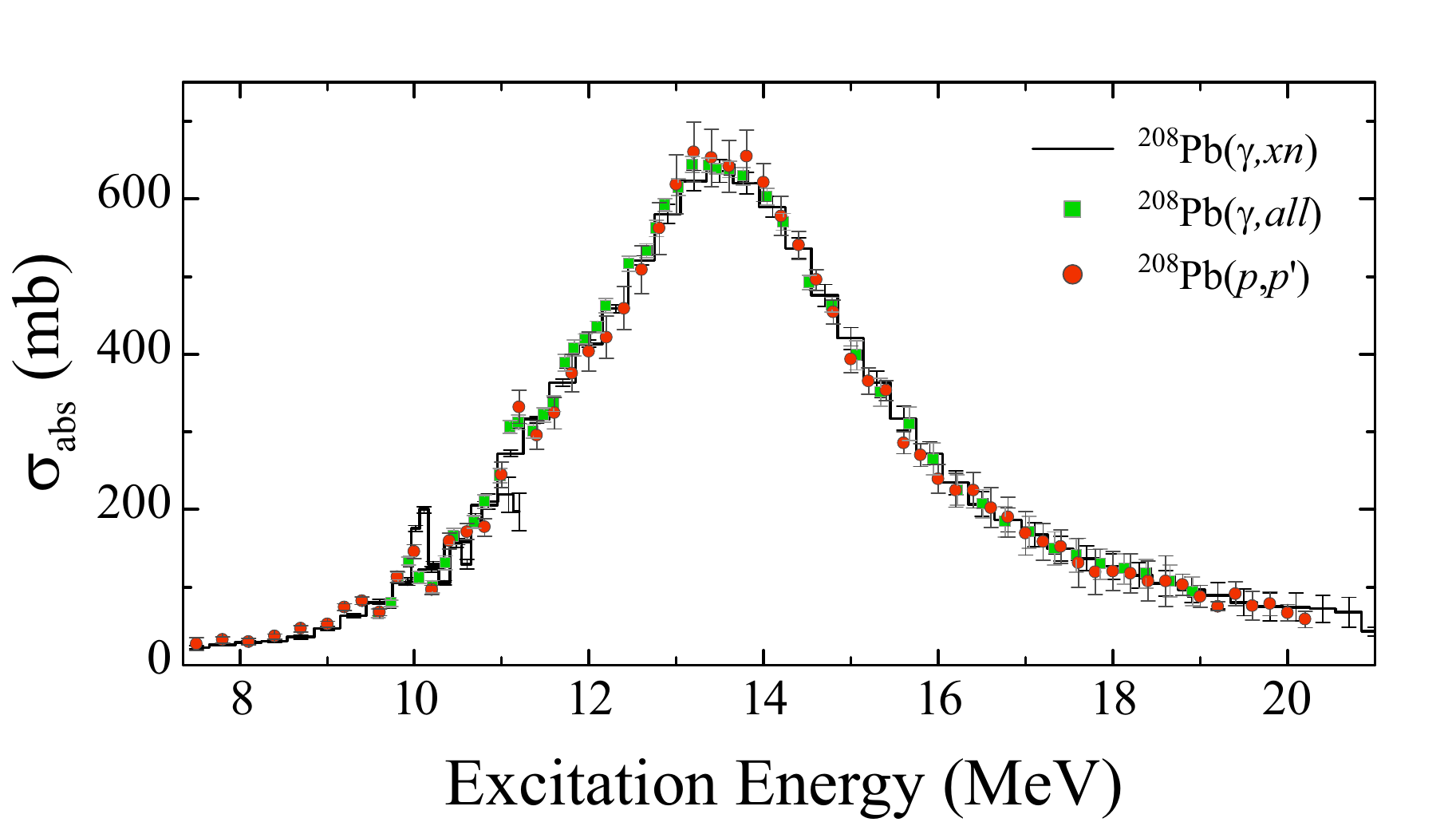}
\includegraphics[width=0.5\linewidth,clip=true]{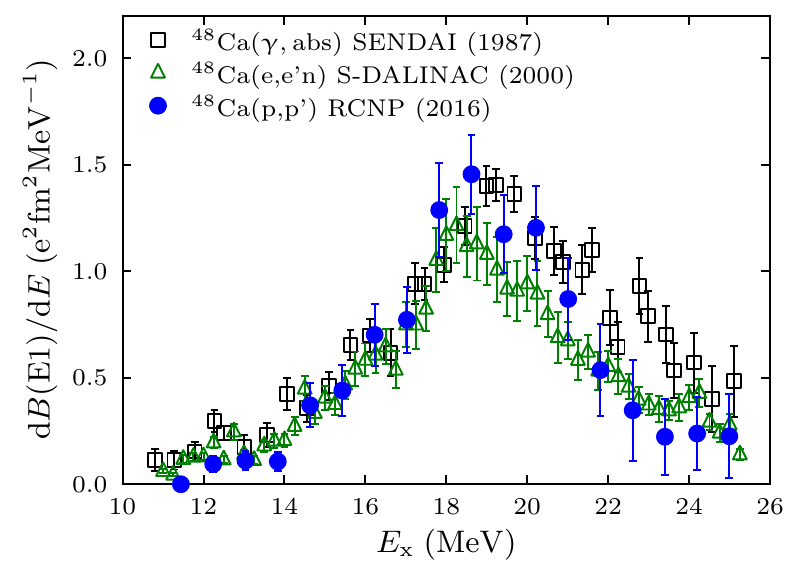}
\caption{\label{fig-pheno-4} Left panel: photoabsorption cross section at forward angles (proportional to the strength function in good approximation corresponding to the IVGDR) in ${}^{208}$Pb as measured in different experiments. Figure taken from Ref.\cite{tamii2014}. Right panel: strength function for the IVGDR in ${}^{48}$Ca with different experimental techniques. Figure taken from Ref.\cite{Birkhan2017}.}
\end{figure}

In general, nuclear shape oscillations are suggested by the fact that some nuclei are deformed in their ground states (most of them) and some have spherical shapes (closed shell nuclei). Therefore, a situation where the shape may show large differences from the ground state one seems plausible. Actually, low-energy states with respect to the ground state energy appear in the energy spectra of most of the known nuclei. Those states are excited by electric quadrupole processes. When dealing with deformed nuclei, those states are known to be part of the so called ground state rotational band. However, for spherical nuclei one is dealing with oscillations in the nuclear shape. At larger energies, there exist other vibration modes that also correspond to shape oscillations of the nucleus as a whole (isoscalar type) and/or fluctuations in which neutrons and protons collectively oscillate out of phase (isovector type). Since nuclei are formed by two types of fermions, also spin and isospin exchange oscillations occur. These higher energy oscillations are commonly known as Giant Resonances (GRs) \cite{bohr69,bortignon1998} and are those in which we are most interested here. As an example, the first vibrational or oscillation mode was measured by photo-absorption. It was latter confirmed that all nuclei show a large increase of the photo-absorption cross section at around 10 to 30 MeV in this type of experiments, the energy depending essentially on the size of the nucleus --that is, on the mass number $A$. This is known as the isovector Giant Dipole Resonance (IVGDR). Two examples are shown in Fig.\ref{fig-pheno-4}. It is worth to note that Migdal realized that an average excitation frequency for dipole absorption could be derived from the nuclear polarizability that is related to the symmetry energy parameters $J$ and $L$ as we will discuss. 

\subsubsection{Excitation energy of Giant Resonances}

In order to provide a general idea on the relation of GRs with the parameters of the nuclear EoS, it is interesting to briefly discuss how GRs can be understood from a macroscopic model. For that, we will closely follow Ref.\cite{bohr69}. We have assumed so far that the nucleus is spherical and in its ground state. A simple extension to that approximation will be to take into account that it may display collective oscillations of the nucleon density as we just briefly introduced. Such an extension can be investigated within the independent-particle picture, where nuclei are modeled to be bound by a one-body potential that, if possible, describes also the phenomenology that we have already discussed in the previous subsection. To stay simple, one can assume that the energy spectra of the nucleus can be approximated by that of an harmonic oscillator and, therefore, density oscillations in nuclei can be described by an harmonic potential \cite{bohr69,ringschuck}.

%\begin{figure}[t!]
%\begin{center}  
%\includegraphics[width=0.5\linewidth,clip=true]{figures/fig-pheno-5.pdf}
%\end{center}
%\caption{\label{fig-pheno-5} Schematic nuclear levels of the harmonic oscillator (left), of the shell model (center) and of the shell model with spin orbit term (right). Figure taken from Ref.\cite{mayer1956}.} 
%\end{figure}

Let us start by reminding some useful expressions for the harmonic oscillator $\mathcal{H}=\frac{1}{2}C\alpha^2+\frac{1}{2}B\dot{\alpha}^2$, where $\alpha$ represents the variation from equilibrium and the conjugate variable of $\alpha$ is defined as $\pi\equiv B\dot{\alpha}$, $C$ corresponds to the restoring force parameter and $B$ to the mass parameter. It is known that the energy spectrum is $E=(n+\frac{1}{2})\hbar\omega$ with $\omega=(C/B)^{1/2}$. In general, one can write the unperturbed Hamiltonian by means of the quanta of excitation as $\mathcal{H}_0= n \hbar\omega_0$ where $n$ is the number operator made of convenient annihilation and creation operators (defined via $\alpha$ and $\pi$ in the usual way). Rewriting the latter expression in terms of $\alpha$, one finds that $\mathcal{H}_0=\frac{1}{2}C_0\alpha^2+\frac{1}{2}B_0\dot{\alpha}^2-\frac{1}{2}\hbar\omega_0$. Adding now a small perturbation $F$ that we assume goes as $\alpha$, the variation of the one-body potential and, therefore, of the density should be linear with the amplitude of the perturbation $\alpha$ and can be written in the form $\delta V = \kappa \alpha F\approx\kappa\alpha^2$. The latter give rise to a potential energy $\delta U=\frac{1}{2}\kappa \alpha^2$. All this allows one to write $\mathcal{H}=\mathcal{H}_0+\delta U=\frac{1}{2}(C_0+\kappa)\alpha^2+\frac{1}{2}B_0\dot{\alpha}^2-\frac{1}{2}\hbar\omega_0$. From this expression one finds that the restoring force has been modified by the perturbation $C=C_0+\kappa$  while the mass parameter stays unaltered $B=B_0$. The energy of the mode will be
\begin{equation}
\hbar\omega = \hbar\left(\frac{C}{B}\right)^{1/2}=\hbar\left(\frac{C_0+\kappa}{B_0}\right)^{1/2}=\hbar\omega_0\left(1+\frac{\kappa}{C_0}\right)^{1/2} \ .
\label{eq-pheno-7}  
\end{equation}  
The unperturbed energy $\hbar\omega_0$ should be chosen in a way to preserve the main features of the ground-state energy spectra. Since we are dealing with a simple model, the energy gap between independent particle shells can be naively expected to decrease as $1/R$. The most commonly used expression for the shell gap is $\hbar\omega_0\approx 41 A^{-1/3}$ MeV.

\begin{figure}[t!]
\includegraphics[width=0.5\linewidth,clip=true]{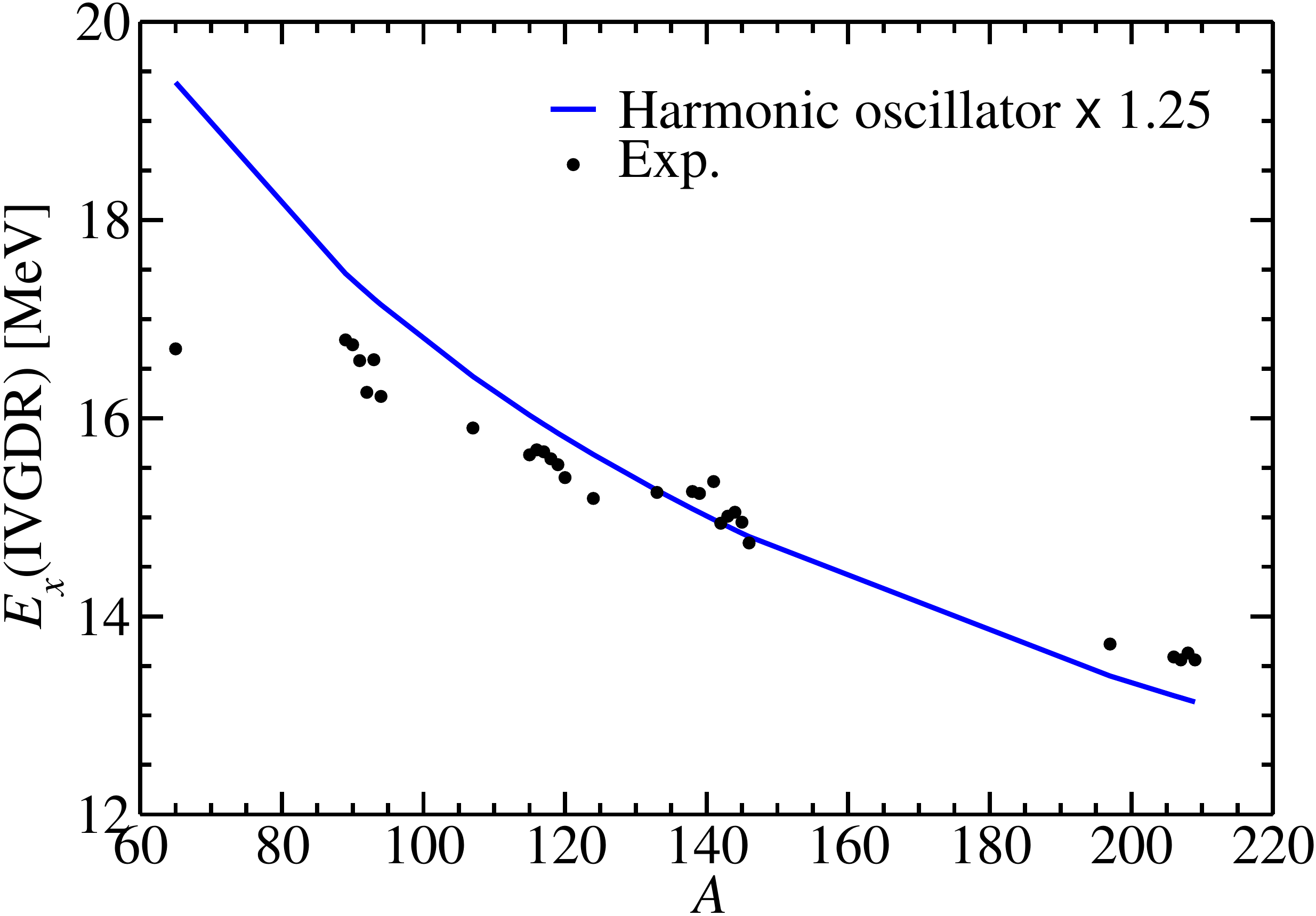}
\includegraphics[width=0.5\linewidth,clip=true]{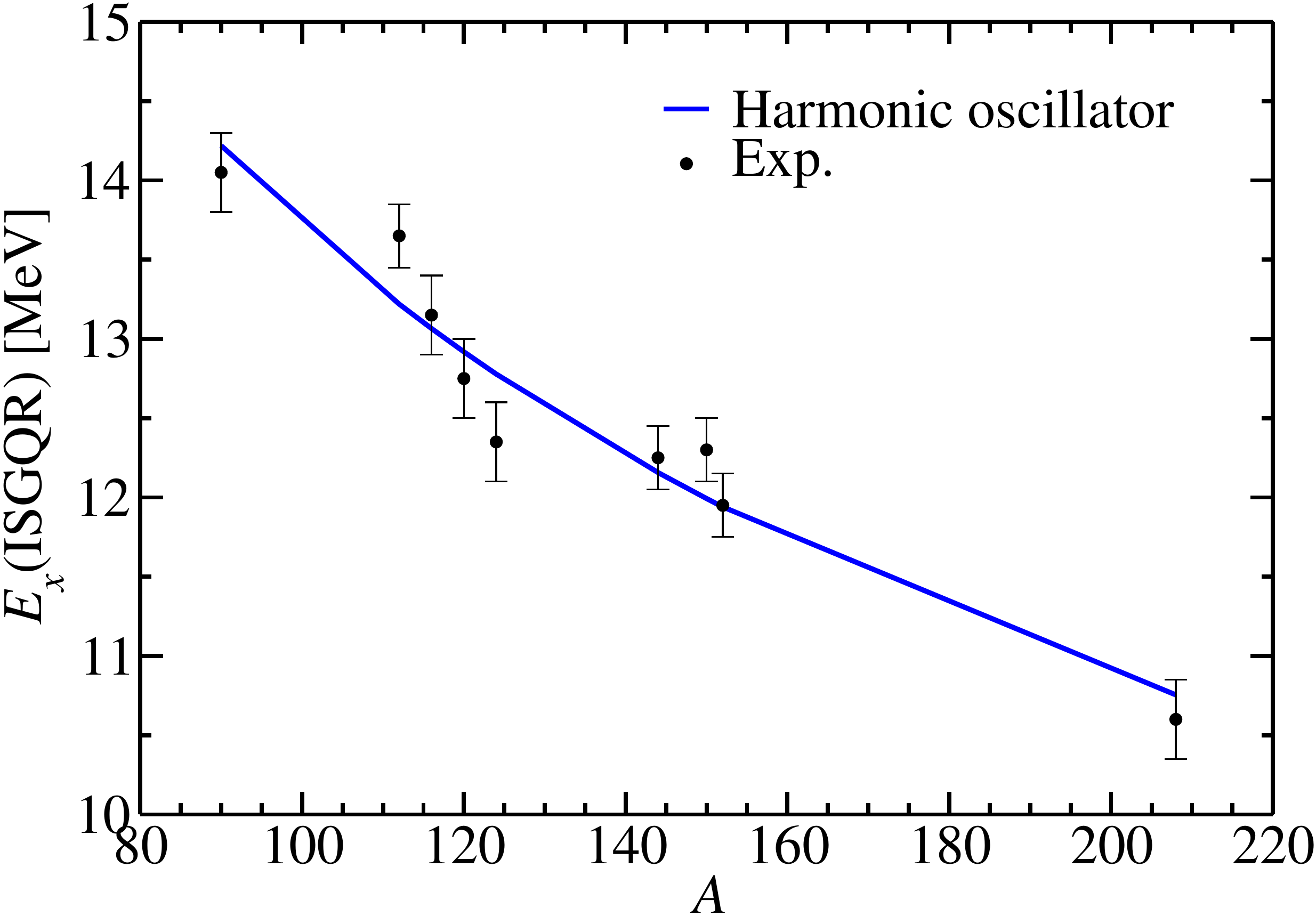}
\caption{\label{fig-pheno-6} Left panel: excitation energies of the IVGDR as a function of the mass number $A$. Measured stable nuclei compiled in Table IV of Ref.\cite{berman1975} and predictions from the harmonic oscillator model are shown (see text). Right panel: excitation energies of the ISGQR for nuclei with $A>90$. Data from Ref.\cite{buenerd1984} and harmonic oscillator results are shown.}
\end{figure}

Now we would like to apply this model to some specific examples. In particular we will briefly discuss the isovector dipole resonance and the isoscalar and isovector giant quadrupole resonance (GQR). Other modes could be understood in a similar way. As mentioned, the IVGDR was the first observed GR. The excitation energy of this resonance can be understood within the simple model discussed here as an interplay of the unperturbed excitations and a collective field proportional to the dipole moment $F(r)=\sum_{i=1}^A r_i \tau_z(i) Y_{10}(\hat{r}_i) $ where $r$ is the radial coordinate and $\tau_z$ is the third component of the Pauli matrices in isospin space. That is $\tau_z=1$ for neutrons and $\tau_z=-1$ for protons. First of all, we determine $C_0$ from the following considerations. In the unperturbed system of $A$ nucleons with mass $m$, one can identify the mass parameter by comparing the kinetic energy of the proposed model $\frac{1}{2}\frac{\pi^2}{B_0}$ with the kinetic energy of an uncorrelated system of particles with average momentum $\langle p^2\rangle=\sum_{i=1}^A\frac{p^2_i}{A}$, that is, $\frac{1}{2}\frac{\langle p^2\rangle}{m}A$. This gives $B_0=m/A$ and, therefore, $C_0 = \omega_0^2B_0=\omega_0^2 m/A$. Next, we evaluate the restoring force $\kappa$. As previously discussed, for small variations $\delta V = k \alpha F$. On the other side, isovector variations of the potential will produce a variation of the isovector density ($\rho_n-\rho_p$), that is, $\delta V \equiv \frac{1}{4} V_{\rm sym} \tau_z \frac{\delta(\rho_n-\rho_p)}{\rho}$ where $V_{\rm sym}$ is the repulsive symmetry potential \footnote{This expression comes from considering that the nuclear potential can be divided in an isoscalar part $V_0$ and an isovector part $V_{\rm sym}$ that can be defined as $V\equiv V_0+\frac{1}{4}V_{\rm sym}\tau_z\frac{N-Z}{A}$. The energy associated to this symmetry potential within the independent particle approximation is $U_{\rm sym}=\frac{1}{8}V_{\rm sym}\frac{(N-Z)^2}{A}$ and the kinetic energy $T_{\rm sym}=\frac{1}{3}\varepsilon_F\frac{(N-Z)^2}{A}$ where $\varepsilon_F$ is the Fermi energy of the symmetric system that will be $\approx 35$ MeV at the nuclear saturation density. All this allows one to connect them with the LDM parameters: $a_A-a_{AS}A^{-1/3} = \frac{1}{3}\varepsilon_F + \frac{1}{8}V_{\rm sym}$. That is $V_{\rm sym}=8 (a_A-a_{AS}A^{-1/3}) - \frac{8}{3}\varepsilon_F$.}. Taking into account that we assume the average amplitude $\langle F\rangle$ equal to $\alpha$ and that $\delta V$ can be approximated by the deformation of the static potential $V$, one can find that $\kappa=\frac{V_{\rm sym}}{4 A \langle r^2\rangle}$ for the assumed excitation operator $F$ within the harmonic approximation \cite{bohr69}. Finally, the excitation energy of the IVGDR is obtained, 
\begin{equation}
\hbar\omega^{\rm IVGDR} = \hbar\omega_0\left(1+\frac{\hbar^2}{4m\langle r^2\rangle}\frac{V_{\rm sym}}{\hbar^2\omega_0^2}\right)^{1/2}\approx\hbar\omega_0\left(1+\frac{\hbar^2}{4m\langle r^2\rangle}\frac{8 (a_A-a_{AS}A^{-1/3}) - \frac{8}{3}\varepsilon_F}{\hbar^2\omega_0^2}\right)^{1/2} 
\label{eq-pheno-8}  
\end{equation}  
where in the last step we have used the relation between $V_{\rm sym}$ and the parameters of the LDM (cf. previous footnote). By using the value of the shell gap modified by the average nucleon effective mass $\hbar\omega_0\rightarrow\sqrt{m/m^*}\hbar\omega_0$ with $m^*\approx 0.7 m$ \footnote{The average effective mass in phenomenological models based on the independent particle picture is in the range $m^*\approx 0.6-0.8 m$.} and the Fermi energy at saturation density $\varepsilon_F\approx 35$ MeV, one can determine $a_{AS}$ by fitting existent experimental data on the IVGDR (e.g., data from Table IV in Ref.\cite{berman1975}) assuming $a_A\approx 28$ MeV as determined from masses. Both, experimental results and fit (re-scaled by a factor $1.25$) are shown in Fig.\ref{fig-pheno-6}. The result for the surface asymmetry coefficient that follows the correct trend of the experimental data is $a_{AS} \approx 9$ MeV, compatible with the one obtained from the neutron skin thickness. We note that due to the approximate nature of Eq.(\ref{eq-pheno-8}), the results had to be shifted by about a 25\% to larger energies as a whole in order to lie on top of the experimental data. Such a systematic error might be expected in this type of macroscopic models. The important result here is that the formula in Eq.(\ref{eq-pheno-8}) follows in good approximation the experimental trends with $A$. Finally, it is interesting to note that Eq.(\ref{eq-pheno-8}) predicts a functional form of the type $E_x^{\rm IVGDR}\approx a A^{-1/3}- bA^{-2/3}$ where $a$ and $b$ should be positive parameters while in the literature a function of the type $E_x^{\rm IVGDR}\approx c A^{-1/3}+ d A^{-1/6}$ has been frequently used \cite{berman1975}. In this model, the term in $A^{-1/3}$ arises from considering the IVGDR as an out-of-phase motion of two fluids with a sharp surface and keeping the total density fixed, the Jensen-Steinwedel model, while the term $A^{-1/6}$ corresponds to a model where neutrons and protons might behave like two separate but interpenetrating density distributions, the Goldhaber-Teller model (cf. Fig.1 of Ref.\cite{myers1977}, see also Eq.(4.12) and Fig.6 therein).

\begin{figure}[t!]
\begin{center}  
\includegraphics[width=0.5\linewidth,clip=true]{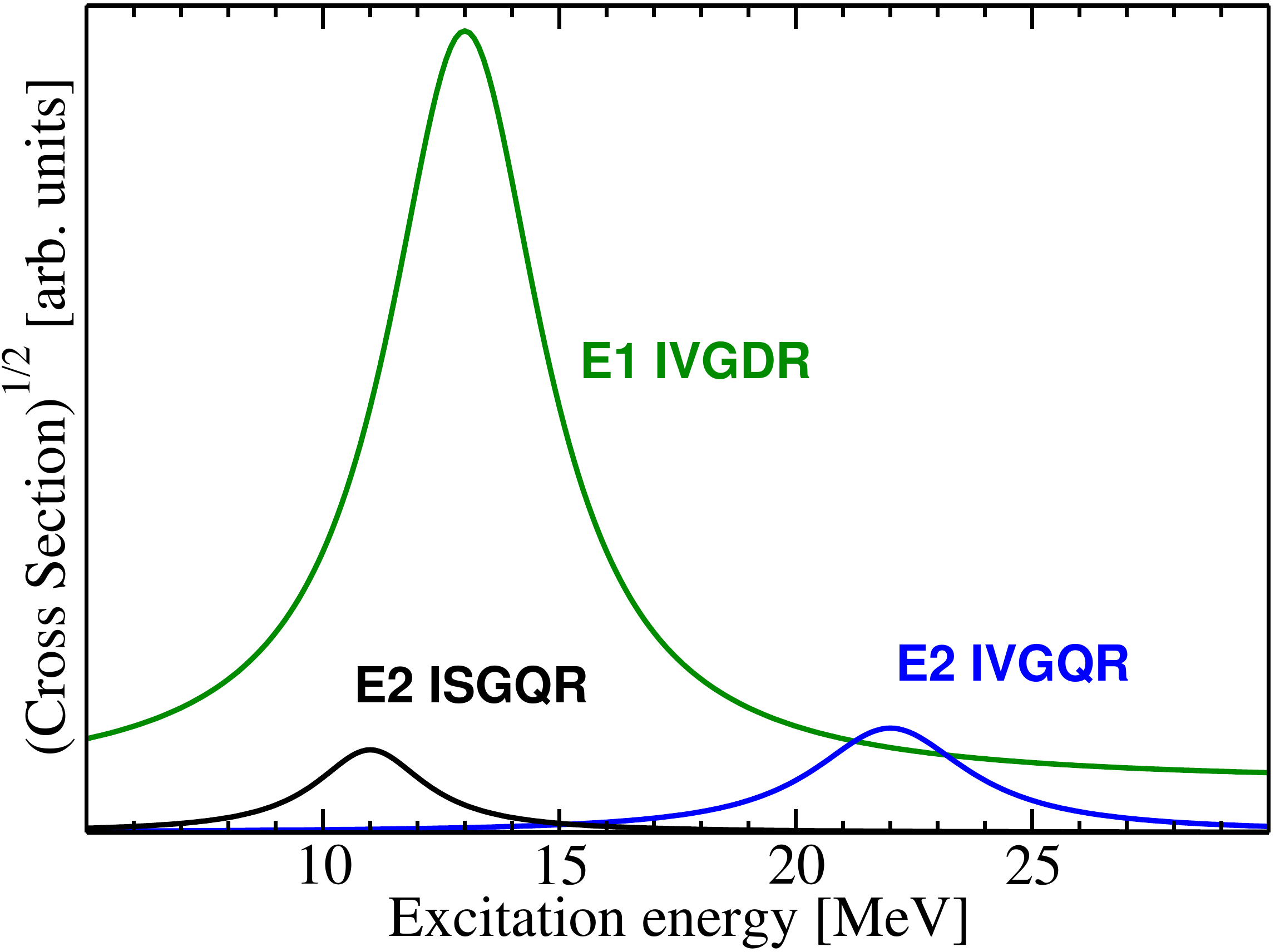}
\end{center}
\caption{\label{fig-pheno-7} Schematic picture for the photon absorption cross section in a nucleus such as ${}^{208}$Pb.}
\end{figure}

The isoscalar giant quadrupole resonance (ISGQR) was discovered in the early 70's in inelastic electron and proton scattering \cite{lewis1972,fukuda1972} and it has been well studied experimentally and theoretically. For example, it has been seen that for nuclei with $A>90$, the strength of this resonance is concentrated in a single peak while, for lighter nuclei, the strength is more fragmented \cite{harakeh01}. The isovector giant quadrupole resonance (IVGQR) is more elusive from the experimental point of view, in part due to the high excitation energy --larger than $2\hbar\omega_0\approx 82 A^{-1/3}$ on the basis of the harmonic oscillator model--, the large width and a relatively small excitation cross section. In Fig.\ref{fig-pheno-7} such difficulties become more evident, specifically, we show a schematic picture for the photon absorption cross section in a nucleus such as ${}^{208}$Pb. Recently, however, there has been some advance and the IVGQR in ${}^{208}$Pb has been measured with unprecedented accuracy \cite{henshaw2011}.   

By adopting the same model but with $F$ now proportional to $r^2Y_{20}(\hat{r})$ --including $\tau_z$ only for the IV mode of excitation-- and very similar considerations as for the dipole case --described in more detail in Ref.\cite{bohr69}--, one can derive an expression for the excitation energy of the ISGQR,
\begin{equation}
\hbar\omega^{\rm ISGQR} = \sqrt{2\frac{m}{m^*}}\hbar\omega_0 \approx 64 A^{-1/3}\quad {\rm MeV}  
\label{eq-pheno-9}  
\end{equation}  
where we have already included the modification of the energy spectra of the harmonic oscillator due to the average effective mass of the nucleon. In this case, we have chosen $m^*= 0.82 m$ since this value allows to better reproduce the experimental data for nuclei with $A>90$ presented in Fig.\ref{fig-pheno-6}. We note in this case that Eq.(\ref{eq-pheno-9}) does not depend (directly) on the parameters of the nuclear EoS since $m^*$, even though one assumes its average value, it is a single-particle property. The correct trend of the experimental data is well reproduced by the presented model.  

In full analogy with the IVGDR, for the case of the IVGQR, one can derive the following expression
\begin{equation}
\hbar\omega^{\rm IVGQR} = \sqrt{2\frac{m}{m^*}}\hbar\omega_0\left(1+\frac{5}{2}\frac{\hbar^2}{2m}\frac{V_{\rm sym}\langle r^2\rangle}{2\frac{m}{m^*}(\hbar\omega_0)^2 \langle r^4\rangle}\right)^{1/2} \ , 
\label{eq-pheno-10}  
\end{equation}  
and adopting $\langle r^4\rangle = \frac{25}{21}\langle r^2\rangle^2$ for a sharp sphere of radius $R$ and $V_{\rm sym}=8 (a_A-a_{AS}A^{-1/3}) - \frac{8}{3}\varepsilon_F$ (cf. previous footnote), 
\begin{equation}
\hbar\omega^{\rm IVGQR} \approx \hbar\omega^{\rm ISGQR}\left(1+\frac{\hbar^2}{2m}\frac{16 (a_A-a_{AS}A^{-1/3}) - \frac{16}{3}\varepsilon_F}{(\hbar\omega^{\rm ISGQR})^2\langle r^2\rangle}\right)^{1/2} \ .  
\label{eq-pheno-11}  
\end{equation}  
We check the trend predicted by this formula for the experimental data in ${}^{40}$Ca and ${}^{208}$Pb: $E_x^{\rm IVGQR}=31.5(1.5)$ MeV \cite{sims1997} and  $E_x^{\rm IVGQR}=23.0(2)$ MeV \cite{henshaw2011}, respectively. By using Eq.(\ref{eq-pheno-11}), we find $\hbar\omega^{\rm IVGQR}\approx 29$ MeV and  $\hbar\omega^{\rm IVGQR}\approx 17$ MeV for ${}^{40}$Ca and ${}^{208}$Pb, respectively, using $a_A\approx 28$ MeV, $a_{AS}\approx 9$ MeV and $m^*/m\approx 0.8$. Also in this case, the model predicts a reasonable trend but underestimates the experimental data.      

\subsubsection{Sum rules}

Sum Rules (SR), or moments of the strength function in nuclei (as that shown in the right panel of Fig.\ref{fig-pheno-4}) provide information on some relevant quantities that have been related to the parameters of the nuclear EoS. The usefulness of SR is that in some cases they are model independent and/or predict very simple expressions that can be exploited to better understand the nuclear phenomenology. This is the reason why we will briefly discuss here some examples adopting a macroscopic approach. Specifically, the isoscalar monopole and dipole resonances as well as the nuclear dipole and quadrupole polarizabilities will be discussed.

So far the strength function has already been mentioned but not defined. Let us use an example to introduce this quantity. We have mentioned that the first GR measured was the IVGDR via photo-absorption. The cross section for the excitation to a final state $\vert \nu\rangle$ with energy $E_{\nu}$ from the ground state $\vert 0\rangle$ with energy $E_0$ by a photon at a given energy $E$ is under some reasonable assumptions \cite{sf1974},
\begin{equation}
\sigma_{\nu}(E) = 4\pi^2 \alpha (E_{\nu}-E_0) \vert\langle\nu\vert F_{\rm dipole}\vert 0\rangle\vert^2\delta(E-E_{\nu}+E_0) \ ,
\label{eq-pheno-12}  
\end{equation}  
where $\alpha$ is the fine structure constant and $F_{\rm dipole}$ is referred to the center of mass (CM) and, thus, differs from the dipole operator defined previously by $\r_i\rightarrow \r_i - \r_{\rm CM}$. The latter change has no influence on the matrix elements as long as the initial and final states do not violate translational invariance. In nuclei, translational invariance is actually violated and, thus, spurious states might appear. Those will be corrected on average only if $F_{\rm dipole}$ is defined with subtracting the CM motion. The total cross section is obtained by summing over all final states $\vert \nu\rangle$ and energies. The result is,
\begin{equation}
\sigma_{\gamma-{\rm abs}} = 4\pi^2 \alpha \sum_{\nu} (E_{\nu}-E_0) \vert\langle\nu\vert F_{\rm dipole}\vert 0\rangle\vert^2 \equiv \frac{m_1}{4\pi^2\alpha}
\label{eq-pheno-13}  
\end{equation}  
where $m_1$ is the energy weighted moment of the strength function (for the dipole case in this example),
\begin{equation}
S(E) \equiv \sum_{\nu} \vert\langle\nu\vert F_{\rm dipole}\vert 0\rangle\vert^2 \delta(E-E_{\nu}+E_0) 
\label{eq-pheno-14}  
\end{equation}  
and
\begin{equation}
m_k \equiv \int dE E^kS(E) = \sum_{\nu} (E_{\nu}-E_0)^k \vert\langle\nu\vert F_{\rm dipole}\vert 0\rangle\vert^2  \ .
\label{eq-pheno-15}  
\end{equation}  
Commonly, the ground state energy $E_0$ is taken as a reference and set to zero. In a similar way, one can derive the corresponding relations for higher multipole electromagnetic transitions \cite{sf1974}. The definition of the strength function and moments remains, in any case, unaltered. Only the transition operator that mimic the experimental process should be carefully defined.

We have just seen that the photo-absorption cross section is proportional to $m_1$ or energy weighted sum rule (EWSR). It is straightforward to show that 
\begin{equation}
m_1 = \sum_{\nu} (E_{\nu}-E_0) \vert\langle\nu\vert F\vert 0\rangle\vert^2 = \langle 0\vert F^\dag[\mathcal{H}, F]\vert 0\rangle
\label{eq-pheno-16}  
\end{equation}  
by using $\mathcal{H}\vert\nu\rangle=E_{\nu}\vert\nu\rangle$. Similarly,
\begin{eqnarray}
  m_0&=&\sum_{\nu} \vert\langle\nu\vert F\vert 0\rangle\vert^2 = \langle 0\vert F^\dag F\vert 0\rangle \nonumber \\
  m_2&=&\sum_{\nu} (E_{\nu}-E_0)^2\vert\langle\nu\vert F\vert 0\rangle\vert^2 = \langle 0\vert F^\dag [\mathcal{H},[\mathcal{H},F]]\vert 0\rangle \nonumber \\
  m_3&=&\sum_{\nu} (E_{\nu}-E_0)^3\vert\langle\nu\vert F\vert 0\rangle\vert^2 = \langle 0\vert F^\dag [\mathcal{H},[\mathcal{H},[\mathcal{H},F]]]\vert 0\rangle\nonumber 
\end{eqnarray}  
and so on and so forth \cite{bohigas1979}. To evaluate, therefore, $m_k$ only information on the ground state is needed. This greatly simplify the calculations. For example, for general operators defined as   
\begin{equation}
  F^{\rm IS} = \sum_{i=1}^A f(r_i)Y_{JM}(\hat{r}_i) \quad {\rm and} \quad
  F^{\rm IV} = \sum_{i=1}^A f(r_i)Y_{JM}(\hat{r}_i)\tau_z(i) 
\label{eq-pheno-17}  
\end{equation}  
where $J$ represents the total angular momentum ($J=1$ corresponds to dipole transitions, $J=2$ to quadrupole transitions, etc.; see below for more details), the EWSR can be found to be assuming a single-particle model
\begin{eqnarray}
  m_1^{\rm IS} &=& \frac{A}{4\pi}\frac{\hbar^2}{2m}\frac{2J+1}{A}\int d\r \left[\left(\frac{df}{dr}\right)^2 + J(J+1)\left(\frac{f}{r}\right)^2\right]\rho(r) \label{eq-pheno-18}  \\
  m_1^{\rm IV}&=&m_1^{\rm IS}(1+\mathcal{K})  
  \label{eq-pheno-19}  
\end{eqnarray}  
where $\mathcal{K}$ is the so called isovector enhancement factor. Experimentally, it has been found to be around 0.2 in the dipole case \cite{harakeh01}. Theoretically, it has a different expression depending on the interaction used for the calculation --that is, $m_1^{\rm IV}$ is model dependent. Specifically, for finite-range or zero-range momentum dependent interactions, the isovector enhancement factor is different from zero. One of the most interesting features comes from the isoscalar EWSR. For isoscalar operators $[V, F^{\rm IS}]=0$ and only the kinetic part contributes to $m_1^{\rm IS}$. This is why we obtain closed, analytic and model independent expression in Eq.(\ref{eq-pheno-18}).

Now we introduce the linear response of a nucleus to an external field $F$ also known as dynamic polarizability. From a macroscopic perspective, this quantity is interesting since it measures in photo-absorption processes the tendency of the nuclear charge distribution to be distorted. In perturbation theory, the linear response can be written as $\alpha_F(E)=2\sum_{\nu}\frac{E_{\nu}-E_0}{(E_{\nu}-E_0)^2-E^2}\vert\langle \nu\vert F\vert 0 \rangle\vert^2$  and, therefore, $\alpha_F(E)\vert_{E\rightarrow 0}= 2(m_{-1}+E^2m_{-3}+ ... )$ \cite{stringari1983}. The static polarizability corresponds to the first term in this expansion, that is, the inverse energy weighted sum rule (IEWSR) $m_{-1}$. Coming back to our initial example, one can see from Eq.(\ref{eq-pheno-12}) that 
\begin{equation}
m_{-1} = 4\pi^2\alpha \int dE \sum_\nu \frac{\sigma_{\nu}(E)}{(E_{\nu}-E_0)^2} \ .  
\label{eq-pheno-20}  
\end{equation}  
That is, the second inverse moment of the photo-absorption cross section is proportional to the static polarizability. As for the EWSR, the calculation of the IEWSR only requires information of the ground state due to the dielectric theorem \cite{bohigas1979}. Consider that the ground state is perturbed by an external field $\lambda F$ assuming $\lambda$ small so that perturbation theory holds. Changes in the expectation value of the Hamiltonian $\mathcal{H}$ can be written as,
\begin{equation}  
  \delta\langle\mathcal{H}\rangle = \lambda^2\sum_{\nu \neq 0}\frac{\vert\langle \nu\vert F\vert 0\rangle\vert^2}{E_{\nu}-E_0} + \mathcal{O}(\lambda^3) = \lambda^2 m_{-1} + \mathcal{O}(\lambda^3)
\label{eq-pheno-21} \ , 
\end{equation}
where standard perturbation theory has been applied, while changes in the expectation value of the operator, 
\begin{equation}  
  \delta\langle F\rangle = -2\lambda\sum_{\nu \neq 0}\frac{\vert\langle \nu\vert F\vert 0\rangle\vert^2}{E_{\nu}-E_0} + \mathcal{O}(\lambda^2) = -2\lambda m_{-1} + \mathcal{O}(\lambda^2)
\label{eq-pheno-22} \ . 
\end{equation}
This leads to, 
\begin{equation}  
  m_{-1} = \frac{1}{2}\frac{\partial^2\langle\mathcal{H}\rangle}{\partial \lambda^2}\Big\vert_{\lambda=0} = -\frac{1}{2}\frac{\partial\langle F\rangle}{\partial \lambda}\Big\vert_{\lambda=0}
\label{eq-pheno-23}  
  \end{equation}
which is nothing but the dielectric theorem. From the relation between $\langle\mathcal{H}\rangle$ and $\langle F\rangle$ one may also write, 
\begin{eqnarray}  
  \frac{\partial^2\langle\mathcal{H}\rangle}{\partial \lambda^2}\Big\vert_{\lambda=0}&=&\frac{\partial}{\partial\lambda}\left(\frac{\partial\langle\mathcal{H}\rangle}{\partial \langle F\rangle}\frac{\partial\langle F\rangle}{\partial \lambda}\Big\vert_{\lambda=0}\right)\nonumber\\
  2m_{-1}&=&(-2m_{-1})^2\frac{\partial^2\langle\mathcal{H}\rangle}{\partial \langle F\rangle^2}\nonumber\\
\frac{1}{2m_{-1}}&=&\frac{\partial^2\langle\mathcal{H}\rangle}{\partial \langle F\rangle^2} \ .
\label{eq-pheno-24}  
\end{eqnarray}

If one now defines a {\it scaled} wave function such that $\vert\eta\rangle \equiv e^{\eta [\mathcal{H},F]}\vert 0\rangle$, where $\eta$ is the scaling parameter, the expectation value of the Hamiltonian proportional to $\eta^2$ in this basis --i.e. $\langle\eta\vert\mathcal{H}\vert\eta\rangle$-- will be $\eta^2\sum_{\nu} (E_{\nu}-E_0)^3\vert\langle\nu\vert F\vert 0\rangle\vert^2$. Therefore, the third moment of the strength function can be evaluated as a second derivative of a small parameter. In the scaling case $m_3=\frac{1}{2}\frac{\partial^2\langle\mathcal{H}\rangle}{\partial \eta^2}\Big\vert_{\eta=0}$. Hence, the third moment of the strength function can also be regarded as a measure of the polarizability of the system: it accounts for the change in the energy of the system when the ground state is deformed according to the scaling defined above. 

With these elements, we will give now some examples. Let us start with the static dipole polarizability or $m_{-1}$ calculated by assuming the previously introduced DM. In Ref.\cite{meyer1982}, this quantity was calculated for the dipole operator leading to the expression\footnote{The expression introduced by Migdal in Ref.\cite{migdal1944} neglected the term in $a_{AS}$.}
\begin{equation}
\alpha_D^{\rm DM} \approx \frac{\pi e^2}{54}\frac{A\langle r^2\rangle}{a_A}\left(1+\frac{5}{3}\frac{a_{AS}}{a_A}A^{-1/3}\right)  
\label{eq-pheno-25}  
  \end{equation}
where we have again identified the asymmetry term in the DM with that in the LDM in order to provide a simple formula with already known parameters. This does not modify the physics contained by the formula but may introduce some differences with respect to the DM result. As an example, this formula with $a_{AS}= 9$ MeV and $a_{A}= 28$ MeV predicts, with remarkable accuracy, that $\alpha_D({}^{208}$Pb$)\approx 19.3$ fm${}^{3}$ (Exp. 19.6(6) fm${}^{3}$ \cite{tamii11}), $\alpha_D({}^{120}$Sn$)\approx 7.8$ fm${}^{3}$ (Exp.  8.6(4) fm${}^{3}$ \cite{hashimoto15}), $\alpha_D({}^{68}$Ni$)\approx 3.1$ fm${}^{3}$ (Exp.  3.9(3) fm${}^{3}$ \cite{rossi13}) and that $\alpha_D({}^{48}$Ca$)\approx 1.8$ fm${}^{3}$ (Exp.  2.1(2) fm${}^{3}$ \cite{Birkhan2017}).

The extension to other multipolarities only changes the power for the expectation value of $r$ and the geometric factors but not the main structure of Eq.(\ref{eq-pheno-25}). For example, for the quadrupole case,  
\begin{equation}
m_{-1}^{\rm DM}(Q) \approx \frac{1}{16\pi}\frac{A\langle r^4\rangle}{a_A}\left(1+\frac{7}{3}\frac{a_{AS}}{a_A}A^{-1/3}\right) \ .  
\label{eq-pheno-26}  
  \end{equation}
These formulas indicate that $a_{AS}/a_A$ depends linearly with the polarizability times $a_A$ for any multipolarity. The linear relation will differ depending on the multipolarity. It will be interesting to confirm if a difference exists in experiment as well. 

Next we will focus on the excitation energy of the ISGMR known as the breathing mode \footnote{From a macroscopic picture it would correspond to an isotropic compression.} and the ISGDR known as the squeezing mode \footnote{From a macroscopic picture it would correspond to a non-isotropic compression.}. It is important to note that excitation operators proportional to $r$ such as the isoscalar dipole operator $rY_{10}(\hat{r})$ produce a center of mass translation and, therefore, does not lead to an excitation of the nucleus. Thus, for the study of the ISGDR the operator should describe second order dipole transitions, that is, it should be proportional to $r^3Y_{10}(\hat{r})$ where mostly $3\hbar\omega$ transitions will be excited. In addition, as in the isovector dipole case, the CM motion should be subtracted. For the ISGMR, the isoscalar monopole operator proportional to $r^0Y_{00}$ will not produce any nuclear excitation for obvious reasons and so, again, one needs to go to second order monopole transitions (2$\hbar\omega$), that is, to use the operator $r^2Y_{00}$.    

We discuss the estimation of the excitation energy of the ISGMR and ISGDR by using the sum rules \cite{stringari1982} since those are more general and subject to less model dependencies than, for example, the harmonic oscillator approach. There are different ways to define an excitation energy by using sum rules. The most common ones are the centroid energy $m_1/m_0$, the constrained energy $\sqrt{m_1/m_{-1}}$ and the scaling energy $\sqrt{m_3/m_1}$. All of them might be very useful but we will now concentrate on the constrained energy.

\begin{figure}[t!]
\begin{center}  
  \includegraphics[width=0.45\linewidth,clip=true]{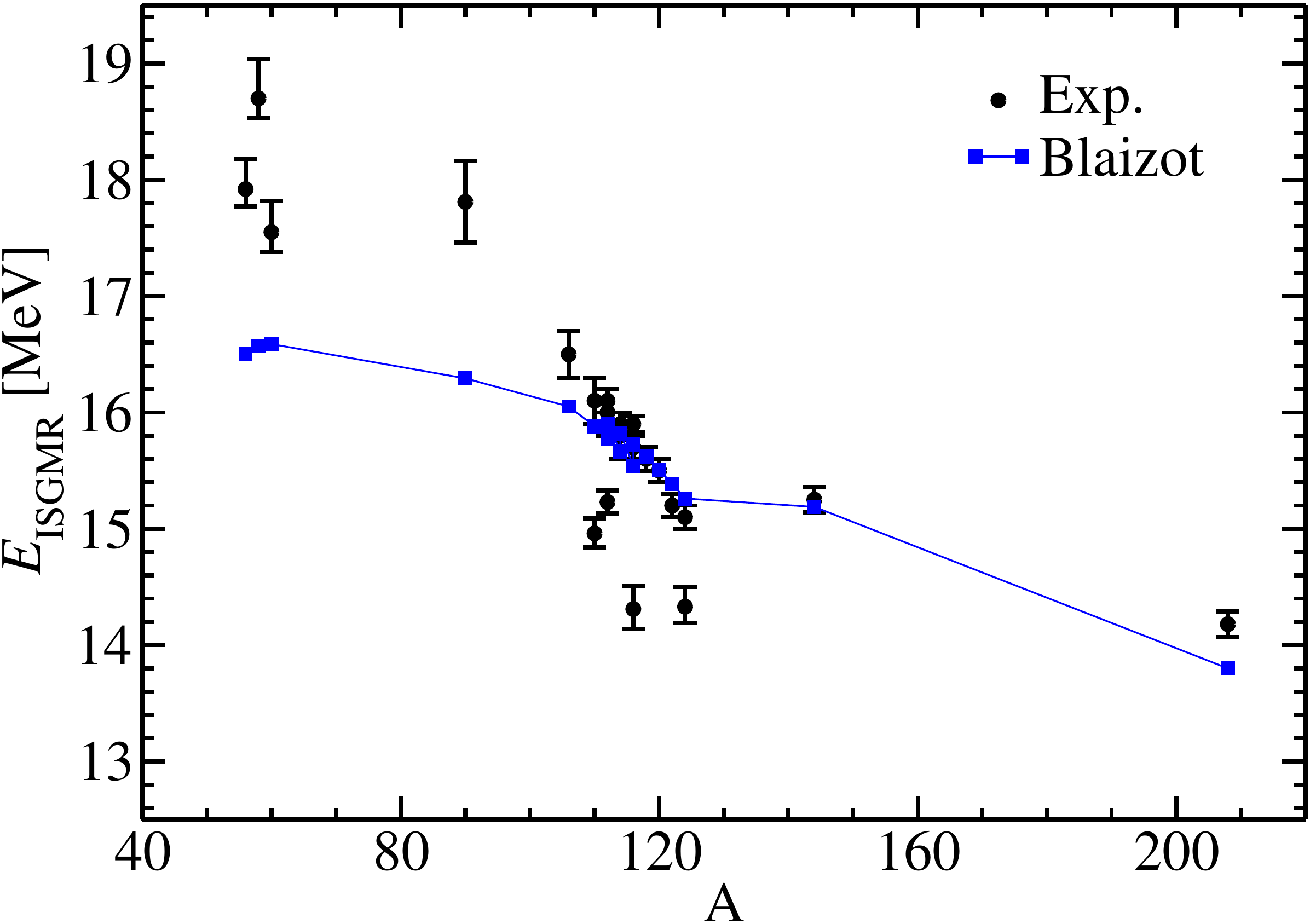}
  \includegraphics[width=0.47\linewidth,clip=true]{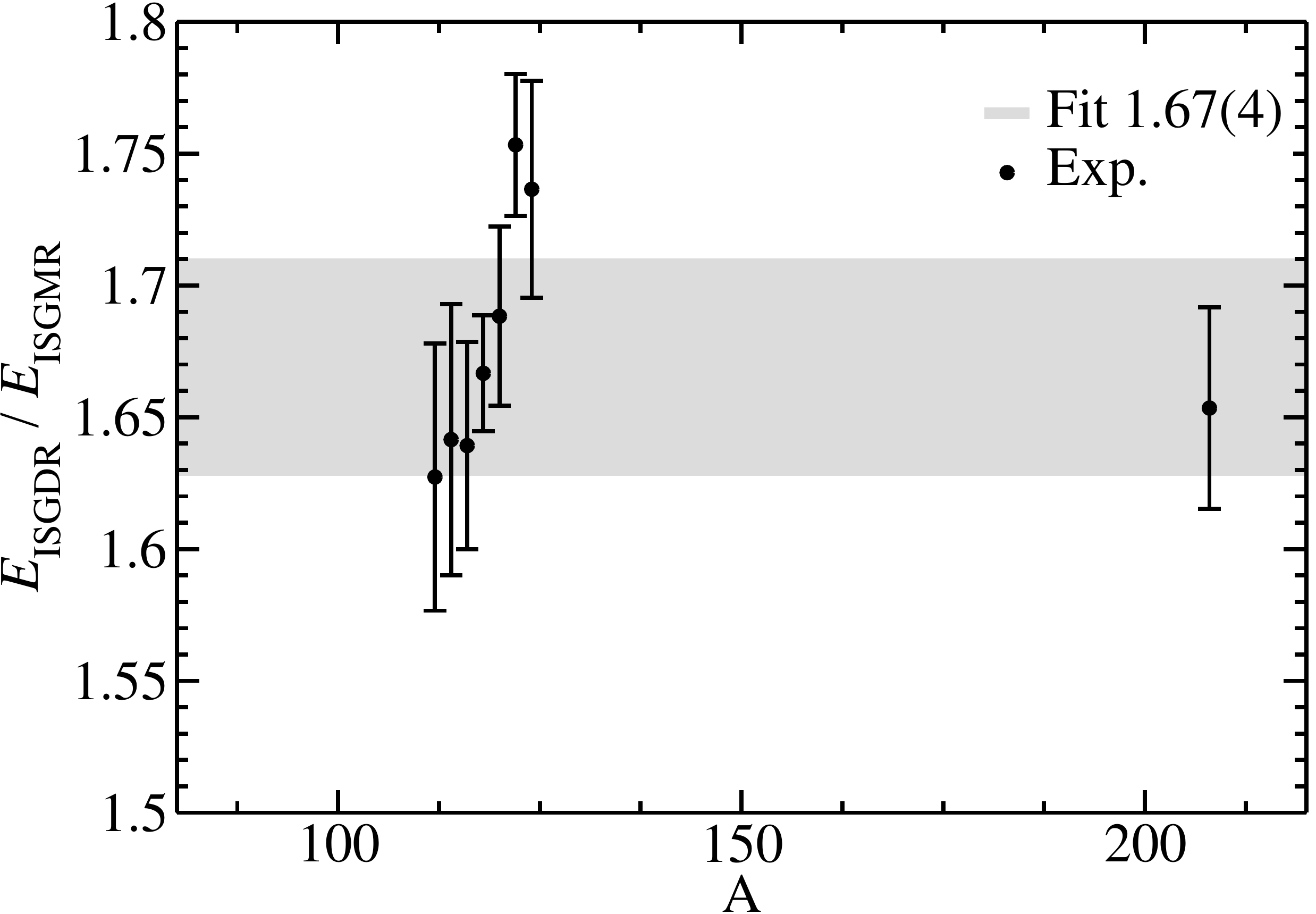}
\end{center}
\caption{\label{fig-pheno-8} Left panel: excitation energy $\sqrt{m_1/m_{-1}}$ of the ISGMR as a function of the mass number. Predictions of Eq.(\ref{eq-pheno-32}) are compared with experiment \cite{harakeh01}. Right panel: ratio between the excitation energies of the ISGDR and ISGMR as a function of the mass number. Experimental data have been fitted to a constant value inspired by the macroscopic model described in the text. See text for details.}
\end{figure}

To start it is instructive to see the ISGMR and ISGDR excitation energies within the harmonic oscillator model of Ref.\cite{bohr69}, as it has been done before for other modes. The prediction is $\hbar\omega^{\rm ISGMR}\approx 5.6 \sqrt{K_0} A^{-1/3}$ and $\hbar\omega^{\rm ISGDR}\approx 8.0 \sqrt{K_0} A^{-1/3}$, that is, the ratio $\frac{\hbar\omega^{\rm ISGDR}}{\hbar\omega^{\rm ISGMR}}\approx 1.43$ is constant within this approximation. In the harmonic oscillator approach, the excitation energy should be proportional to the square root of the restoring force. In this case, macroscopically both modes can be understood as volume compression of the nucleus and, thus, the restoring force has to be related to the incompressibility of the finite nucleus of mass $A$. In the previous expression $K_0$ has been used instead; this corresponds to a rough approximation justified by the fact that employed macroscopic model aims at a qualitative picture. It would be, therefore, interesting to define the incompressibility of the finite nucleus. The concept can be extended from the infinite system of nucleons to the finite system as follows. For a given nucleus with mass number $A$, using Eqs.(\ref{eq-pheno-18}) and (\ref{eq-pheno-24}), considering the monopole operator ($J=0$) and that $\langle H\rangle = E$, one finds  
\begin{equation}  
(E_x^{\rm ISGMR})^2=\frac{m_1}{m_{-1}}= 4\frac{\hbar^2}{m}\langle r^2\rangle\frac{\partial^2 E}{\partial \langle r^2 \rangle^2}=4 A\frac{\hbar^2}{m\langle r^2\rangle}\langle r^2\rangle^2\frac{\partial^2 (E/A)}{\partial \langle r^2 \rangle^2}\equiv K_A \frac{\hbar^2}{m\langle r^2 \rangle}\ ,
\label{eq-pheno-27}  
\end{equation}
where the latter expression is commonly used to define the finite nucleus incompressibility $K_A\equiv 4 A \langle r^2\rangle^2\frac{\partial^2 (E/A)}{\partial \langle r^2 \rangle^2}$. $K_A$ should be related to the compressibility of the nucleus if a macroscopic picture is adopted, but which is the relation between $K_A$ and $K_0$? Or, in other words, are these definitions in harmony?. First of all let us answer the latter question. From the thermodynamical definition of the compressibility $\chi = \frac{1}{V}\left(\frac{\partial P}{\partial V}\right)^{-1}$ one may write, assuming spherical symmetry, 
\begin{equation}
\frac{1}{\chi} = \frac{r}{3}\left(-rP +\frac{1}{4\pi r^2}\frac{\partial^2 E}{\partial r^2}\right)  \ .
\label{eq-pheno-28}  
\end{equation}
Imposing that the nucleus is like a liquid drop at equilibrium in its ground state and, therefore, its pressure at any point can be taken as zero \footnote{We remind that this is just a simplified picture to justify the definition in Eq.(\ref{eq-pheno-27}).} and rewriting the incompressibility of infinite symmetric nuclear matter in terms of the compressibility $K_0 = 9/(\rho_0 \chi)$, we can define in analogy  
\begin{equation}
K_A = \frac{9 V}{\chi} = 9\frac{r^2}{9}\frac{\partial^2 E}{\partial r^2}=Ar^2\frac{\partial^2 (E/A)}{\partial r^2} \ .
\label{eq-pheno-29}  
\end{equation}
If one does the variations with respect to $r^2$ (instead of $r$), it leads to the expression  
\begin{equation}
  K_A = 4 A (r^2)^2\frac{\partial^2 (E/A)}{\partial (r^2)^2} \ ,
\label{eq-pheno-30}  
\end{equation}
so one can say that $K_A$ and $K_0$ are comparable quantities. Next we briefly comment on how these two quantities can be related \cite{Blaizot1980,chossy1997}. For this purpose, we will use a simple macroscopic model. Assuming that the nucleus is a sharp sphere of radius $R$ and neutron excess $I$, the energy per particle of the nucleus can be written as the energy per particle of the infinite system (EoS) plus surface and Coulomb corrections. That is, $E/A = e(\rho,\delta) + \frac{4\pi R^2\sigma}{A}+\frac{3}{5}\frac{Z^2e^2}{A R}$ where $\sigma$ is the so called surface tension. By deriving the energy per particle with respect to the density (taking into account that $R$ will change accordingly), equating the result to zero and expanding the EoS around saturation ($\rho_0$) and for small asymmetries ($I\rightarrow 0$), one finds at lowest order in the density expansion the equilibrium density in the finite system $\rho_{\rm eq}=\rho_0+\frac{3\rho_0}{K_0}\left(1-LI^2+2\frac{4\pi\sigma R^2}{A}-\frac{3}{5}\frac{e^2}{R}\frac{Z^2}{A}\right)$. One, of course,  can improve this solution to take into account also higher order terms in the density. Substituting the expression for the equilibrium density (taking into account terms up to second order in the density expansion) of the finite nucleus into the second derivative of the energy per particle multiplied by $9\rho_{\rm eq}$ in analogy with the definition of $K_0$ one finds \cite{Blaizot1980,chossy1997,xavier} 
\begin{equation}
  K_A = 9\rho_{\rm eq}\frac{\partial^2 (E/A)}{\partial \rho^2}\Big\vert_{\rho=\rho_{\rm eq}} = K_0 +\left(22-2\frac{K^\prime}{K_0}\right)b_SA^{-1/3} + \left[K_{\rm sym} +L\left(\frac{K^\prime}{K_0}-6\right)\right]I^2 + b_C\left(\frac{K^\prime}{K_0}-8\right)\frac{Z^2}{A^{4/3}}
\label{eq-pheno-31}  
\end{equation}
where terms in $\sigma^\prime$ are zero, terms in $\sigma^{\prime\prime}$ have been neglected and parameters $b_S$ and $b_C$ conveniently collect some constants \footnote{Such as the surface tension evaluated in a semi-infinite nuclear matter calculation of a symmetric system evaluated at $\rho_0$. This is encoded in $b_S$.}. Primes stand for density derivatives evaluated at $\rho_0$. This expression is analogous to the liquid drop mass formula. Many attempts have been done to fit the parameters of this type of function to available experimental data on the excitation energy of the ISGMR (see Ref.\cite{stone2014}). In this way one can provide important qualitative information on the relation of the excitation energy of the ISGMR and the different parameters of the EoS. In order to show here an example, we will use a simplified model, known as the Blaizot formula \cite{Blaizot1980}, commonly used in the literature and that resembles more the liquid drop model,  
\begin{equation}
  K_A \equiv K_V +K_SA^{-1/3} + K_{\tau}I^2 + K_C\frac{Z^2}{A^{4/3}}
\label{eq-pheno-32}  
\end{equation}
where different parameters ($K$'s), to be fitted to the experimental $E_x^{\rm ISGMR}$ via the relation in Eq.(\ref{eq-pheno-27}) --that is, within the constrained approximation-- have been defined in an obvious way. Analogous to the LDM energies, the volume term here is identified with the corresponding parameter of the EoS, that is, $K_0$. The latter is, of course, an assumption that depends on the reliability of Eq.(\ref{eq-pheno-32}). The result of the fit to 18 experimentally known $E_x^{\rm ISGMR}$ \cite{harakeh01} (see also Ref.\cite{stone2014} and references therein) is $K_0=326(150)$ MeV, $K_S=-877(293)$ MeV and $K_{\tau}=-417(664)$ MeV if the Coulomb term is fixed to $K_C=-5.2$ MeV (assuming no error) which has been estimated from microscopic models \cite{sagawa2007} \footnote{Errors in $K_0$, $K_S$ and $K_\tau$ correspond to one standard deviation and are underestimated since $K_C$ has been assumed to be well known.}. From this fit it is clear that available experimental data is not enough to constrain the parameters of this formula \cite{Vinas2015}. In Fig.\ref{fig-pheno-8} the excitation energy of the ISGMR ($\sqrt{m_1/m_{-1}}$) as a function of the mass number is shown. Predictions of Eq.(\ref{eq-pheno-32}) are compared with experimental data \cite{harakeh01} used in the fit. 

Finally, for the case of the ISGDR, one can follow similar considerations and find the constrained energy as we have done before for the ISGMR. The result will be again proportional to $\sqrt{K_A}$. In Ref.\cite{stringari1982} it was derived theoretically that the relation between the ISGMR and ISGDR excitation energies should roughly scale by a constant factor $\frac{E_x^{\rm ISGDR}}{E_x^{\rm ISGMR}}\approx \sqrt{\frac{15}{7}}\approx 1.46$ (very close to the HO prediction) meaning that the information content of both observables is, in first approximation, redundant. In the right panel of Fig.\ref{fig-pheno-8}, the rate between these energies for available experimental data in ${}^{208}$Pb\cite{hedden2002} and Sn isotopes \cite{li2010} \footnote{Instead of a rate between excitation energies calculated as $\sqrt{m_1/m_{-1}}$, energies from a Lorentzian fit to the experimental strength function are used.} are shown. A fit to a constant value is also depicted being the result equal to 1.67(4). The main discrepancy between the experimental fit and the theoretical prediction is probably due to the fact that the ISGDR is splitted into two peaks and in the last figure we have only used the excitation energy of the high energy peak while, to be fully consistent in the comparison with theory, one has to estimate from experiment $m_1/m_{-1}$ for the full energy range. On the other side, although the results for the Sn isotopes are consistent with a constant value of $\frac{E_x^{\rm ISGDR}}{E_x^{\rm ISGMR}}$, it seems that there is some dependence on the neutron excess.

In summary, we have seen a way to connect the excitation energy of the ISGMR and ISGDR with $K_A$ and, thus, with $K_0$ and $K_{\rm sym}$. Similar expressions could be derived for the scaling energy $\sqrt{m_3/m_1}$ using instead the relation derived above for the third moment of the strength function $m_3=\frac{1}{2}\frac{\partial^2\langle\mathcal{H}\rangle}{\partial \eta^2}\Big\vert_{\eta=0}$. 

%% file: theory.tex
\section{Theoretical framework: self-consistent mean field models}
\label{theo}
The nuclear EoS describes a system that does not exist isolated in nature. Therefore, to access the parameters that characterize the EoS one needs to compare theory with experimental data on finite nuclei, heavy ion reactions or with astrophysical observations, especially those coming from neutron stars. Here, we will concentrate on the information one can gain from (bulk) ground and collective excited state properties of nuclei. In general, it is important to keep in mind that the separation between bulk and surface properties of nuclei is model dependent and, thus, to establish the relation between the properties of finite nuclei and the EoS will suffer from model intrinsic uncertainties difficult to evaluate.
 
As we have seen in the previous section, bulk properties of finite nuclei such as masses, radii and integrated moments of the strength function, directly related to the linear response of the nucleus to small perturbations, are clearly connected with the main parameters of the nuclear EoS. It is, therefore, crucial to analyze available experimental data within a microscopic theoretical framework that can reliably describe all these observables and then predict the EoS on equal footing. Nowadays, the most suitable framework is that of self-consistent mean field models (SCMF) \cite{skyrme1}. From a theoretical point of view, SCMF can be regarded as a possible way to derive an approximate EDF. The importance of such a relation is that EDF theory is based on the Hohenberg-Kohn theorems that ensure the existence of an exact energy density functional. By exact here one means that both total energy and density as well as the expectation value of any one-body operator would correspond to an exact prediction (since any one-body operator can be reabsorbed in the one-body potential in the Kohn-Sham realization of an EDF). As it will be seen in Sec.\ref{gs}, nuclear EDF derived from SCMF calculations have shown a very high accuracy on the overall description of nuclear masses (below the 0.5\% for medium and heavy nuclei) and charge radii (about few \%). Regarding moments of the strength function, EDFs are successful in the description of excitation energies of GR and polarizabilities (see Sec.\ref{excitations}). The latter can also be justified from an EDF perspective because, whenever the external perturbation is an isoscalar and one-body operator, the calculation of $m_k$ will stay within the Kohn-Sham shceme of the Hoheberg-Kohn theorems.

In what follows we will briefly review the most commonly used SCMF models and their extension to deal with the linear response of nuclei when perturbed by an external field. For further details we refer the reader to the literature herein. 

\subsection{Nuclear effective forces}
In the ground state of a nucleus, or at low excitation energies (MeV), nucleons do not feel the bare nucleon-nucleon interaction due to the presence of many other nucleons. It is, therefore, natural to consider a {\it screened} bare interaction that can be parameterized in terms of an effective interaction in the medium and that may greatly differ from the bare one. Opposite to the bare interaction derived from nucleon-nucleon scattering data in the vacuum, effective interactions can be well behaved at all distances and many-body methods can be applied more easily. Commonly used effective interactions depend on about ten parameters and are fitted to many-body data such as masses or charge radii, so a many-body scheme has to be fixed before the determination of the optimal parameters. That is, the use of these models is restricted to the many-body scheme employed in the fitting of the parameters.

Current SCMF models are built based on effective density-dependent two-body interactions. The ansatz can be chosen freely, however, only three types of interactions are commonly used nowadays. The main features that need to be ensured are the short-range character of the interaction, the saturation mechanism, the isospin invariance, and the fermionic nature of the nucleon. Those successful interactions are known as the non-relativistic zero-range Skyrme \cite{skyrme1958} and finite-range Gogny \cite{decharge1980} interactions as well as the relativistic SCMF models of zero-range and finite-range as well as including self-interacting meson terms or density dependent coupling constants \cite{meng2016}.  

\subsubsection{Non-relativistic effective interactions}
The Skyrme interaction is the most widely used interaction in nuclear structure calculations. The reason is simple: it is a zero-range (but momentum dependent) interaction that greatly simplifies calculations in many-body systems. It was introduced by T. H. R. Skyrme in Ref.\cite{skyrme1958} as the lowest order momentum expansion of a (finite-range) Yukawa interaction plus a three-body contact term that is usually approximated by a density dependent two-body contact term as well as with a spin-orbit term that has its origin in covariant theories,
\begin{eqnarray}
  V_{\rm Skyrme}({\bf r})&=&t_0(1+x_0P_{\sigma})\delta({\bf r})+\frac{1}{2}t_1(1+x_1P_{\sigma})\left[{\bf k}^{\dag 2}\delta({\bf r})+\delta({\bf r}){\bf k}^2\right]+t_2(1+x_2P_{\sigma}){\bf k}^{\dag}\cdot\delta({\bf r}){\bf k}\nonumber\\
&+& \frac{1}{6}t_3(1+x_3P_{\sigma})\delta({\bf r})\rho^\alpha({\bf R}) +iW_0{\bf \sigma}\cdot{\bf k}^\dag\times\delta({\bf r}){\bf k} \ ,  
\label{eq-theo-33}  
\end{eqnarray}  
where tensor contributions have not been included; $t$'s, $x$'s, $\alpha$ and $W_0$ are the parameters to be fitted\footnote{$\alpha=1$ would correspond to the reduction of a pure three-body force to a two-body density dependent one, thus, $\alpha$ different from 1 and 0 mimic a many-body interaction term.}; ${\bf r}\equiv{\bf r}_1-{\bf r}_2$ is the relative coordinate between particles 1 and 2; ${\bf R}\equiv\frac{{\bf r}_1+{\bf r}_2}{2}$ is the CM coordinate considering two equal (mass) nucleons; ${\bf k}\equiv {\bf k}_1-{\bf k}_2$ in momentum space or ${\bf k}=\frac{i}{2}\left({\bf \nabla}_1-{\bf \nabla}_2\right)$ in coordinate space correspond to the relative momentum operator between the interacting particles labeled with 1 and 2; ${\bf \sigma}\equiv{\bf \sigma}_1+{\bf \sigma}_2$ is the sum of the Pauli matrices referred to particle 1 and 2, respectively; $P_\sigma\equiv\frac{1+{\bf \sigma}_1\cdot{\bf \sigma}_2}{2}$ is the spin-exchange operator between particles 1 and 2. 

Although there are hundreds of parameterisations of this effective interaction in the literature, only some of them are considered to be accurate enough to be used for a general purpose. In Ref.\cite{dutra2012} 240 parameterizations were analyzed based on some criteria and a short list of preferred parameterizations was given. Other comparisons are available in the literature as well (see for example Refs.\cite{erler2012,goriely2013}).

Another type of commonly used effective interaction in non-relativistic models is the so called Gogny interaction \cite{decharge1980} that has two finite-range terms of Gaussian type, a zero-range density dependent term and a spin-orbit term, being the last two terms equal to those of the Skyrme interaction,  
\begin{eqnarray}
  V_{\rm Gogny}({\bf r})&=&\sum_{i=1}^{2}e^{\frac{r}{\mu_i}}\left(W_i+B_iP_\sigma-H_iP_\tau-M_iP_\sigma P_\tau\right) \nonumber\\
&+& \sum_{i=1}^2t_0^i(1+x_0^iP_{\sigma})\delta({\bf r})\rho^{\alpha_i}({\bf R}) +iW_0{\bf \sigma}\cdot{\bf k}^\dag\times\delta({\bf r}){\bf k} \ ,  
\label{eq-theo-34}  
\end{eqnarray}  
where $P_\tau$ is the isospin exchange operator (analogous to the spin-exchange one previously defined) and $W$'s, $B$'s, $H$'s, $M$'s, $t_3$, $x_3$, $\alpha$, and $\mu$'s are the parameters that have to be fitted to experimental data. Much less parameterizations of this force are  available in the literature. Nevertheless, its accuracy in the description of masses, charge radii and excitation energies of GR is comparable to that of the Skyrme models \cite{goriely2009}.

\subsubsection{Relativistic effective interactions}
Within the relativistic framework, the effective interaction is carried by heavy mesons. The Walecka model \cite{serot1984} represents the first attempt to propose a relativistic theory of the nucleus. It has been later improved including meson self-interacting terms or density dependent meson-nucleon vertex functions. Those interactions are finite-range since the bosons carrying the interaction are massive. However, there are also zero-range versions of relativistic interactions based on the heavy-mass meson limit that allow to neglect the momentum transfer in the scattering of two nucleons in a nucleus within a good approximation \cite{sulaksono2003,Liang2012a}.

These theories start from a Lagrangian density usually composed of three parts, the nucleonic free Lagrangian, the mesonic free Lagrangian and the Lagrangian describing the interactions. The first one corresponds to the Dirac equation for the free nucleon fields while the second one corresponds to its equivalent in the case of bosons, that is, the Klein-Gordon equations for the free meson fields. The interaction Lagrangian can be generally written as,
\begin{eqnarray}
{\mathcal L}_{int} &=& {\bf \bar{\Psi}}{\Gamma}_{\sigma}({\bf \bar{\Psi}},{\bf  \Psi}){\bf  \Psi}\Phi_{\sigma} 
                   +  {\bf \bar{\Psi}}{\Gamma}_{\delta}({\bf \bar{\Psi}},{\bf  \Psi})\mbox{\boldmath$\tau$}{\bf  \Psi}{\bf \Phi}_{\delta}\nonumber\\ 
                  & & -{\bf \bar{\Psi}}{\Gamma}_{\omega}({\bf \bar{\Psi}},{\bf  \Psi})\gamma_{\mu}{\bf\Psi}A^{(\omega) \mu}
                  - {\bf \bar{\Psi}}{\Gamma}_{\rho}({\bf \bar{\Psi}},{\bf  \Psi})\gamma_{\mu}{\mbox{\boldmath$\tau$}}{\bf \Psi}{\bf A}^{(\rho) \mu}
%                  & & -e{\bf \bar{\Psi}}\hat{Q}\gamma_{\mu}{\bf\Psi}A^{(\gamma) \mu}
   \label{eq-theo-35}
\end{eqnarray}
where the neutrons and the protons are represented as Dirac spinors ${\bf \Psi}=\left(\begin{array}{c} \psi_p \\ \psi_n\end{array}\right)$ and the four mesons ($\sigma$, $\omega$, $\rho$ and $\delta$) carrying the effective nuclear strong interaction represented by the fields $\Phi_{\sigma}$, $A^{(\omega)}_{\mu}$, ${\bf A}^{(\rho)}_{\mu}$ and ${\bf \Phi}_{\delta}$. The index $\mu$ indicates the time- and space-like components of the vector fields and the bold face indicates the vector nature of a field in the isospin space. The Pauli matrices in isospin space are represented by $\mbox{\boldmath$\tau$}$ and the Dirac matrices by $\gamma_\mu$. The $\Gamma$'s are the couplings associated to each meson field and they may depend on the nucleon field ${\bf \Psi}$, however, since covariance is required, the dependence should reduce to a scalar operator of the form $\hat{\rho}({\bf \bar{\Psi}},{\bf \Psi})$. Hence, one should take a prescription for the specific expression of this Lorentz-invariant operator. The prescription $\hat{\rho}^2 = \hat{j_{\mu}}\hat{j^{\mu}}$ ---where $\hat{j_{\mu}}={\bf  \bar{\Psi}}\gamma_{\mu}{\bf \Psi}=\hat{\rho}\hat{u}_{\mu}$ is the nucleon vector current and $\hat{u}_{\mu}$ a four velocity ($\hat{u}_{\mu}\hat{u}^{\mu}={\bf 1}$)--- has been chosen frequently in modern versions of these models. The reason relies on the fact that its expectation value at the ground state is just the nucleon density $\rho$. Hence, the connections with other theories will be more natural. 

In the original Walecka model, the $\Gamma$'s where simple constants (no dependence on the nucleon field) and no $\delta$ or $\rho$ meson fields were taken into account. This was enough to reproduce the saturation mechanism in a Hartree (mean-field) calculation since $\sigma$ simulate the short range attraction produced by two-pion exchange and $\omega$ produces the needed repulsion. It is, however, not successful in reproducing the incompressibility of infinite symmetric nuclear matter or isovector properties in general since, looking at Eq.(\ref{eq-theo-35}), it is evident that $\sigma$ is a scalar-isoscalar meson and $\omega$ is a vector-isoscalar meson while their isovector counterparts are $\rho$ and $\delta$ (i.e. terms involving $\mbox{\boldmath$\tau$}$). It is important to note that the pion is not explicitly included in theories that do not go beyond the mean-field Hartree approach since, for symmetry reasons, the expectation value of the pion-field (pseudo-vector) in a Hartree calculation is zero. Interesting features of these models are that they preserve Lorentz invariance (as in the underlying QCD) allowing for a natural description of the spin-orbit coupling, imposing restrictions on the number of parameters, and preserving causality among other relevant properties. 

As mentioned, relativistic interactions employed nowadays are different from the original Walecka model. The first type corresponds to non-linear Walecka model and includes the $\rho$ meson field and meson self-interacting terms up to some given order in the expansion of the meson fields \cite{furnstahl1997}. The minimal refinement and one of the most commonly used is just to include $\sigma$ self-interacting terms up to the fourth order modifying the interaction Lagrangian by adding terms as $-\frac{g_2}{3}\Phi_{\sigma}^3-\frac{g_3}{4}\Phi_{\sigma}^4$ \cite{NL3}. Such a modification allows to better reproduce, e.g., the incompressibility. 

The other, even more successful approach is based on effective Lagrangians where the interaction couplings in Eq.(\ref{eq-theo-35}) are considered to be dependent on the baryon density. In these models, there is the freedom to choose the explicit density dependence of the coupling constants. The most extended strategy is to map or mimic the density dependence derived from Dirac-Brueckner-Hartree-Fock calculations in infinite nuclear matter as done, e.g., in Refs.\cite{typel1999,hofmann2001,roca-maza2011}. Finally, zero-range reduction of the theory by assuming the heavy-meson mass limit have been also shown to be successful \cite{burvenich2002}.     

In the relativistic case, there exist also hundreds of parameterizations of the effective interaction. Most of them have also been examined in the study of Ref.\cite{dutra2014}, similar to the one previously mentioned for the Skyrme interaction \cite{dutra2012}. Other comparisons are available in the literature as well (see for example Ref.\cite{afanasjev2013}).

\subsection{Self-consistent mean-field approach}
Nuclear phenomenology justify the assumption that nucleons move independently in an average one-body potential or mean-field produced by all nucleons. To derive such a potential from two and three body interactions, it is customary to apply the variational principle using Slater determinants as a very simple trial wave functions. In practice such a procedure is realized by the Hartree or Hartree-Fock approximations based on a nuclear effective interaction such as the ones discusses so far. As mentioned, these types of models allows for a description of nuclei by an EDF dependent on generalized densities, i.e., the neutron and proton densities, spin densities, kinetic energy densities, currents, etc. For simplicity, it is common to encode all this information keeping a simple notation: $E[\rho]$. 

The variational equation can be written as $\delta E[\Psi] = 0$, where $E[\Psi]\equiv \langle\Psi\vert \mathcal{H}\vert\Psi\rangle / \langle\Psi\vert \Psi\rangle $ is equivalent to the Schroedinger equation $\langle\delta\Psi\vert \mathcal{H}-E\vert\Psi\rangle=0 \rightarrow \mathcal{H}\vert\Psi\rangle=E\vert\Psi\rangle$. As long as one does not know the exact wave function $\vert\Psi\rangle$, trial wave functions $\vert\Phi\rangle$ should be assumed. The variational principle tells us that for any trial wave function $E[\Phi]\geq E_0$ being trivially equal if $\vert\Phi\rangle=\vert\Psi\rangle$ and being $E_0$ the ground state energy. The latter inequality is very easy to prove \cite{ringschuck}. Not knowing $\vert\Psi\rangle$ will always result in an approximate solution for $E_0$. For the calculation of excited states on the bases of the trial (approximate) wave function, one solves the variational equation $\delta E[\Phi_\nu] = 0$ imposing that the excited state $\nu$ should be orthogonal to the ground state $\langle\Phi_\nu\vert \Phi_0\rangle=0$ (and to all other excited states).

In what follows we will briefly describe the main idea behind the HF method in order to give an overall idea to the non-conversant reader. For a deeper insight we refer the reader to, e.g., Refs.\cite{skyrme1,ringschuck} and for recent developments to Refs.\cite{drut2010, dobaczewski2011, ring2011,erler2012,afanasjev2013,goriely2013}.

\subsubsection{Hartree-Fock ground state energy}
As previously discussed, it is justified to assume that a one-body potential can describe reasonably well the nuclear phenomenology. The HF theory assumes a Slater determinant
%, that is, 
%\begin{equation}
%  \Phi^{\rm HF}({\bf x_1,x_2.....x_A})= \frac{1}{\sqrt{A!}}
%\begin{vmatrix}
%\phi_1({\bf x_1}) & \phi_1({\bf x_2})& \cdots & \phi_1({\bf x_A})\\ 
%\phi_2({\bf x_1}) & \phi_2({\bf x_2})& \cdots & \phi_2({\bf x_A})\\ 
%\cdots             &  \cdots          & \ddots & \cdots \\ 
%\phi_A({\bf x_1}) & \phi_A({\bf x_2})& \cdots & \phi_A({\bf x_A})
%\end{vmatrix} \ ,
%  \label{eq-theo-36}
%\end{equation}  
%where $\phi_i({\bf x_j})$ represent the $i-$th single-particle orbital for the coordinates of the $j-$th particle,
as a trial wave function to calculate the one-body potential from the expectation value of a two-body interaction. Then, the ground state energy of the system within the HF approximation can be found by applying the variational principle that will give as a result the optimal $\vert\Phi_0^{\rm HF}\rangle$ and, thus, the ground state density $\rho^{\rm HF}=\langle\Phi_0^{\rm HF}\vert\Phi_0^{\rm HF}\rangle$. Hence, one needs to calculate first the HF energy from an effective Hamiltonian\footnote{In the relativistic framework one can obtain the Hamiltonian from the Lagrangian by a Legendre transformation $\mathcal{H}=T^{00}$ where $T^{\mu \nu} = \sum_{i} \frac{\partial {\mathcal L}}{\partial(\partial_{\mu} \phi_{i})}\partial^{\nu}\phi_{i} - g^{\mu \nu} {\mathcal L}$ is the energy-momentum tensor, $g^{\mu\nu}$ is the metric, and $\phi_i$ runs over all the nucleon and meson fields.}  
%\begin{eqnarray}
%E^{\rm HF}&=&\int \mathcal{H}^{\rm HF} d{\bf r}\nonumber\\  
%\mathcal{H}^{\rm HF}&=& \langle\Phi^{\rm HF}\vert\mathcal{H}\vert\Phi^{\rm HF}\rangle=\langle\Phi^{\rm HF}\vert T\vert\Phi^{\rm HF}\rangle+\langle\Phi^{\rm HF}\vert V\vert\Phi^{\rm HF}\rangle \ ,
%\label{eq-theo-37}
%\end{eqnarray}
and, then, to perform variations with respect to the wave function keeping the orthonormality of the bases in order to find the ground state energy. After doing so, one is able to simplify the A-body problem into a 1-body problem. That is, one finds a set of equations for each orbital with the form of a Schroedinger equation $h({\bf x})\phi_i({\bf x})=\varepsilon_i\phi_i({\bf x})$ where $\varepsilon_i$, $\phi_i$ and $h({\bf x})$ represent the single-particle energy, wave function and Hamiltonian for a given particle labeled with $i$. The single-particle Hamiltonian relates with the HF Hamiltonian as follows, $\mathcal{H}^{\rm HF}=\sum_{k=1}^A h({\bf x}_k)$. The HF equations are commonly solved iteratively until consistency between the one-body potential and the single-particle wave functions is reached. It is important to notice that we just highlight some of the very general features of the ground state energy calculations via the HF approach. In actual calculations, relevant details like nuclear deformations or paring correlations among others have to be taken into account.

Independently of the effective interaction adopted, it is important to note that some differences between the relativistic and non-relativistic frameworks arise from the kinetic energy. The non-relativistic expression for the kinetic energy in the HF approximation can be written as $-\frac{\hbar^2}{2m}\sum_{j=1}^A\langle\phi_j\vert\nabla_j^2\vert\phi_j\rangle$ while the relativistic one is $\sum_{j=1}^A\langle\phi_j\vert\left(i\gamma_{\mu}\partial^{\mu}-M\right)\vert\phi_j\rangle$ \footnote{Adopting the no-sea approximation where negative-energy states are not taken into account, that is, vacuum polarization is explicitly neglected.} giving quantitative differences in actual calculations.

\subsubsection{Infinite asymmetric nuclear matter}
Within the Hartree or Hartree-Fock approximations, one can now numerically estimate the energy and density of the ground state as well as the sum rules discussed in Sec.\ref{pheno} for any nucleus adopting the preferred effective interaction. Some results will be presented in the following sections. On the other hand, by neglecting the Coulomb interaction and assuming plane waves since the system is translationally invariant, i.e. $\phi = \frac{1}{\sqrt{\Omega}}e^{i{\bf k}\cdot{\bf r}}$, one can also address the problem of calculating the uniform and spin-saturated nuclear matter equation of state for arbitrary neutron and proton densities at zero temperature, that is, what we have called so far the nuclear EoS. The limiting cases, i.e., the symmetric matter EoS will correspond to $\delta=0$ ($\rho_n=\rho_p$), and the pure neutron matter EoS will correspond to $\delta=1$ ($\rho=\rho_n$).      

{\bf Skyrme.} For the Skyrme interaction, one finds that the nuclear EoS can be written as \cite{dutra2012},
\begin{eqnarray}
  e(\rho,\delta)&=&\frac{3}{5}\frac{k_F^2}{2m}f_{5/3}+\frac{1}{8}t_0\rho\left[2(x_0+2)-(2x_0+1)f_2\right] + \frac{1}{48}t_3\rho^{\alpha+1}\left[2(x_3+2)-(2x_3+1)f_2\right]\nonumber\\
&+& \frac{3}{40}k_F^2\rho\left\{\left[t_1(x_1+2)+t_2(x_2+2)\right]f_{5/3}+\frac{1}{2}\left[t_2(2x_2+1)-t_1(2x_1+1)\right]f_{8/3}\right\}  
\label{eq-theo-42}
\end{eqnarray}
where $k_F=\left(\frac{3\pi^2}{2}\rho\right)^{1/3}$ and $f_m=\frac{1}{2}\left[(1+\delta)^m+(1-\delta)^m\right]$. 

{\bf Gogny.} For the Gogny interaction, one finds \cite{shellaewa2014} 
\begin{eqnarray}
  e(\rho,\delta)&=&\frac{3}{5}\frac{k_F^2}{2m}f_{5/3}+\frac{1}{2}\sum_{i=1}^2\left\{\left[\frac{\pi^{3/2}\mu_i^3}{4}\left(4W_i+2B_i-2H_i-M_i\right)+\frac{3}{4}t_0^i\rho^{\alpha_i}\right]\rho\right.\nonumber\\
  &&\hspace{3.1cm}\left.+\left[-\frac{\pi^{3/2}\mu_i^3}{4}\left(2H_i+M_i\right)-\frac{1}{4}t_0^i(1+2x_0^i)\rho^{\alpha_i}\right]\rho\delta^2\right\}\nonumber\\
  &-&\frac{1}{2}\sum_{i=1}^2\left\{\frac{1}{\sqrt{\pi}}\left(W_i+2B_i-H_i-2M_i\right)\left[\frac{1+\delta}{2}g(\mu_ik_{F_n})+ \frac{1-\delta}{2}g(\mu_ik_{F_p})\right]\right\}\nonumber\\
  &+&\frac{1}{\sqrt{\pi}}\left(H_i+2M_i\right)h(\mu_ik_{F_n},\mu_ik_{F_p})
\label{eq-theo-43}
\end{eqnarray}
where $k_{F_q}$ with $q=n$ or $p$ is the Fermi momentum of neutrons and protons respectively. It relates with $k_F$ as follows $k_{F_q}=k_F(1+\tau_q\delta)^{1/3}$ where $\tau_n=-1$ and $\tau_p=1$. The two functions have been defined: $g(q)\equiv\frac{2}{q^3}-\frac{3}{q}-\left(\frac{2}{q^3}-\frac{1}{q}\right)e^{-q^2}+\sqrt{\pi}{\rm erf}(q)$; and $h(s,t)\equiv 2\frac{s^2-st+t^2-2}{s^3+t^3}e^{-\frac{(s+t)^2}{4}}-2\frac{s^2+st+t^2-2}{s^3+t^3}e^{-\frac{(s-t)^2}{4}}-\sqrt{\pi}\frac{s^3-t^3}{s^3+t^3}{\rm erf}\left(\frac{s-t}{2}\right)+\sqrt{\pi}{\rm erf}\left(\frac{s+t}{2}\right)$.

{\bf Non-linear Walecka model.} In this case, as an illustrative example, we give the version of the model with just the self-interacting $\sigma-$meson terms up to fourth order \cite{NL3}. An extended model can be found in Ref.\cite{furnstahl1997}. Within the Hartree approximation, the solutions of the Dirac equations for protons and neutrons are the usual plane-wave Dirac spinors. The infinite matter reduction of the meson field equations of motion are, within the same approximation,
\begin{equation}
\Phi_{\sigma} = \frac{\Gamma_{\sigma}}{m_{\sigma}^2 }\rho^s  ;\quad
A^{(\omega) 0} = \frac{\Gamma_{\omega}}{m_{\omega}^2 }\rho   ;\quad
\Phi_{\delta} = \frac{\Gamma_{\delta}}{m_{\delta}^2 }\rho^s_3 ;\quad {\rm and}\quad
A^{(\rho) 0} = \frac{\Gamma_{\rho}}{m_{\rho}^2 }\rho_3 \ ,
\label{eq-theo-44}
\end{equation}
where the nucleon and scalar densities can be computed as follows,  
\begin{eqnarray}
\rho_{_q} &=& \frac{2}{(2\pi)^3}\int_{|k|< k_{F_{q}}}d^3k=\frac{{k_{F_{q}}}^3}{3\pi^2} 
\nonumber\\
\rho^s_{_q} &=& \frac{2}{(2\pi)^3}\int_{|k|< k_{F_{q}}}\frac{m^*_{_q}}{E_{_q}}d^3k\nonumber\\
           &=&\frac{m^*_{_q}}{2\pi^2}\left[ k_{F_{q}}E_{F_{q}} -
  {m^*_{_q}}^2 \ln\left(\frac{k_{F_{q}}+E_{F_{q}}}{m^*_{_q}}\right)\right]
\label{eq-theo-45}
\label{infros}
\end{eqnarray}
and where the Fermi energy of a proton ($q=p$) or a neutron ($q=n$) is $E_{F_{_q}}=\sqrt{{{\bf k}_{F_{_q}}}^2 +{m^*_{_q}}^2}$; $k_{F_q}$ is the Fermi momentum and $m^*_q=m-\Gamma_{\sigma}(\rho)\Phi_{\sigma}-\tau_{_q}\Gamma_{\delta}(\rho)\Phi_{\delta}$ is the Dirac effective mass of neutrons or protons; $\tau_p=-1$ while $\tau_n=+1$. With these elements, one can calculate the energy density ($\mathcal{E}=\frac{E}{V}=\rho\frac{E}{A}=\rho e$)   
\begin{eqnarray}
\mathcal{E}(\rho_n,\rho_p) = \langle 0|T^{00}|0\rangle  &=& \frac{1}{4}\left[3E_{F_n}\rho_n +  m^*_n\rho^s_n \right]+
                                   \frac{1}{4}\left[3E_{F_p}\rho_p +  m^*_p\rho^s_p \right] \nonumber\\
                & & + \frac{1}{2}\left[ m_{\sigma}^2{\Phi_{\sigma}}^2 + m_{\omega}^2{A^{(\omega) 0}}^2 
                    + m_{\delta}^2{\Phi_{\delta}}^2 + m_{\rho}^2{A^{(\rho) 0}}^2+\frac{g_2}{3}\Phi_\sigma^3+\frac{g_3}{4}\Phi_\sigma^4\right]
\label{eq-theo-46}
\end{eqnarray}    
where the coupling constants $\Gamma$ do not depend on the density and enter into the latter equation via the meson-fields [cf. Eq.(\ref{eq-theo-44})]. 

{\bf Density dependent relativistic models.} For these models the coupling constants $\Gamma$ explicitly depend on the density \cite{typel1999,ddme}. The expression for the energy density is equal to that of the previous model but with no $g_2$ or $g_3$ term. In addition, only for the case of zero-range (or point coupling models) the coupling constants should be redefined based on the heavy-meson mass approximation, that is, $\frac{\Gamma_i(\rho)}{m_i}\rightarrow \alpha_i(\rho)$ \cite{niksic2008}. These type of models as well as the non-linear Walecka model have been traditionally solved at the H level but HF calculations are also available \cite{meng2016}.

In the case of relativistic models, the equations need to be solved numerically --due to the expression of the scalar density-- by fixing the neutron and proton densities. In addition, it is straight forward to check the thermodynamic consistency of this theory by comparing the pressure as calculated through the energy-momentum tensor with the thermodynamical definition: $p=\frac{1}{3}\sum_{i=1}^{3}\langle 0|T^{ii}|0\rangle = \rho^2[\partial (e/\rho)/\partial\rho]$ and that the energy-momentum tensor is conserved $\partial_\mu T^{\mu \nu} = 0$ which implies that the four-momentum is a constant of motion as expected.

The parameters of the nuclear equation of state defined in Eqs.(\ref{eq-pheno-5}) and (\ref{eq-pheno-6}) can be easily calculated from the expressions above. Explicit formulas can be found in \cite{dutra2012} for the Skyrme, \cite{shellaewa2014} for the Gogny, and \cite{dutra2014} for relativistic models. 

\subsection{Linear response}
The aim of this section is to give some basic definitions and concepts that may help the non-conversant reader to better understand the results given in the following sections (for further details see, e.g., \cite{ringschuck} or \cite{nakatsukasa2016}).

The linear response of a nucleus under the action of an external and time-dependent oscillating field  of the form $F(t)= Fe^{-i\omega t} + F^\dag e^{+i\omega t}$, where the amplitude $F$ is assumed to be small, will be described. In the second quantization, one can write the operator $F(t)$ assuming it is one-body as $F(t)=\sum_{nm}f_{nm}(t)a_n^\dag a_m$ where $a^\dag$ and $a$ are creation and annihilation operators acting on the HF vacuum. Now the wave function of the system is time-dependent $\vert\Phi(t)\rangle$ and a time-dependent one-body density can be defined as $\rho_{nm}(t)=\langle\Phi(t)\vert a^\dag_ma_n\vert\Phi(t)\rangle$. By working within the HF bases, $\rho_{nm}$ will correspond to a Slater determinant (with the property $\rho^2=\rho$) and, therefore, one will be extending HF or the independent particle picture to its time-dependent version and one can refer to a time dependent EDF \cite{nakatsukasa2016}. As we have previously discussed, we assume that the density (and, therefore, the HF single-particle potential) oscillates with the external field. Then, when the frequency $\omega$ is close to the excitation energy  of a nuclear excited state one will obtain a resonance such as a giant resonance studied before.

Within these approximations, the one-body density will follow the time-dependent Schroedinger equation that might be written in the operator form as \cite{ringschuck},
\begin{equation}
i\hbar\frac{d\rho}{dt}=[h(\rho)+f(t),\rho]  
\label{eq-theo-47}  
\end{equation}
where $h(\rho)$ is the HF single-particle Hamiltonian or {\it field} previously defined and we denote the HF ground state density as $\rho^0$, i.e., $[h(\rho^0),\rho^0]=0$. It is important to note that $\rho^0$ and $h(\rho^0)$ are diagonal in the HF basis. For a small perturbation, the density is linear with the external field and could be written as,
\begin{equation}
\rho = \rho^0 + \delta\rho(t) \quad {\rm where}\quad \delta\rho(t)=\delta\rho e^{-i\omega t}+\delta\rho^\dag e^{i\omega t} \ ,
\label{eq-theo-48}  
\end{equation}
and $\delta\rho$ is the so called transition density $\delta \rho_{nm}=\langle 0\vert a^\dag_ma_n\vert\nu\rangle$ between the ground state $\vert 0\rangle$ and an excited state $\vert\nu\rangle$. Inserting now Eq.(\ref{eq-theo-48}) into Eq.(\ref{eq-theo-47}) and keeping the terms linear in $f$,  
\begin{equation}
i\hbar\frac{d\rho}{dt}=[h(\rho^0),\delta\rho] + \left[\frac{\partial h}{\partial\rho}\delta\rho,\rho^0\right] + [f,\rho^0] \ . 
\label{eq-theo-49}  
\end{equation}
Particle-particle ($pp$) and hole-hole ($hh$) matrix elements will be zero by construction\footnote{Due to the Slater approximation, it can be seen by imposing $\rho^2=\rho$.} and only particle-hole ($ph$) or hole-particle ($hp$) matrix elements will contribute.
%After some algebra, one arrives at the linear response equation in matrix form
%\begin{equation}
%\left\{
%\begin{pmatrix}
%A&B\\ 
%A^*&B^*\\ 
%\end{pmatrix}
%-\hbar\omega
%\begin{pmatrix}
%1&0\\ 
%0&-1\\ 
%\end{pmatrix}
%\right\}
%\begin{pmatrix}
%\delta\rho_{ph}\\ 
%\delta\rho_{hp}\\ 
%\end{pmatrix}
%=
%\begin{pmatrix}
%f_{ph}\\ 
%f_{hp}\\ 
%\end{pmatrix}
%\label{eq-theo-50}
%\end{equation}  
%where the elements $A$ and $B$ can be written as, 
%\begin{eqnarray}
%  A_{ph,p^\prime h^\prime}&=&(\varepsilon_p-\varepsilon_h)\delta_{pp^\prime}\delta_{hh^\prime}+\frac{\partial h_{ph}}{\partial \rho_{ph}} \nonumber\\
%  B_{ph,p^\prime h^\prime}&=& \frac{\partial h_{ph}}{\partial \rho_{hp}} \ .
%\label{eq-theo-51}  
%\end{eqnarray}
%Eq.(\ref{eq-theo-50})
Eq.(\ref{eq-theo-49}) leads to the so called Random Phase Approximation (RPA) if one assumes a zero amplitude of the external field;
%rewriting $\delta\rho_{ph}=\langle 0\vert a_h^\dag a_p\vert\nu\rangle\equiv X_{ph}^\nu$ and $\delta\rho_{hp}=\langle 0\vert a_p^\dag a_h\vert\nu\rangle\equiv Y_{ph}^\nu$ which are probability amplitudes whose square gives the probability of finding the states $a^\dag_p a_h\vert 0\rangle$ and $a^\dag_h a_p\vert 0\rangle$ in the excited state $\vert\nu\rangle$;
and as long as the antisymmetrized effective interaction ($\bar{V}$) employed to solve the HF equations is equal to $\bar{V}^{qq^\prime}=\frac{\partial h_q}{\partial \rho_{q^\prime}}=\frac{\partial^2 E^{\rm HF}}{\partial\rho_{q}\partial\rho_{q^\prime}}$ where $q$ denotes neutrons or protons. The latter expression is very important in nuclear physics since it allows for the implementation of density dependent forces in the RPA approximation.
%With all these, the RPA equation  will, thus, read 
%\begin{equation}
%\begin{pmatrix}
%A&B\\ 
%A^*&B^*\\ 
%\end{pmatrix}
%\begin{pmatrix}
%X_{ph}\\ 
%Y_{ph}\\ 
%\end{pmatrix}
%=\hbar\omega
%\begin{pmatrix}
%1&0\\ 
%0&-1\\ 
%\end{pmatrix}
%\begin{pmatrix}
%X_{ph}\\ 
%Y_{hp}\\ 
%\end{pmatrix} \ .  
%\label{eq-theo-52}
%\end{equation}  
%The solution of this eigenvalue problem determines the energy $\hbar\omega_\nu$ (or $E_\nu$) of the excited state $\vert\nu\rangle=\sum_{ph}(X^\nu_{ph}a^\dag_pa_h+Y^\nu_{ph}a^\dag_ha_p)\vert 0\rangle$.
The eigenvalues and eigenvectors obtained by solving the RPA equations give access to the excited spectrum and, thus, one can explicitly evaluate the strength function [Eq.(\ref{eq-pheno-14})] as well as the different sum rules [Eq.(\ref{eq-pheno-15})]. Such an approximation has been shown to be successful in the description of giant resonances.

Recently, a practical method for solving the RPA equations has been proposed in Ref.\cite{nakatsukasa2007} and successfully applied to different nuclear effective models. This method is known as Finite Amplitude Method (FAM) that allows for an RPA calculation with a simple extension of a HF ground state calculation. This is very convenient from the numerical point of view. For details, we refer the reader to the original reference.  

\subsection{Statistical theoretic errors}

Most of the nuclear models available in the literature omit theoretical error estimations. This gives a limited reliability on its predictions. When proposing a model, one usually tries to adopt the simplest possible model with a minimal number of the effective interaction terms and associated parameters. Although adding more parameters may improve the quality measure \footnote{e.g. the $\chi^2$ value or the root mean square difference between experimental data and model predictions.} it does not always mean that the overall quality of the fit has been improved. As a simple example, when adding a new term to the model, it can happen that large changes are produced in already existing parameters. This is a clear signature that the model with the new parameter is introducing uncontrolled and/or unphysical correlations and, thus, error estimation in the predictions will not be reliable either. This problem is very important in order to meaningfully constrain the parameters of the EoS. 

It is important to mention that there exists also the uncertainty associated to the choice of the model: the systematic uncertainty. In the absence of an exact EDF for finite nuclei, systematic uncertainties associated to a given prediction can only be estimated by comparing different kinds of nuclear EDFs, or nuclear models more in general. This is part of the scope of this review and the comparison between models will be essential in the following sections. As an example, in Ref.\cite{erler2012}, statistical and systematic uncertainties in nuclear masses were investigated within a set of different parameterizations of the Skyrme EDFs finding that systematic uncertainties govern the total uncertainty for the studied observables. 

Coming back to statistical uncertainties and correlations, there are several strategies one can follow. Here, we will briefly overview the covariance analysis that allows to study the statistical errors within a given model as well as the correlations between parameters and between observables such as those derived (analytically) from macroscopic models in the previous section. More details on the covariance analysis can be found in Refs.\cite{bevington,dobaczewski14}.

\subsubsection{Covariance analysis}

Consider a model characterized by $n$ parameters $\bm{p} = (p_1 , . . . , p_n)$. Those parameters define the model space and are usually coupling constants of an effective model. Observables ($\mathcal{O}$) are, therefore, functions of the parameters $\mathcal{O}(\bm{p})$. Hence, the assumption of a model implies correlations between computed quantities. 

{\bf $\chi^2$ definition}. The $\chi^2$ defines a {\it quality measure} and is commonly used in information theory. In low-energy nuclear physics, as in many other fields, it has been exploited in the calibration of theoretical effective models such as EDFs. In the latter case,
\begin{equation}
\chi^2(\bm{p}) = \sum_{\imath=1}^{m}
 \left(\frac{\mathcal{O}_\imath^{\rm theo.}(\bm{p})-\mathcal{O}_\imath^{\rm ref.}}
 {\Delta\mathcal{O}_\imath^{\rm ref.}}\right)^2 \,
\label{eq-theo-53}
\end{equation}
where ``theo.'' stands for calculated values, and ``ref.'' may refer to experimental, observational and/or {\it pseudo-data} that sometimes is used as a simple way to {\it guide} the models. The use of derived, not directly observable, {\it pseudo-data} should be understood as a temporary benchmark never substituting real data and just helping model extrapolations. It, therefore, influences the value of the quality measure and the results should be taken with great care. $\Delta\mathcal{O}^{\rm ref.}$ stands for the {\it adopted errors}. In principle, this quantity should stand only for the experimental standard deviation $\sigma_{\rm exp}$, however, such a choice is not realistic since the error of the model $\sigma_{\rm theo}$ is, in most of the cases in nuclear physics, larger than the experimental one. Therefore, $\Delta\mathcal{O}^{\rm ref.}$ should be chosen to be equal to $\sigma_{\rm exp} + \sigma_{\rm theo}$. The value of $\sigma_{\rm theo}$ is unknown {\it a priori}. It can be fixed for each type of observable by imposing $\chi^2({\bf p})=1$ at the optimal value of ${\bf p}$ (see below). This type of situation will guide model extrapolations, commonly leading to a {\it semi}$-$objective construction of the $\chi^2$. In addition, there exist huge amount of data that, if fitted, would require a high computational cost. Being part of the data redundant for the fitting of the parameters\footnote{For example, there is no need to fit the more than two thousand measured masses to obtain a good parameterization of the model.}, some freedom exists in choosing a convenient set of $\mathcal{O}_\imath^{\rm ref.}$ and $\Delta\mathcal{O}^{\rm ref.}$ that may characterize the nucleus. So practitioners use to ``equilibrate'' different terms in the $\chi^2$ according to their physical understanding of the model and to the final purpose of the fit. 

{\bf Parameters and observables.} Assuming that the $\chi^2$ is a well behaved function of the parameters around their optimal value $\bm{p}_0$: $\partial_{\bm{p}}\chi^2(\bm{p})\mid_{\bm{p}=\bm{p}_0} = 0$; and that near the minimum can be approximated by a Taylor expansion up to the second order in the parameter space; one finds 
\begin{equation}
\chi^2(\bm{p})-\chi^2(\bm{p}_0) \approx 
 \frac{1}{2}\sum_{\imath, \jmath}^n(p_{\imath}-p_{0\imath})
 \partial_{p_\imath}\partial_{p_\jmath}\chi^2(p_{\jmath}-p_{0\jmath}) \ ,
\label{eq-theo-54}
\end{equation}
the curvature matrix, $\mathcal{M}_{ij}\equiv \partial_{p_\imath}\partial_{p_\jmath}\chi^2$. The estimate for the errors ($\bm{e}$) on the fitted parameters can be calculated as follows, 
\begin{equation}
e_\imath \equiv e(p_\imath) = 
 \sqrt{\left(\mathcal{M}^{-1}\right)_{\imath\imath}} \equiv 
 \sqrt{\mathcal{E}_{\imath\imath}} \ \ , 
\label{eq-theo-55}
\end{equation}
where the covariance (or error) matrix $\mathcal{E}$ has been defined. It is clear from this definition that if the curvature matrix around the minimum takes a large (small) value along the $p_i$ direction, it means that a small (large) change in this parameter will already produce a large (small) change in the value of the $\chi^2$, that is, $p_i$ will be associated to a small (large) error. The covariance matrix can be used also to study correlations. Actually, the correlation matrix ($\mathcal{C}$) is defined as,  
\begin{equation}
\mathcal{C}_{\imath\jmath} \equiv \frac{\mathcal{E}_{\imath\jmath}}
 {\sqrt{\mathcal{E}_{\imath\imath}\mathcal{E}_{\jmath\jmath}}} 
\label{eq-theo-56}
\end{equation}
where $\mathcal{C}_{\imath\jmath}$ takes values form $-1$ to $1$. $\mathcal{C}_{\imath\jmath}\approx 1$ indicates a large correlation and $-1$ a large anti-correlation between parameters $p_\imath$ and $p_\jmath$, respectively, while $\mathcal{C}_{\imath\jmath}$ around zero means that no correlation holds at all between parameters $p_\imath$ and $p_\jmath$. This indicates that both parameters are needed for the description of the set of observables used for the fit.

To estimate the correlation between two predicted observables $A$ and $B$, assuming a smooth behavior when evaluated around of the optimal parameterization,
\begin{eqnarray}
A(\bm{p}) &=& A(\bm{p}) + (\bm{p}-\bm{p}_0)
 \partial_{\bm{p}}A(\bm{p})\mid_{\bm{p}=\bm{p}_0}\nonumber\\
&\equiv& A_0 + (\bm{p}-\bm{p}_0)\bm{A_0} \ , 
\label{eq-theo-57}
\end{eqnarray}
and the same for $B$, one needs to estimate the covariance between them ($C_{AB}$). This calculation leads, within the approximations adopted so far and assuming that $\chi^2(\bm{p})-\chi^2(\bm{p}_0) = 1$, to
\begin{equation}
C_{AB} \approx \sum_{\imath\jmath}^n 
\left.\frac{\partial A(\bm{p})}{\partial p_\imath}\right\vert_{\bm{p}=\bm{p}_0}
\mathcal{E}_{\imath\jmath}
\left.\frac{\partial B(\bm{p})}{\partial p_\jmath}\right\vert_{\bm{p}=\bm{p}_0} \ .
\label{eq-theo-58}
\end{equation}
The root square of $C_{AA}$ estimates the error in $A$ and the correlation between $A$ and $B$ can be estimated by the the Pearson-product moment correlation coefficient as,           
\begin{equation}
c_{AB} \equiv \frac{C_{AB}}{\sqrt{C_{AA}C_{BB}}} \ . 
\label{eq-theo-59}
\end{equation}
$c_{AB}=1$ means complete correlation between observables $A$ and $B$, $-1$ means complete anti-correlation and $c_{AB}=0$ means no correlation.

\subsubsection{Examples}
In the following, we will examine different examples on some correlations of interest for the EoS as predicted by different EDFs of common use in nuclear physics. This is important since it allows to better understand and validate those correlations already discussed from a macroscopic perspective such as the correlation of the neutron skin thickness, the IVGDR or the dipole polarizability with the symmetry energy parameters; the excitation energy of the ISGMR with the nuclear matter incompressibility; or the excitation energy of the ISGMR with the effective mass.
\begin{figure}[t!]
\begin{center}  
\includegraphics[width=0.45\linewidth,clip=true]{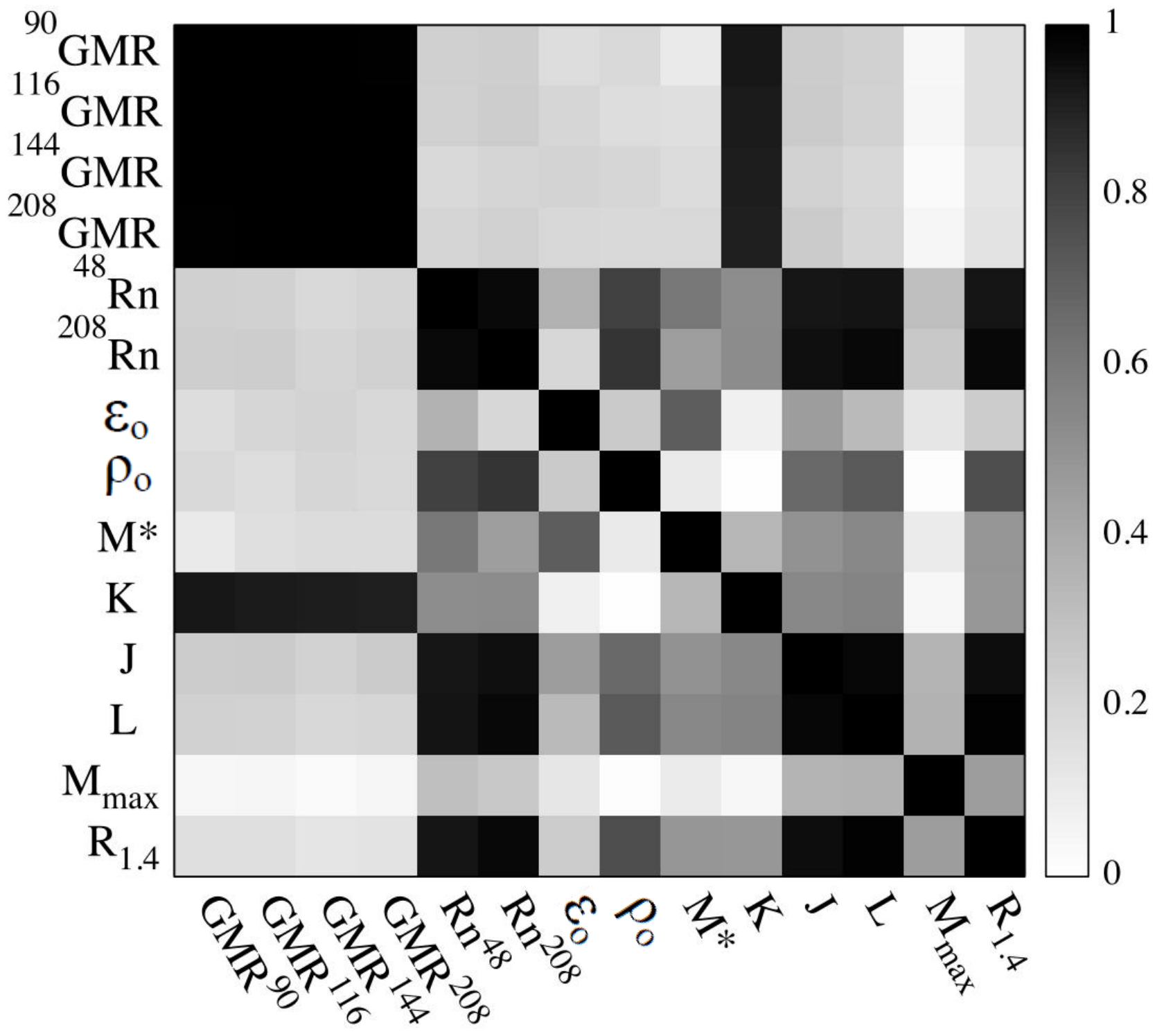}
\hspace{1cm}  
\includegraphics[width=0.4\linewidth,clip=true]{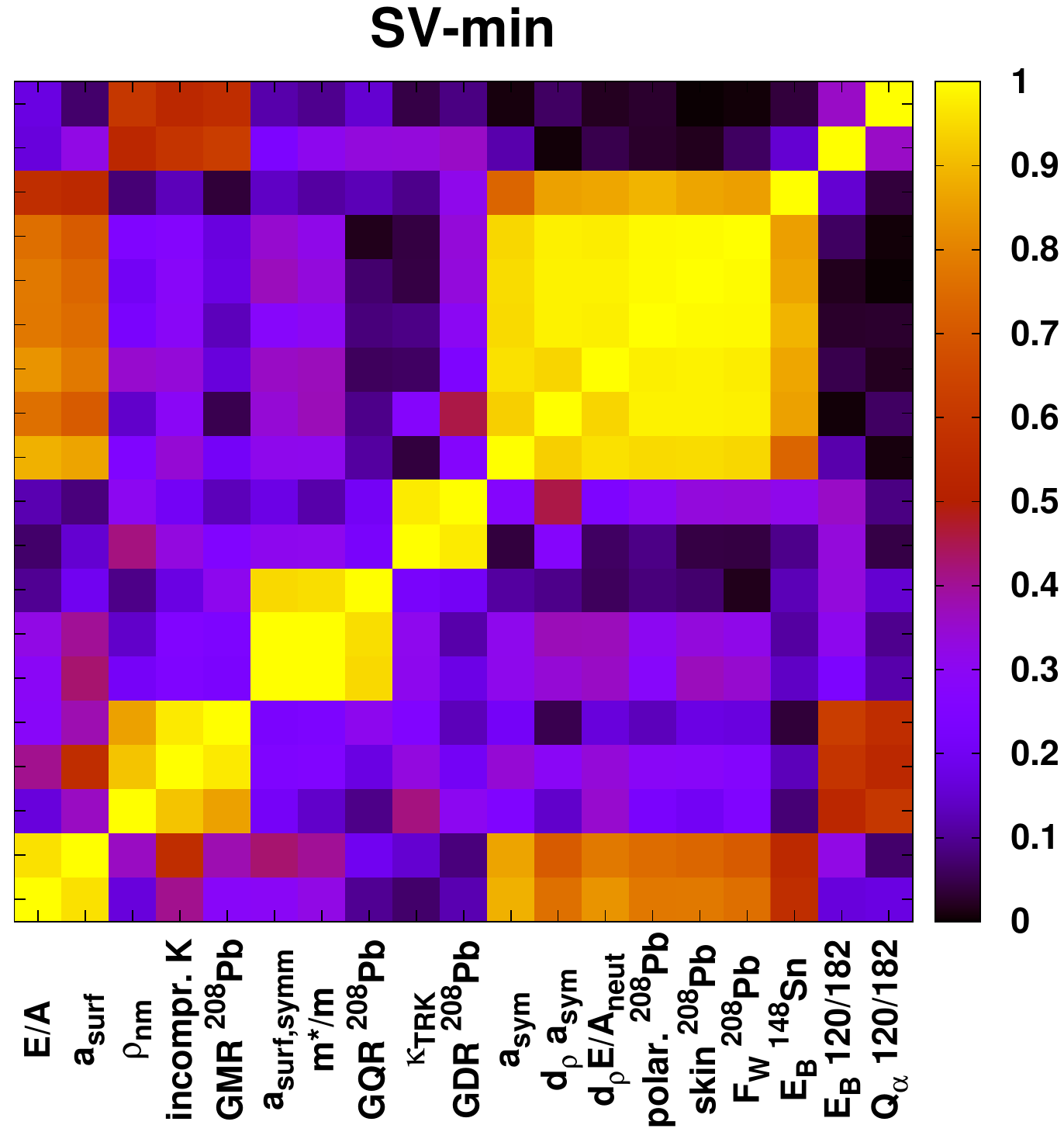}
\includegraphics[width=0.45\linewidth,clip=true]{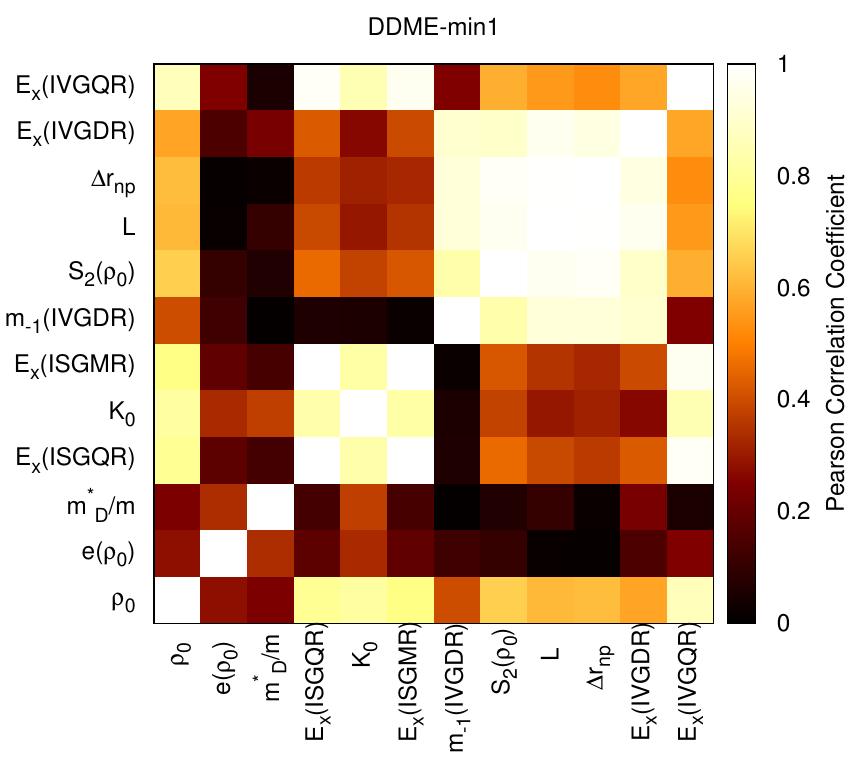}
\includegraphics[width=0.45\linewidth,clip=true]{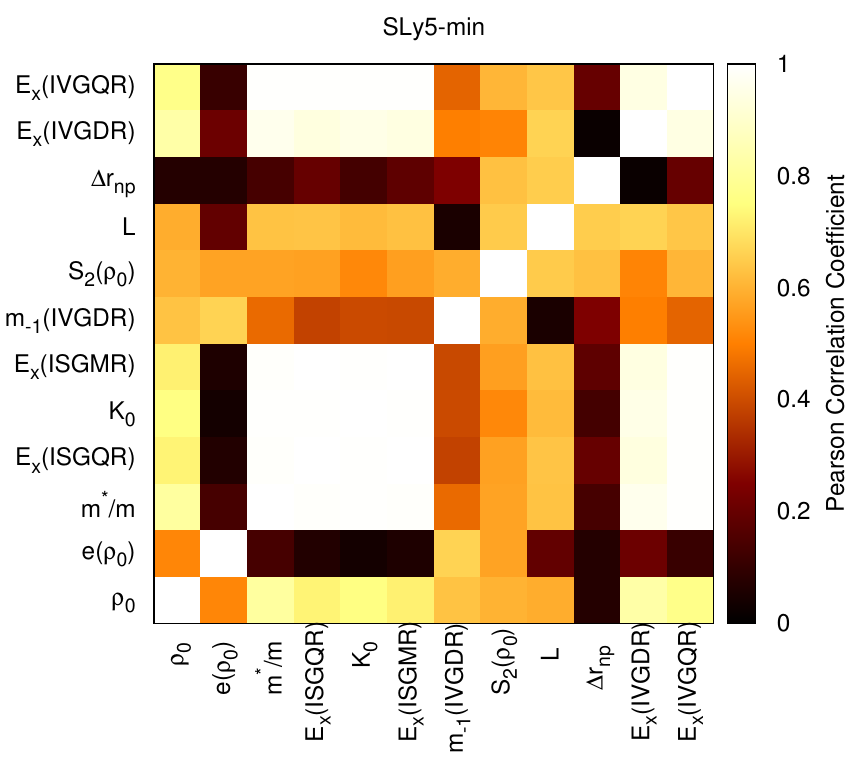}
\caption{\label{fig-theo-9} Absolute value of the Pearson-product moment correlation coefficients between different observables for finite nuclei and EoS parameters as predicted by different EDFs. Observables on GR refer to its excitation energy. Whenever not specified, observables for finite nuclei refer to the example case of ${}^{208}$Pb. $S_2(\rho_0)\equiv J$. The meaning of the different quantities relevant for the discussion are given in the text, for a detailed information on the other quantities, we refer the reader to the original references. Upper-left panel: results obtained with the relativistic non-linear Walecka model parameterization FSUGold 2 \cite{todd-rutel2005,chen2014} are shown. Figure taken from Ref.\cite{piekarewicz2015}. Upper-right panel: results obtained with the Skyrme parameterization SV-min \cite{kupfel2009} are shown. Figure taken from Ref.\cite{erler2015}. Lower-left panel: results obtained with the relativistic density dependent meson exchange parameterization DDME-min1 \cite{niksic2002,roca-maza2015} are shown. Figure taken from Ref.\cite{roca-maza2015}. Lower-right panel: results obtained with the Skyrme parameterization SLy5-min \cite{Chabanat1998,roca-maza2015} are shown. Figure taken from Ref.\cite{roca-maza2015}.}
\end{center}
\end{figure}

In Fig.\ref{fig-theo-9}, absolute values of the Pearson-product moment correlation coefficients between different finite nuclei observables and EoS parameters are shown as predicted by different EDFs. Observables on GR refer to its excitation energy. Whenever not specified, finite nuclei observables refer to the example case of ${}^{208}$Pb. Note that $S_2(\rho_0)\equiv J$. For a detailed information and definition on the different quantities, we refer the reader to the original references \cite{piekarewicz2015, erler2015, roca-maza2015}, here we will just discuss some of them in what follows. In the upper-left panel of Fig.\ref{fig-theo-9} are shown results obtained with the relativistic non-linear Walecka model parameterization FSUGold 2 \cite{todd-rutel2005,chen2014}; in the upper-right panel are shown results obtained with the Skyrme parameterization SV-min \cite{kupfel2009}; in the lower-left panel are shown results obtained with the relativistic density dependent meson exchange parameterization DDME-min1 \cite{niksic2002,roca-maza2015}; and in the lower-right panel are shown results obtained with the Skyrme paramterization SLy5-min \cite{Chabanat1998,roca-maza2015}. In all panels the lightest colors indicate strong correlation while dark colors indicate weak correlation except for the upper-left panel in which the latter scale is inverted. 

First of all, we can compare the correlations predicted by the different EDFs for the excitation energy of the ISGMR in ${}^{208}$Pb (labeled as ${}^{208}$GMR, $E_x({\rm ISGMR})$ or GMR ${}^{208}$Pb in the different panels) with the nuclear matter incompressibility (labeled as $K_0$ or $K$) as it is suggested by the constrained energy $\sqrt{m_{1}/m_{-1}}$ derived from the sum-rules [cf. Eqs.(\ref{eq-pheno-27}) and (\ref{eq-pheno-31})]. It is quite evident that the correlation coefficient is very high in all cases confirming our previous discussions. One could conclude then, that: {\it the excitation energy of the ISGMR (at least in ${}^{208}$Pb) determines the curvature of the symmetric matter EoS around saturation}. The question now is how accurately it can be determined. This question will be answered by supplementing the statistical error with the systematic spread of values obtained by calculations from different models for the quantity of interest. We will discuss this issue in more detail in the following subsections. As an example on how to proceed is to confirm that all these models are accurate in the overall description of bulk data in nuclei and, specifically, also accurate enough in the description of the $E_x({\rm ISGMR})$ in ${}^{208}$Pb. The $E_{x}^{{\rm ISGMR}}$ is predicted to be 14.00$\pm$0.36 MeV (SLy5-min), 13.87$\pm$0.49 MeV (DDME-min1), 13.76$\pm$0.08 MeV (FSUGold 2) and 13.5$\pm$0.2 (SV-min) while the experimental value is $14.2\pm 0.3$ MeV \cite{Youngblood1999}. Therefore, being compatible with experimental data the SLy5-min, DDME-min1 and FSUGold 2 EDFs and knowing that ${{K}_{0}}$ is predicted to be 230.5$\pm$9.0 MeV (SLy5-min), 261$\pm$23 MeV (DD-ME-min1) and 238$\pm$2.8 MeV (FSUGold 2) one can say that $K_0$ should lie between 220 MeV and 280 MeV (if no other model enlarges this range). This range is large in comparison to the statistical uncertainties obtained from these functionals.  

Next, one can consider the correlations between the excitation energy of the ISGQR (labeled $E_x({\rm ISGQR})$ or GQR ${}^{208}$Pb in the different panels) with the non-relativistic effective mass (labeled as $m^*/m$). This quantity is connected to the single-particle properties of the model under consideration and not to bulk properties but, to look at this correlation, may also serve to exemplify the relevance of the covariance analysis. For the results shown in Fig.\ref{fig-theo-9}, it is also clear that a large correlation is found between the $E_x({\rm ISGQR})$ in ${}^{208}$Pb and the effective mass for the non-relativistic models\footnote{Note that the relativistic effective mass \cite{jaminon1989} is different from the non-relativistic effective mass. The latter coincides with the one discussed in previous sections while the former does not.}. This again confirms our previous considerations from the simple HO model when non-relativistic and momentum dependent forces are analyzed (such as the Skyrme one). As before, the $E_x({\rm ISGQR})$ is predicted to be 10.7$\pm$0.6 by SV-min and $12.6\pm 0.6$ MeV by SLy5-min while, experimentally, it has been determined to be 11.0$\pm$0.3 MeV \cite{harakeh01} and, thus, only SV-min will be compatible. The effective mass predicted by SV-min is $m^*/m=0.95\pm 0.15$.

Let us now consider the excitation energy of the IVGDR (labeled $E_x({\rm IVGDR})$ or GDR ${}^{208}$Pb in the different panels) and how it is correlated with the parameters $J$ and $L$ of the EoS (labeled also as $S_2(\rho_0)$ or $a_{\rm sym}$ and $d_\rho a_{\rm sym}$, respectively). Here the picture is not as clear as in the previous cases. There is a large correlation of these two quantities for the DDME-min1 model, a modest correlation for SLy5-min model and, finally, a low correlation when inspecting the SV-min model. As suggested by the HO approach, this might be due to the fact that it is a combination of the asymmetry surface ($a_{AS}$ that can be related with $L$ within a local density approximation) and volume ($a_A$ that can be identified with $J$) parameters leading this excitation energy. Therefore, one should be more careful in this case since the connection of $E_x({\rm IVGDR})$ with the parameters of the nuclear EoS is not one to one but, possibly, with a combination of $J$ and $L$. Nevertheless, one can also compare the obtained results with experimental data. For the $E_x({\rm IVGDR})$ it is found 11$\pm$5 MeV for SV-min, 13.9$\pm$1.8 MeV for SLy5-min and 14.64$\pm$0.38 MeV for DDME-min1 while the experimental value is 13.25$\pm$0.10 MeV \cite{ryezayeva2002}. Thus SV-min and SLy5-min are compatible with experimental data although SV-min is rather unconstrained on the properties that determine $E_x({\rm IVGDR})$. The values of $J$ and $L$ for SLy5-min are 32.60$\pm$0.71 MeV and 47.5$\pm$4.5 MeV, respectively. Summarizing, the information presented in this subsection does not allow to understand the relation between $E_x({\rm IVGDR})$, $J$ and $L$. However, the macroscopic model presented in Sec.\ref{pheno}, give us some hints that one may follow in order to unravel the possible relation between this observable and the parameters of the EoS. 

The dipole polarizability, proportional to  $m_{-1}$ of the IVGDR [labeled as polar. ${}^{208}$Pb or $m_{-1}($IVGDR$)$] has been seen to be correlated to the ratio $a_{AS}/a_A$ (or qualitatively to $L/J$). In Fig.\ref{fig-theo-9}, the polarizability is strongly correlated with $J$ for DDME-min1 and SV-min and modestly correlated in the case of SLy5 while for $L$ only DDME-min1 and SV-min show correlation. The situation here is similar to that of the $E_x({\rm IVGDR})$: there is not enough information in what has been presented in Fig.\ref{fig-theo-9} that points towards a one to one relation between $m_{-1}(IVGDR)$ and $J$ or $L$. To further investigate $m_{-1}(IVGDR)$ or $E_x({\rm IVGDR})$, one can explore trends inspired by macroscopic formulas using well calibrated EDFs (see for example Ref.\cite{roca-maza13a,colo15}). This has been done in many papers and some light has been shed into the relation of these observables with the parameters of the EoS (cf. Sec.\ref{excitations}). Another strategy has also been used by different groups with a similar scope. It is as follows, after obtaining an optimal parameterization, one may generate around slightly modified models by fixing some parameters of the EoS while varying others. This might be useful because it allows to isolate the effect of changing one parameter of the EoS into the rest of predictions of the model in a very systematic way. This is commonly referred in the literature as generating a {\it family of systematically varied interactions} and it will be discussed in the next subsection. 

Finally, one of the most paradigmatic correlations in the recent years is that between the neutron skin thickness in ${}^{208}$Pb --or the neutron rms radius since the proton rms radius is usually well fixed in the fits-- (labeled as $R_n^{208}$, skin ${}^{208}$Pb, or $\Delta r_{np}$) and the slope parameter $L$ of the symmetry energy. It was first reported in Ref.\cite{brown00} and it is also seen in Eq.(\ref{eq-pheno-3}) if we assume, as before, that $a_{AS}/a_A \rightarrow L/J$ and that the variation of $J$ as compared to that of $L$ can be neglected\footnote{For the models shown in Fig.\ref{fig-theo-9}, the value of $J$ is 30.7$\pm$1.4 MeV for SV-min, 37.62$\pm$1.11 MeV for FSUGold 2, 32.60$\pm$0.71 MeV for SLy5-min, and 33.0$\pm$1.7 MeV for DDME-min1 while $L$ takes the values 112.8$\pm$16.1 MeV for FSUGold 2, 47.5$\pm$4.5 MeV for SLy5-min, and 55$\pm$16 MeV for DDME-min1 ($L$ was not reported for SV-min in Ref.\cite{kupfel2009}).}. Looking at the predictions shown in Fig.\ref{fig-theo-9}, all the EDFs agree with the existence of such a correlation. One may conclude, therefore, that such a correlation, justified from a macroscopic picture, is well founded since all microscopic EDF models also agree with it. Specifically, the neutron skin thickness predicted by these models is 0.1655$\pm$0.0069 fm for SLy5-min, 0.20$\pm$0.03 fm for DDME-min1, 0.287$\pm$0.020 for FSUGold 2 and 0.17$\pm$0.13 fm for SV-min. From the experimental side, there is a lack of accuracy in the determination of the neutron skin thickness. For example, experiments using hadronic probes ($\sim$ 0.15$-$0.2 fm) do not seem to agree with the central value predicted by experiments using electroweak probes ($\sim$0.3 fm) although their error bars largely overlap. The predicted values for $L$ are 47.5$\pm$4.5 MeV for SLy5-min, 55$\pm$16 MeV for DDME-min1 and 112.8$\pm$16.1 MeV for FSUGold 2 ($L$ was not reported for SV-min in Ref.\cite{kupfel2009}). There is, thus, the need to better constrain this parameter of the EoS and the neutron skin thickness of ${}^{208}$Pb seems to be a very good candidate if accurately and precisely measured.

As we will discuss in Sec.\ref{excitations}, the excitation energy of the IVGDR and IVGQR among others as well as the dipole polarizability are also good candidates to constrain the nuclear matter EoS parameters, being related to a combination of $J$ and $L$ rather than to a one to one correlations as it happens in good approximation for the neutron skin thickness.

\subsection{Systematically varied effective interactions}
Families of systematically varied interactions have been commonly used in the recent literature \cite{niksic2008,kupfel2009,agrawal2010, piekarewicz2011, roca-maza13}. These families are commonly build as follows. After obtaining an optimal parameterization of a given EDF, one fixes one parameter of the EoS at a time refitting the functional using the same quality measure, e.g. $\chi^2$, used to find the optimal (unconstrained) parameterization. This might be useful because it allows to isolate the effect of changing one parameter of the EoS into the rest of predictions of the model in a very systematic way.

\begin{figure}[t!]
\begin{center}  
\includegraphics[width=0.6\linewidth,clip=true,angle=-90]{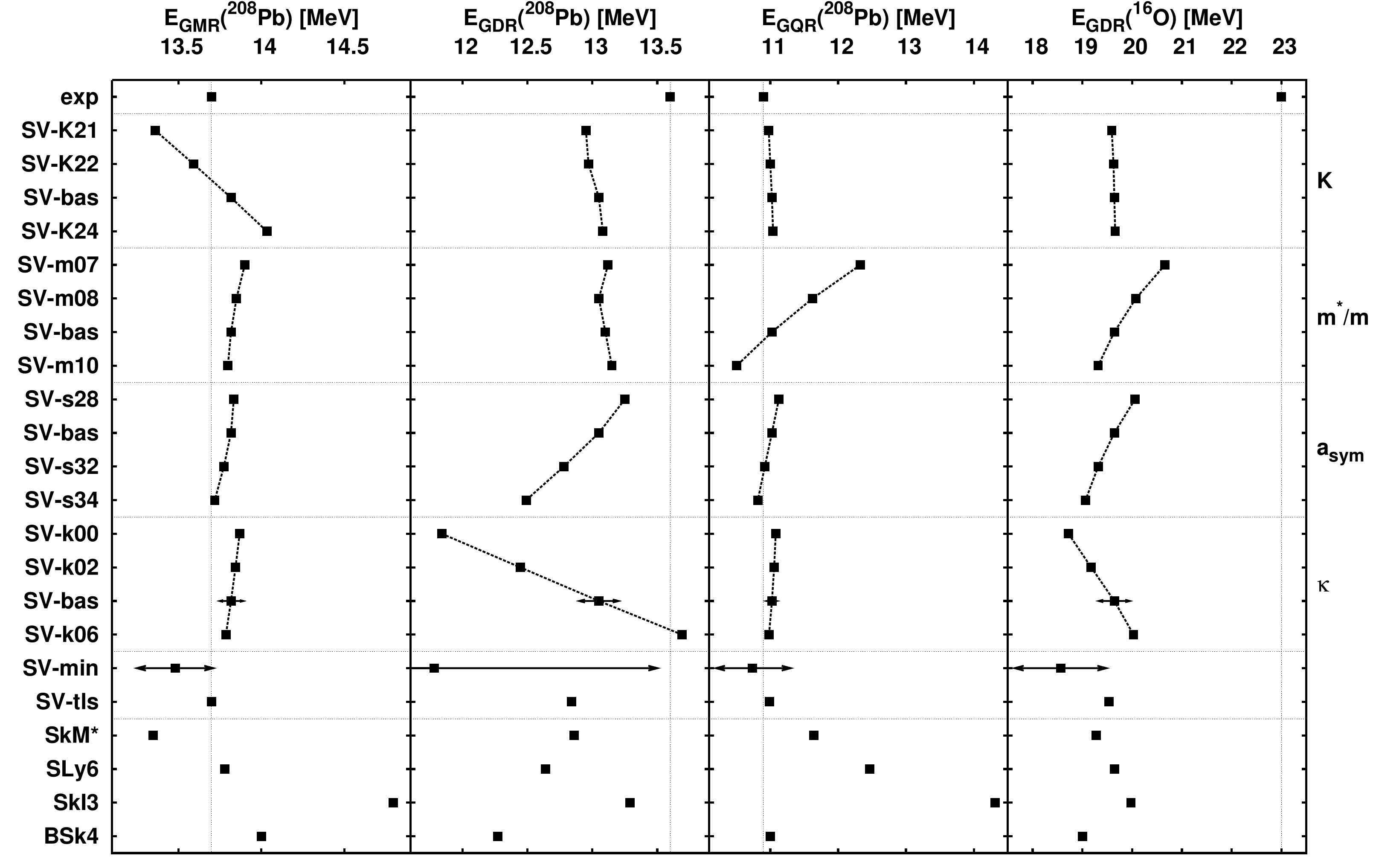}
\caption{\label{fig-theo-10} Excitation energy of the ISGMR, IVGDR, ISGQR in ${}^{208}$Pb as well as the IVGDR in ${}^{16}$O are shown as predicted SV-min, some families of interactions based on SV-min: SV-K family correspond to a variation of $K_0$; SV-m correspond to a variation of $m^*/m$; SV-s to a variation of $J$; and SV-k to a variation of the isovector enhancement factor $\mathcal{K}$. Other models are also shown. For further details we refer the reader to the original work. Figure taken from Ref.\cite{kupfel2009}}
\end{center}
\end{figure}

The usefulness of producing a {\it family} is exemplified in Fig.\ref{fig-theo-10} where SV-min is the optimal parameterization of the Skyrme functional used in this case. After that, SV-K parameterization fixes the value of the nuclear matter incompressibility at different values different from the optimal one (i.e. that predicted by SV-min); SV-m corresponds to a variation on the effective mass; SV-s on the symmetry energy parameter $J$; while SV-k on the isovector enhancement factor (or isovector effective mass). For details on other models see Ref.~\cite{kupfel2009}. The predictions of all these families for the excitation energy of the ISGMR, IVGDR, ISGQR in ${}^{208}$Pb as well as the IVGDR in ${}^{16}$O are shown. The figure shows that a variation on $K_0$ produces a clear variation on $E_x({\rm ISGMR})$ leaving almost constant the value of the other excitation energies corresponding to other multipolarities. Similarly, a change on $m^*/m$ impacts clearly on the ISGQR, as expected. The variation of the IVGDR in ${}^{16}$O with the effective mass indicates that the $E_x({\rm IVGDR})$ in light nuclei might be more sensitive to the shell structure than in heavy nuclei (cf. for the case of ${}^{208}$Pb that there is no variation with $m^*$). Finally, the variation of the $E_x({\rm IVGDR})$ is most influenced by the isovector parameters $J$ (labeled $a_{\rm asym}$) and $\mathcal{K}$ (the isovector enhancement factor labeled as $\kappa$) as it could also be expected.

Hence, from this simple exercise, one can learn how the different parameters of the nuclear EoS impact different observables when predicted by a given model. If this exercise is repeated for different families of interactions based on different EDFs, one can have a more complete understanding on the EoS parameters.

%% file: eos.tex
\section{Equation of State}
\label{eos}

Being the EoS [Eq.(\ref{eq-pheno-4})] at the heart of any nuclear model, all experimental and observational data sensitive to the nucleon degrees of freedom will be related to it. As already discussed in Secs.~\ref{pheno} and ~\ref{theo}, there are specific observables that alone have the potential to put stringent constraints on the EoS. The present review is devoted to {\it what can we learn on the EoS from ground state and collective excited states in finite nuclei}. In Sec.\ref{pheno}, we have seen that the equilibrium density of a finite nucleus estimated by using a very simple model is approximately equal to 
\begin{equation}
\rho_{\rm eq}\approx \rho_0+\frac{3\rho_0}{K_0}\left(1-LI^2+2a_SA^{-1/3}-a_C\frac{Z^2}{A^{4/3}}\right)
%\nonumber \ , 
\end{equation}
which means, that properties on finite nuclei and collective excitations (small perturbations to the ground state or equilibrium density) will be sensitive to densities at and around $\rho_{\rm eq}$. According to this formula, the corrections to $\rho_0$ are not large for a reasonable value of the parameters. This implies that one should expect to probe the EoS from ground state and collective excited states in finite nuclei not far from $\rho_0$. This justifies the study of the parameters that characterize the EoS at and around $\rho_0$ [cf. Eqs.(\ref{eq-pheno-4})-(\ref{eq-pheno-6})],
\begin{equation}
e(\rho,\delta) = e(\rho_0,0) + \frac{1}{2}K_0(\frac{\rho-\rho_0}{3\rho_0})^2 + \left[J + L\frac{\rho-\rho_0}{3\rho_0} + \frac{1}{2}K_{\rm sym}(\frac{\rho-\rho_0}{3\rho_0})^2\right]\delta^2 + \mathcal{O}[\rho^3] \ ,
%\nonumber
\end{equation}
from models that accurately reproduce masses, radii, and collective excitations along the nuclear chart. To be more quantitative, it has been seen that the symmetry energy of a heavy nucleus such as ${}^{208}$Pb equals the symmetry energy of the infinite system, i.e. the EoS, at about $2\rho_0/3\approx 0.1$ fm${}^{-3}$ in EDFs\cite{centelles09}. Therefore, one may expect the latter density to be proben by the observables reviewed here among others \cite{tsang12,s-epja,Horowitz:2014bja}. It is not the aim of this review to cover all of them. We will only mention, for the sake of comparison, some illustrative results derived from neutron star observations and heavy ion collisions without being exhaustive. Theoretical predictions coming from the {\it ab initio} approaches will also be briefly addressed. The following section is thought to serve, in part, as a brief introduction to the main constraints to the EoS that will be discussed in more detail in Secs.\ref{gs} and \ref{excitations}.

\subsection{Summary on available constraints on the EoS}

Recent studies have taken care of collecting different constraints on the nuclear EoS. Most of them have been concentrated on the symmetry energy parameters \cite{tsang12,lattimer2013,li2013,s-epja,Baldo2016,oertel2017} and some of them to the nuclear matter incompressibility and $K_\tau\equiv K_{\rm sym} - L(6-K^\prime/K_0)$ [see Eqs.(\ref{eq-pheno-31} and (\ref{eq-pheno-32})] \cite{shlomo2006,garg2007,centelles09,li2010,stone2014,patel2012}. 
\begin{figure}[t!]
\begin{center}  
  \includegraphics[width=0.40\linewidth,clip=true]{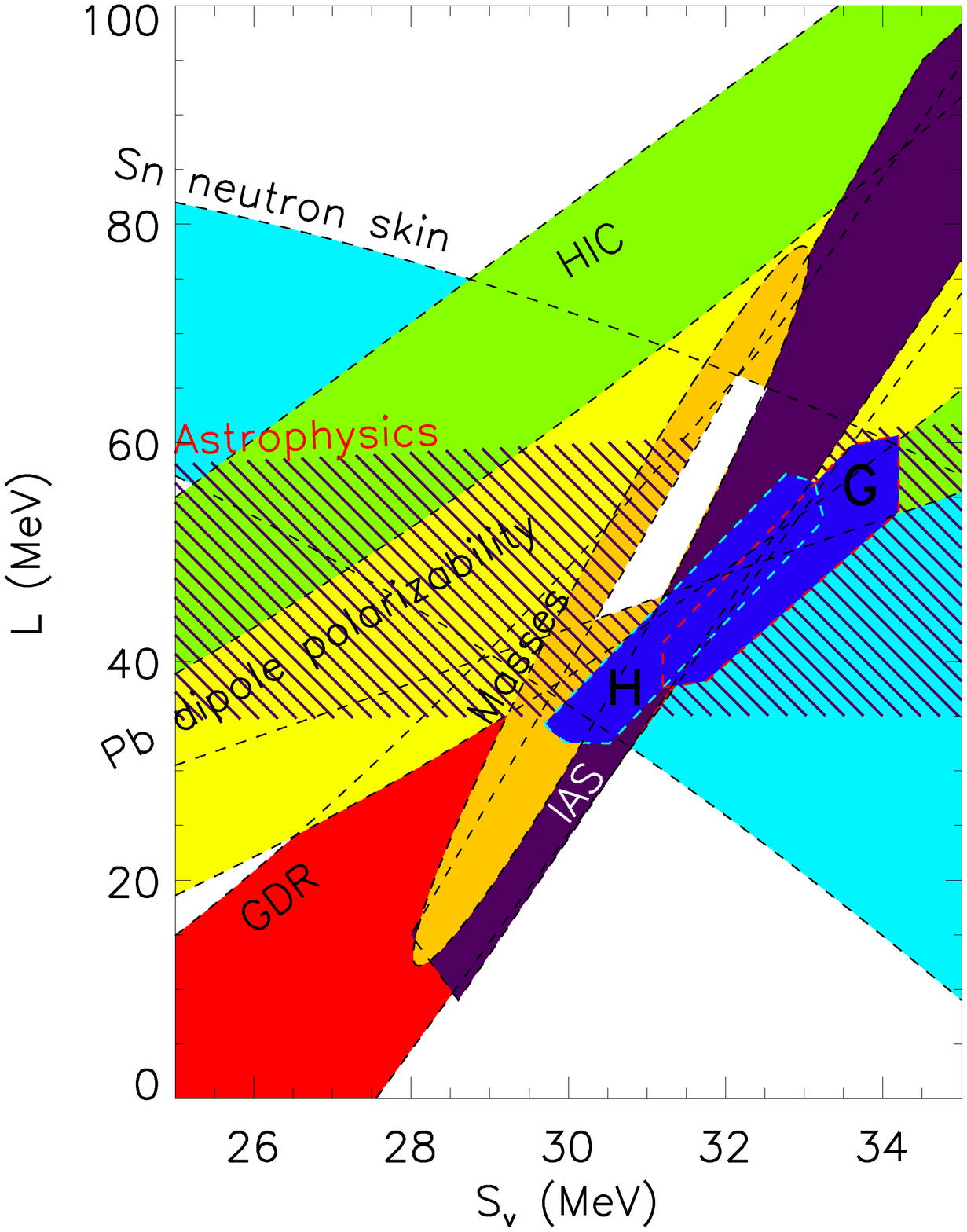}
  \includegraphics[width=0.58\linewidth,clip=true]{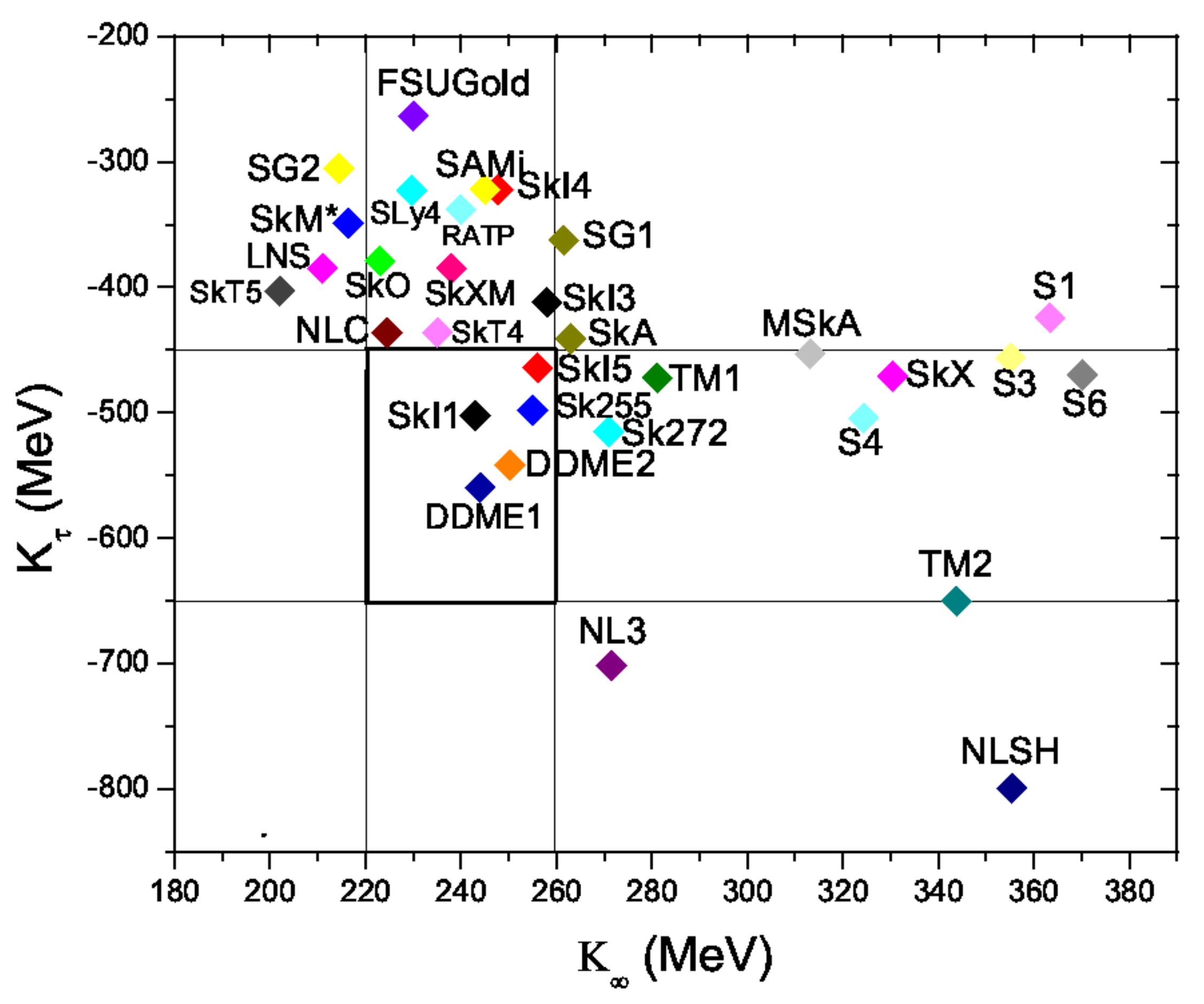}
  \caption{\label{fig-eos-11} Left panel: experimental constraints for the symmetry energy parameters $J\equiv S_v$ and $L$, from the analysis of the neutron skin thickness, heavy ion collisions (HIC), astrophysical observations, dipole polarizability, masses, isobaric analog state (IAS) and neutron matter studies of Gandolfi et al. \cite{gandolfi2012} and Hebeler et al. \cite{hebeler2010}, denoted by G and H, respectively. For details on each analysis we refer to the original work. Figure taken from Ref. \cite{lattimer2014}. Right panel: Values of $K_0\equiv K_\infty$ and $K_\tau$ calculated from different EDFs. The vertical and horizontal lines enclose the experimental ranges as reported in Ref.\cite{colo15}. For details on the EDFs used we refer the reader to the same reference. Figure taken from Ref.\cite{colo15}.}
\end{center}
\end{figure}

Regarding the symmetry energy parameters, in Ref.\cite{tsang12}, available results on laboratory experiments and commonly used theoretical models have been overviewed and discussed. As a conclusion, a constraint centered around $J\approx 32.5$ MeV and $L\approx 70$ MeV was given (errors where not estimated in detail). Later on, in Ref.\cite{lattimer2013} a compilation on experimental, theoretical, and observational analyses lead to the ranges on the symmetry energy parameters $29.0 <J<32.7$ MeV and $40.5<L<61.9$ MeV. In Refs.\cite{li2013,oertel2017} two very systematic and complete studies on available constraints in the literature lead to a global average of $J=31.6\pm 0.9$ MeV and $L=58.9\pm 16.5$ MeV (cf. Table 1 and 2 of Ref.\cite{li2013}) and $J= 31.7 \pm 3.2$ MeV and $L= 58.7\pm 28.1$ MeV [cf. Table II, Fig. 8 and text in Ref.\cite{oertel2017}]. A recent update \cite{kong2017} of Ref.\cite{li2013}, also taking into account nuclear excitations narrowed these values. 
%Being the last compilations in good agreement between them and also with many other works (cf. references herein and within the previously quoted articles), 
Based on last compilations (cf. references herein and within the previously quoted articles), one may conclude that $J$ is much better constrained than $L$ by available experimental and observational data and that $J$ should be close to 31-32 MeV while $L$ is very likely to lie within 30 and 90 MeV. As an illustrative example, in Fig.\ref{fig-eos-11} from Ref.\cite{lattimer2014}, some constraints in the $J$ versus $L$ plane are shown. Specifically, Fig.\ref{fig-eos-11} shows the experimental constraints coming from the analysis of the neutron skin thickness, heavy ion collisions (HIC), astrophysical observations, dipole polarizability, masses, isobaric analog state (IAS) and two theoretical neutron matter studies \cite{gandolfi2012, hebeler2010} labeled as G and H. 
%Such a correlation in all models is due to the following. 
The reason to show the $J$ versus $L$ plane is based on the positive correlation expected between them. It can be understood as follws. The nuclear symmetry energy around an average nuclear density of $0.1$ fm${}^{-3}$ is well determined in effective models fitted to masses and charge radii as compared to the uncertainties in $J$ and $L$. As already discussed, $S(\rho=0.1$fm$^{-3})$ corresponds in good approximation to the symmetry energy of the finite nucleus \cite{centelles09} and, hence, approximately to the LDM asymmetry term. Thus, from Eq.\ref{eq-pheno-6} keeping up to linear order in the density: $S(\rho=0.1$fm$^{-3})=J-L/8$; that is {\it $J$ and $L$ are expected to be linearly correlated} if higher order terms in the expansion of Eq.\ref{eq-pheno-6} are negligible. By considering Fig. \ref{fig-theo-9} one can conclude that in all models, $J$ and $L$ are linearly correlated and, hence, the interest of showing the correlations in such a way. We refer the reader to Secs.~\ref{gs} and \ref{excitations} for more details about recent constraints on the EoS and symmetry energy
based on ground and excitation properties in nuclei.
%Some of these constraints will be discussed in the following sections, for details on the other analyses or theoretical calculations we refer the reader to the original work. 

Regarding the parameters characterizing the curvature of the asymmetric matter EoS at saturation density, there is consensus on $K_0=240\pm 20$ MeV \cite{shlomo2006,colo2008,colo2017} from the analysis of the ISGMR in closed shell nuclei such as ${}^{208}$Pb. Recent analysis on available open-shell nuclei points toward a larger range of possible values quoting $250<K_0<315$ MeV \cite{stone2014} or slightly smaller central value that would be closer to 200 MeV \cite{avogadro2013}. So there is still some controversy about this parameter. As long as $K_\tau$ is regarded, there is much less experimental information. An analysis of the neutron skins of different nuclei from anti-protonic atoms (those in Fig.\ref{fig-pheno-3}) lead to $K_\tau=-500^{+120}_{-100}$ MeV \cite{centelles09} which is in reasonable agreement with analysis of the ISGMR in Sn isotopes $K_\tau= -395 \pm 40$ MeV \cite{garg2007} and $K_\tau=-550\pm 100 $ MeV \cite{li2010} and the ISGMR in Cd isotopes $K_\tau=-555\pm 75 $ MeV \cite{patel2012}. The overall analysis made in \cite{stone2014}, reported a range $-840<K_\tau<-350$ MeV. As an additional example, we show in Fig.\ref{fig-eos-11} the values of $K_0$ and $K_\tau$ as calculated from different EDFs and compare them with the experimental ranges (vertical and horizontal lines) as reported in Ref.\cite{colo15}.
%(complied from previous literature). 
%For details on the EDFs used we refer the reader to the same refernce. 

\begin{figure}[t!]
\begin{center}  
  \includegraphics[width=0.45\linewidth,clip=true]{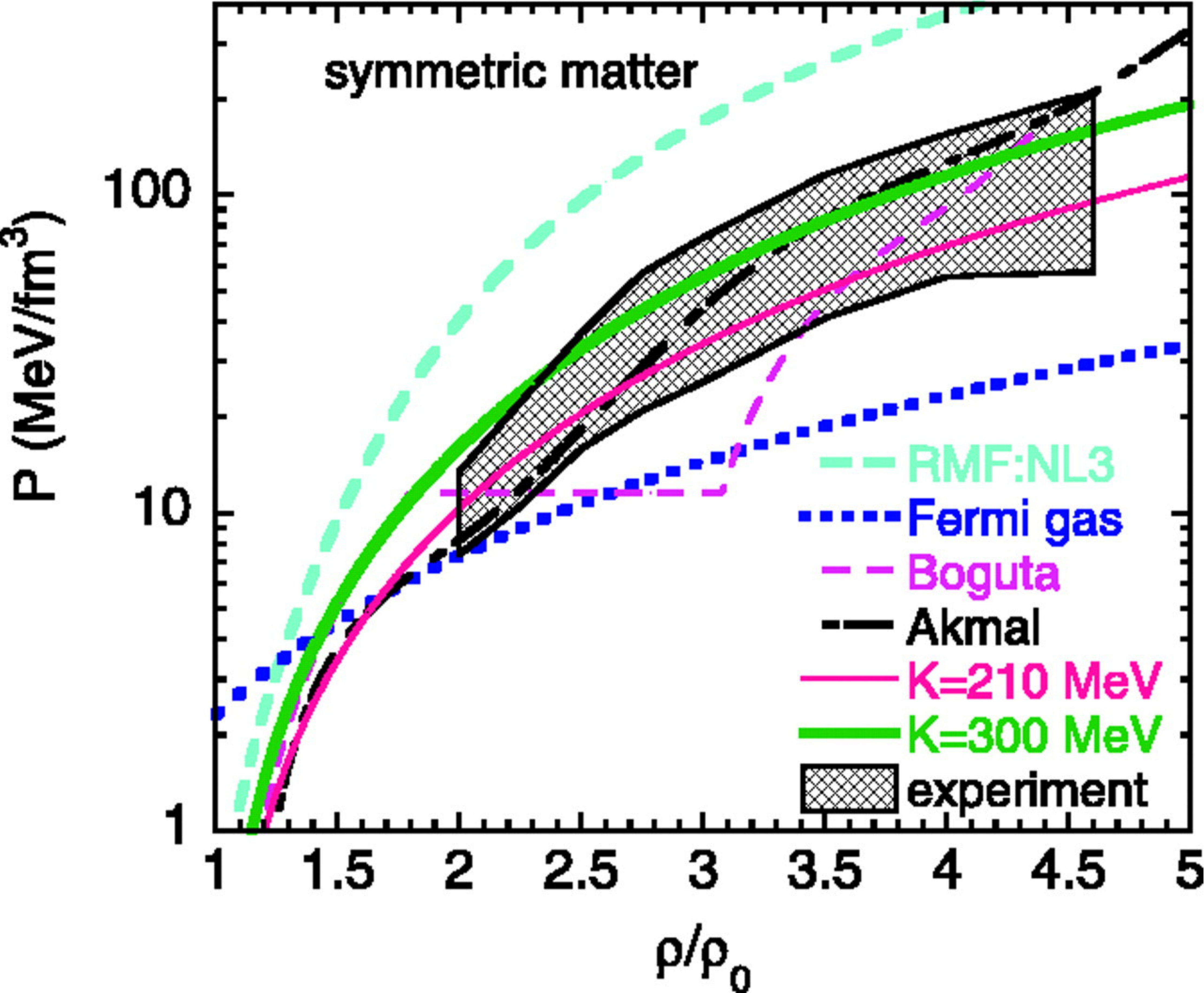}
  \includegraphics[width=0.45\linewidth,clip=true]{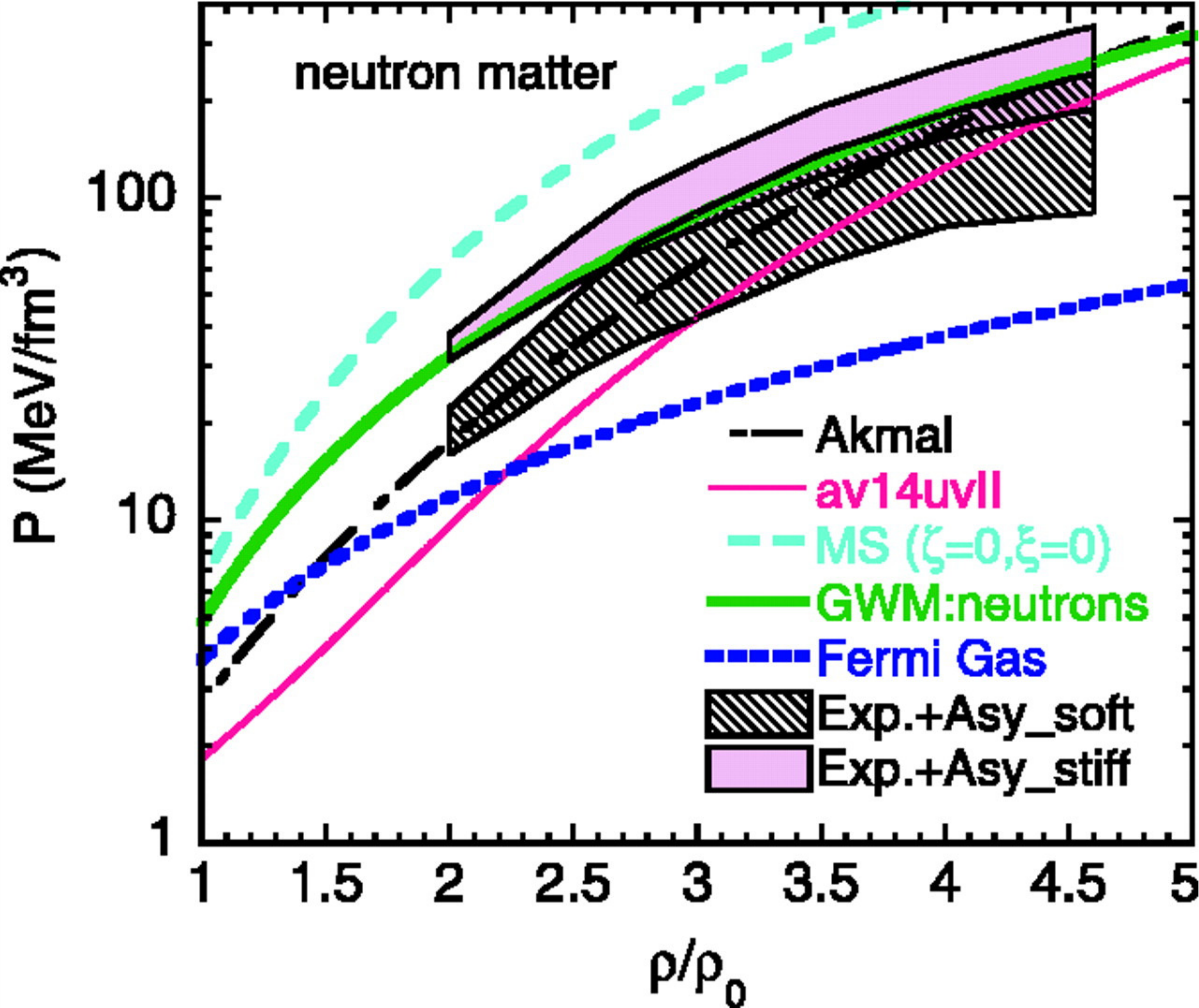}
  \caption{\label{fig-eos-12} Zero-temperature EOS (pressure as a function of density) for symmetric nuclear matter (left panel) and neutron matter (right panel). The shaded region corresponds to the region of pressures consistent with the experimental data. The different curves correspond to the prediction of different models. For details on these models we refer the reader to the original reference. Figures taken from Ref.\cite{danielewicz2002}.}
\end{center}
\end{figure}

In addition, we would also like to highlight here a study where the density behaviour of the EoS was derived form heavy ion collisions \cite{danielewicz2002} where matter can be compressed in a very short period of time to a very high densities only reachable in nature in compact stellar systems such as neutron stars. Therefore, such experiments probe the EoS at different densities than those probed by experiments measuring giant resonances, masses or nucleon distributions. Such type of studies, might be very useful for a better characterization of the nuclear EoS. We will not discuss this analysis here since it goes beyond the scope of the present review.
% that is focused on bulk ground and excited state properties of nuclei. 
However, for the sake of completeness we show in Fig.\ref{fig-eos-12} the experimental constraints (depicted as bands) derived from such work on the symmetric matter EoS (left panel) and neutron matter EoS (right panel) compared with some theoretical calculations. In this figure the pressure instead of the energy per particle is shown which at zero temperature contains the same information. For details on the  models shown in the figures, we refer the reader to the original reference \cite{danielewicz2002}. It is interesting to note for the purpose of this section that the experimental bands would be compatible with incompressibilities ranging from $210<K_0<300$ MeV as it can be seen in the left panel (green and magenta solid lines). In the right panel some external constraints on the symmetry energy were imposed in order to derive the two bands corresponding to limiting cases for the value of $J$ of about 30-35 MeV (cf. Refs.\cite{danielewicz2002, prakash1988}).

\begin{figure}[t!]
\begin{center}  
  \includegraphics[width=0.4\linewidth,clip=true]{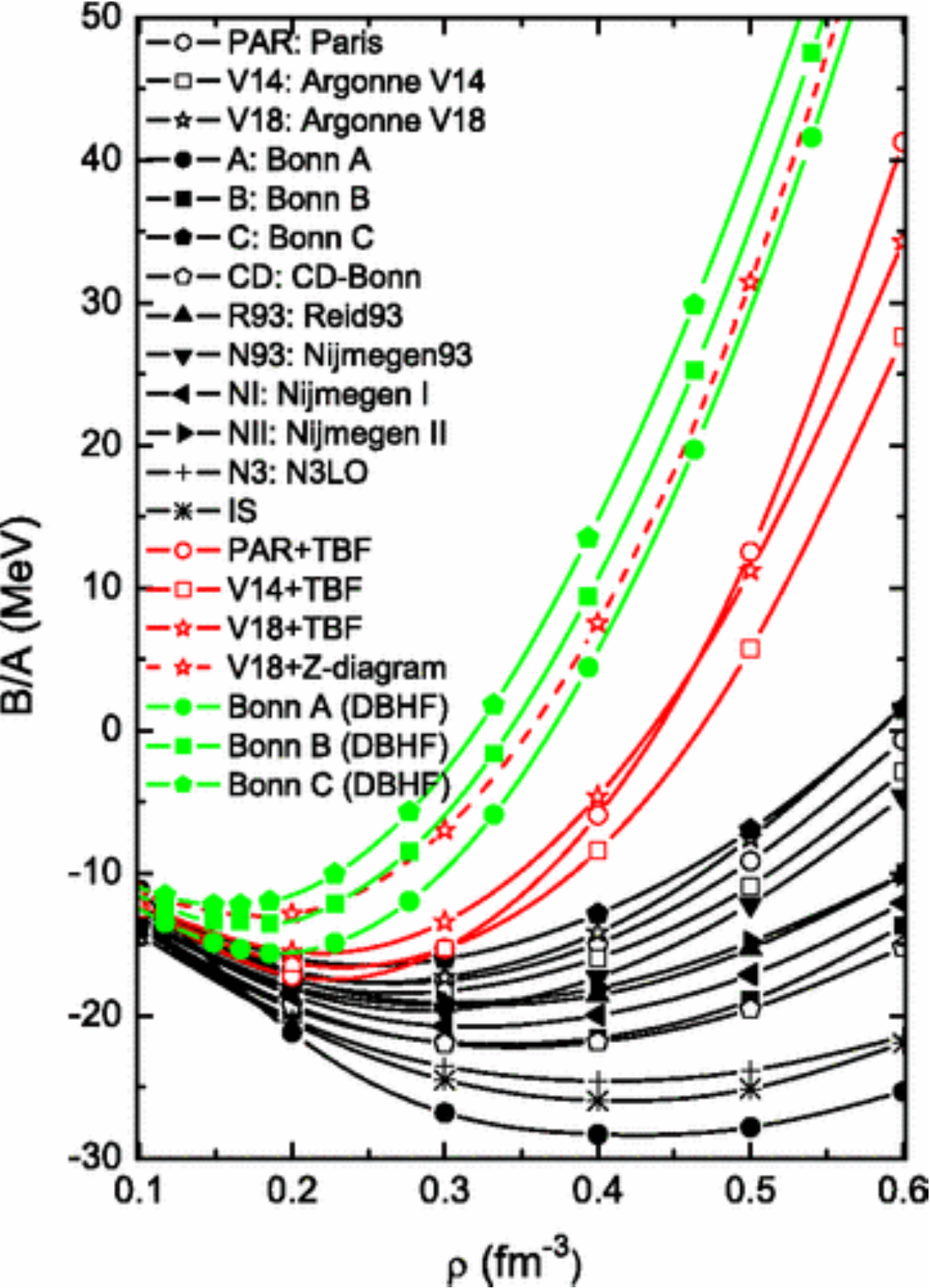}
  \includegraphics[width=0.55\linewidth,clip=true]{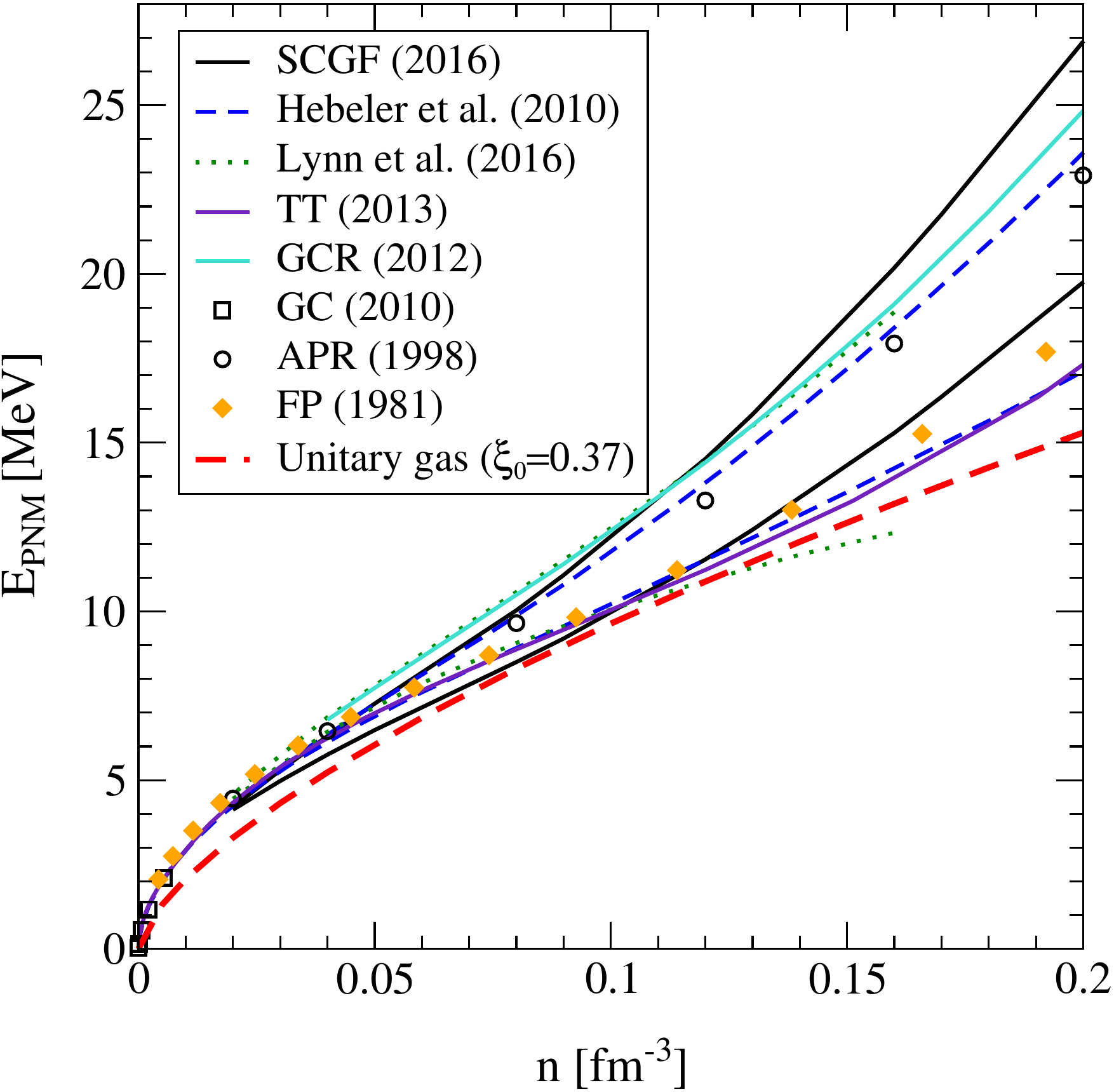}
  \caption{Left panel: Brueckner and Dirac-Brueckner Hartree-Fock calculations for the symmetric matter EoS as a function of the density. Figure taken from Ref.\cite{li2006}. Right panel: unitary gas bound compared to some {\it ab-initio} calculations for pure neutron matter EoS as a function of the density. See text and Ref.\cite{tews2017} for details. Figure taken from Ref.\cite{tews2017}.\label{fig-eos-13}}
\end{center}
\end{figure}

\subsection{Model predictions}
In this section we overview some of the latest works based on different theoretical frameworks that address the problem of studying the nuclear EoS. This will give the reader a richer idea on the systematic theoretical errors discussed so far. 

\subsubsection{Ab initio models}
In low-energy nuclear physics, we refer to {\it ab initio} approaches to all models that fit the nuclear effective potential from few-body data and, then, solve the many-body problem based on a given many-body method that will commonly be approximated. Some examples on nuclear effective potentials fitted to nucleon-nucleon scattering data in the vacuum are the Paris, Argonne, Bonn \cite{paris,argonne,bonn} or chiral effective potentials \cite{machleidt2013,hagen2015}. There are also efforts in deriving the nuclear strong interaction directly from QCD \cite{savage2012}. Therefore, the term {\it ab initio} should be understood in a very broad and flexible way. The relevance of these type of models is that they constitute our unique bridge between hadron physics (or QCD) and low-energy nuclear physics. Hence, {\it ab initio} EoS are thought to represent our best attempt to better understand the underlying physics of the nuclear problem.     

The first {\it ab initio} studies of nuclear matter were based on the Brueckner approximation within the Brueckner or Dirac-Brueckner Hartree-Fock (BHF or DBHF) and within variational approaches. The current status in the EoS established within these models are shown in Fig.\ref{fig-eos-13} (left panel) \cite{li2006}. The difference in the results can be seen not only if two-body (black) forces or two plus three-body (red) forces (TBF) are employed but also if relativistic (green) or non-relativistic (black and red) frameworks are adopted. A similar situation is extensive also for other {\it ab initio} approaches (cf. Fig.4 of Ref.\cite{summarruca2015} or Fig.4 of Ref.\cite{drischler2016b}). 

In Fig.\ref{fig-eos-13} (right panel), unitary gas lower bound on the pure neutron matter EoS \cite{tews2017} is compared to some state-of-the-art {\it ab initio} calculations based on different interactions and many-body methods (see original reference for details). Most of these models respect the lower bound due to repulsive three-body forces and show discrepancies between them of few MeV at saturation density. This agrees well with the uncertainty derived from phenomenology on the value of $J$. At the moment, knowledge on the EoS from {\it ab initio}, that is, from realistic potentials is limited. Taking this warning into account, one can inspect recent results on the neutron matter EoS from Refs.\cite{gandolfi2012} labeled as G in Fig.\ref{fig-eos-11} or labeled as H in the same figure~\cite{hebeler2010}. Those results are very narrow as compared to other estimations for the values of $J$ and $L$ coming from different experimental analysis but nicely overlap with them. 

\subsubsection{Macroscopic-microscopic models}
Macroscopic-microscopic (mac-mic) models are phenomenological models that have been devised to reproduce as accurately as possible nuclear masses. Indeed, those models are the most accurate in the literature displaying root mean square deviations with respect to experimental masses of only few hundreds of keV (300-600keV) \cite{frdm,myers1996,pomorski2003,bhagwat2010,moller2011,wang2014}. Opposite to {\it ab initio} approaches, mac-mic models can be applied to the study of nuclear masses, deformations, radii, etc. along the whole nuclear chart. The disadvantage is that any connection with a more fundamental theory such as QCD is completely lost and that macroscopic and microscopic terms are not derived based on the same theoretical grounds.

The main idea behind mac-mic models is that masses can be already qualitatively well understood from a macroscopic picture. If this picture is later complemented by the addition of ``small'' microscopic effects (shell corrections and pairing energies), mac-mic models become a quantitative approach to nuclear masses. As we have already discussed, masses provide information on the nuclear EoS such as the saturation energy of symmetric nuclear matter or the symmetry energy around saturation. This is why the input from this type of models will be of relevance in studying the nuclear EoS. In this sense the difficulty is to extract from these models some of the parameters defined in Eqs.\ref{eq-pheno-5} and \ref{eq-pheno-6}. In Ref.\cite{wang2015} this has been done for two mac-mic models, giving as a result (cf. Table II of the original reference): $e(\rho_0,0)=-15.494\pm0.004$ MeV, $K_0=230\pm 11$ MeV $J=29.2\pm 0.2$ MeV and $L=41.6\pm 7.6$ MeV for the model of Ref.\cite{pomorski2003}; and $e(\rho_0,0)=-15.583\pm0.007$ MeV, $K_0=235\pm 11$ MeV, $J=29.7\pm 0.3$ MeV and $L=51.5\pm 9.6$ MeV for the model of Ref.\cite{wang2014}. The relevant information here is that regardless of the reliability of the attached errors that will depend on the way they were extracted, the values predicted by mac-mic models seem to be in consonance with the predictions from other type of models.    

\subsubsection{EDFs}

\begin{figure}[t!]
\begin{center}  
  \includegraphics[width=0.45\linewidth,clip=true]{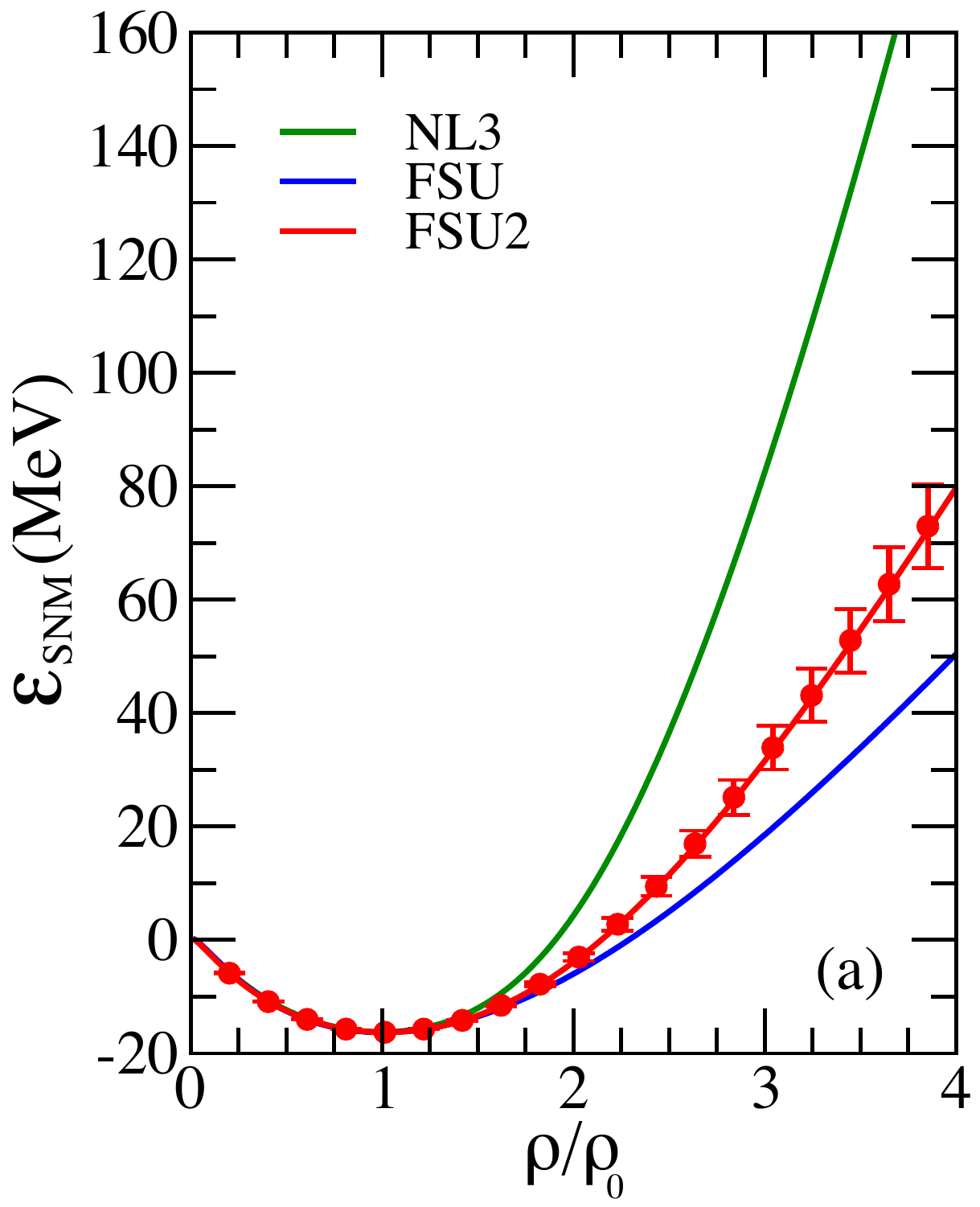}
  \includegraphics[width=0.45\linewidth,clip=true]{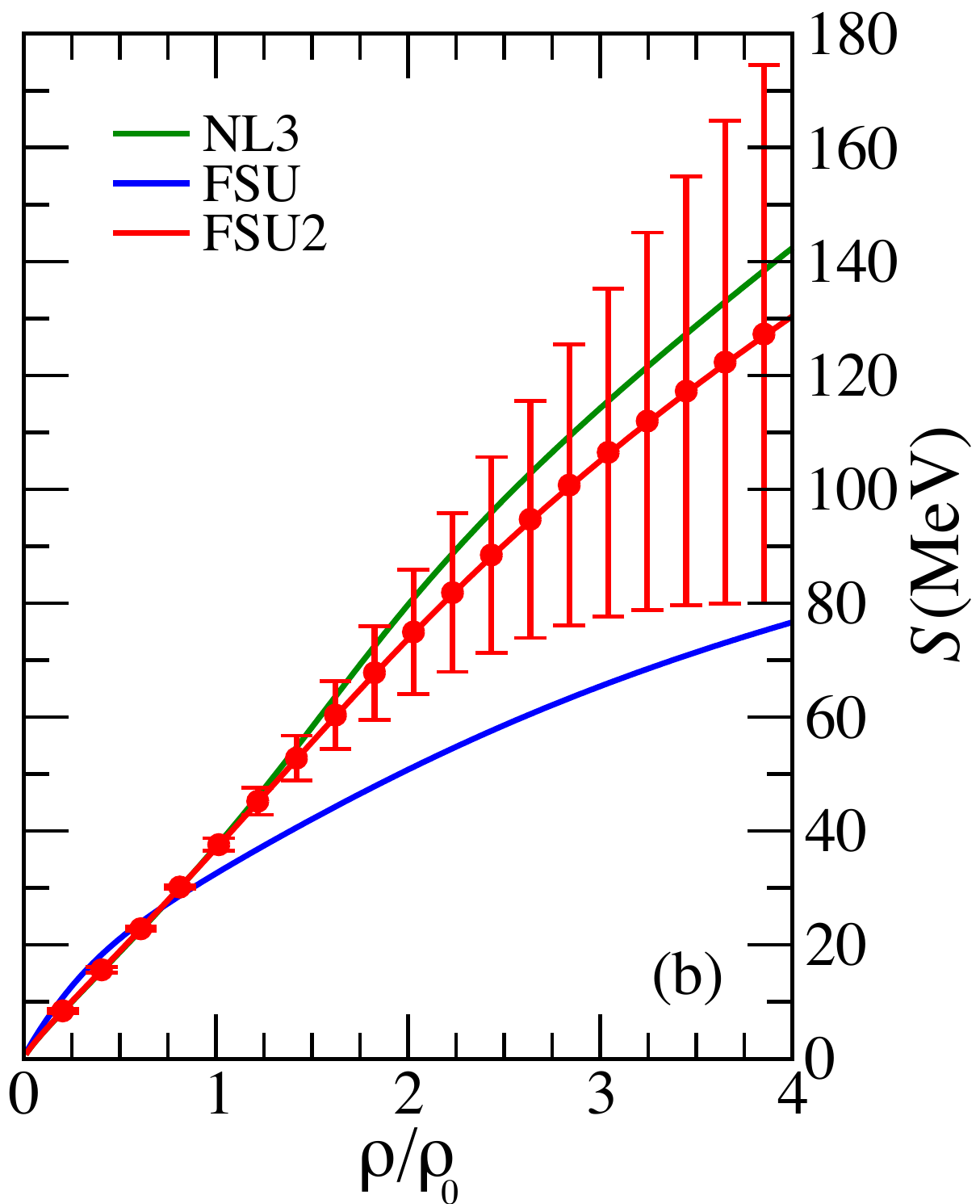}
  \caption{\label{fig-eos-14} Symmetric nuclear matter (left panel) and symmetry energy (right panel) as a function of $\rho_0$. Predictions of three different covariant models are shown. The FSU2 model contains an estimate of the statistical theoretical errors. Figures taken from Ref.\cite{chen2014}.}
\end{center}
\end{figure}

Energy density functionals are commonly based on effective Hamiltonians solved at the mean-field (Hartree or Hartree-Fock) level. These types of models are fully microscopic and essentially different from the {\it ab initio} models in what follows. The EDFs are fitted to many-body data (and no to few-body data) requiring self-consistency between the HF solution and the experimental data on finite nuclei. The performance of these models on the description of bulk properties of nuclei such as density distributions, masses, deformations or the excitation energies of giant resonances along most of the nuclear chart is remarkable taking into account the theoretical and numerical simplicity of the approach. Since the nucleus is not a {\it pure mean-field system}, the weakness of EDFs derived from a HF calculation is that the parameters of the model will contain many-body correlations that go beyond the mean-field approach adopted and, thus, the connection with any realistic nucleon-nucleon or three nucleon interaction is lost. Therefore, extrapolations of the EoS toward the densities not probed by the data used in the fit should be taken with great care.

Over the past years, the parameters characterizing the EoS around saturation density have been extensively discussed\cite{skyrme1}. Two recent and quite complete references are \cite{dutra2012} where 240 Skyrme functionals were analyzed and \cite{dutra2014} where a similar study was made for 263 relativistic mean-field models of different types.  In table II of Ref. \cite{dutra2012} and Table VII of Ref.\cite{dutra2014} most of the relevant parameters of the nuclear EoS at saturation for those models ($> 500 !$) are given. Specifically, in Refs.\cite{dutra2012,dutra2014}, the analysis consisted in imposing some empirical constraints on the EoS to be fulfilled by the models giving a list of well behaved EDFs as a result. The selection of used empirical constraints as a filter for all models might be misleading or at least controversial: not all constraints are of the same nature and theoretical systematic errors coming from the modelization of the strong interaction needed for the analysis of experimental data dealing with strongly interacting probes is not always realistically assessed. Proof of this is that some models considered by the authors to give the optimal nuclear matter properties (non-observable) are known not to be as accurate as other non selected models in the description of masses or charge radii (observables). This situation should foster such type of global studies but for the case of observable quantities in order to better constrain the nuclear EoS \cite{afanasjev2016}. That is, {\it going from successful models in the description of observables to the EoS is a well defined and safe strategy while ensuring reasonable parameters of the EoS do not necessarily lead to a good reproduction of the data on finite nuclei.} Note that this situation will be very different if exact {\it ab initio} calculations were available as in condensed matter physics.

From a different perspective, in Ref.\cite{agbemava2014}, the global performance of some state-of-the-art covariant EDFs on some nuclear observables was analyzed. The authors focused on binding energies, deformations, charge radii, neutron skin thicknesses, among other quantities. That is, observable quantities were used to qualitatively assess the reliability of the models. Some comparison with non-relativistic Skyrme functionals is also provided. In a different work \cite{kortelainen2010}, a Skyrme EDF was presented together with an exhaustive evaluation of statistical theoretical errors. The confidence ellipsoid labeled as ``Masses'' in Fig.\ref{fig-eos-11} where the $L$ versus $J$ correlation is shown comes from this study. Although masses give some information on the symmetry energy parameters, it still remains an open question whether the isovector properties of EDF's can accurately be defined from masses alone. Proof of this is that the plotted ellipsoid only contains the information on the above mentioned correlation as predicted by a single model --no systematic errors were evaluated-- and it does not allow to provide a narrow window for the values of $L$. In Ref.\cite{chen2014}, a covariant model (FSU2) was proposed including also a careful analysis of the statistical errors. Such a model was fitted to reproduce different properties of finite nuclei and neutron stars. In Fig.\ref{fig-eos-14} the symmetric matter EoS (left panel) and the symmetry energy (right panel) as a function of $\rho/ \rho_0$ are shown. Predictions of two other covariant models (NL3 and FSU) of common use are also shown for comparison. The FSU2 model contains an estimate of the statistical theoretical errors. In the figure it is evident that the model cannot be applied with confidence in isospin asymmetric systems at high density. A similar situation is found for other types of EDFs and in {\it ab initio} calculations. The authors of this work conclude that also at normal densities there is still a meaningful constraint missing in the isovector sector (cf. Table IV in Ref.\cite{chen2014} where the error in $L$ is larger than 15 MeV and where the values predicted by FSU2 for $J$ and $L$ are very large as compared to the other EDFs of common use in the field). 

Some efforts have been devoted to refine existent EDFs in order to build a {\it universal} functional \cite{erler2012,kortelainen2012,goriely2013} that can be regarded to be as accurate as mac-mic models in the description of masses. This has only been achieved by the model presented in Ref.\cite{goriely2013} and in subsequent refinements. To pursue such an accuracy from a purely microscopic model might be instrumental in the study of unexplored areas of the nuclear chart and might also be very informative: first because they provide an understanding of nuclear masses from a consistent microscopic point of view (shell effects and the bulk part of the functional are consistent, opposite to what happens in mac-mic models); and, second, because the EDF theory is rooted on the Hohenberg-Kohn theorems that ensure the existence of an exact energy density functional (and, therefore, one may learn how to approach to such an exact functional or if is it possible). 

There has also been some effort in building EDFs inspired by DFT developed in condensed matter physics where the EoS is based on {\it ab initio} calculations. In the nuclear case, the latter calculations are not as accurate as in Coulomb systems but, in this way one may fix an important part of the functional by more fundamental calculations and learn if EDFs are compatible with them and, at the same time, with experimental data on finite nuclei. This is a way to test both EDFs and {\it ab initio} results on the EoS. Such a strategy has been adopted in Refs.\cite{drut2010, baldo2017, roca-maza2011} and it seems to be feasible as long as the {\it ab initio} EoS is fitted within some errors and not exactly fixed --this is mainly related to the fact that {\it ab initio} approaches do not exactly saturate at the empirical values.

Finally, with a more phenomenological perspective, other types of the EDFs have been proposed \cite{margueron2017}: the starting point is an EDF parameterized directly with the parameters of the nuclear EoS [as in Eqs.(\ref{eq-pheno-4})-(\ref{eq-pheno-6})], and where a phenomenological surface term that can account for the properties of finite nuclei is added. Such an strategy has been shown to be successful although the information content of the nuclear observables used in the fit are not able to constrain the parameters of the EoS (cf. Table I of Ref.\cite{margueron2017}). So, as expected, this strategy is equivalent to a direct fit of the parameters of the nuclear effective interaction as it is commonly done in EDFs.

%% file: gs.tex
\section{Ground state properties}
\label{gs}

Nowadays, EDFs represent the only approach that consistently and microscopically can address the description of ground state and excited state properties in nuclei along the nuclear chart and of the EoS. In this section some representative results that impact omn the EoS properties around saturation on masses and rms radii from the neutron and proton density distributions will be given. For the case of masses, some mac-mic model results will also be presented for comparison since these models are the most accurate in describing nuclear masses. Finally, in this subsection, we will focus on the overall performance of EDFs not discussing in detail effects such as shell effects or deformations among others. This is because the EoS properties should not be influenced by such effects. 
\subsection{Nuclear masses}

\begin{figure}[t!]
\begin{center}  
  \includegraphics[width=0.45\linewidth,clip=true]{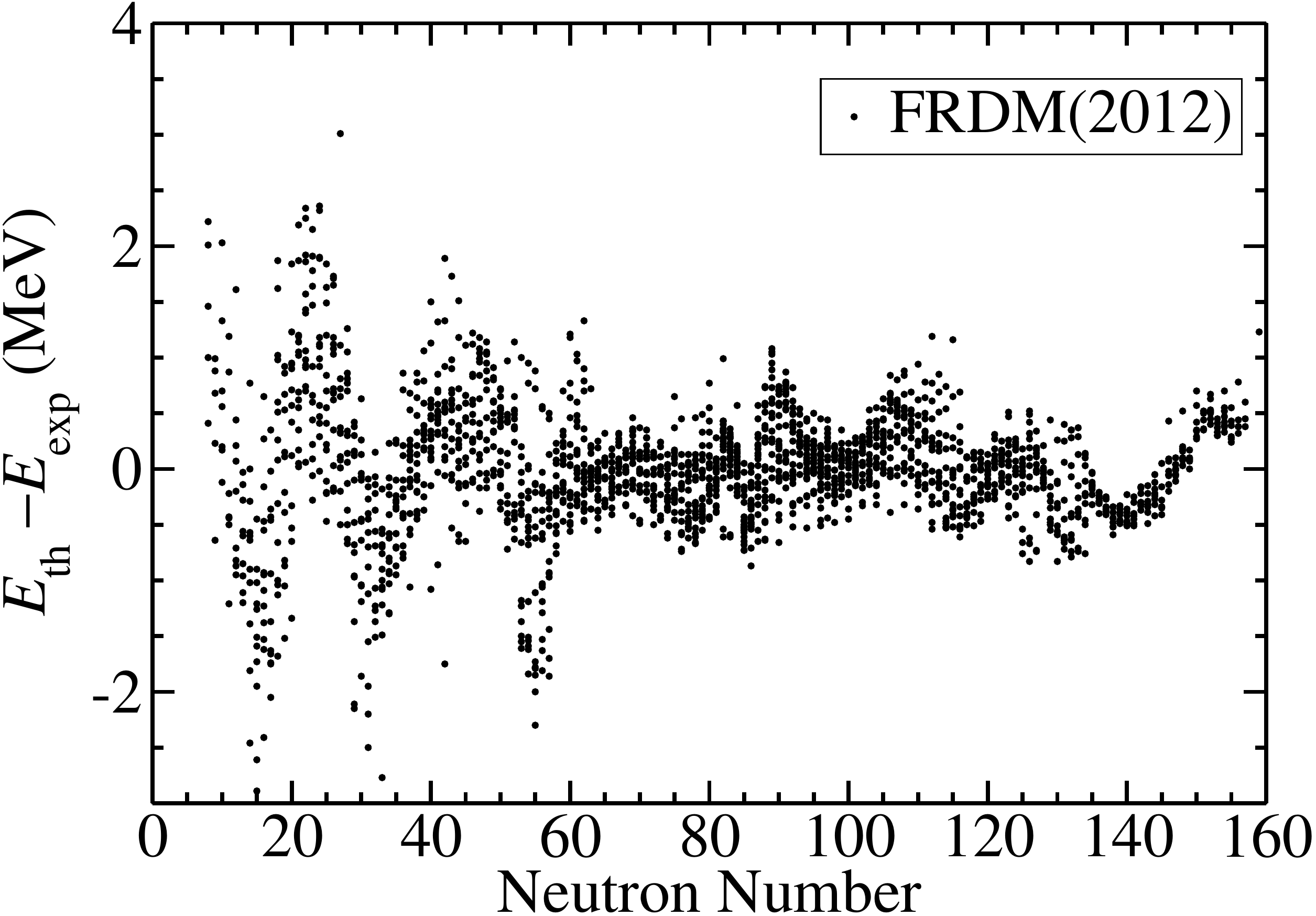}
  \includegraphics[width=0.45\linewidth,clip=true]{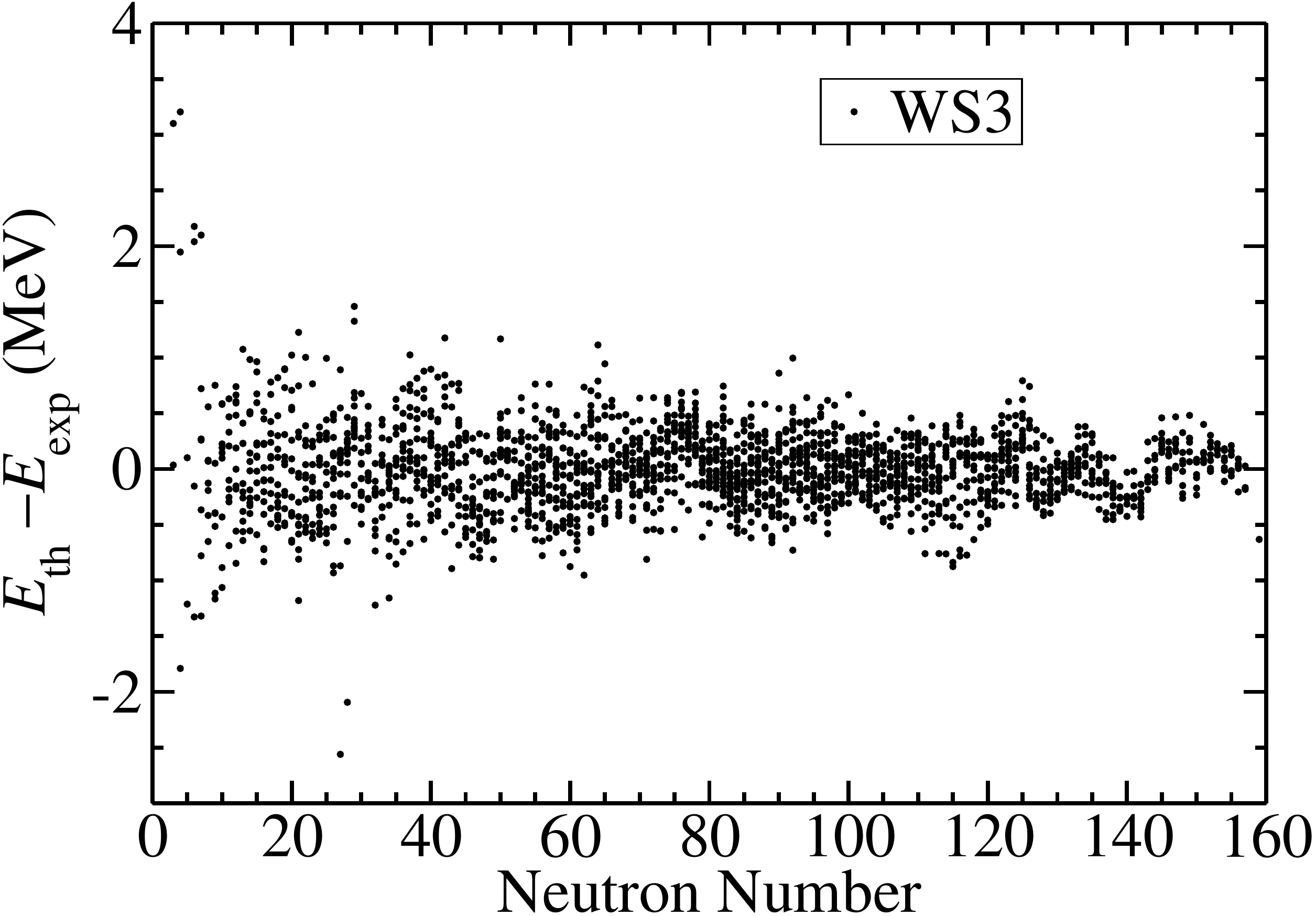}
  \includegraphics[width=0.45\linewidth,clip=true]{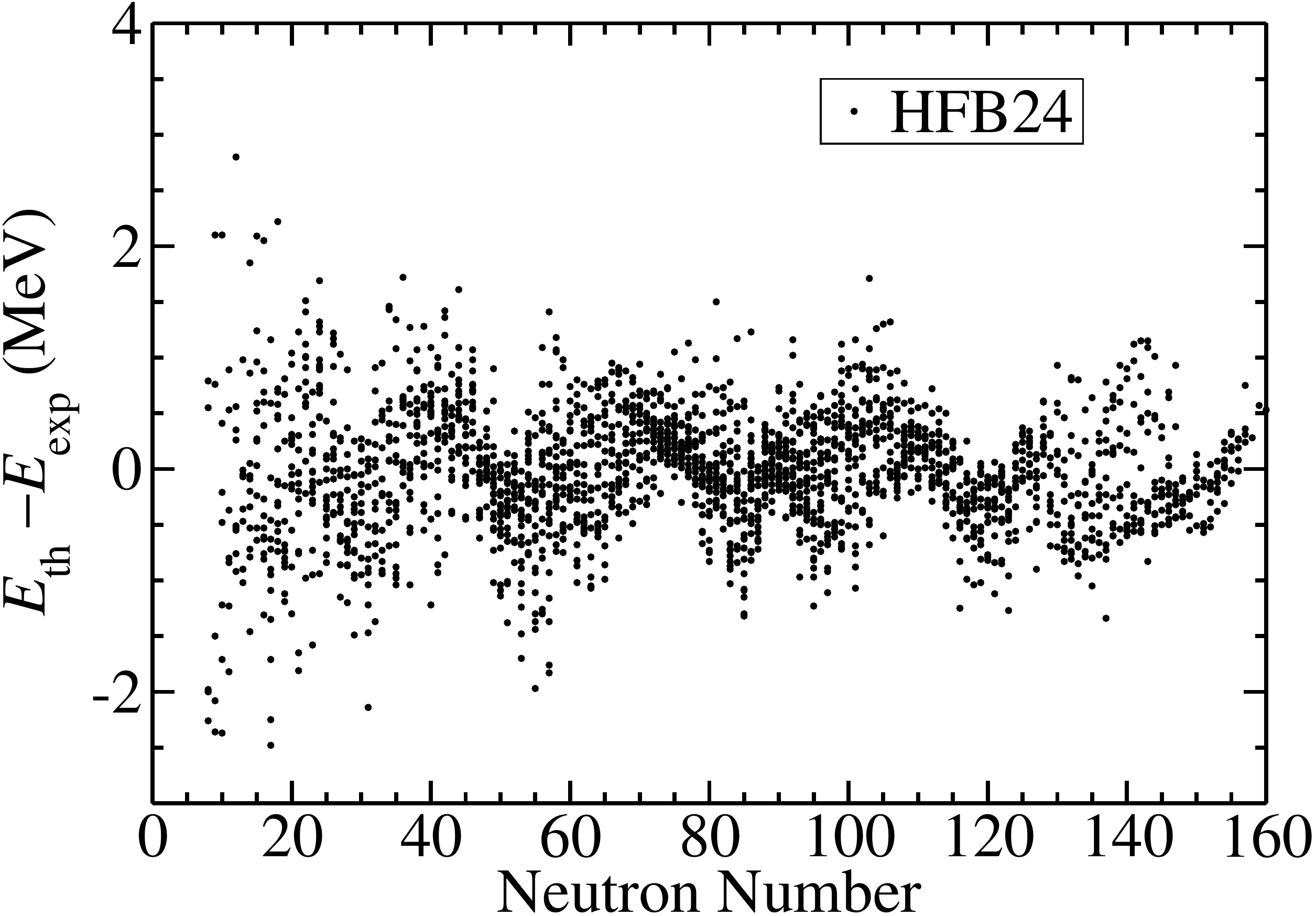}
  \includegraphics[width=0.45\linewidth,clip=true]{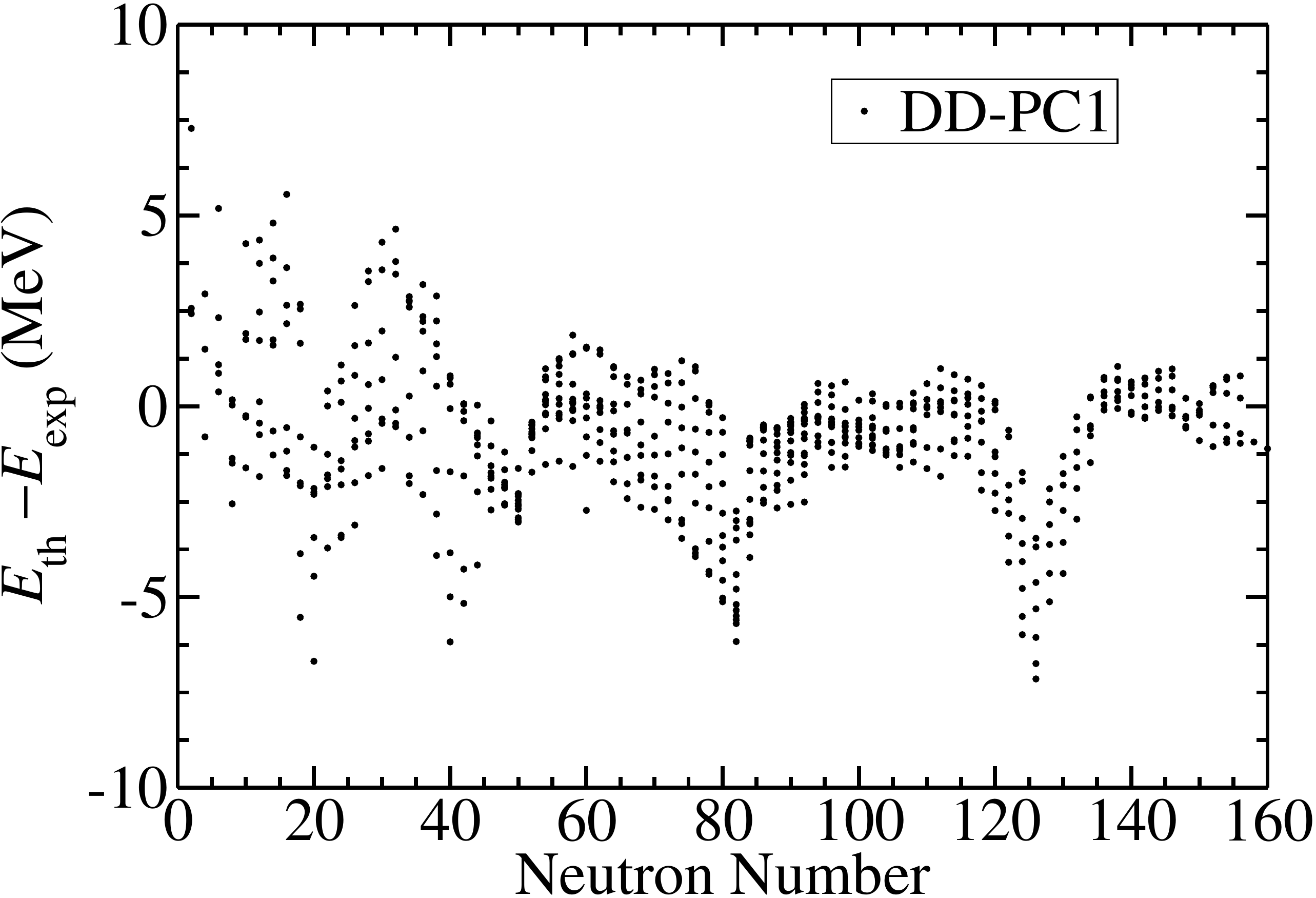}
  \caption{\label{fig-gs-15} Mass differences as a function of the neutron number for the mac-mic model [FRDM(2012)] of Refs.\cite{moller2011,moller2012} (left upper panel); mac-mic model [WS3] of Ref.\cite{wang2011} (right upper panel); Skyrme model [HFB24] of Ref.\cite{goriely2013} (left lower panel); and relativistic density dependent point coupling model [DD-PC1] of Ref.\cite{niksic2008} (right lower panel).}
\end{center}
\end{figure}

Figure~\ref{fig-gs-15} shows the mass differences (or residuals) as a function of the neutron number for the mac-mic model [FRDM(2012)] of Refs.\cite{moller2011,moller2012} (left upper panel); mac-mic model [WS3] of Ref.\cite{wang2011} (right upper panel); Skyrme model [HFB24] of Ref.\cite{goriely2013} (left lower panel); and relativistic density dependent point coupling model [DD-PC1] of Ref.\cite{niksic2008} (right lower panel). EDFs and mac-mic models are based on different theoretical grounds. Nevertheless the accuracy in the prediction of nuclear masses is of the same quality when comparing FRDM(2012) and HFB24, both containing a similar number of constants to be determined from experiment (cf. Table\ref{tab-gs-1}). The mac-mic model WS3 (and newer version WS4\cite{wang2014}) is giving the best description of nuclear masses, reducing the arch structure in the residuals that is clearly present in all the other models, indicating that some physics is missing in such models\footnote{Residuals of an exact model should follow a gaussian distribution if an histogram is constructed.}. Finally, the most accurate relativistic EDF model in the description of nuclear masses has been shown to be the point coupling model DD-PC1 \cite{agbemava2014}. Opposite to most of available EDFs, this model was fitted to reproduce the binding energy of well deformed nuclei (most of the known nuclei are deformed). Deformed nuclei have also been used in constraining non-relativistic functionals, e.g., in Ref.~\cite{baldo2017}. In addition, EDFs based on the mean-field approach are more suitable for the description of not light, open-shell and deformed nuclei since correlation energies that go beyond the mean-field are only due to rotational energies (relatively easy to correct) while closed shell spherical nuclei contain relevant correlations such as surface vibrations that goes beyond the mean-field approach and that are not easy to incorporate \cite{bohr69,roca-maza2017}. The non-relativistic interaction HFB24 also contained in the fitting protocol deformed nuclei.

\begin{figure}[t!]
\begin{center}  
  \includegraphics[width=0.8\linewidth,clip=true]{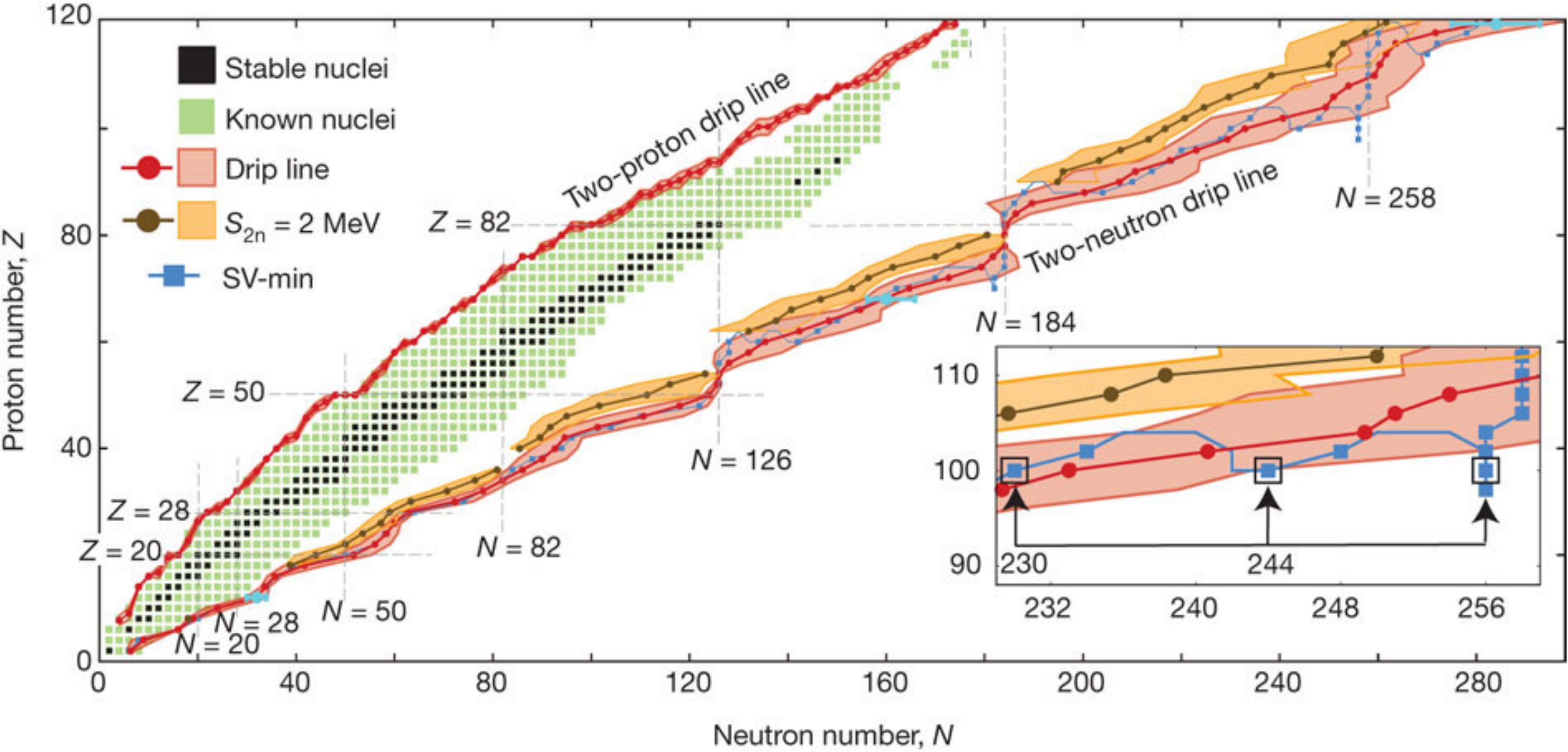}
  \includegraphics[width=0.37\linewidth,clip=true,angle=-90]{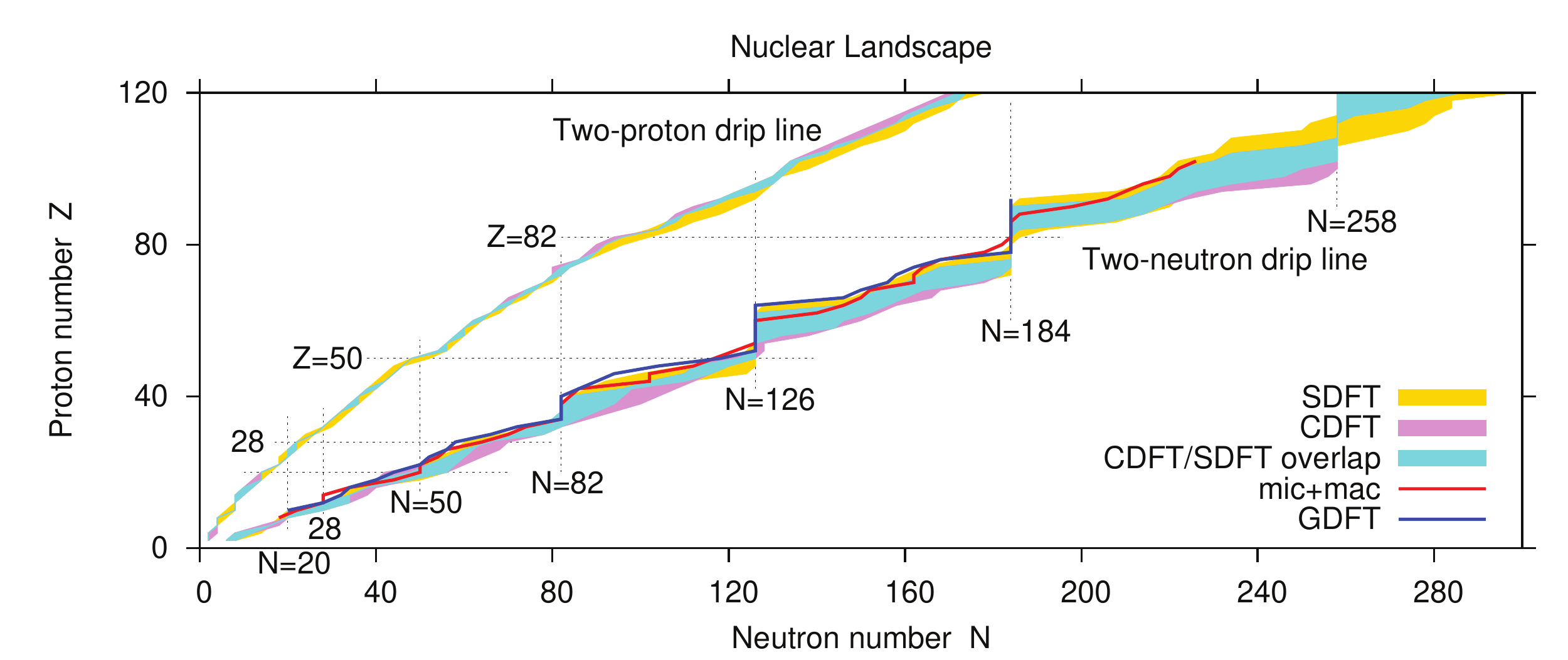}
  \caption{\label{fig-gs-16} Nuclear landscape ($Z$ versus $N$ plot). In the upper panel known (2012) even-even nuclei are shown. Stable nuclei are depicted in black squares and radioactive ones with green squares. Mean drip lines and their uncertainties (red) are also shown. The two-neutron separation energy $S_{2n}=2$ MeV line is also shown (brown) together with its systematic uncertainty (orange). The inset shows the irregular behavior of the two-neutron drip line around $Z= 100$. Figure taken from Ref.\cite{erler2012}. In the lower panel a comparison of the estimated uncertainties in the definition of two-proton and two-neutron drip lines obtained in relativistic (CDFT) and non relativistic (SDFT) EDFs is depicted. The blue shaded area shows the overlap between them. The two-neutron drip lines obtained by mic-mac FRDM(1995) \cite{moller1995} and Gogny (GDFT) EDFs \cite{delaroche2010} are shown by red and blue lines, respectively.Figure taken from Ref.\cite{afanasjev2013}.}
\end{center}
\end{figure}

In Table\ref{tab-gs-1}, the root mean square deviation with respect the experimental data on nuclear masses ($\sigma_{M}$) of the models shown in Fig.\ref{fig-gs-15} and discussed above are given. Other details on the predicted parameters of the EoS and the number of constants that compose each model are also included. The Skyrme UNEDF1 \cite{kortelainen2010,kortelainen2012} functional has been added to the table since it constitutes one of the major efforts to build a universal EDF--also fitted to deformed nuclear masses. From this table, one 
can observe that UNEDF1 and DD-PC1 interactions have a relatively small number of parameters when compared to the other models and that provide a less accurate description of nuclear masses ($\sigma_{M}\sim 2$ MeV) while the other models with a larger number of parameters reach a much better overall description ($0.3$ MeV$<\sigma_{M}<0.6$ MeV). Here a word of caution is important. Although having (adding) more parameters in (to) a reasonable model may improve the quality measure ($\sigma_{M}$) it does not necessarily mean that the model itself has a larger predictive power and, thus, provides more reliable extrapolations \cite{roca-maza2015}. On this regard, one of the most outstanding questions in physics, that has not been answered yet, is related to the limits of 
existence of nuclei, i.e., to establish the so-called proton and neutron drip lines where atomic nuclei will decay by the emission of a proton or a neutron respectively \cite{erler2012,afanasjev2013}. Many studies have been devoted to such a question (see for example \cite{erler2012,afanasjev2013}). The drip-line position is strongly affected by pairing correlations that make nuclei with even number of neutrons and protons more bound than their odd counterparts. Therefore, the one particle emission drip line is reached earlier than the two particle emission drip line. In Fig.\ref{fig-gs-16}, the two neutron and proton drip lines have been calculated by using different EDFs in order to estimate these limits of the nuclear landscape. Specifically, in the upper panel \cite{erler2012} known (2012) even-even nuclei are shown. From those, stable nuclei are depicted in black squares and radioactive ones with green squares. Mean value for the drip lines and their uncertainties (red) as predicted by the studied models are also shown. The two-neutron separation energy $S_{2n}=2$ MeV line is shown (brown) together with its systematic uncertainty (orange). The inset shows the irregular behavior of the two-neutron drip line around $Z= 100$. In the lower panel \cite{afanasjev2013} a comparison of the estimated uncertainties in the definition of two-proton and two-neutron drip lines obtained in relativistic (CDFT) and non relativistic (SDFT) EDFs is depicted. The blue shaded area shows the overlap between them. The two-neutron drip lines obtained by mac-mic FRDM(1995) \cite{moller1995} and Gogny (GDFT) EDFs \cite{delaroche2010} are shown by red and blue lines, respectively. It is evident from this figure that proton drip-lines, much closer to the stability valley due to the effect of the Coulomb interaction, show much less spread in all calculations while neutron drip lines are less well determined. This result is intimately connected to the fact that EDFs are not well constrained in the description of systems with large neutron excess. That is, available data does not seem to contain enough information to precisely determine the limits of existence of neutron-rich nuclei. A signature of this can also be seen by inspecting the values for $J$ and $L$ as predicted by some of the most accurate models in the description of nuclear masses presented in Fig.\ref{fig-gs-15} and Table\ref{tab-gs-1}. In this table, all models predict a relatively narrow band for the values for $\rho_0$ ($0.154$ fm${}^{-3}<\rho_0< 0.159$ fm${}^{-3}$), $e_0$ ($-16.2$ MeV$<e_0< -15.6$ MeV) and $K_0$ ($220$ MeV$<K_0<240$ MeV) and a clearly larger band for the possible values of $J$ ($29$ MeV$<J<35$ MeV) and $L$ ($46$ MeV$<L<113$ MeV). These ranges are orientative, the spread (or lack of) in these predictions give an idea on the sensitivity to fitted data--commonly masses and charge radii--of each one of these parameters. In addition to this spread, statistical theoretical error bars (in Table\ref{tab-gs-1} are given within parenthesis when available) may produce a further increase on the above discussed ranges. The most sensitive parameter to the reproduction of masses is $e_0$. In particular, the volume binding energy grows with $e_0A$, then if energy differences do also grow or decrease with $A$ means that $e_0$ has not been correctly determined. In Fig.\ref{fig-gs-15} it is quite clear that all models show no slope of the mass differences with the neutron number (it is the same when plotted against the total mass number). Therefore, it is reasonable to assume that these state-of-the-art results provide the best estimation on the central value of $e_0$. Such range is, as reported above, $-16.2$ MeV$< e_0 < -15.6$ MeV. 

\begin{table}[t!]
\vspace{-0.5cm}
\caption{Example of different model predictions on some nuclear matter EoS parameters. Estimated errors, when available, are reported within parenthesis and the overall relative variation is also given in the last row. Number of total fitted parameters as well as root mean square deviation $\sigma_M$ with respect to the experimental masses are also given. See the text for details.}
\begin{center}
\begin{tabular}{lccccccccc}
\hline\hline
Model & Type &N$^{\rm o}$ par.& $\rho_0$ [fm$^{-3}$]& $e_0$ [MeV] & $K_0$ [MeV] & $J$ [MeV] & $L$ [MeV] &$\sigma_{\rm M}$ [MeV]\\
\hline
FRDM(2012)&Mac-Mic&38${}^{a}$ & --    &$-$16.195&240  &32.5(5) &53(15)&0.559$^{b}$\\
WS4$^{c}$ &Mac-Mic&18& --    &$-15.583(7)$&$235(11)$&$29.7(3)$ &59(10) &0.298$^{d}$\\
HFB24     &EDF&30${}^{e}$&0.1578 &$-$16.048&245.5&30.0  &46.4&0.549$^{f}$\\
UNEDF1    &EDF&12&0.1587(4)&$-$15.800&220.0&29.0(6)&40(13)&1.88${}^{g}$\\
DD-PC1    &EDF&9&0.154  &$-$16.12 &238  &35.6  &113&2.01${}^{h}$\\
\hline
Rel. var. & & &3\% & 4\%& 9\% & 20\% & 80\%& \\
\hline\hline
\end{tabular}
\end{center}
\label{tab-gs-1}

{\tiny
${}^{a}$ 21 fixed from other considerations than fit to masses;
${}^{b}$ With respect to AME2003\cite{ame2003};
${}^{c}$ Estimated properties\cite{wang2015};
${}^{d}$ With respect to AME2012\cite{ame2012};
${}^{e}$ Some of them fixed {\it manually} \cite{goriely2013};
${}^{f}$ Only even-even nuclei with $N,Z>8$ have been considered and compared with AME2003\cite{ame2003};
${}^{g}$ Only even-even nuclei with $N,Z>8$ have been considered;
${}^{h}$ Only even-even nuclei with $Z\leq 104$ have been considered and compared to AME2012 \cite{ame2012}.
}
\end{table}

\begin{figure}[t!]
\begin{center}  
  \includegraphics[width=0.5\linewidth,clip=true]{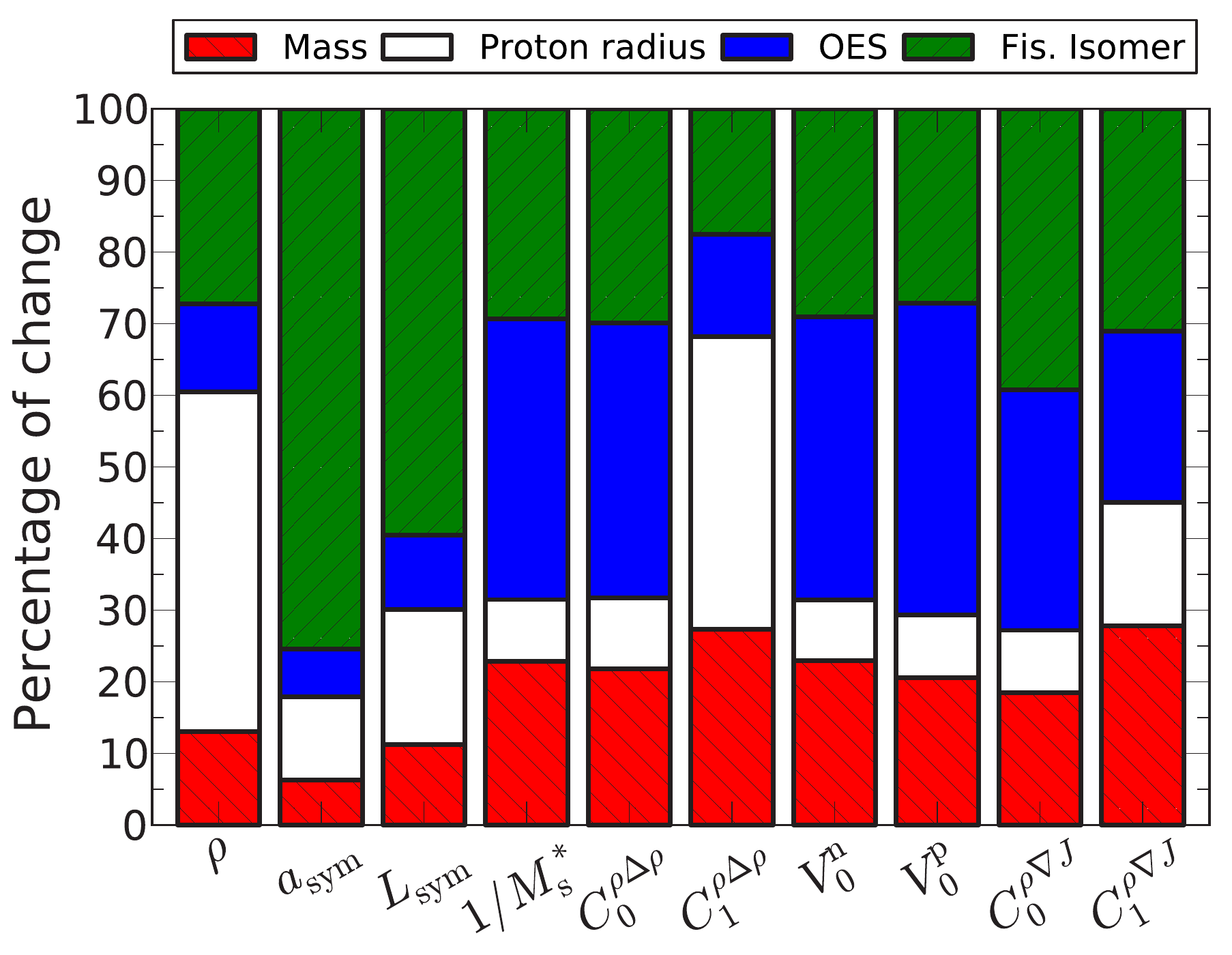}
  \caption{\label{fig-gs-17} Sensitivity of the UNEDF1 interaction to different types of data entering the quality measure ($\chi^2$). Figure taken from Ref.\cite{kortelainen2012}}.
\end{center}
\end{figure}

Next, as a quantitative example, the impact of each type of observable used for the fit of UNEDF1 \cite{kortelainen2012} on some parameters of the nuclear EoS is shown in Fig.\ref{fig-gs-17}. Focusing on the three first columns on this figure (for further details see the original publication): saturation density (``$\rho$'' same as $\rho_0$); symmetry energy at saturation (``$a_{\rm sym}$'' same as $J$); and slope of the symmetry energy at saturation (``$L_{\rm sym}$'' same as $L$); one sees that the effect (weight) of masses (``Mass'') on these parameters is small. The saturation density is mainly determined by experimental proton radii (``Proton radius'') included in the fit while the symmetry energy parameters seem to be much more sensitive to fission isomer excitation energies (``Fis. Isomer''). The latter information was included in the fitting of UNEDF1 interaction because the authors aimed to produce a model optimized for fission. On the one hand, this information does not tightly constrain $J$ nor $L$ (statistical theoretical errors are large, see Table\ref{tab-gs-1}) and, on the other hand, the experimental determination of fission isomer excitation energies \cite{singh2002} depends on a model dependent analysis for which systematic errors might be also large.

Finally, it is important to mention that some observations on neutron stars such as the mass and radius will have impact on the nuclear EoS \cite{lattimer2007,steiner2013}. However, the unknowns on the structure and composition of a neutron star--mostly on the inner core but not only--prevent us to put stringent constraints on the EoS. It seems, however, that for sufficiently light neutron stars ($\sim 0.5 M_\odot$), the neutron matter equation of state at normal densities might be constrained in a more transparent way (see \cite{carriere2003} and following subsection). In connection with nuclear masses, there exist also indirect observational data, for example, on crustal modes in strongly magnetized neutron stars that may be influenced by the composition and structure of the neutron star crust (see e.g. \cite{piro2005} or \cite{sotani2016}). Regarding the outer crust, it is believed to be made of fully equilibrated matter where ionized neutron rich nuclei form a Coulomb crystal embedded in an electron gas \cite{chamel2008}. The main unknown in such a model corresponds to the binding energy of neutron rich nuclei present in the lattice. Part of the nuclei believed to be present in the outer crust have been already measured in the laboratory. However, experimental information is not sufficient to provide a full prediction for the composition of this neutron star outermost layer and one should rely on model predictions. This has fostered experimental studies on mass measurements of exotic nuclei. In Ref.\cite{wolf2013} the mass of ${}^{82}$Zn was measured for the first time and the composition of the outer crust was altered with respect to previous model predictions and constrained by experimental data deeper into the crust than before. Then, if convenient astrophysical observations are clearly identified, the mass of very neutron rich nuclei not reachable in earth laboratories might eventually be inferred and, thus, model predictions on masses far from the stability valley constrained.

\subsection{Nuclear radii}
In this section, the focus is put on one of the main bulk properties in nuclei: its size. Specifically, the nuclear size is customary studied in terms of the second moment of the corresponding density distribution of a nucleus in its ground state
\begin{equation}
  \langle r^2_q\rangle = \int d{\bf r}\rho_q({\bf r})r^2
  \label{eq-gs-60}
\end{equation}  
where $q=n,p$ indicate neutrons or protons. This quantity is also referred as the root mean-square (rms) radius of neutrons or protons. Also the total matter density ($\rho_n+\rho_p$) have been studied via the matter rms radius. In previous sections, we have seen that measured proton radius essentially constrain the possible values of the nuclear matter saturation density $\rho_0$. On the other side, it has also been discussed that the neutron skin thickness (or equivalently the neutron rms as long as the proton rms is known) of a heavy nucleus correlates well with the symmetry energy and its density dependence around saturation density. Here, more details on microscopic calculations will be given. As in the previous section, the EDFs are our basic tool to consistently access at the same time nuclear distributions and the EoS parameters. Results based on other theoretical frameworks will also be given for comparison. 

\begin{figure}[t!]
\begin{center}  
  \includegraphics[width=0.45\linewidth,clip=true]{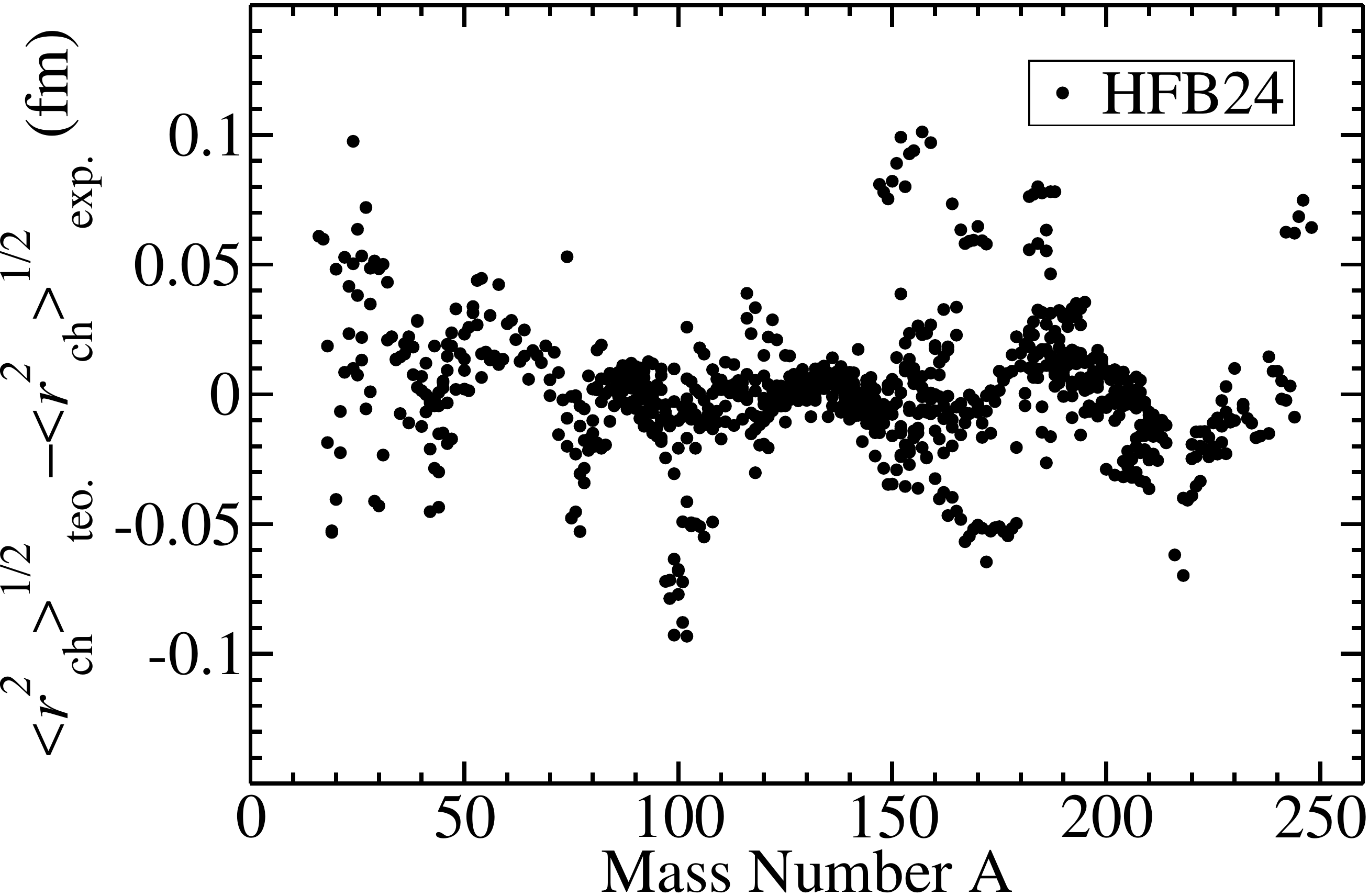}
  \includegraphics[width=0.45\linewidth,clip=true]{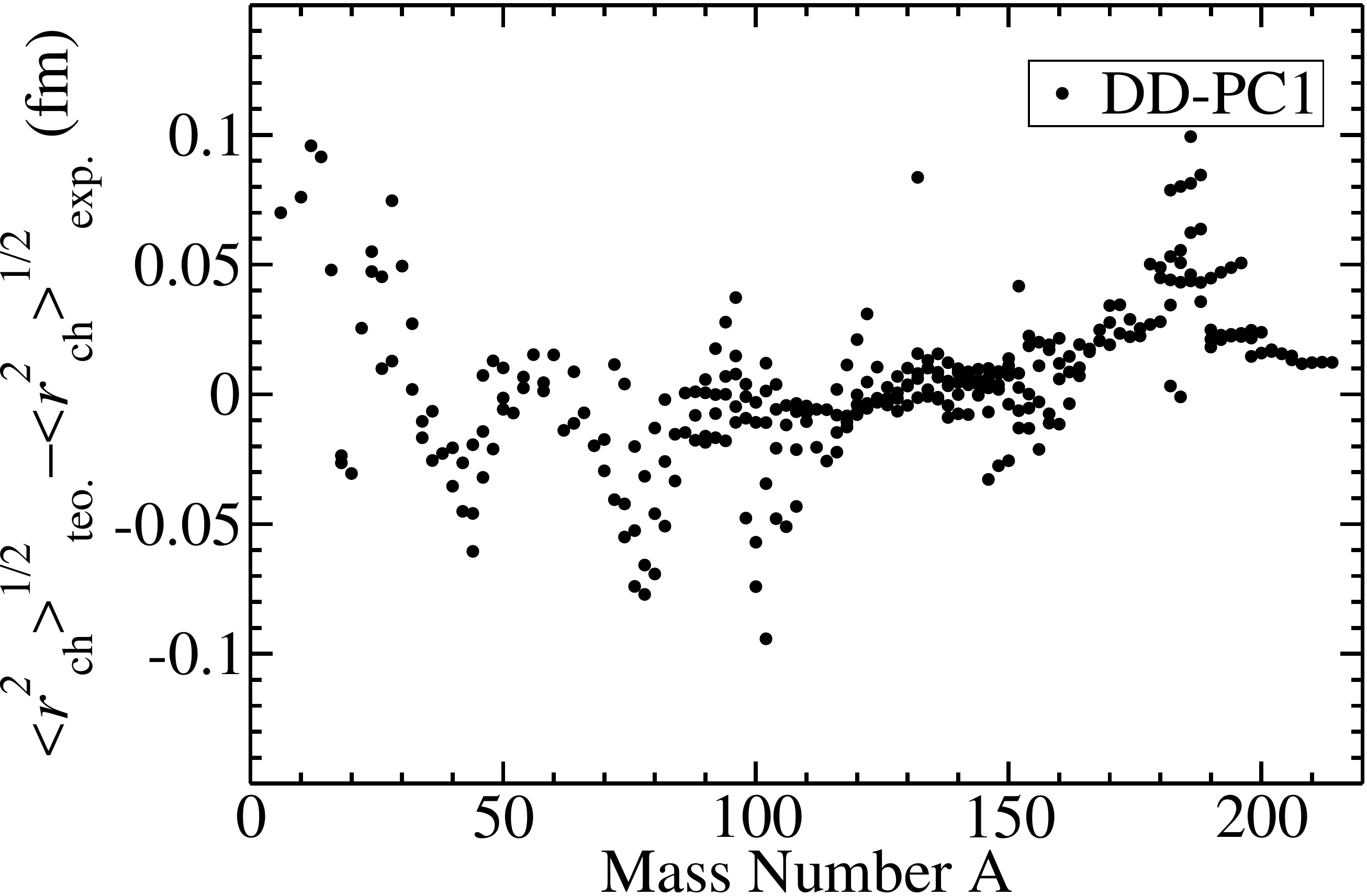}
  \caption{\label{fig-gs-18} Electric charge radius differences as a function of the mass number for the Skyrme model [HFB24] of Ref.\cite{goriely2013} (left panel); and relativistic density dependent point coupling model [DD-PC1] of Ref.\cite{niksic2008} (right panel).}
\end{center}
\end{figure}

\subsubsection{Proton density distributions}
Elastic electron (and muon) scattering off nuclei is a model independent technique to investigate the size and shape of stable nuclei \cite{hofstadter1956,devries1987,drechsel1989} and unstable nuclei \cite{ohnishi2015, antonov2011, karataglidis2017}. This is because electrons interact via the electroweak interaction with the nucleons. Being the weak charge density of the nucleus much harder to probe\footnote{The weak neutral current interaction is mediated by the $Z^0$ boson while the electromagnetic interaction is mediated by the photon. As an example, for 1GeV electron beam scattered by a ${}^{208}$Pb target at forward angles, 1 electron out of $10^6$ interact via the weak interaction exchanging a $Z^0$ boson with a neutron instead of a photon with a proton.}, experiments on elastic electron scattering off nuclei safely neglect such a contribution. If  the  energy  of  the incident electrons is high enough (hundreds of MeV), they become an optimal probe of the internal structure of nuclei. Specifically, the analysis of electron-scattering data provides information about the electric charge distribution in atomic nuclei. Since the proton is not an elementary particle, in order to extract the proton distribution one needs to rely on single-proton electromagnetic form factors. The determination of the latter is accurate enough for the purpose of studying the proton distribution in nuclei. Magnetic effects and neutron contributions to the electric charge distribution can also be accounted for. Specifically, the proton rms radius can be written in good approximation, assuming HF single-particle orbitals $\rho_q({\bf r})=\sum_{nljq}v_{nljq}\phi^*_{nljq}({\bf r})\phi_{nljq}({\bf r})$ where $v_{nljq}$ represents the occupation number, as
\begin{equation}
\langle r^2_{\rm p}\rangle \approx\langle r^2_{\rm ch}\rangle - \langle r^2\rangle_{\rm prot}-\frac{N}{Z}\langle r^2\rangle_{\rm neut}-\frac{1}{Z}\left(\frac{\hbar}{M}\right)^2\sum_{nljq} v_{nljq}(2j+1)\mu_q\langle\bm{\sigma}\cdot{\bf l}\rangle   
\label{eq-gs-61}
\end{equation}
where ${nlj}$ follow the usual notation for the principal, orbital and total angular momentum quantum numbers, respectively; $\bm{\sigma}$ are the spin Pauli matrices; $\mu_q$ corresponds to the magnetic moment of a neutron/proton ($\mu_p=2.793$ and $\mu_n=-1.913$ in units of the nuclear magneton); and $\langle r^2\rangle_{\rm prot}^{1/2}=0.8783(86)$ fm and $\langle r^2\rangle_{\rm neut}=-0.1149(27)$ fm$^2$ are the single-proton/neutron electric rms and mean square radii respectively \cite{angeli2013}. The last term in Eq.\ref{eq-gs-61} is the electromagnetic spin-orbit effect. It is customary in available EDFs to adopt as a recipe $\langle r^2_{\rm p}\rangle = \langle r^2_{\rm ch}\rangle - 0.8^2$[fm$^2$] since the expected global accuracy of the EDFs in the description of charge rms radii in nuclei is of the order of a few \% and the latter approximation will not affect the overall accuracy, at least for medium and heavy mass nuclei.

\begin{figure}[t!]
\begin{center}  
  \includegraphics[width=0.39\linewidth,clip=true]{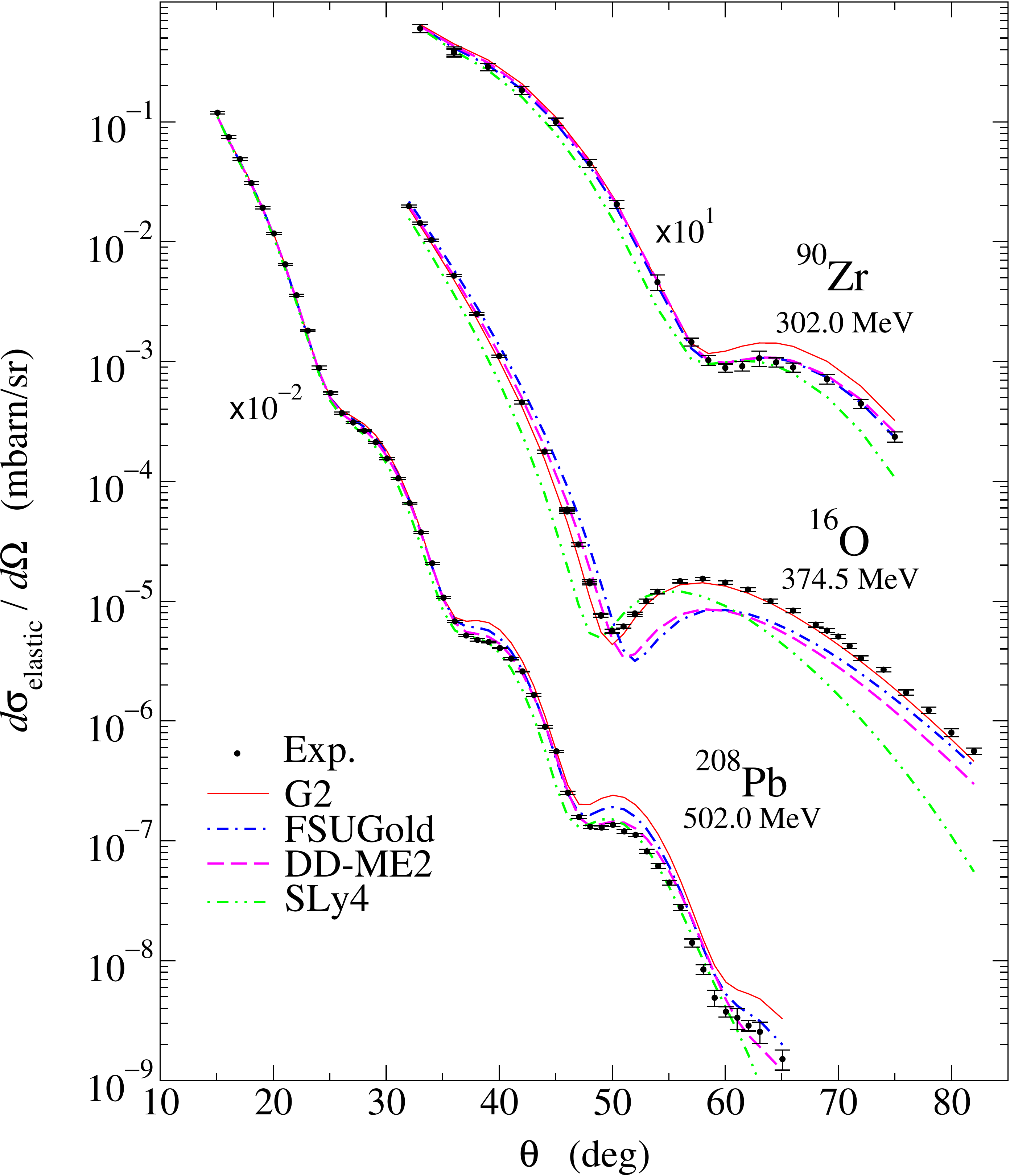}
  \includegraphics[width=0.45\linewidth,clip=true]{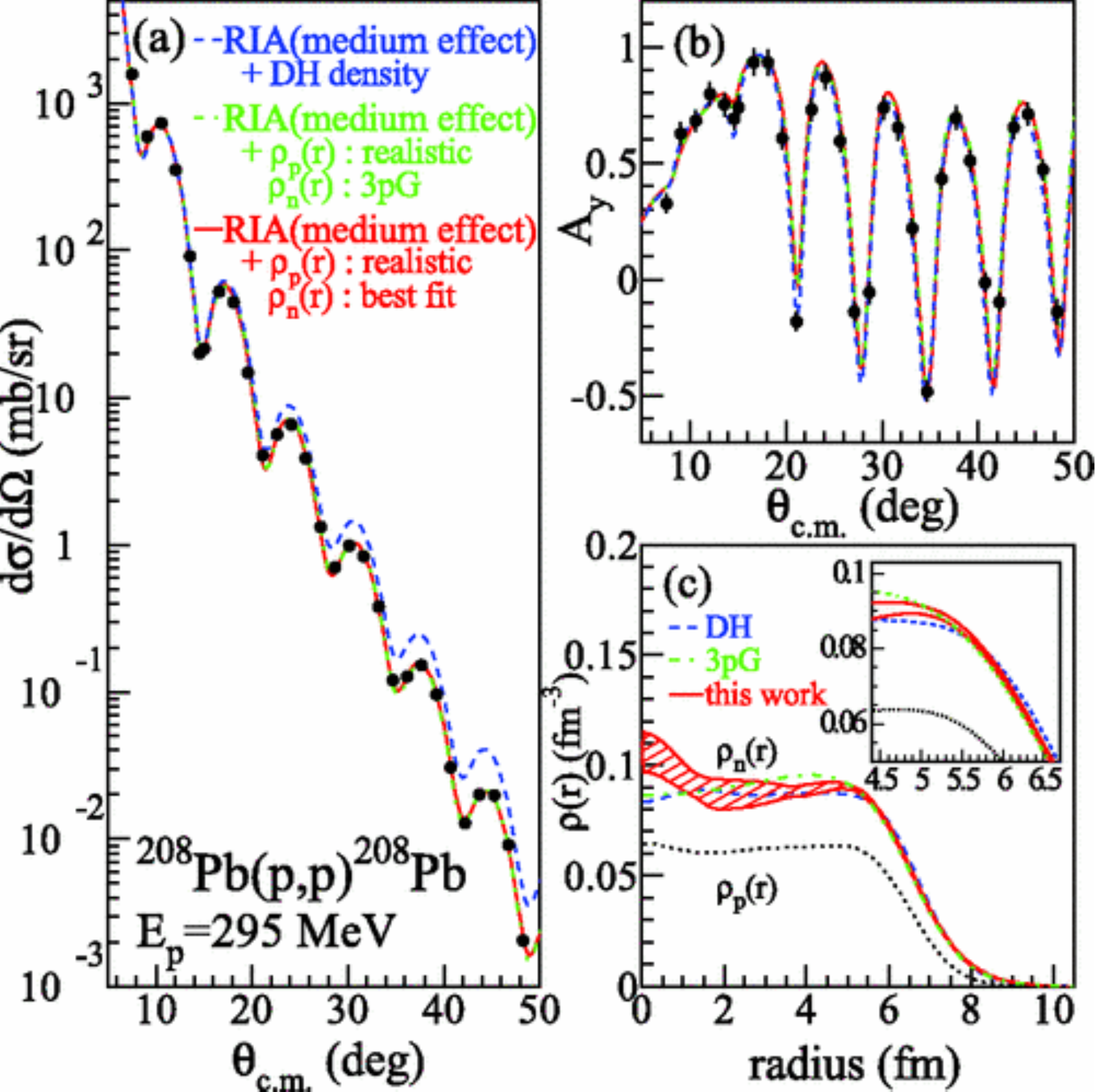}
  \caption{\label{fig-gs-19} Elastic differential cross section (DCS) for electron scattering on ${}^{16}$O, ${}^{90}$Zr, and ${}^{208}$Pb as a function of the scattering angle $\theta$, at the electron beam energies shown (left most panel). The results from the EDFs indicated in the legend are compared with the measured DCS. For details see Ref. \cite{roca-maza2008}, figure taken from the same reference. Experimental elastic proton-nucleus scattering DCS [panel (a)], analyzing power [panel (b)] and extracted neutron density of ${}^{208}$Pb with error envelope [panel (c)]. The dashed and dash-dotted lines are theoretical calculations including medium effects and realistic proton density distributions. Figure taken from \cite{zenihiro2010}.}
\end{center}
\end{figure}

In Fig.\ref{fig-gs-18}, the electric charge radii deviations with respect to the experiment are shown for two of the models that are within the most accurate in the description of nuclear masses. It is remarkable that DD-PC1 with much less parameters than HFB24 is able to reproduce charge radii at the same level of accuracy. Indeed, both show an overall relative deviation from experiment which is below 1\% and an rms deviation for nuclei with $N$ and $Z$ larger than 8 of $\sigma_{r_{\rm ch}}=0.025$ fm for HFB24 and of $\sigma_{r_{\rm ch}}=0.028$ fm for DD-PC1. Assuming that $\delta r_{\rm ch}\sim 1$ \% in well calibrated EDFs, this implies $\delta\rho_0\sim 1$ \% if one implements the simple relation $\langle r^2\rangle \sim \sqrt{3/5}r_0A^{1/3}$ where $r_0\equiv (3A/4\pi \rho_0)^{1/3}$. This result is of the same order found in the presented EDFs, that is, around a 3\% error in $\rho_0$ (cf. Table\ref{tab-gs-1}). Since $\rho_0$ is very sensitive to proton radii (charge radii), it is quite reasonable to assume the spread of this two state-of-the-art models in the prediction of $\rho_0$ as the best estimate from EDFs for the central value of the saturation density of nuclear matter (note that statistical theoretical errors have not been estimated for HFB24 and DD-PC1). The resulting range is, $0.154$ fm$^{-3} < \rho_0 < 0.158$ fm$^{-3}$.  

Regarding stable nuclei, in the left most panel of Fig.\ref{fig-gs-19}, elastic differential cross section (DCS) is shown for electron scattering on ${}^{16}$O, ${}^{90}$Zr, and ${}^{208}$Pb nuclei as a function of the scattering angle $\theta$, at several hundreds of MeV electron beam energies (indicated in the figure). The results from the EDFs (G2, FSUGold, DD-ME2 and SLy4) indicated in the legend are compared with the measured DCS (see Ref. \cite{roca-maza2008} for further details). This figure shows the accuracy of the EDFs in describing the experimental DCS and, in direct connection, the electric charge distribution in stable nuclei. Data is reproduced in most cases up to the first minimum, thus ensuring a good reproduction of the so called ``sharp radius'' which identifies the mean position of the surface. In some cases, the description is very good also for larger scattering angles, providing the information on the surface fall off of the density. Both quantities together are essential to properly reproduce the experimental charge rms radius. Theoretical studies on elastic electron scattering by exotic nuclei \cite{roca-maza2008, chu2010, meucci2013, roca-maza2013, karataglidis2017} have recently been fostered by projects such as SCRIT \cite{ohnishi2015} or ELISe \cite{antonov2011}.  

Finally, in a very recent work \cite{brown2017}, the charge radius difference in mirror nuclei has been postulated as a model independent observable that is very sensitive to neutron matter EoS and, in particular, to the slope of the symmetry energy around saturation. Keeping in mind that the neutron skin thickness of a medium-heavy or heavy nucleus is related to the $L$ parameter (cf. Fig.\ref{fig-gs-18}), this correlation might be understood as follows. Assuming exact isospin symmetry (EIS) in nuclei, one can write that $\Delta r_{\rm np}^{\rm EIS}(N,Z) = \langle r_{\rm n}^2(N,Z)\rangle^{1/2}-\langle r_{\rm p}^2(N,Z)\rangle^{1/2}=\langle r_{\rm p}^2(Z,N)\rangle^{1/2}-\langle r_{\rm p}^2(N,Z)\rangle^{1/2}$. Considering now that the electric charge rms radius can be written in good approximation as $\langle r^2_{\rm ch}\rangle = \langle r^2_{\rm p}\rangle + 0.8^2$[fm$^2$] (see previous discussion), one finds that, in exact isospin symmetry, the neutron skin thickness of a nucleus with $N$ neutrons and $Z$ protons should correlate with the difference between its charge radius and the charge radius of its mirror nucleus: $\Delta r_{\rm np}^{\rm EIS}(N,Z) = \langle r_{\rm ch}^2(Z,N)\rangle^{1/2}-\langle r_{\rm ch}^2(N,Z)\rangle^{1/2}$. Isospin symmetry, however, is broken in nuclei. The leading isospin symmetry breaking (ISB) term is due to the Coulomb interaction but also the neutron-proton mass difference or ISB terms in the nuclear strong interaction contribute. In Ref.\cite{brown2017} it has been shown on the basis of different EDFs that the correlation holds also when the Coulomb interaction is taken into account. So it has to be confirmed that other ISB terms do not spoil this relation. If this is confirmed, the challenge is to individuate nuclei and its mirror counterpart that have been (or could be in the future) accurately measured. It is important to note, that the physics behind the differences in the charge radii in mirror nuclei is related to the physics of the Isobaric Analog State excitation energy (see Sec.\ref{excitations})  \cite{roca-maza2017b}. The advantage of this other observable is that has been accurately measured in many nuclei \cite{antony1997}.

\subsubsection{Neutron density distributions and neutron skin thickness}

\begin{figure}[t!]
\begin{minipage}{0.6\linewidth}
    \includegraphics[width=0.9\linewidth,clip=true]{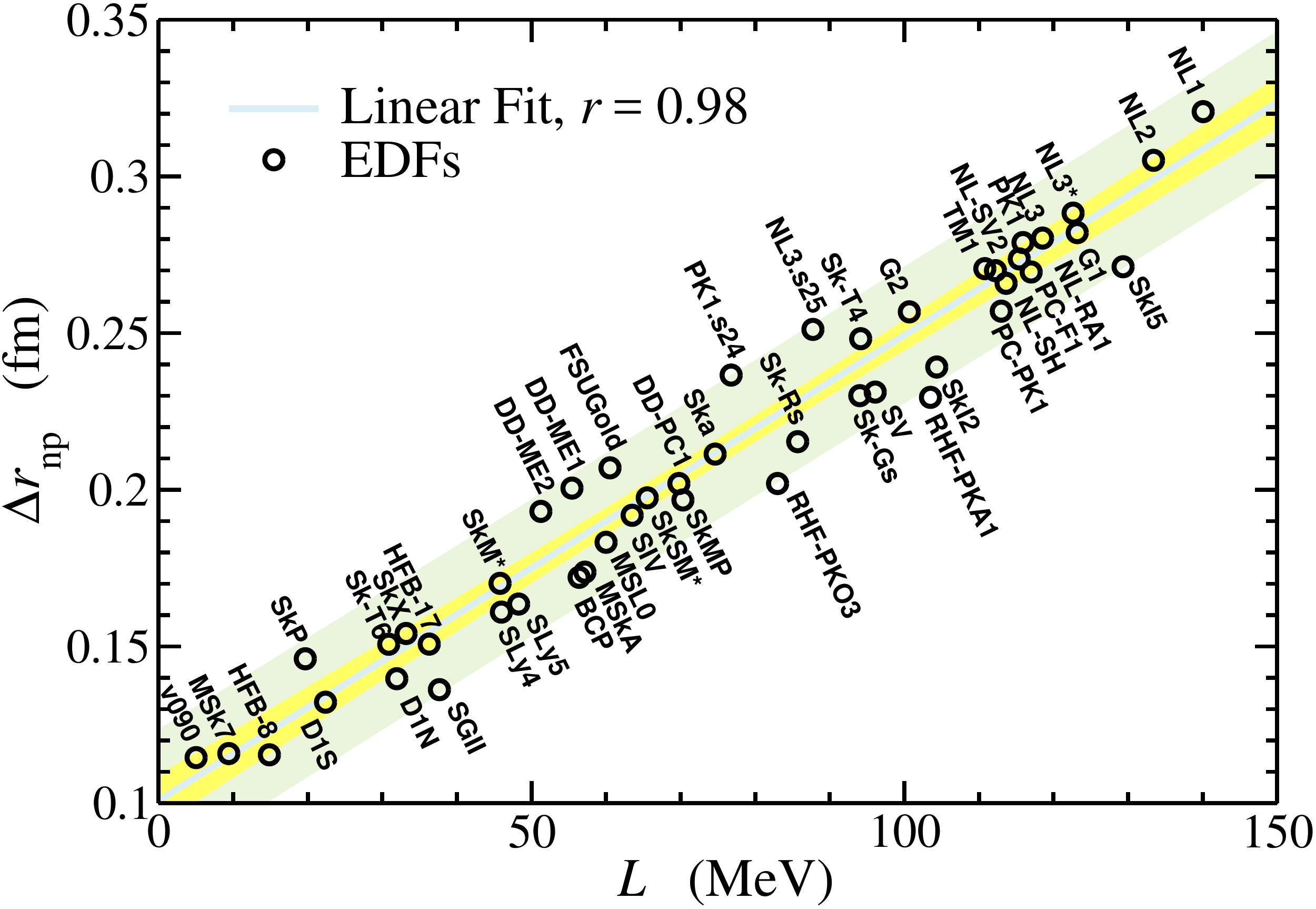}
    \includegraphics[width=0.9\linewidth,clip=true]{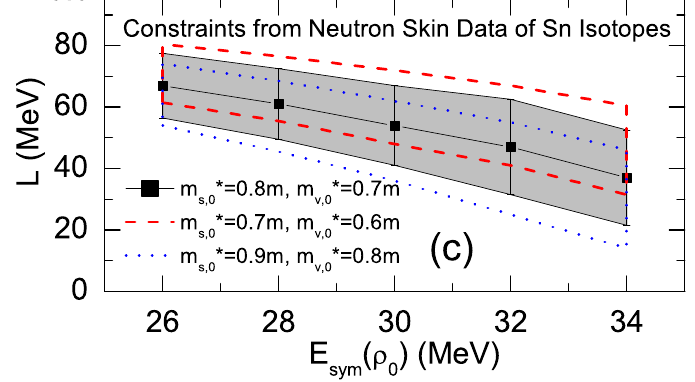}
  \end{minipage}
  \begin{minipage}{0.35\linewidth}  
    \includegraphics[width=\linewidth,clip=true]{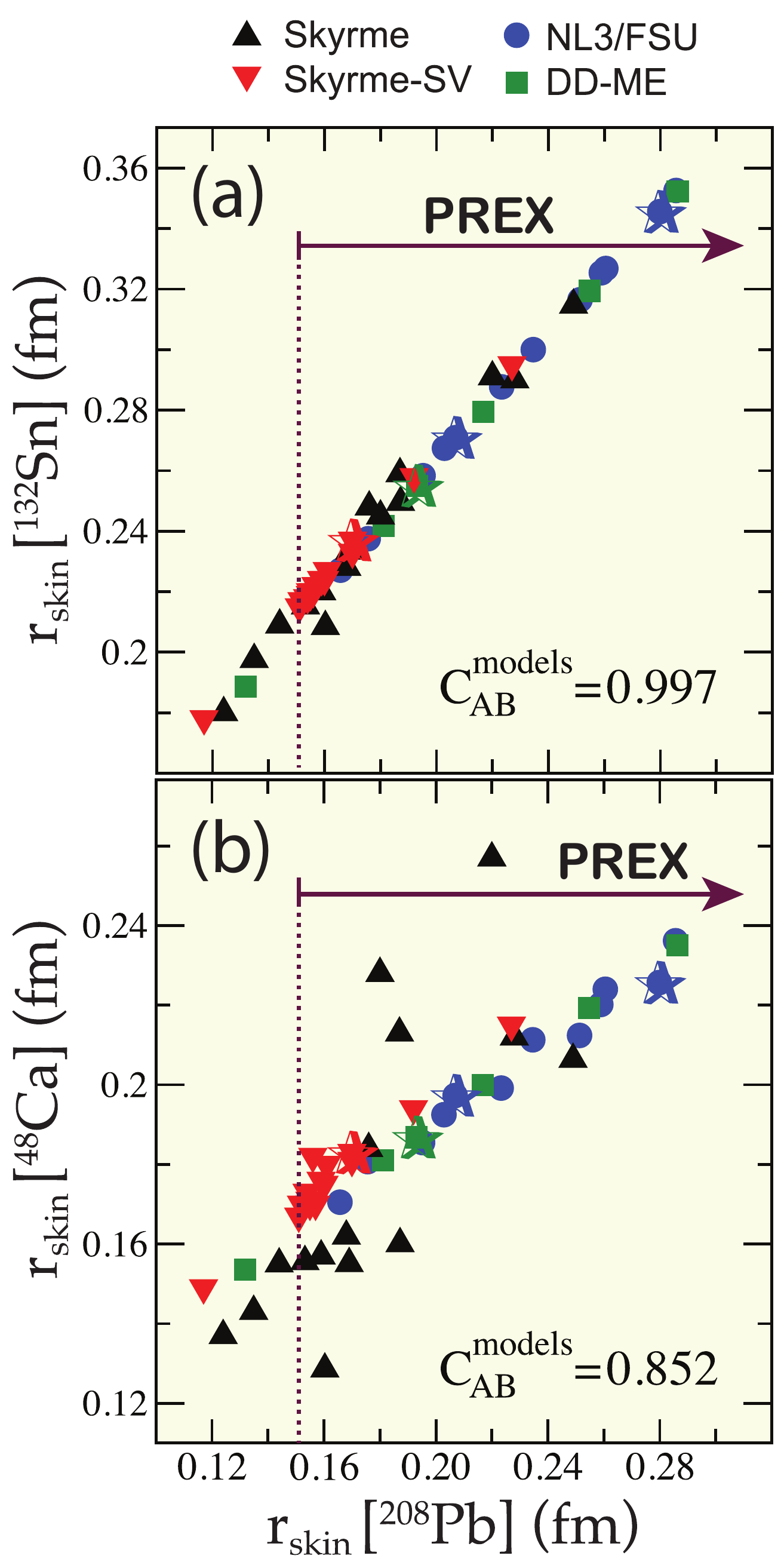}
  \end{minipage}  
  \caption{\label{fig-gs-20} Left upper panel: neutron skin thickness of ${}^{208}$Pb as predicted by a large set of EDFs against the $L$ parameter. The inner (outer) colored regions depict the loci of the 95\% confidence (prediction) bands of the regression. A linear fit is also shown. Figure modified from Ref.\cite{roca-maza11}. Right panels: predictions from several EDFs for the neutron skin thickness of ${}^{208}$Pb and ${}^{132}$Sn (a) and ${}^{48}$Ca (b). Constraints on the neutron skin thickness from PREX\cite{prex} have been incorporated into the plot. Figure taken from Ref.\cite{piekarewicz12}. Left lower panel: Constraints from a theoretical analysis on $L$ and $J$ (here written as $E_{\rm sym}(\rho_0)$) from a $\chi^2$ analysis of available data from proton scattering, $\alpha$ scattering, and antiprotonic atoms on the neutron skin thickness of some tin isotopes. Bands and dashed as well as dotted lines represent an estimation of the error. Figure taken from Ref.\cite{Chen2010}.}
\end{figure}

The experimental determination of neutron density distributions is limited even for stable nuclei. As the neutron total electric charge is zero, neutron densities have been probed mostly by using strongly interacting particles \cite{hoffman1980,zenihiro2010,Aumann2017} and nuclear effects in exotic atoms \cite{klos2007,friedman2009}. Even if some of these experiments report small errors, all hadronic probes require model assumptions to deal with the strong force introducing systematic uncertainties. By the contrary, parity violating elastic electron scattering is a sensitive and model independent probe of the weak charge in nuclei that is essentially determined by the neutron density \cite{prex}. However, the latter technique is very challenging experimentally and can only be thought to be applied to stable nuclei at the moment. 

Since the proton rms radius of many nuclei is well known from parity conserving electron elastic scattering, to study the neutron rms radius or the neutron skin thickness ($\Delta r_{\rm np} = \langle r_{\rm n}^2\rangle^{1/2}-\langle r_{\rm p}^2\rangle^{1/2}$) can be regarded as equivalent. Recently, many efforts have been devoted to the theoretical and experimental studies of the neutron skin thickness in medium and heavy nuclei \cite{tsang12,s-epja,Horowitz:2014bja} since it has been shown to be linearly correlated with the $L$ parameter of the EoS (see Refs.\cite{furnstahl02,brown00} and cf. Fig.3 of Ref.\cite{roca-maza11}). In the left uppermost panel of Fig.\ref{fig-gs-20}, the neutron skin thickness of ${}^{208}$Pb as predicted by a large set of EDFs against the $L$ parameter is shown. It is interesting to understand if the neutron skin thickness of other nuclei also contains the same physical information. This can be qualitatively understood by inspecting the right most panels in Fig.\ref{fig-gs-20} where predictions from several EDFs for the neutron skin thickness of ${}^{208}$Pb and ${}^{132}$Sn (a) and ${}^{48}$Ca (b) are displayed \cite{piekarewicz12} (constraints on the neutron skin thickness from the parity violating elastic electron scattering ${}^{208}$Pb Radius Experiment (PREX) \cite{prex} have been incorporated into the plot). These panels clearly show that, in the basis of EDF, the same information can be accessed if the neutron skin thickness of ${}^{208}$Pb or ${}^{132}$Sn are measured. This is not exactly the case for ${}^{48}$Ca: the neutron skin thickness of ${}^{48}$Ca and ${}^{208}$Pb (or ${}^{132}$Sn) do not account for exactly the same information--their do not correlate one to one--and, thus, measuring both can be of relevance. This indicates that $L$ should be expected to correlate well only with the neutron skin of heavy nuclei, something that is well justified by macroscopic models of the nucleus (cf. Ref.\cite{centelles09} and Eq.\ref{eq-pheno-3}). Next, in left lower panel of the same figure, constraints from a theoretical analysis on $L$ and $J$ (written as $E_{\rm sym}(\rho_0)$ in the figure) from a $\chi^2$ analysis of available data from proton scattering, alpha scattering, and antiprotonic atoms (see Ref.\cite{Chen2010} for details) on the neutron skin thickness of some tin isotopes are depicted. This analysis corresponds to the band labeled ``neutron skin'' in Fig.\ref{fig-eos-11} and it is important to spend some words on these results. The authors of this work found an anti-correlation between $J$ and $L$. This is against all other analysis done so far in the literature (see for example \cite{s-epja,Horowitz:2014bja}) so one should take this results with great care. Without going into detail, the simplest consideration one may follow to understand the linear correlation between $J$ and $L$ is to explore the EoS around saturation (the neutron skin thickness is a ground state property) without assuming a model. That is to write $S(\rho)\approx J - L\frac{\rho_0-\rho}{3\rho_0}$. We know that for heavy nuclei like Sn isotopes the average density, to which the EoS is most sensitive, is around 0.1 fm${}^{-3}$ \cite{centelles09}. Therefore, within a local density approximation, the symmetry energy of the finite nucleus will approximately correspond to $S(\rho\approx 0.1$ fm${}^{-3})\approx J - L/8$ and, thus, $J$ should be positively correlated with $L$ and show a slope not far from $0.125$ [cf. Eqs.(12-14) in Ref.\cite{roca-maza2015b}]. 

An example of the latest attempts to experimentally determine the neutron skin thickness of ${}^{208}$Pb from the measurement of the ground state neutron distribution and, therefore, learn about the density dependence of the symmetry energy around saturation is presented in the right most panels of Fig.\ref{fig-gs-19}~\cite{zenihiro2010}. In this work polarized proton elastic scattering off nuclei at intermediate energies ($\sim$300 MeV) has been performed. This technique is not exempt from uncertainties related to the nucleon-nucleon interaction but medium effects have been calibrated in a transparent way and, thus, it is believed to be one of our best means to experimentally access the neutron distributions in heavy nuclei by using hadronic probes. Specifically, in the right most panels of Fig.\ref{fig-gs-19}, the experimental elastic proton-${}^{208}$Pb scattering DCS [panel (a)], analyzing power [panel (b)] and extracted neutron density of ${}^{208}$Pb with error envelope [panel (c)] are shown. The dashed and dash-dotted lines are theoretical calculations including medium effects and realistic proton density distributions taken from the literature on elastic electron scattering. The obtained value for the neutron skin thickness of ${}^{208}$Pb is $\Delta r_{\rm np} = 0.211^{+0.054}_{-0.063}$ fm. This result supports a range of values of $40$ MeV$\lesssim L\lesssim 110$ MeV which is quite large (cf. from left uppermost panel in Fig.\ref{fig-gs-20}). The same technique was previously applied to some tin isotopes in Ref.\cite{terashima2008}, for the results see Table III of this reference. 

Another example corresponds to the global analyses of antiprotonic and pionic atoms that show a reasonably good agreement of the root-mean-square radii of the neutron distributions \cite{friedman2009}. Techniques based on antiprotonic atoms \cite{klos2007,antip1} have been recently revitalized\footnote{See PUMA: anti-Proton Unstable Matter Annihilation EU project at \url{http://cordis.europa.eu/project/rcn/212041_en.html}.}. When a slow anti-proton is captured by an atom it emits X-rays until the final annihilation with a proton or a neutron of the atomic nucleus occur. This gives information on the tail of the neutron and proton distributions since anti-matter is very sensitive to the presence of matter. This technique is not exempt from strong interaction uncertainties. In Ref.\cite{antip1}, the neutron skin of 26 stable nuclei from Ca to U was reported by using this technique (see Fig.\ref{fig-pheno-3}). It was empirically demonstrated that the neutron skin thickness in stable nuclei grows linearly with the neutron excess (as expected also from macroscopic considerations, see Eq.\ref{eq-pheno-3}).  

Information on the neutron skin has also recently been extracted from coherent pion photo-production cross sections \cite{tarbert2014,gardestig2015}. In this technique, the target nucleus remains in its ground state. For photon energies of around 200 MeV, the $\Delta$ excitation is dominant and the amplitudes for neutron and proton $\Delta$ excitations are expected to be the same. Therefore $(\gamma,\pi^0)$ cross section is equally sensitive to both nucleon distributions. Compared to other hadronic probes, this reaction is not complicated by initial state interactions but for final state interactions instead. Assuming the proton distribution as known, the neutron distribution can be parameterized and the constants fitted to reproduce the above mentioned cross section. By doing this, the neutron skin thickness of ${}^{208}$Pb found was $\Delta r_{\rm np}=0.15\pm0.03({\rm stat.})^{+0.01}_{-0.03}({\rm sys.}) $ fm that implies $0$ MeV$\lesssim L\lesssim 60$ MeV which is also quite large (cf. from left uppermost panel in Fig.\ref{fig-gs-20}).

Parity violating elastic electron scattering is the only experimental technique up to now that has allowed to measure, in a model independent way, the weak charge rms radius of a heavy nucleus such as ${}^{208}$Pb \cite{prex}. The weak charge radius essentially determines the rms of the neutron distribution \cite{donnelly1989,horowitz01c,horowitz12,roca-maza11}. This experimental technique exploits the fact that ultra-relativistic electrons interact with the nucleus by the Coulomb plus or minus the weak interaction depending only on their helicity of the incident electrons. Therefore, by measuring the so called longitudinal asymmetry (or parity violating asymmetry), which is the difference between the DCS of electrons with the spin aligned in the direction of the beam with those anti-aligned with it and divided by the sum of them, one is sensitive to the weak charge distribution. That is, the parity violating asymmetry is a kind of interference observable that gives access to the very small effect due to the weak interaction. In the only parity violating experiment done so far\footnote{further experiments are planed at the Jefferson Laboratory in USA and MESA at Mainz.}, the neutron skin of ${}^{208}$Pb was determined to be $\Delta r_{\rm np}=0.33^{+0.16}_{-0.18}$ fm. Unfortunately, this type of experiment is very challenging and the achieved error was too large, preventing from a tight constraint on $L$ (cf. left uppermost panel in Fig.\ref{fig-gs-20}). 

In this section, we have overviewed some recent experiments that probe the ground state density distributions in the atomic nucleus. The neutron skin thickness can also be accessed via the measurement of collective excitations that probe nuclear restoring forces after the ground state densities have been slightly perturbed by some convenient external probe. We will focus on collective excitations in the next section. In addition, proton and neutron density distributions can be deduced from heavy-ion collisions as well. We refer the reader to Refs.\cite{baran2005,gulminelli2006,li2008,s-epja,tsang12,Horowitz:2014bja} since this topic will not be covered in this review. 

Finally, from the astrophysical side, low-mass ($\sim 0.5 M_\odot$) neutron stars with central densities close to $\rho_0$ provide information about the EoS at such densities. Those stars are rare and, thus, difficult to observe but a strong correlation between the neutron radius of ${}^{208}$Pb and the radius of a $\sim 0.5 M_\odot$ neutron star seems to be reasonable \cite{carriere2003,lattimer2007}. Thus, the radius of a $\sim 0.5 M_\odot$ neutron star can be inferred from a measurement of the neutron radius of ${}^{208}$Pb and, eventually, the neutron skin thickness in ${}^{208}$Pb (or $L$) might be deduced from observational data \cite{fattoyev2018}. Recently, neutron star physics have received a new strong boost, as the LIGO-Virgo collaboration announced the first detection of gravitational waves from a binary neutron star merger, setting a new type of constraint on the radius of a neutron star \cite{LIGO2017}.

%% file: excitations.tex
\section{Excitations in nuclei and EOS}
\label{excitations}
% general introduction on the modes
% mention modes that we don not cover (IS,IV M1,IV monopoles - large uncertainties)
In this section, the properties of different collective excitations (so called Giant Resonances) that have been shown to be sensitive to the parameters of the nuclear EoS will be presented. We do not aim at covering all possible excitations in nuclei or discuss features that are not related to the aim of this review. Time dependent density functional theory (and its small amplitude limit: RPA) is the only theoretical tool available in the literature that can be systematically applied for a reliable study of the excitation energy and sum rules of collective excitations along the nuclear chart. Hence, we will focus in this section mainly on the predictions based on the EDFs.  

%Constraining simultaneously nuclear symmetry energy and neutron-proton effective mass splitting with nucleus giant resonances using a dynamical approach
%By: Kong, Hai-Yun; Xu, Jun; Chen, Lie-Wen; et al.
%PHYSICAL REVIEW C   Volume: 95   Issue: 3     Article Number: 034324   Published: MAR 29 2017 

\subsection{Isoscalar modes}
\subsubsection{Isoscalar giant monopole resonance}

The excitation energies of isoscalar monopole resonance are essential to constrain the incompressibility of nuclear matter $K_0$. Early review on the relation between the compressibility of nuclear matter and the frequencies of the collective monopole vibrations of nuclei is given in Ref.~\cite{Blaizot1980}.
Over the past years, a large number of theoretical 
studies, supplemented with the measurements based on advanced experimental 
methods, have determined the limits for $K_0$ (see also discussion in Sec.\ref{pheno}).  Recent theoretical studies are mainly
based on microscopic approaches based on nuclear energy density functionals 
(relativistic or non-relativistic) and macroscopic models.

From the experimental side, inelastic $\alpha$ scattering at small angles became a standard
approach to measure the ISGMR excitation spectra~\cite{Youngblood1999,Youngblood1999b,Youngblood2004,Youngblood2007,Youngblood2009,Youngblood2013,Youngblood2015,Li2007,Li2007b,li2010,Liu2011,Patel2013}. Recently, the ISGMR in $^{116}$Sn and $^{208}$Pb  have been
investigated using small-angle inelastic scattering of deuterons and multipole decomposition analysis, indicating that this approach could be employed to study the ISGMR in radioactive isotopes in inverse kinematics~\cite{Patel2014}.
In order to reduce 
uncertainty in the prediction for $K_0$, it is necessary to employ fully self-consistent
theory frameworks. The extracted values of $K_0$ are model dependent~\cite{Colo2004}.
While in previous studies there was a discrepancy between predictions for $K_0$ from
non-relativistic and relativistic approaches~\cite{Colo2004b}, more recently, models based on 
Gogny, Skyrme and relativistic functionals reach a satisfactory level of consistency in $K_0$.
A stringent constraint has been established from the microscopic theory based on the
RPA and experimental data on isoscalar 
giant monopole and dipole resonances in closed shell nuclei, $K_0=(240 \pm 20)$ MeV~\cite{shlomo2006}. However this result is still devated since analysis on open shell nuclei points towards lower values of $K_0$ \cite{Avogadro2011,colo2017}. 
In the following we review a selection of recent advances in addressing compressional modes in nuclei to constrain $K_0$.

% macroscopic model analysis J. Stone
In Ref.~\cite{stone2014} the  incompressibility
of finite nucleus $K_A$ is parameterized in the form of a leptodermous expansion in powers
of $A^{-1/3}$~\cite{Blaizot1980} (cf. Eq.\ref{eq-pheno-32}). Leptodermous expansion represents a parameterized description
that allows a connection between the experimental data and parameters of the expansion,
$K_{vol}$,$K_{surf}$,$K_{curv}$,$K_{\tau}$, and $K_{coul}$, that is, volume, surface,curvature, isospin and Coulomb contributions, respectively. The contributions from these terms have been determined from a fit to values of $K_{A}$, obtained from the experimental
excitation energies of the GMR (cf. Eq.\ref{eq-pheno-27}), and assuming that the coefficient of the volume term $K_{vol}$ corresponds to $K_0$. Within this model, that neglects microscopic effects, the result from Ref.~\cite{shlomo2006}, $K_0=(240 \pm 20)$ MeV was reproduced with the assumption that the ratio of the surface and volume coefficients $c=K_{surf}/K_{vol}$ equals -1, that is consistent with most of the energy density functionals~\cite{stone2014}. However, by releasing the ratio $c$, it was shown that the fits significantly improve
and acquire larger values of $K_0$, that is $250 < K_0 < 315$ MeV~\cite{stone2014} (see also discussion in Sec.\ref{pheno}).
 
% see references of ISGMR in Ref.~\cite{stone2014}.

\begin{figure}[t!]
\centering
\includegraphics[width=0.46\linewidth,clip=true]{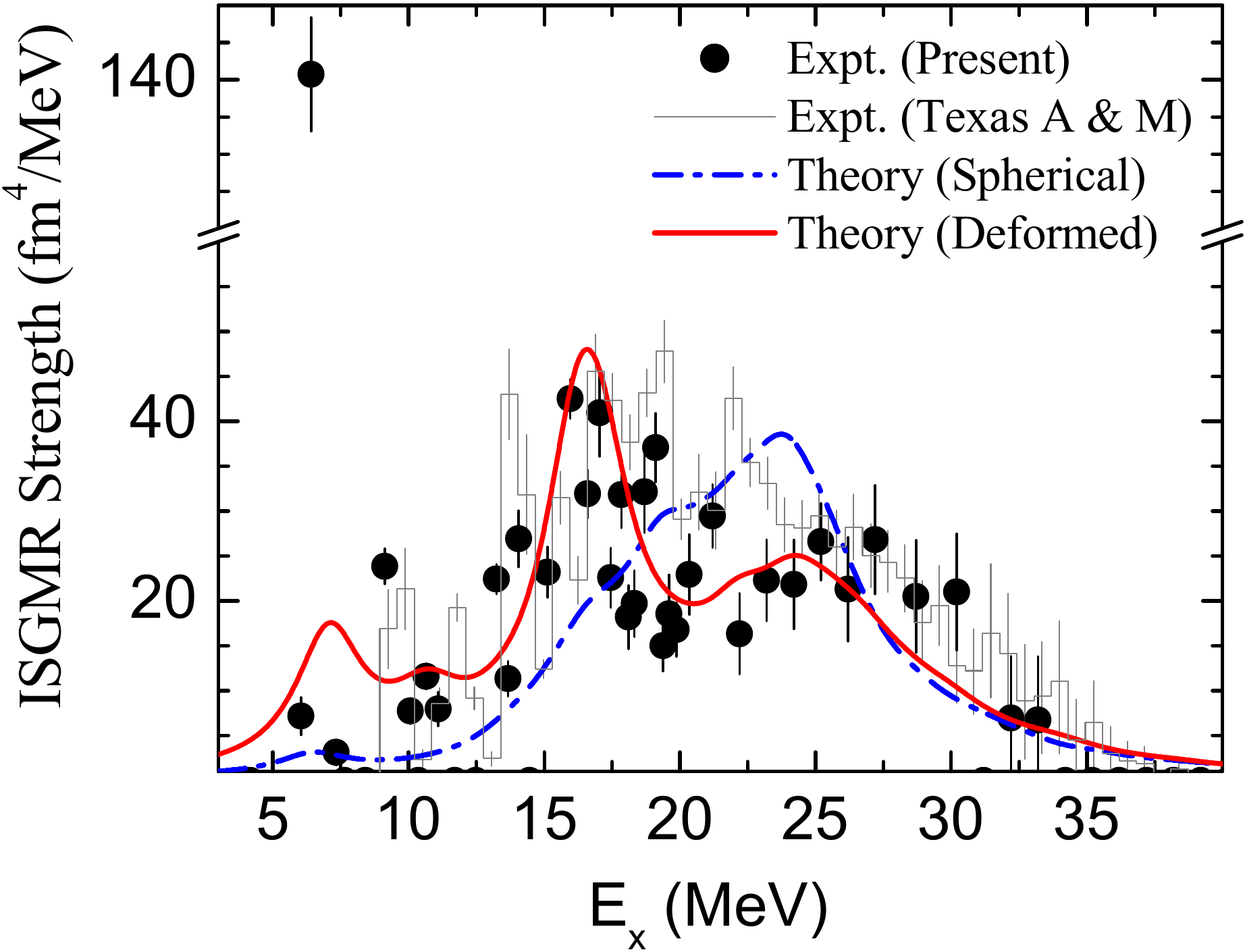}
\includegraphics[width=0.45\linewidth,clip=true]{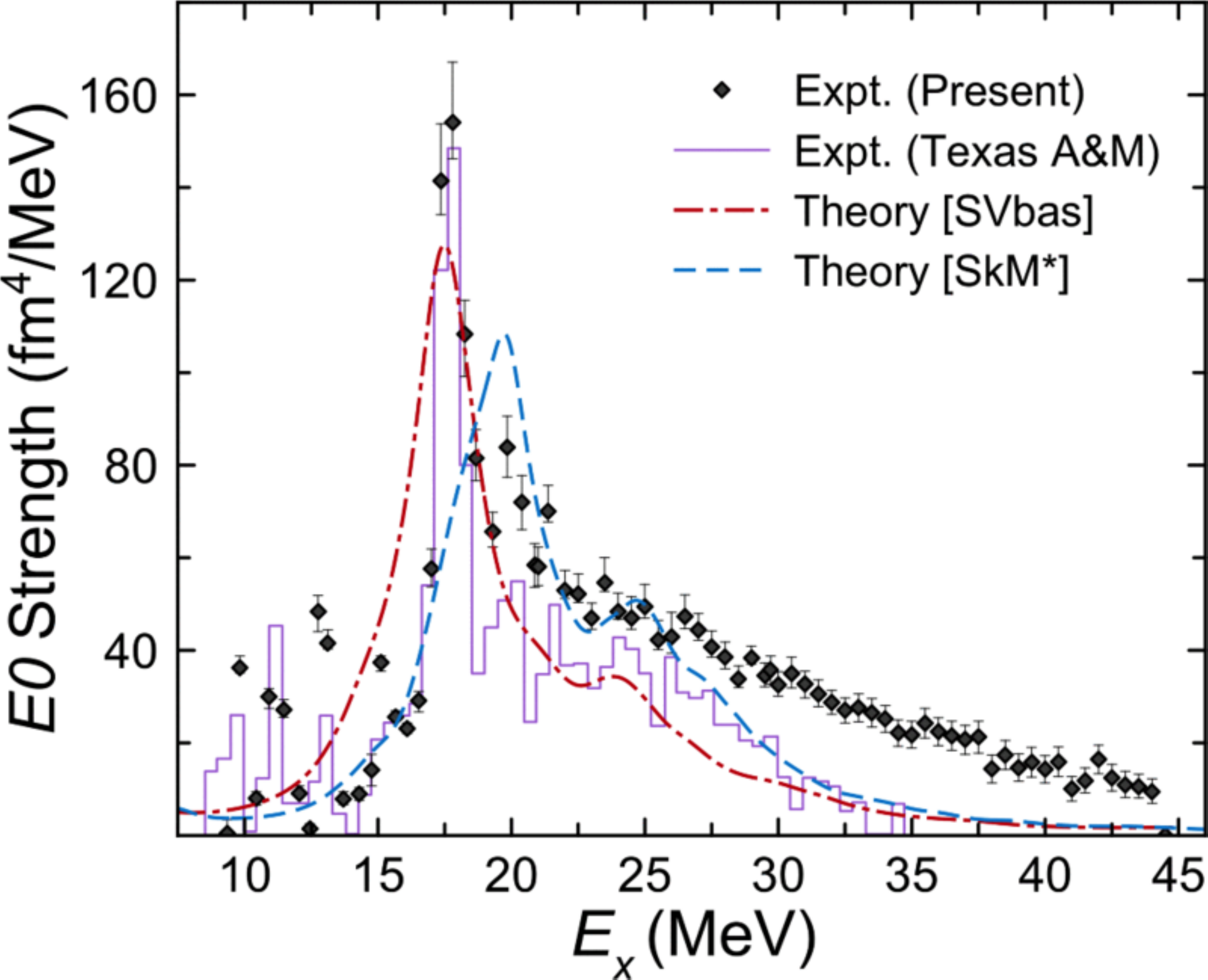}
\caption{\label{fig-mono-peach} 
Left Panel: The ISGMR strength distribution in $^{24}$Mg from RCNP~\cite{Gupta2015} and Texas A$\&$M ~\cite{Youngblood2009} inelastic
$\alpha$ scattering, in comparison with the results of spherical and deformed HFB+QRPA calculations for the SkM* functional. Figure taken from  Ref.~\cite{Gupta2015}.
Right panel: The ISGMR strength distributions from
RCNP~\cite{Peach2016} and by Texas A$\&$M~\cite{Youngblood2007} data, in comparison with the results from 
deformed QRPA calculations with Skyrme functionals SkM* and SVbas ~\cite{Peach2016,Kvasil2016}. Figure taken from  Ref.~\cite{Peach2016}. }
\end{figure}

%Effect of ground-state deformation on GMR
The properties of ISGMR, as well as other multipoles, are sensitive to the ground state
deformation of nucleus. The ISGMR strength distributions have been measured
in light deformed nuclei $^{24}$Mg and $^{28}$Si using inelastic scattering of $\alpha$ particles \cite{Youngblood1999, Youngblood2009,Gupta2015} and $^{6}$Li particles at small angles \cite{Chen2009}, resulting in considerable fragmentation of the spectra and splitting of the ISGMR. 
However, only recently it has been shown that a two-peak structure of the ISGMR strength distribution in $^{24}$Mg can be explained by the deformation of the ground state \cite{Gupta2015}. Fig.~\ref{fig-mono-peach} (left panel) shows the ISGMR strength distribution in $^{24}$Mg from inelastic $\alpha$ scattering at Texas A$\&$M ~\cite{Youngblood2009} and RCNP~\cite{Gupta2015}, resulting in fragmented spectra with a pronounced double-hump structure in the latter case. Similar structure has previously been observed in $^{154}$Sm, nucleus with a deformed ground state~\cite{Youngblood1999b,Youngblood2004,Itoh2003}. As shown in Fig.~\ref{fig-mono-peach} comparison of the ISGMR strength distributions for $^{24}$Mg obtained in Hartree-Fock-Bogliubov (HFB) plus quasi-particle RPA calculations (QRPA) for spherical and prolate-deformed cases clearly demonstrate the importance of deformation effects to reproduce measured two-peak structure. Similar study of ISGMR $^{28}$Si~\cite{Peach2016} showed the effect of the ground state deformation, and the QRPA calculations based on the SVbas functional for an oblate-deformed ground state could reproduce the experimental data (right panel in Fig.~\ref{fig-mono-peach}). Giant resonances,
including ISGMR, have also been explored in axially-symmetric-deformed QRPA based on Gogny D1S effective interaction in deformed $^{26-28}$Si and $^{22-24}$Mg, demonstrating the impact of 
the intrinsic nuclear deformation on collective nuclear excitations~\cite{Peru2008}. The ISGMR displays splitting in deformed nuclei that is not present if the QRPA is based on spherical HFB ground state. Deformation induced splitting of the ISGMR has also been systematically studied in a wide range of masses covering medium, rare-earth,
actinide, and superheavy axial deformed nuclei, in the framework of fully self-consistent QRPA based on the Skyrme functional~\cite{Kvasil2016}.

\begin{figure}[t!]
\centering
\includegraphics[width=0.35\linewidth,clip=true]{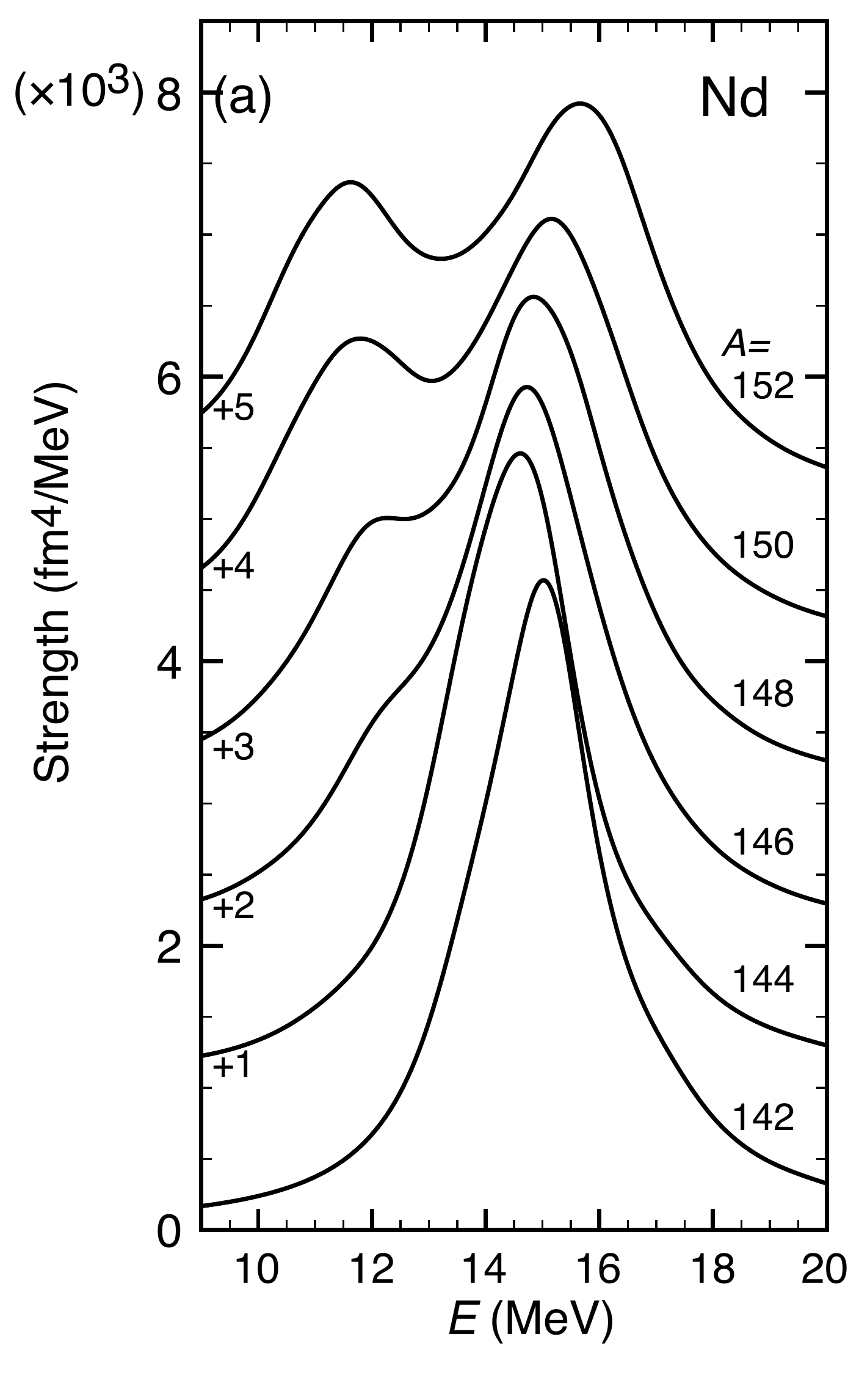}
\includegraphics[width=0.276\linewidth,clip=true]{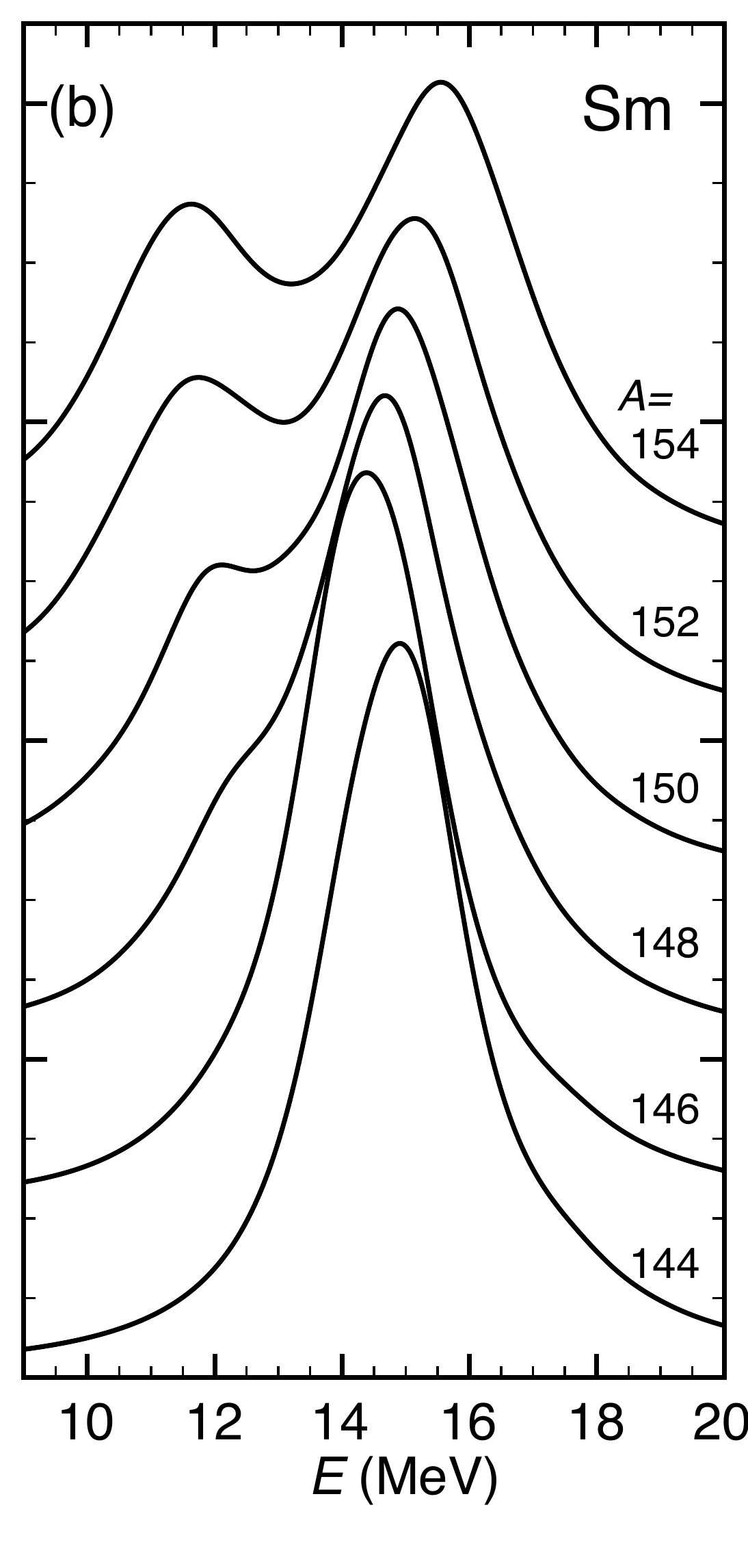}
\caption{\label{fig_mono_yoshida} 
The evolution of ISGMR strength distributions in Nd (left panel) Sm (right panel) isotopes, based on deformed HFB+QRPA (SkM*) calculations. Strength distributions are shifted for different isotopes as denoted in figures. Figure taken from  Ref.~\cite{Kenichi2013}.}
\end{figure}

% Effect of deformation on ISGMR in Sm and Nd isotopes
In the implementation of deformed HFB+QRPA based on Skyrme energy density functional with SkM*,SLy4,SkP parameterizations, the evolution of strength distributions for giant resonances has been studied in Nd and Sm isotopes, as shown in Fig.~\ref{fig_mono_yoshida} for the case of ISGMR~\cite{Kenichi2013}. While in spherical isotopes a sharp peak around 15 MeV is obtained, deformed nuclei have a double-peak structure. In the case of $^{154}$Sm, lower and higher peak exhaust 31.4$\%$ and 60.6$\%$ of the EWSR, respectively~\cite{Kenichi2013}.
The magnitude of the peak splitting and the fraction of the EWSR in the lower peak of the ISGMR reflect the nuclear deformation effect. 
Model calculations of the ISGMR and other compression modes showed a consistency of the results with the nuclear matter incompressibility $K\simeq$210-230 MeV and the effective mass $m^*_0/m \simeq$ 0.8-0.9. Only the high-energy octupole resonance resulted in some deviation, indicating a smaller effective mass, and need for further precise measurements of this mode.

% Mo and Zr isotopes exp.
The ISGMR has also been systematically explored by inelastic scattering of $\alpha$ particles at small angles including 0${}^\circ$ in isotopes $^{92,96,98,100}$Mo and$^{90,92,94}$Zr~\cite{Youngblood2013,Youngblood2015,Krishichayan2015}. The HF+RPA calculations with KDE0v1 Skyrme functional reproduce the main ISGMR peak, however, with $~$2-3 MeV higher excitation energies, and high-energy tail could not be reproduced~\cite{Youngblood2015}. Further theoretical studies including pairing correlations, deformation, and other effective interactions are necessary to clarify the respective experimental data.
% The role on compressibility

\begin{figure}[t!]
\centering
\includegraphics[width=0.5\linewidth,clip=true]{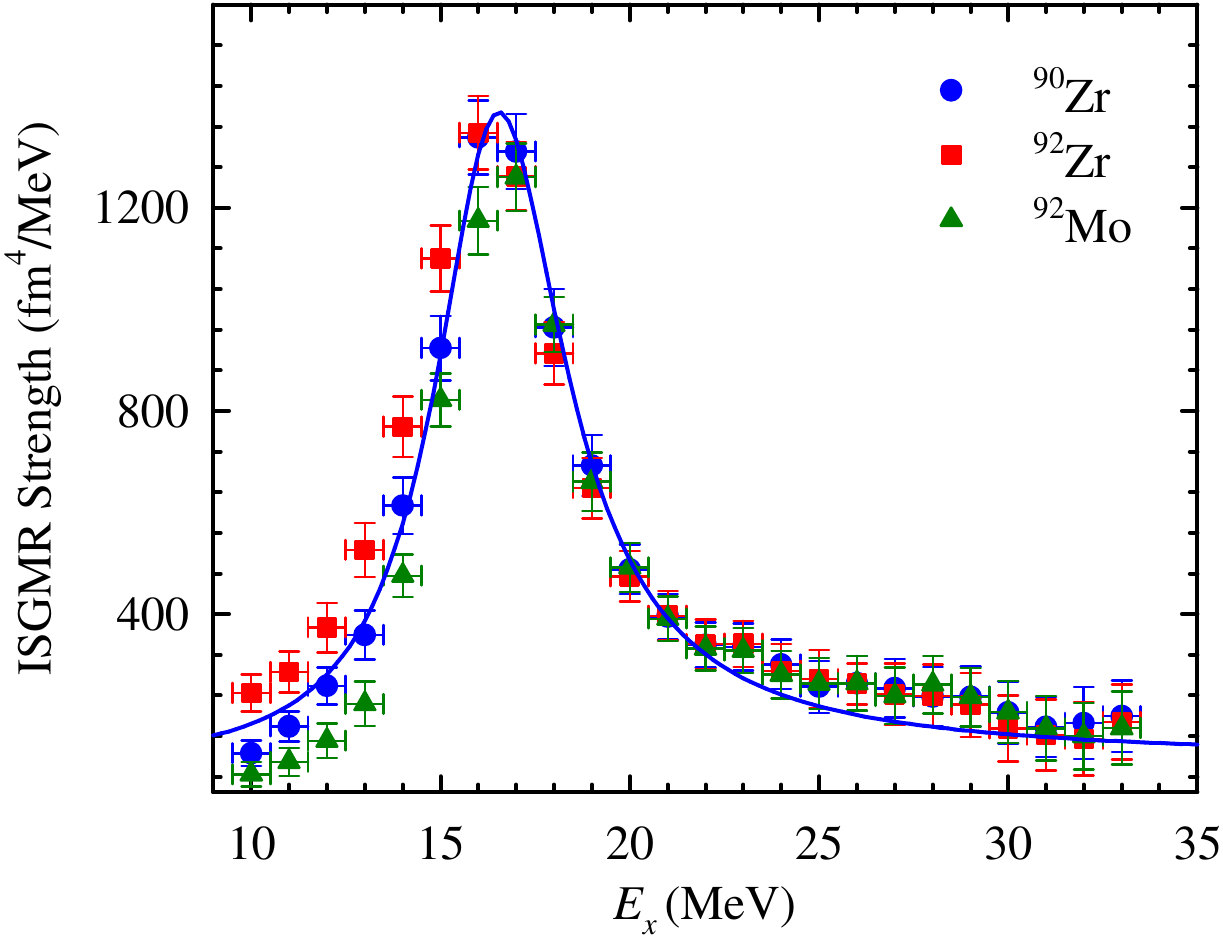}
\caption{\label{fig_mono_gupta} 
The ISGMR strength distributions for $^{90,92}$Zr and $^{92}$Mo from inelastic $\alpha$-particle scattering at extremely forward angles at RCNP. Figure taken from  Ref.~\cite{Gupta2016}.}
\end{figure}

As pointed out in Ref.~\cite{Youngblood2013}, the basic assumption used to determine incompressibility of nuclear matter from the 
energy of the ISGMR has been that its energy is not affected by the details of the nuclear structure beyond the general features contained
in the calculation of the ISGMR energy for a specific nucleus. This assumption has been tested in the analysis of Zr and Mo isotopes~\cite{Youngblood2013}, showing that some difficulties exist. The ISGMR energies in A=92 nuclei lead to nuclear compressibility
considerably above those of other nuclei nearby, thus raising a question on the impact of nuclear structure effects on the ISGMR energy and 
the corresponding incompressibility of nuclear matter~\cite{Youngblood2013}. In contrast with Ref.~\cite{Youngblood2013}, more recent study of inelastic $\alpha$-particle scattering on $^{90,92}$Zr and $^{92}$Mo at RCNP resulted in the ISGMR strength distributions that coincide for these nuclei, as shown in Fig.~\ref{fig_mono_gupta}. Therefore, in this case the incompressibility of nuclear matter is not influenced by the nuclear shell structure~\cite{Gupta2016}. It is argued that the origin of discrepancies in the results may be in the background subtraction approach. In Ref.~\cite{Gupta2016}, the background subtraction leaves the physical continuum as part of the excitation spectra, while in Ref.~\cite{Youngblood2013} additional assumptions are imposed on the
shape of the background, that may also subtract the physical continuum~\cite{Gupta2016}.

%FAM
Recent implementation of the finite amplitude method (FAM) in the construction of the QRPA, using iterative algorithms with adopted
generalized conjugate residual method, provides a novel approach to describe excitation strength functions in open-shell nuclei~\cite{Avogadro2011}. The advantages of this approach are coding feasibility and reasonable computational cost, with equal accuracy as the traditional methods. The first application of this method includes a study of the ISGMR in $^{174}$Sn, demonstrating the feasibility and usefulness of the FAM approach~\cite{Avogadro2011}. The implementation of the FAM-QRPA based on Broyden's iterative procedure, has been first demonstrated in the case of isoscalar and isovector monopole strengths for strongly deformed configurations
in $^{100}$Zr and $^{240}$Pu~\cite{Stoitsov2011}. In this way,  large-scale calculations of strength distributions in well-deformed superfluid nuclei become feasible across the nuclear landscape.

\begin{figure}[t!]
\centering
\includegraphics[width=0.7\linewidth,clip=true]{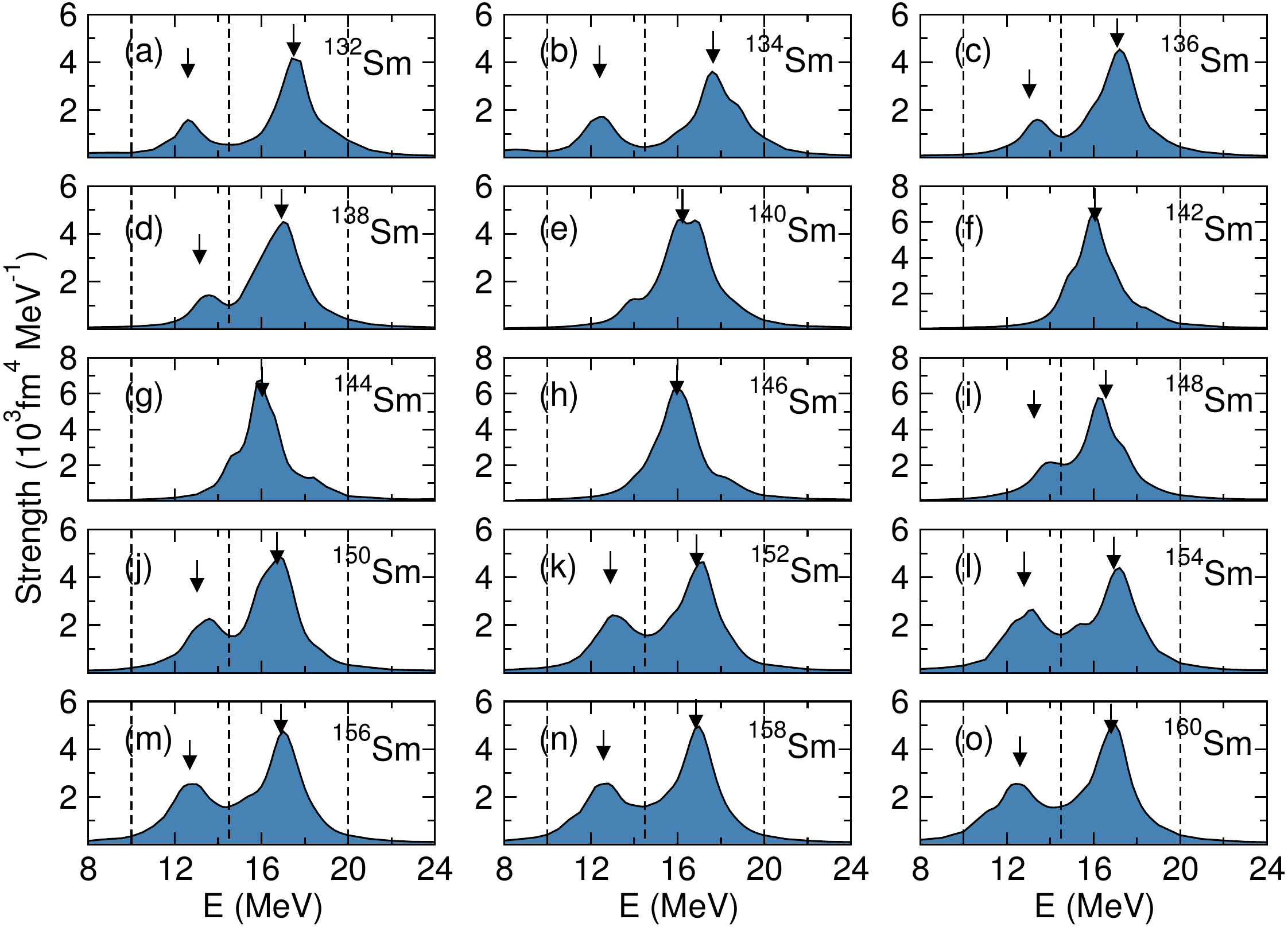}
\caption{\label{fig_mono_fam_niksic} 
The $K^{\pi}=0^+$ strength functions for $^{132-160}$Sn calculated with the FAM-RQRPA method. The centroid energies of the low- and high-energy components are denoted by arrows. Figure taken from  Ref.~\cite{Niksic2013}.}
\end{figure}

In Ref.~\cite{Hinohara2013} the FAM-QRPA method has been derived under the assumption
of axial and mirror symmetries, based on the Skyrme energy density functionals, and employed in the description of the isoscalar monopole strength for the oblate configuration of $^{24}$Mg. The FAM-RPA has also been developed in the framework of the covariant density functional theory, and its feasibility is demonstrated in the cases of
the ISGMR for $^{132}$Sn and $^{208}$Pb~\cite{Liang2013}. One of the advantages of this implementation is
that the rearrangement terms due to density dependent couplings are implicitly calculated without extra
computational cost. In Ref.~\cite{Niksic2013} the relativistic QRPA has been developed in the FAM framework for
deformed nuclei, and tested in a calculation of the monopole response in $^{22}$O. Its application in the chain of Sm isotopes demonstrates the splitting of the ISGMR in axially deformed systems. Fig.~\ref{fig_mono_fam_niksic}
shows the evolution of $K^{\pi}=0^+$ transition strength functions for $^{132-160}$Sn and splitting between
two main peaks for deformed isotopes. In order to explore the multipole excitations in triaxially deformed superfluid nuclei, latest developments include  the implementation of the FAM-QRPA in three-dimensional Cartesian coordinate space~\cite{Washiyama2017}.

\begin{figure}[t!]
\centering
\includegraphics[width=0.5\linewidth,clip=true]{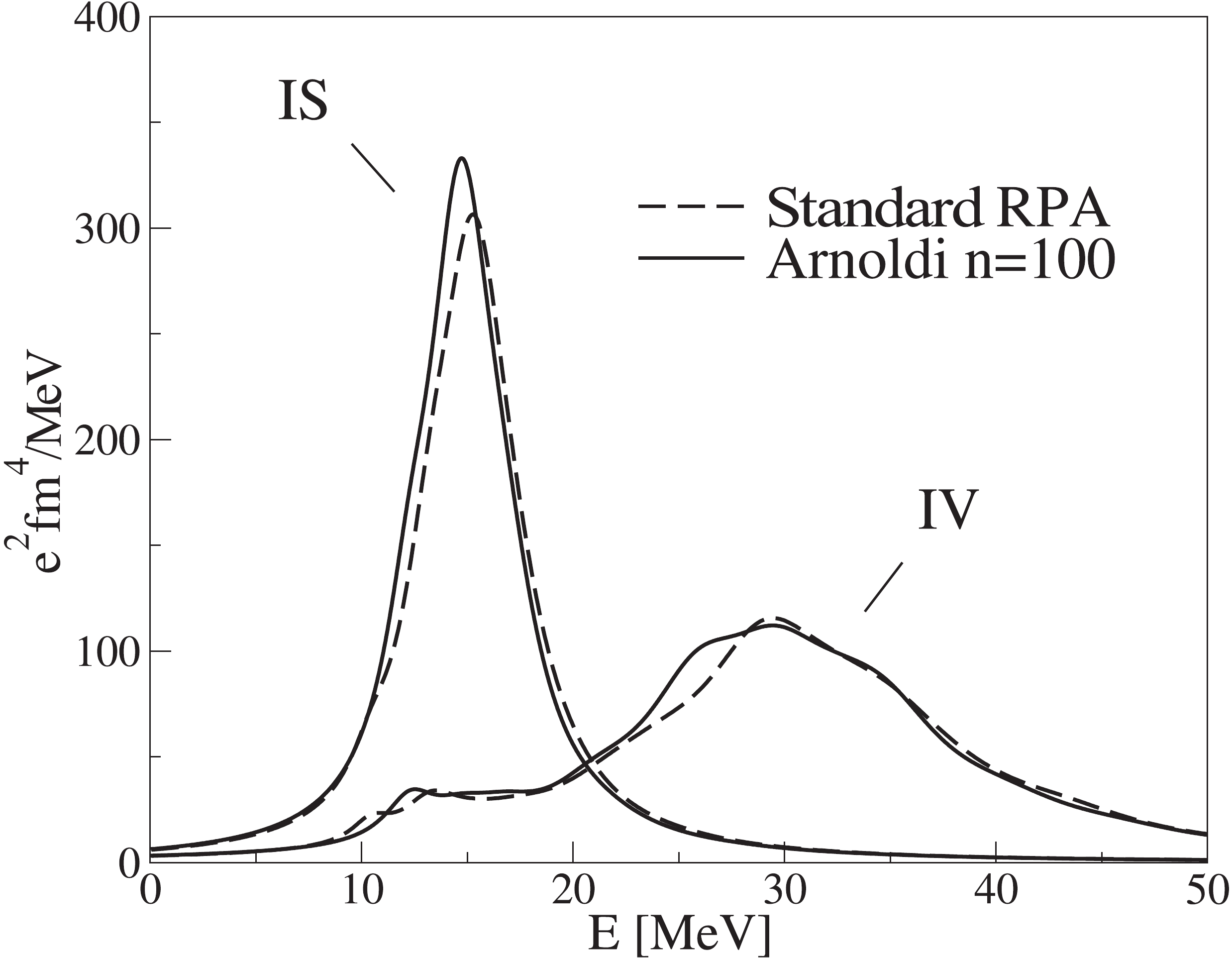}
\caption{\label{fig_mono_toivanen} The isoscalar and isovector monopole strength distributions in $^{132}$Sn
calculated using 100 Arnoldi iterations, compared with the standard RPA result (SLy4 functional). Figure taken from  Ref.~\cite{Toivanen2010}.}
\end{figure}

% Arnoldi
The ISGMR transition strength functions have also been studied in the implementation of another method to solve the RPA, using iterative non-Hermitian Arnoldi diagonalization method, which does not explicitly calculate and store the RPA matrix \cite{Toivanen2010}. As shown in Fig. ~\ref{fig_mono_toivanen}, the ISGMR strength distributions in $^{132}$Sn, 
calculated using SkM* functional and 100 Arnoldi iterations, result in excellent agreement with the standard RPA.

\begin{figure}[t!]
\centering
\includegraphics[width=0.6\linewidth,clip=true]{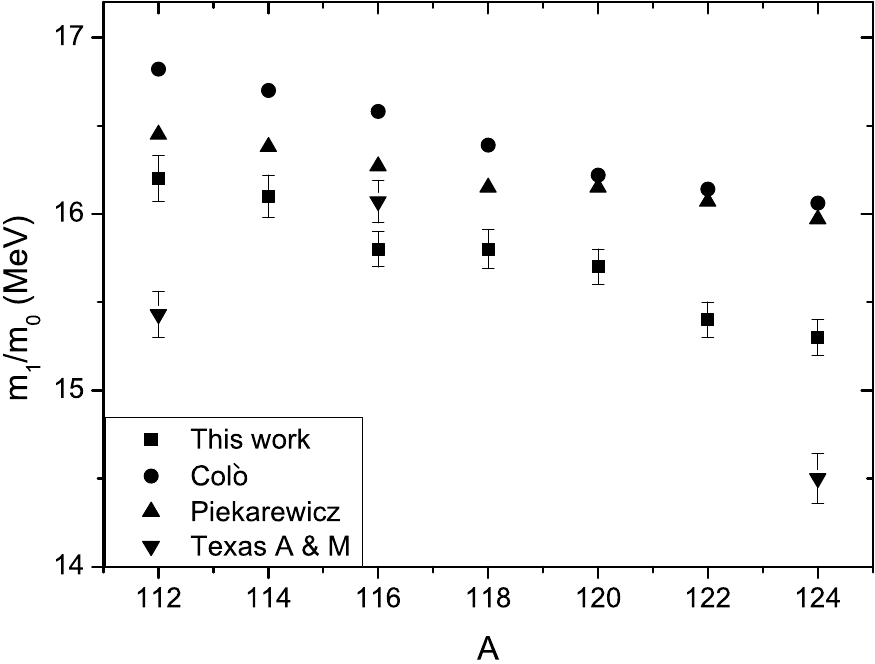}
\caption{\label{fig_mono_Li_Sn} The centroid energies of the ISGMR
calculated using non-relativistic~\cite{Li2007} and
relativistic Q(RPA)~\cite{Piekarewicz2007}, compared with the experimental
data~\cite{Li2007,Youngblood2004,Lui2004}
Figure taken from  Ref.~\cite{Li2007}.}
\end{figure}

%ISGMR Tin isotopes problem
One of the challenges on the properties of ISGMR is the question why the theoretical studies 
that accurately describe the ISGMR in $^{208}$Pb overestimate measured ISGMR excitation energies
in the Sn isotopes~\cite{Li2007,garg2007}. As shown in Fig.~\ref{fig_mono_Li_Sn}, the ISGMR centroid energies calculated both in the non-relativistic and relativistic RPA appear systematically above the measured values for all even-even $^{112-124}$Sn isotopes~\cite{Li2007}.
More recent study ~\cite{li2010} with inelastic scattering 386-MeV $\alpha$ particles on even-even isotopes $^{112-124}$Sn at extremely forward angles, including $0^{\circ}$, resulting in "background-free" spectra. Similar as in Ref.~\cite{Li2007}, the centroid energies of the 
ISGMR in tin isotopes appear considerably lower than the results of theoretical studies~\cite{li2010}.

In Ref.~\cite{Vesely2012}, the HFB+QRPA, based on Arnoldi method to solve the linear-response problem,  have been employed in a study of the role of finite-range and separable  pairing  interactions on monopole strength functions. It has been shown that the pairing forces cannot
resolve discrepancy of the calculated ISGMR energies in tin isotopes with the experimental 
data.
In Ref.~\cite{Li-Gang2012} the ISGMR has been studied in Cd, Sn and Pb isotopes within the self-consistent HF plus Bardeen-Cooper-Schrieffer(BCS\footnote{Well known approximation to deal with pairing correlations in open shell nuclei \cite{ringschuck}}.) approach and QRPA, by employing Skyrme interactions SLy5, SkM* and SkP, spanning a range of $K_0$=230, 217, and 202 MeV, respectively, supplemented with three types of pairing interaction.  The SkP cannot reproduce the ISGMR strength distribution for all isotopes due to low $K_0$. On the other side, SLy5, supplemented with the pairing interaction, provides reasonable description of the ISGMR in Cd and Pb isotopes, but it underestimates the ISGMR energies for Sn isotopes. The SkM* interaction, with a softer value of the nuclear incompressibility, improves description of the ISGMR energies for Sn isotopes. However, SkM* fails to reproduce the centroid energy of ISGMR in $^{208}$Pb.
The ISGMR in Sn isotopes has also been investigated in a relativistic mean field formalism and RPA based on NL3 and FSUGold parameterizations~\cite{Piekarewicz2009}. Using the general expression for the incompressibility coefficient of infinite neutron-rich matter, hybrid model has been constructed, that improves description of the
ISGMR strength distributions in Sn isotopes, however, similar to other models, it underestimates the respective strength distribution in $^{208}$Pb~\cite{Piekarewicz2009}.
One attempt to resolve the question "why the Sn isotopes are soft" is based on possible mutually enhanced magicity (MEM) effect~\cite{Khan2009,Patel2014}. The MEM effect refers to strong under-binding observed in HF mass formulas for all doubly magic nuclei and their nearest neighbors. Accordingly, doubly magic nuclei should 
be stiffer than open-shell nuclei, and the respective effect should be visible in the ISGMR excitation energies.
Specifically, the ISGMR energy in $^{208}$Pb would be higher about 600 keV than in the cases of $^{204,206}$Pb.
However, the experimental study of these isotopes showed that the ISGMR centroid energies are similar, in 
contradiction to the expectation from the model calculations, thus the effect of MEM in nuclear incompressibility has been ruled out~\cite{Patel2014}. 

\begin{figure}[t!]
\centering
\scalebox{0.35}{\includegraphics{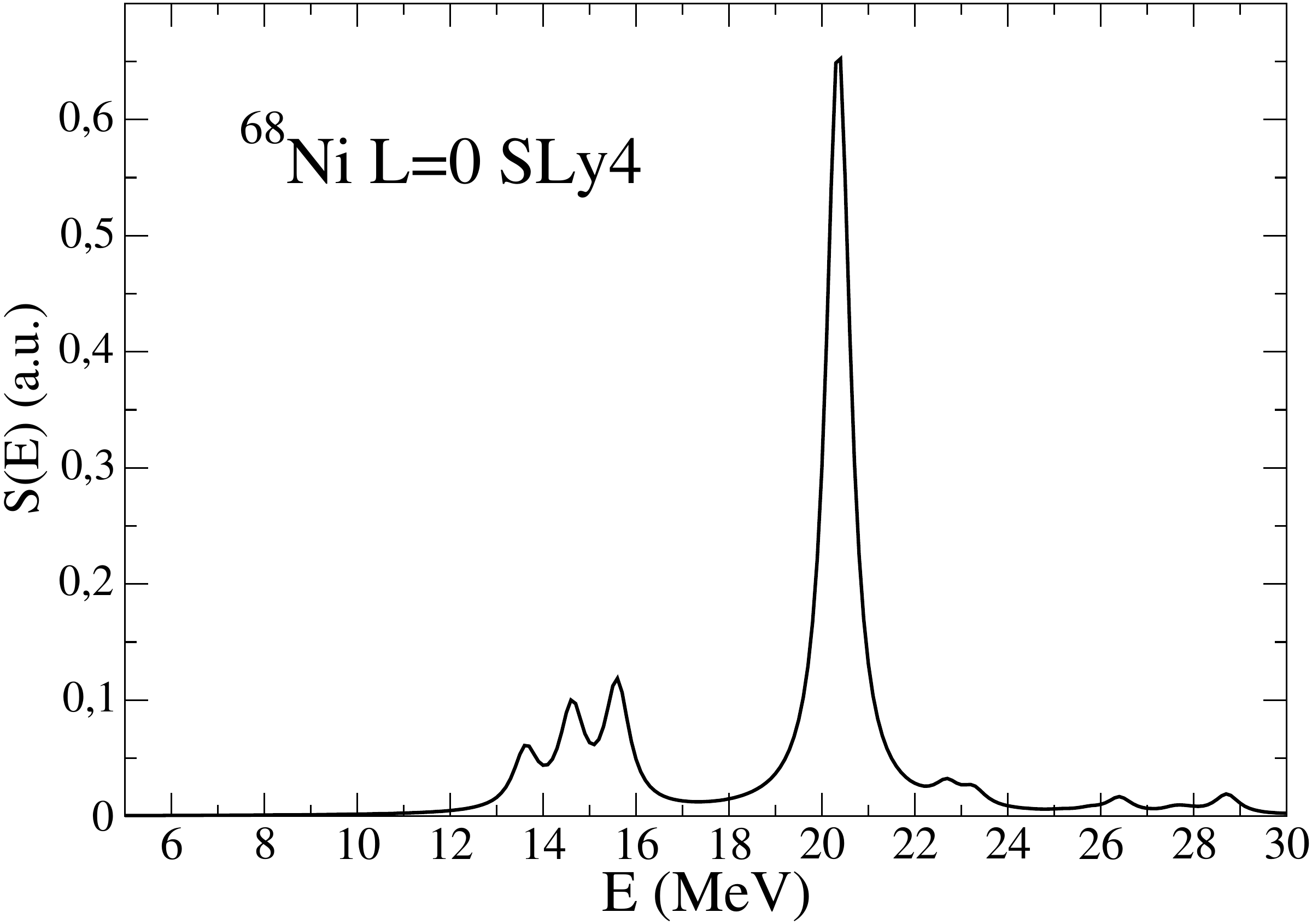}\includegraphics{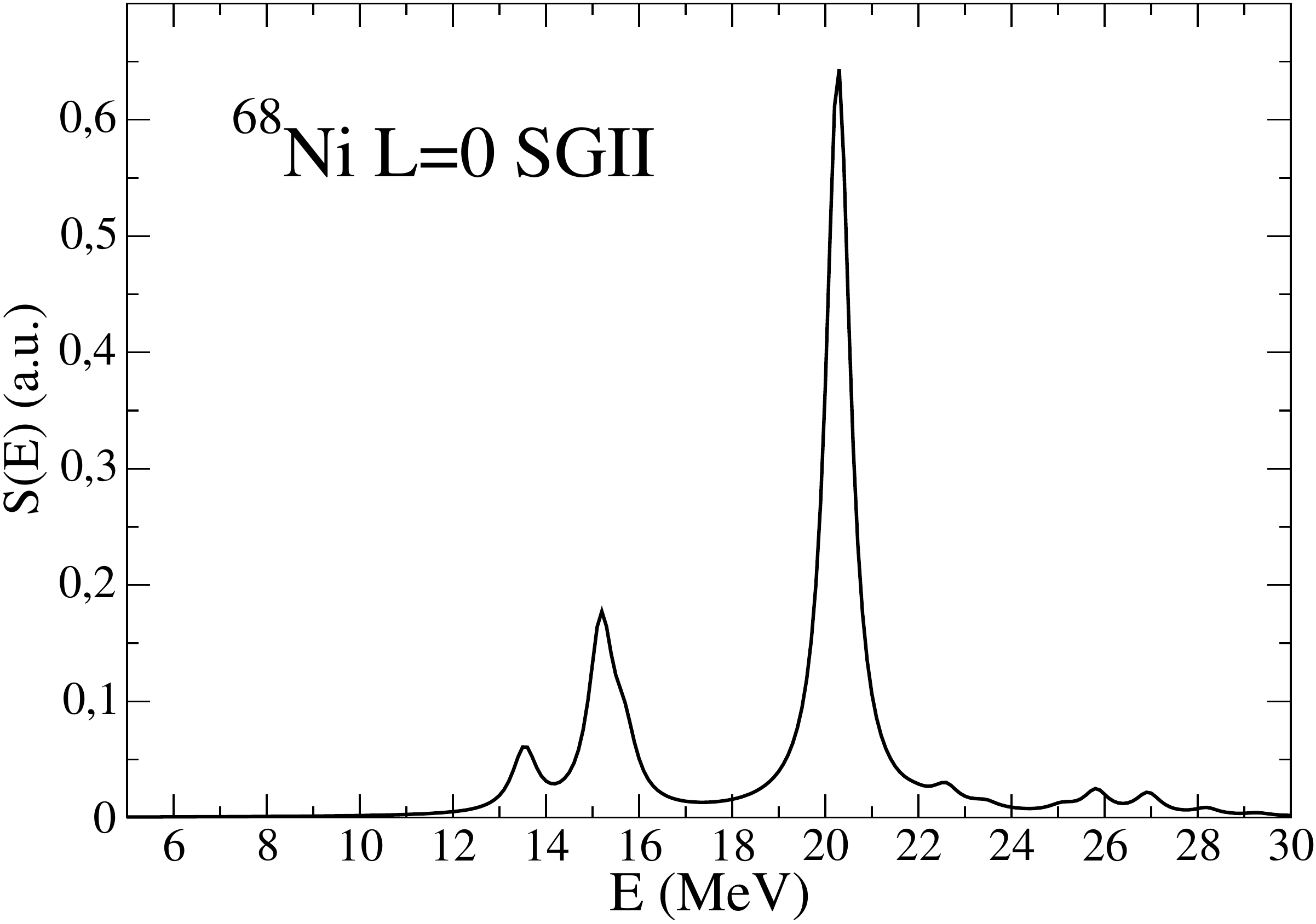}}
\caption{\label{fig_mono_khan} The isoscalar monopole strength distributions in $^{68}$Ni
calculated for (a) SLy4 and (b) SGII functionals. Figure taken from  Ref.~\cite{Khan2011}.}
\end{figure}

In the focus of recent studies is also the evolution of the ISGMR properties in unstable nuclei. The first measurement of the ISGMR in short-lived nucleus is reported for $^{56}$Ni, using deuterons as isoscalar probe and resolving difficult conditions in inverse kinematics, using active target (Maya) and set-up with an angular coverage close to $4\pi$~\cite{Monrozeau2008}. It was shown that the ISGMR in $^{56}$Ni exhausts a large fraction of the EWSR.
In Ref.~\cite{Centelles2005} 
the average ISGMR energies have been calculated in several isotopic chains from the proton to the neutron 
drip lines. It has been shown that the description based on sum rule approach, simplified by implementing the scaling method, provides the ISGMR energies in agreement with the calculation based on the RPA~\cite{Centelles2005}.
Calculations show that while approaching the neutron drip line, the ISGMR energies decrease and the resonance width increases.
Calculations performed using the microscopic Skyrme HF + RPA and relativistic RHB + RQRPA predicted the occurrence of additional low-energy monopole states for neutron-rich Ni isotopes, well separated from the isoscalar giant monopole resonance~\cite{Khan2011}. Fig. ~\ref{fig_mono_khan} shows the monopole transition strength distributions for $^{68}$Ni, calculated with Skyrme functionals Sly4 and SGII, resulting in pronounced low-energy monopole strengths between 10 and 15 MeV, below the main ISGMR peaks. In general, theoretical analyses predicted gradual enhancement of low-energy monopole strength with neutron excess~\cite{Khan2011}. Inelastic alpha and deuteron scattering in inverse kinematics for $^{68}$Ni support theoretical predictions, providing indications for soft isoscalar monopole mode~\cite{Vanderbrouck2015}. Enhanced soft monopole strength has
also been demonstrated in the application of the FAM-QRPA approach in modelling
collective excitations of weakly bound deformed Mg isotopes~\cite{Pei2014}.

% Ktau vs. K_inf - check other references
The question if the leptodermous expansion of the finite nucleus incompressibility (Blaizot's formula) can be fitted using the available experimental data of ISGMR has been addressed in a number 
of studies (see also discussion in Sec.\ref{pheno}). The parameter $K_{\tau}$ of the asymmetry term has been investigated in Refs.~\cite{Li2007,Li2007b}, using inelastic scattering of 400-MeV $\alpha$-particles at extremely forward angles to measure the ISGMR energies. Fig.~\ref{fig-eos-11} (right panel) shows the ranges of $K_{\tau}$ and $K_{0}=K_{\infty}$ values obtained from the expansion of $K_A$ in volume, asymmetry and Coulomb terms, and by implementing the ISGMR energies~\cite{Li2007b}. For comparison, the $K_{\tau} - K_{\infty}$ values are shown for a number of non-relativistic and relativistic functionals. In most of the cases, the results of theoretical predictions remain outside the experimental range of values, except for the density-dependent meson-exchange interactions DD-ME1 and DD-ME2 interactions~\cite{ddme2}. The study based on measured ISGMR energies for even-even isotopes $^{112-124}$Sn resulted in $K_{\tau}=-550\pm100$ MeV for the asymmetry term 
in the nuclear incompressibility.
%
%\begin{figure}[t!]
%\centering
%\includegraphics[width=0.7\linewidth,clip=true]{figures/fig_Li_Ktau_Kinf_2007.pdf}
%\caption{\label{fig_Li_Ktau_Kinf_2007}  The values of $K_{\tau}$ vs. $K_0=K_{\infty}$ from microscopic calculations based on nonrelativistic and relativistic functionals. The experimental ranges from the work ~\cite{Li2007,Li2007b} are denoted by horizontal and vertical lines.Figure taken from  Ref.~\cite{Li2007b}.}
%\end{figure}
%
%
Possible determination of the bulk symmetry incompressibility from the ISGMR has been revisited in
Ref.~\cite{Vinas2015}. In the analysis using RMF model with the NL3 effective interaction as a benchmark, the incompressibility of finite nuclei has been calculated for a set of nuclei in the scaling approach. By fitting these values to Blaizot's formula and implementing the covariance analysis, it is shown that it does not seem possible to use the coefficients of Blaizot's formula fitted to the ISGMR data to accurately constrain EDFs~\cite{Vinas2015}.
%
% overtones
In addition to the ISGMR, $L=0$ overtone for the transition operator $(r^4-\xi r^2)Y_{00}(\hat {r})$ has also been
investigated. In Ref.~\cite{shlomo2003}, transition density associated with the ISGMR overtone
has been derived using HF+RPA and semi-classical Fermi-liquid approach, resulting in qualitative
agreement between the two methods.

In order to explore the multipole excitations in triaxially deformed superfluid nuclei, latest developments include
 the implementation of the FAM-QRPA in three-dimensional Cartesian coordinate space~\cite{Washiyama2017}.
 The ISGMR has also been studied within the approaches going beyond the (Q)RPA, e.g., by including
the quasi-particle-phonon coupling at the level of the time-blocking approximation ~\cite{Lyutorovich2015},
and second RPA based on Skyrme and Gogny functionals~\cite{Gambacurta2016}. Since the implementation
of complex configurations results in modifications of the excitation spectra, their role on determination
of the nuclear matter incompressibility still needs to be investigated in detail.
Within the {\it ab initio} inspired approaches,
the isoscalar monopole and other giant resonances have been investigated on the
basis of unitarily transformed two-nucleon plus phenomenological three-nucleon interactions,
resulting in only slightly overestimated resonance centroid energies~\cite{Gunther2014}. However, it is expected that the
implementation of higher-order configurations would shift resonances to lower energies~\cite{Gunther2014}. The
feasibility study of large-scale second RPA calculations, based on realistic interaction derived from the Argonne
V18 potential, has been reported in Ref.~\cite{Papakonstantinou2010}.
 
%%%%%%%%%%%%%%%%%%%%%%%%%%%%%%%%%%%%%%%%%%
\subsubsection{Isoscalar giant dipole resonance}
%%%%%%%%%%%%%%%%%%%%%%%%%%%%%%%%%%%%%%%%%%

In addition to the ISGMR, another compression mode also provides constraint on the nuclear matter incompressibility
coefficient $K_0$: isoscalar giant dipole resonance (ISGDR). It is a second order effect, built on $3\hbar \omega$
or higher configurations, and in a macroscopic picture it corresponds to a compression wave traveling back and forth
through the nucleus along a definite direction.  More details and respective references on earlier studies
of the ISGDR are given in Ref.~\cite{Paar2007}. 
There exists also available experimental data on the ISGDR in $^{90}$Zr,$^{116}$Sn,$^{144}$Sm,$^{208}$Pb from 
inelastic $\alpha$-scattering, that can constrain the value of $K_0$ from this 
mode~\cite{Clark1999,Clark2001,Youngblood2004b,Uchida2003,Uchida2004,Nayak2006}.
In Ref.~\cite{Garg2004} an overview of the ISGDR measurements using small-angle inelastic $\alpha$-particle scattering  at RCNP, Osaka, and implications on the determination of
nuclear incompressibility are given in detail.

Over the past years, one of the open problems was 
the question of inconsistent results on $K_0$ deduced from the ISGMR and ISGDR, i.e. discrepancy between the theoretical predictions and experimental data on the ISGDR excitation spectra~\cite{shlomo2003b}.
As shown in Ref.~\cite{Dumitrescu1983}, the HF-RPA calculations that reproduce the ISGMR excitation energies, overestimate the ISGDR energies by more than 3 MeV. In the first studies, several approximations have been introduced, in particular, the effective interaction in the HF was not fully consistent with the residual interaction
in the RPA (with Coulomb and spin-orbit interaction neglected) together with the limitations in the $ph$ configuration space. As a result, spurious state contamination could appear in the ISGDR excitation spectra. Self-consistent calculations based on a unified effective interaction at the HF and RPA level resolve this problem, resulting in a clear separation of
$1^-$ spurious state close to the zero energy. From the experimental side, improvements
include precise and instrumental-background-free measurements of the isoscalar dipole strength
distributions using inelastic scattering of $\alpha$ particles~\cite{Itoh2003,Uchida2003,Uchida2004,Nayak2006}. As pointed out in Ref.~\cite{Nayak2009},
the value of the nuclear incompressibility, obtained from the ISGDR data is
consistent with that from the ISGMR data. 

\begin{figure}[t!]
\centering
\includegraphics[width=0.5\linewidth,clip=true]{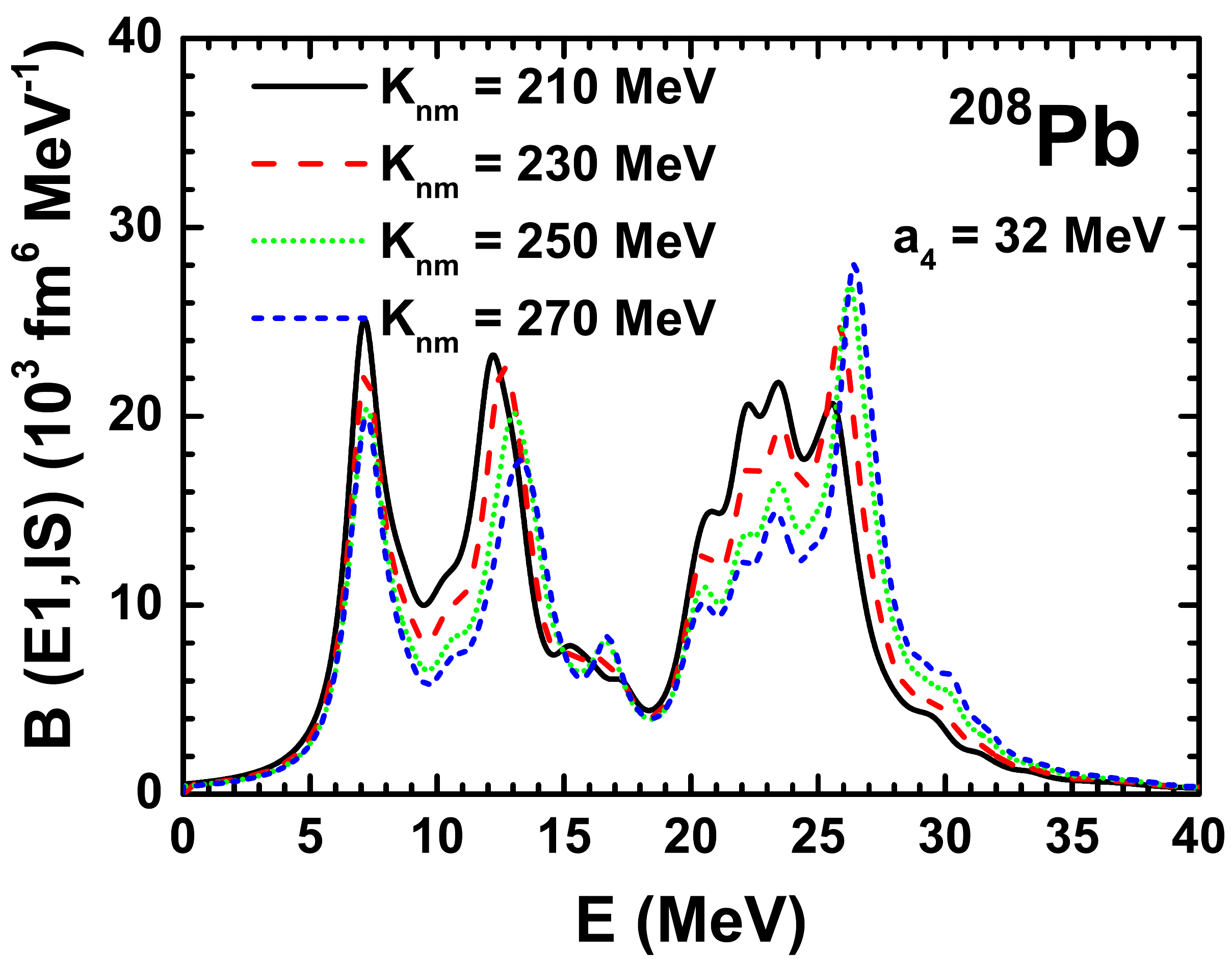}
\caption{\label{fig_isgdr_Pb208_vretenar} Isoscalar dipole transition strength distributions in $^{208}$Pb calculated with the relativistic RPA for the set of effective interactions that differ in the nuclear matter incompressibility coefficient as denoted in the legend.}
\end{figure}

The isoscalar excitation spectra are composed from two separate regions: a low-energy part, and a high-energy part that corresponds to the dipole compression mode (ISGDR)~\cite{Colo2000}. Fig.~\ref{fig_isgdr_Pb208_vretenar} shows the isoscalar dipole transition strength distribution for $^{208}$Pb, calculated with relativistic RPA using a set of four density-dependent meson-exchange effective interactions that differ only in  the value of nuclear matter incompressibility, $K_0$=210,230,250, and 270 MeV~\cite{Vretenar2012}. As expected, the high-energy component is rather sensitive to the choice of the nuclear matter incompressibility, with increasing $K_0$, the energy is shifted towards higher energies. On the other side, the effect on low-energy transitions is rather small; by varying $K_0$ the energies do not change, and only the B(E1) strength is somewhat reduced with increased $K_0$. Similar dependence is obtained within non-relativistic models using Skyrme functionals~\cite{Colo2000}. As a possible solution for understanding the low-energy dipole strength, it has been suggested that it may correspond to the toroidal resonance that would include nucleon currents within a torus~\cite{Vretenar2002}. For more details see the review in Ref.~\cite{Paar2007}.

\begin{figure}[t!]
\centering
\includegraphics[width=0.5\linewidth,clip=true]{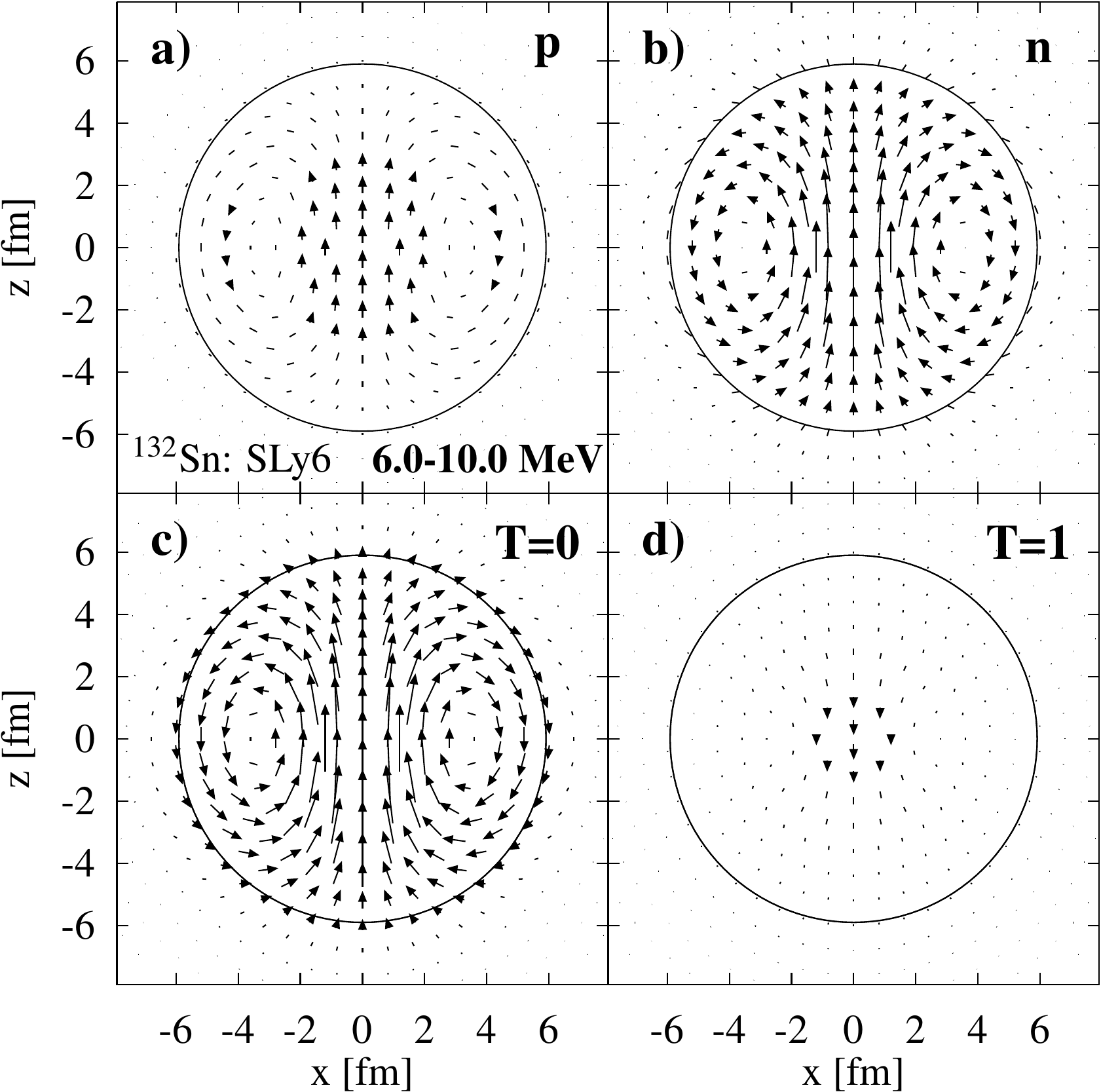}
\caption{\label{fig_toroidal_nesterenko}   The current fields for the low-energy dipole strength in the interval 6-10 MeV for $^{132}$Sn, 
calculated with the RPA: (a) proton, (b) neutron, (c) isoscalar, and (b) isovector current fields. Figure taken from  Ref.~\cite{Nesterenko2016}.}
% corresponds to Fig.4. in Nesterenko2016.
\end{figure}

The bi-modal structure of the isoscalar dipole transition strength and the interplay of dipole compressional and vortical nuclear 
currents have also been confirmed within semi-classical nuclear Fermi-fluid dynamic approach~\cite{Misicu2006}.
In Ref.~\cite{Nesterenko2016} the toroidal resonance has been explored in relation to the pygmy dipole strength (see subsection below) 
since both appear in the same low-energy region. Fig. ~\ref{fig_toroidal_nesterenko} shows the proton, neutron, 
isoscalar and isovector current fields in the $z-x$ plane for the low-energy excitations in the interval 6-10 MeV, 
calculated using the RPA with the Skyrme functional SLy6 for $^{132}$Sn~\cite{Nesterenko2016}. Clearly, the low-energy transitions
correspond to isoscalar toroidal nuclear motion (see Fig.~\ref{fig_toroidal_nesterenko} c)).
Nuclear vorticity in isoscalar dipole response in $^{208}$Pb has also been 
explored in Ref.~\cite{Reinhard2014}, by inspecting two basic concepts, hydrodynamical (HD) and Rawenthall-Wambach (RW)~\cite{Wambach1987}. In the HD case vorticity is defined as a curl of the velocity, while in the RW case just a given component of the current is proposed as an indicator of vorticity~\cite{Wambach1987,Reinhard2014}.
Model calculations based on self-consistent RPA using
Skyrme functional SLy6 showed that RW concept is not robust and the vorticity is
better characterized by the toroidal strength corresponding to HD treatment~\cite{Reinhard2014}.

\begin{figure}[t!]
\centering
\includegraphics[width=0.8\linewidth,clip=true]{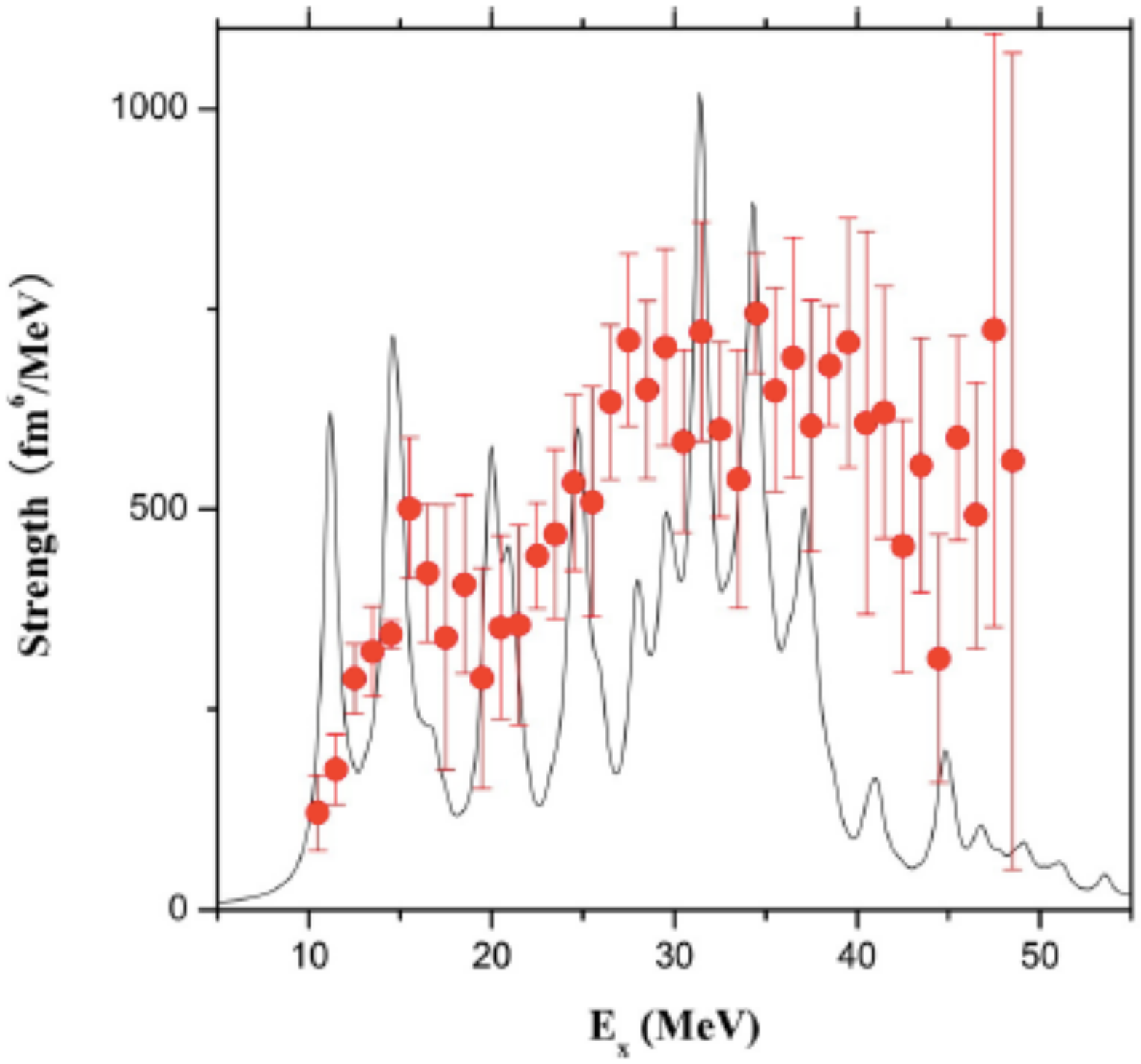}
\caption{\label{fig_isoscalar_dipole_nayak}The ISGDR strength distribution in $^{58}$Ni from $\alpha$ particle scattering at extreme forward angles in comparison to the QRPA calculation. Figure taken from  Ref.~\cite{Nayak2006}.}
% corresponds to Fig.5. in Nayak2006.
\end{figure}

Measurement of the ISGDR in $\alpha$-particle scattering on $^{58}$Ni represents the first confirmation
of the bi-modal structure of isoscalar dipole strength in A$<$90 nucleus~\cite{Nayak2006}. In Fig.~\ref{fig_isoscalar_dipole_nayak} the measured isoscalar dipole strength distribution is compared with
those from the QRPA calculation. The experimental and theoretical strength distributions are in
qualitative agreement, except at the highest energies, where the experimentally extracted 
transition strength is compromised by the limitations of the multipole decomposition analysis~\cite{Nayak2006}. There are also limitations present at the theoretical side, because contributions 
due to complex configurations are not taken into account. In order to improve description of high-energy
spectra, the excitation and proton-decay of the ISGDR have been measured using
$^{208}$Pb$(\alpha,\alpha'p)^{207}$Tl reaction at 400 MeV~\cite{Nayak2009}. It was shown that
the ISGDR strength, previously observed at highest excitation energies, does not appear
in the coincidence spectra, indicating that this excess strength was spurious and originates from 
other, non-resonant phenomena~\cite{Nayak2009}. The isoscalar multipole strengths have also been measured in exotic doubly-magic $^{56}$Ni using inelastic $\alpha$ particle scattering in inverse kinematics~\cite{Bagchi2015}. This measurement confirmed $\alpha$-particle scattering as
an appropriate method for exciting the ISGMR and ISGDR modes in radioactive isotopes.

%
% SRPA, phonon coupling model
%

\subsubsection{Isoscalar giant quadrupole resonance}

\begin{figure}[t!]
\centering
\includegraphics[width=0.5\linewidth,clip=true]{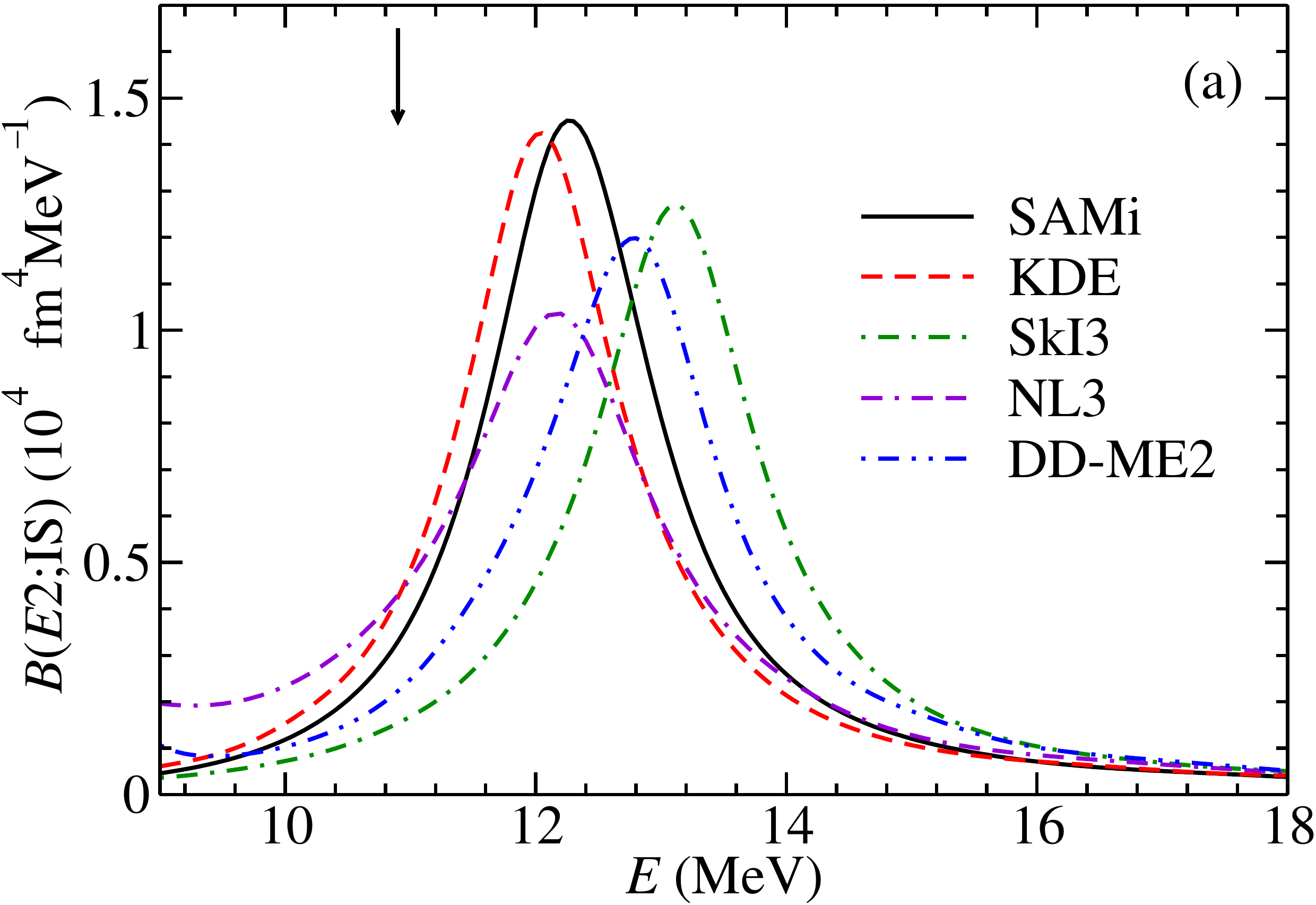}
\caption{\label{fig_isgqr_roca-maza} The ISGQR strength functions in $^{208}$Pb calculated using the RPA with Skyrme (SAMi~\cite{roca-maza12b}, KDE~\cite{kde0},SkI3~\cite{SkI3}) and relativistic (NL3~\cite{NL3}, DD-ME2~\cite{ddme2}) functionals. The experimental ISGQR energy is denoted by arrow~\cite{Youngblood2004}.Figure taken from  Ref.~\cite{roca-maza13}.}
% corresponds to Fig.1(a). in roca-maza13.
\end{figure}

The isoscalar quadrupole response is composed from the two main structures in the low- and high-energy region. In addition to a pronounced low-energy $2^+$ state, a collective ISGQR strength appears at higher
energies. The energy of the transition from the ground to the first excited $2^+$ state is strongly sensitive on the nuclear shell effects~\cite{Raman2001} and depends on the number of particles 
outside the closed shell~\cite{Tsoneva2011}. On the other side, the ISGQR as a collective resonant mode is expected to vary smoothly with the mass number $A$. The ISGQR provides a constraint on the effective mass as discussed in Sec.\ref{pheno}, it is also sensitive to the nuclear matter incompressibility, and contributes in determining the nuclear
symmetry energy and the neutron skin thickness as we will discuss below~\cite{roca-maza13}. Fig.~\ref{fig_isgqr_roca-maza}
shows the ISGQR transition strength distributions for $^{208}$Pb, calculated using RPA with Skyrme type functionals, SAMi~\cite{roca-maza12b}, KDE~\cite{kde0}, SkI3~\cite{SkI3}, and relativistic functionals, 
NL3~\cite{NL3} and DD-ME2~\cite{ddme2}. Model calculations overestimate measured ISGQR energy (denoted in figure by an arrow) $\approx 1-3$ MeV. The energy of the ISGQR is closely related to the effective
nucleon mass $m^*/m$~\cite{Blaizot1980} (decreases with increasing effective mass), and the model calculations indicate that empirical ISGQR
energy in $^{208}$Pb favors $m^*/m \approx 1$ (see Fig.22 of Ref.\cite{Blaizot1980}), that is above the respective values in the effective interactions used in Fig.~\ref{fig_isgqr_roca-maza}. 

\begin{figure}[t!]
\centering
\includegraphics[width=0.6\linewidth,clip=true]{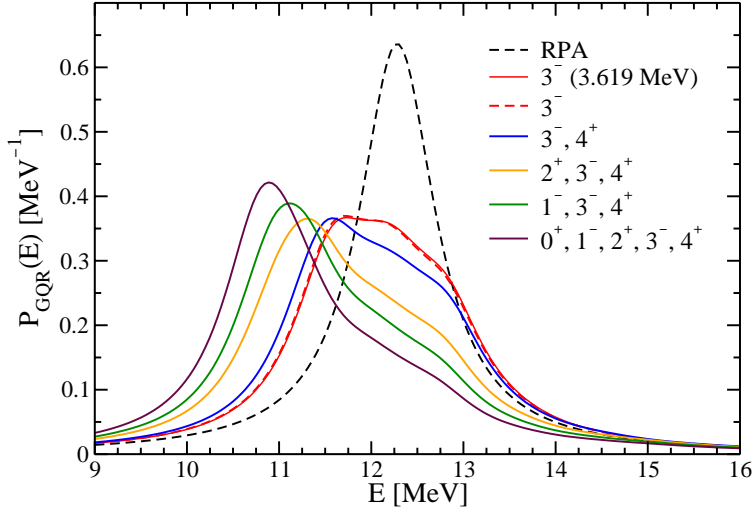}
\caption{\label{fig_isgqr_brenna}. Probability to find the ISGQR state at an
energy E. The RPA result is compared with different curves corresponding
to the implementation of the phonons denoted in the legend as intermediate states.
Figure taken from  Ref.~\cite{Brenna2012}.}
% corresponds to Fig.5 in Brenna2012.
\end{figure}

Another important effect comes from beyond
mean-field correlations. In Ref.~\cite{Brenna2012} the width of the resonance has been calculated
in a microscopic theory of the $\gamma$-decay of the ISGQR, based on the Skyrme functional, treating the ground-state decay within the fully self-consistent RPA and the decay to low-lying states at the lowest order beyond RPA. Fig.~\ref{fig_isgqr_brenna} shows the probability of finding the ISGQR, calculated by including an increasing number of intermediate phonons with
multipolarity ranging from 0 to 4 and with natural parity~\cite{Brenna2012}. As shown in Fig.~\ref{fig_isgqr_brenna}, the main contribution to the spreading width of the order of 2 MeV comes from the low-lying $3^-$ state, while the other phonons have the effect on the ISGQR energy shift toward lower energies.

\begin{figure}[t!]
\centering
\includegraphics[width=0.5\linewidth,clip=true]{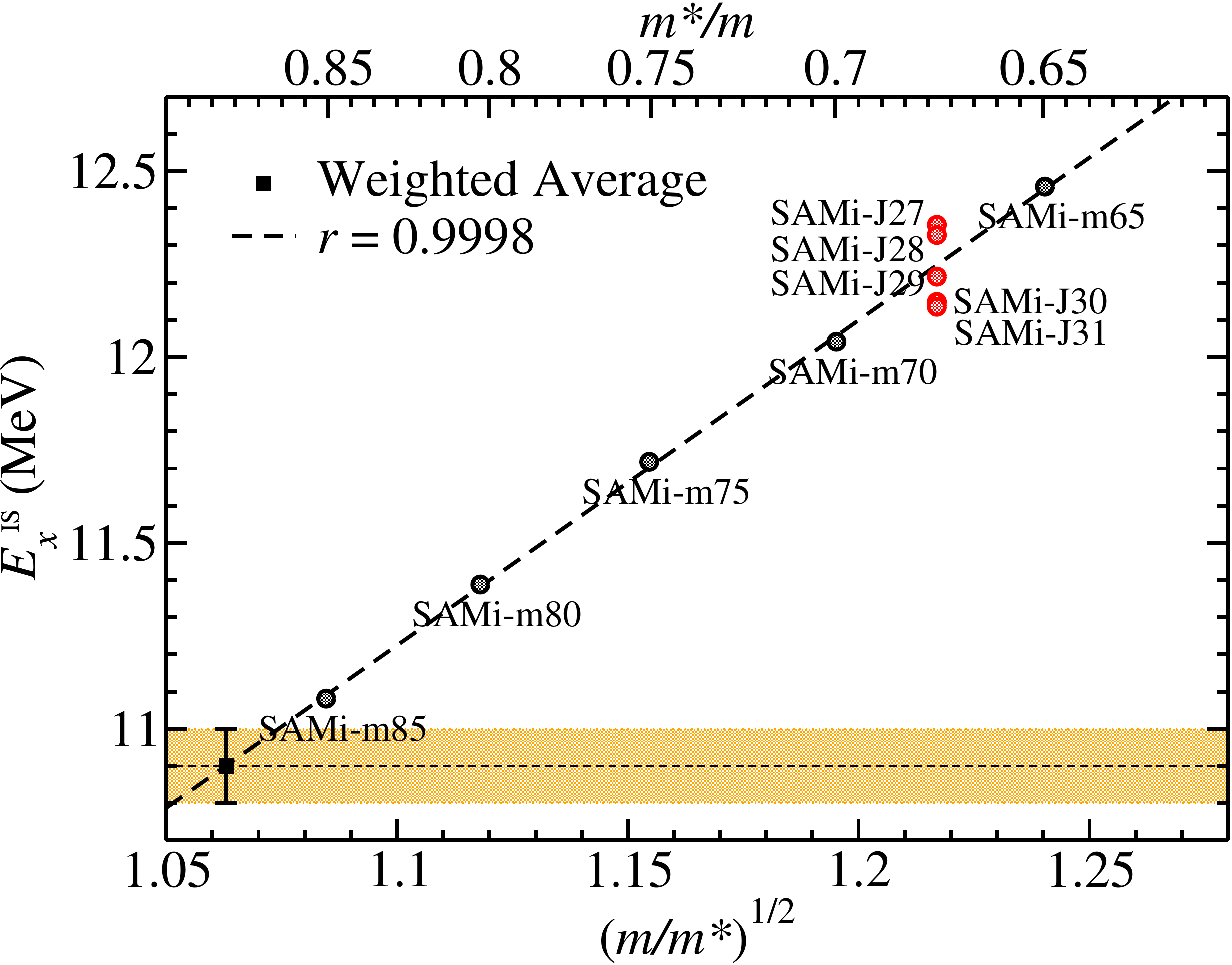}
\caption{\label{fig_isgqr_effmass_roca-maza} The ISGQR excitation energy in $^{208}$Pb as a function of the effective mass (upper x-axis, $m^*/m$. Calculations are based on the RPA with the SAMi-m and SAMi-J families of Skyrme functionals. Figure taken from  Ref.~\cite{roca-maza13}.}
%
%corresponds to Fig.4. in roca-maza13.
%
\end{figure}

As mentioned above, the ISGQR provides important information on the nucleon effective 
mass $m^*$~\cite{Blaizot1980}. Fig.~\ref{fig_isgqr_effmass_roca-maza} shows the
ISGQR excitation energies in $^{208}$Pb as a function of ${\sqrt {m/{m^*}}}$, calculated 
with two families of Skyrme functionals (see Sec.\ref{theo}), SAMi-m and SAMi-J~\cite{roca-maza12b}, 
that systematically vary the effective mass $m^*$ and  the symmetry energy at saturation 
density $J$, respectively~\cite{roca-maza13}. For the SAMi-m interactions, the figure 
demonstrates the correlation between the ISGQR energies and ${\sqrt {m/{m^*}}}$.
For the range of $J=27-31$ MeV, the variation of ISGQR energy is rather small~\cite{roca-maza13}.

\begin{figure}[t!]
\centering
\includegraphics[width=0.5\linewidth,clip=true]{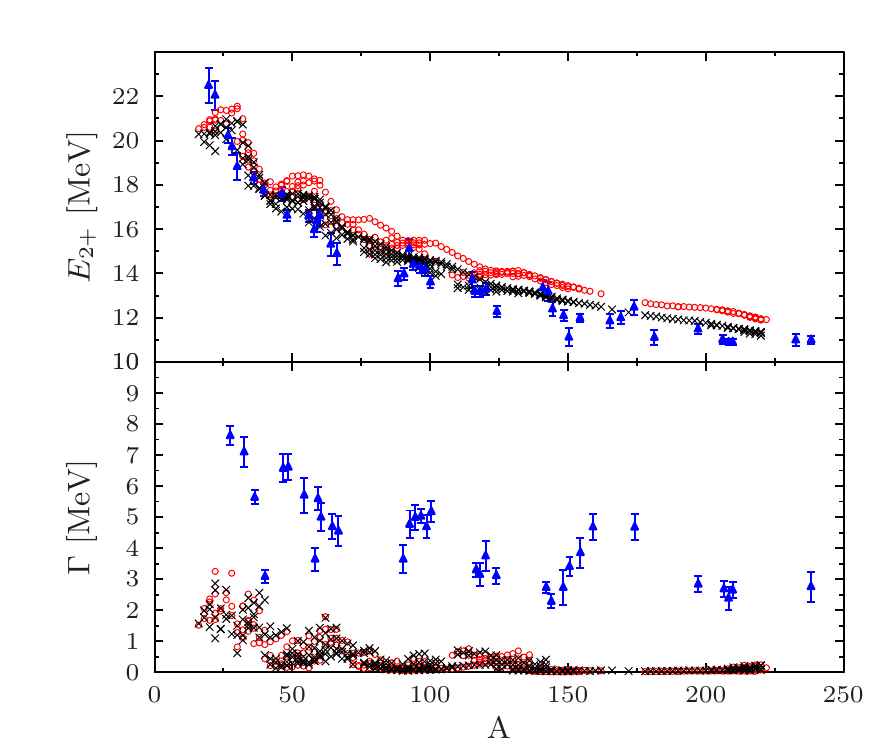}
\caption{\label{fig_isgqr_scamps} The ISGQR excitation energy (upper panel) and widths (lower panel) for a set of spherical nuclei. Calculations are based on TDHF+BCS approach with the SkM* (crosses) and SLy5 (circles) functionals. Experimental data are denoted by triangles~\cite{Bertrand1981}. Figure taken from  Ref.~\cite{Scamps2013}.}
% corresponds to Fig.9. in Scamps2013.
\end{figure}

\begin{figure}[t!]
\centering
\includegraphics[width=0.5\linewidth,clip=true]{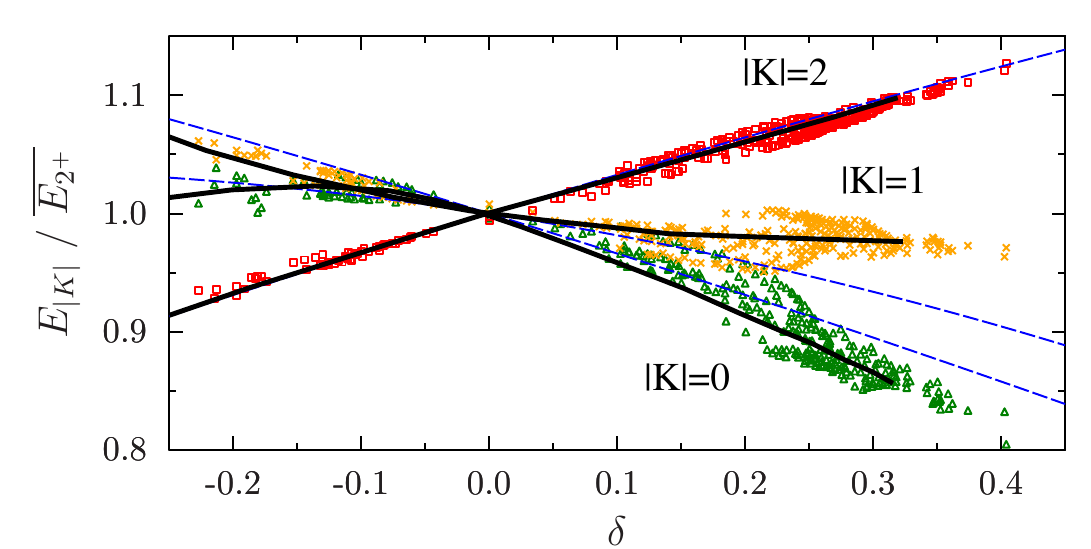}
\caption{\label{fig_isgqr_axial_scamps} The main peak energies of the ISGQR as a function of deformation parameter $\delta$, for the components $|K|$=0,1,2. The results from the fluid-dynamical
model~\cite{Nishizaki1985} are shown as solid black lines.  
Figure taken from  Ref.~\cite{Scamps2014}.}
% corresponds to Fig.8. in Scamps2014.
\end{figure}

In the framework of fully self-consistent QRPA with Skyrme interaction and density-dependent
pairing functionals, the isoscalar and isovector response in the deformed $^{24-26}$Mg and $^{34}$Mg have been investigated~\cite{Losa2010}. The analysis of low-lying isoscalar $2^+$ state in open shell nucleus $^{26}$Mg showed that the main contributions originate from 
$pp$ transitions, while in the case of $^{24}$Mg $ph$ states play the dominant role.
The ISGQR and IVGQR responses have been systematically investigated in spherical~\cite{Scamps2013} and deformed~\cite{Scamps2014} nuclei using time-dependent HF+BCS approach. The calculated  ISGQR energies and the corresponding widths are shown for a set of spherical nuclei in Fig.~\ref{fig_isgqr_scamps}, for SkM* and SLy5 Skyrme functionals~\cite{Scamps2013}. The calculated mean-energy of the collective high-energy ISGQR state are in the agreement with experimental data~\cite{Bertrand1981}, however, due to the missing two-body effects their widths could not be reproduced~\cite{Scamps2013}.
Various deformation effects on collective ISGQR modes have been explored in Ref.~\cite{Scamps2014}.
For axially symmetric nuclei, a splitting of the ISGQR is observed in three components $|K|$=0,1,2. As shown in Fig.~\ref{fig_isgqr_axial_scamps}, the main peak energies of the three components are 
clearly separated with increasing the deformation parameter $\delta$. The results from TDHF+BCS calculations are in qualitative agreement with the fluid-dynamical model~\cite{Nishizaki1985}.
By increasing the deformation, the ISGQR in medium- and heavy-nuclei acquires a non negligible spreading width for all $|K|$ components that significantly contributes to the overall fragmentation of the strength~\cite{Scamps2013}.
The analysis in Ref.~\cite{Scamps2014} also includes triaxial nuclei, resulting with increased complexity in the fragmentation and splitting of the collective GQR modes.

The properties of lowest $K=2^+$ states have been explored in axially deformed rare earth and actinide nuclei using separable RPA based on Skyrme functional~\cite{Nesterenko2016b}. Although in some cases agreement with the experimental data
has been achieved, the calculations indicated that further progress of self-consistent models is necessary for a systematic quantitative description of $2^+$ states.
Fully consistent axially-symmetric deformed QRPA, based on the Gogny D1S effective interaction, has been employed in a study of  giant resonances in a heavy nucleus $^{238}$U~\cite{Peru2011}. It results with strongly fragmented quadrupole response, especially in the low-energy region below 10 MeV.

In Ref.~\cite{Peach2016} the ISGQR has been studied in inelastic $\alpha$ scattering at extremely
forward angles on $^{28}$Si, with a focus on the effect of ground-state deformation on the resonance properties. It has been shown that excitation energy of the ISGQR at 17.7 MeV appears consistent with the calculations for an oblate-deformed ground state.

\begin{figure}[t!]
\centering
\includegraphics[width=0.65\linewidth,clip=true]{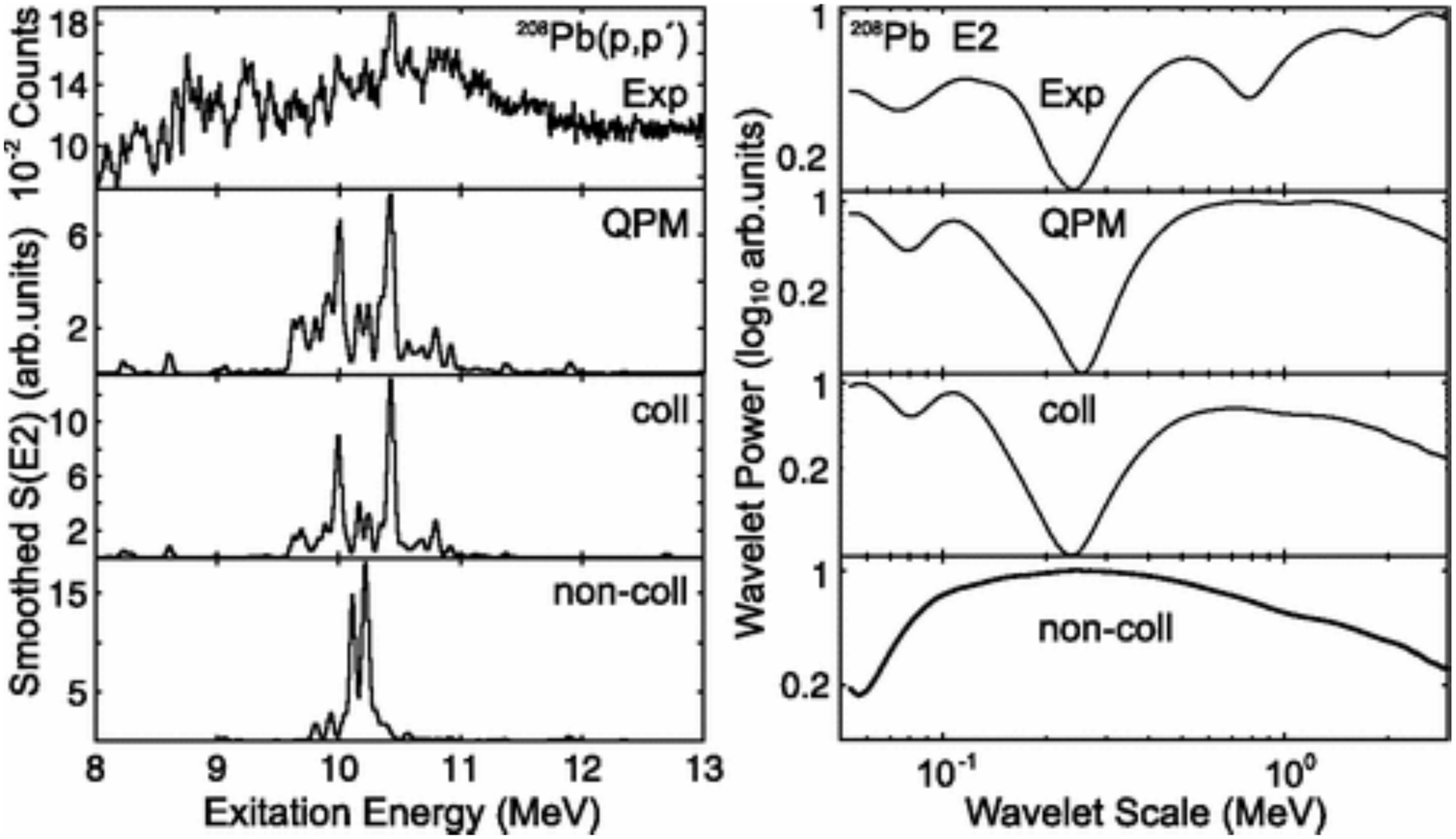}
\caption{\label{fig_isgqr_scales} (Left panel) The measured spectrum from $^{120}$Sn$(p,p')$ reaction at the maximum of the ISGQR cross section in comparison with the QPM and ETDHF 
calculations. (Right panel) Corresponding wavelet power spectra.
Figure taken from  Ref.~\cite{Shevchenko2009}.}
% corresponds to Fig.17. in Shevchenko2009
\end{figure}

The fine structure of the ISGQR has been investigated experimentally in high-resolution proton scattering 
on $^{58}$Ni, $^{90}$Zr, $^{120}$Sn, and $^{208}$Pb, and wavelet analysis has been employed for the extraction of scales characterizing the fine structure~\cite{Shevchenko2004}. Fig.~\ref{fig_isgqr_scales} shows a comparison between the measured ISGQR spectrum in $^{120}$Sn and related quasi-particle phonon model (QPM), and extended time-dependent Hartree-Fock (ETDHF) calculations, together with the corresponding wavelet power spectra that vary between the experiment and model calculations~\cite{Shevchenko2009}.
From the comparison with the model calculations including $2p2h$ degrees of freedom, it has been shown that collective coupling of the ISGQR to low-energy surface vibrations represent the main source of the observed scales.  
In Ref.~\cite{Shevchenko2009}
several models have been employed in the study of ISGQR in $^{208}$Pb, the QPM, ETDHF, second RPA, and extended theory of finite Fermi systems (ETFFS), resulting in significant differences for the values of scales. Possible origin of these differences could be implementation of different effective interactions, truncation schemes of the model spaces, coupling to the continuum and couplings of complex configurations involved. More systematic and consistent studies are needed to benchmark theoretical description of the fine structure of the ISGQR. Recently,
the studies of the fine structure of ISGQR using proton inelastic scattering have been extended toward lower mass nuclei, i.e., $^{40}$Ca~\cite{Usman2011}. It has been shown that characteristic scales for $^{40}$Ca are already present at the mean-field level pointing to their origin in Landau damping, different than in heavier nuclei and Second RPA (SRPA) calculations based on phenomenological effective interactions, where the fine structure is explained by the coupling to two-particle two-hole (2p-2h) states~\cite{Usman2011}.

Within the experiment with inelastic scattering of $\alpha$ particles at small angles, the
ISGQR has been measured at small angles in $^{48}$Ca~\cite{Liu2011}.
In Ref.~\cite{Monrozeau2008} the ISGQR has been measured in unstable nucleus
$^{56}$Ni using $^{56}$Ni$(d,d')$ reaction, resulting with the centroid energy 
16.2$\pm$0.5 MeV, that is comparable to those from $(\alpha,\alpha')$ scattering 
on $^{58}$Ni and $^{56}$Fe. 

Novel developments of the FAM-QRPA solver have also been exploited in description 
of the ISGQR in heavy deformed nucleus $^{240}$Pu, without any truncations in
the quasi-particle space~\cite{Kortelainen2015}. The FAM-QRPA in the three-dimensional
Cartesian coordinate space has been implemented to study isoscalar quadrupole strength 
functions in  axially deformed superfluid nuclei, $^{110}$Ru and $^{190}$Pt~\cite{Washiyama2017}.

%Quasiparticle random-phase approximation with interactions from the Similarity Renormalization %Group, Hergert Phys. Rev. C 83, 064317 
%
% SRPA Papakonstantinou, Roth, Gambacurta, Grasso
%
%~\cite{Hergert2011}

%%%%%%%%%%%%%%%%%%%%%%%%%%%%%%%%%%%%%%%%%%%
\subsection{Isovector modes}
%%%%%%%%%%%%%%%%%%%%%%%%%%%%%%%%%%%%%%%%%%%
\subsubsection{Isovector giant dipole resonance}
%%%%%%%%%%%%%%%%%%%%%%%%%%%%%%%%%%%%%%%%%%%

The isovector giant dipole resonance (IVGDR) corresponds to a collective vibration
mode of protons oscillating against neutrons~\cite{Dumitrescu1983}. 
Within the original studies based on a macroscopic model, the giant dipole resonance
has been described as a combination of the Goldhaber-Teller displacement mode 
and the Steinwedel-Jensen acoustic mode, with the restoring forces from 
the droplet model~\cite{myers1977} (see also Sec.\ref{pheno}). The GDR was found to contain a large
component of the Goldhaber-Teller motion, but in heavy nuclei the Steinwedel-Jensen
mode also contributes.

Covariance analysis of energy density functionals, that provides insight into correlations between different observables and nuclear properties (see Sec.\ref{theo}), indicates strong
correlations between the IVGDR excitation energy, neutron-skin thickness, and symmetry energy parameters, or more specifically, a combination of them. The value of the symmetry energy $S(\rho)$ associated
with the Skyrme functionals, at sub-saturation densities around 0.1fm$^{-3}$, is strongly 
correlated with the centroid energy of the IVGDR in spherical nuclei~\cite{trippa08}. Specifically, in the latter work, it was found that $23.3$ MeV$<S(\rho=0.1$ fm${}^{-3})<24.9$ MeV. Since the IVGDR provides essential information to constrain the nuclear matter symmetry energy, detailed experimental and theoretical studies of the IVGDR are required. The link between measurements of the IVGDR properties and neutron skin thickness has been established in Ref.~\cite{Krasznahorkay1991} in inelastic scattering of $\alpha$ particles by $\alpha'-\gamma$ coincidence measurements. The
analysis of experimental data for $^{116}$Sn, $^{124}$Sn, and $^{208}$Pb, with an 
assumption of Goldhaber-Teller picture of the IVGDR, resulted in ${\Delta r_{\rm np}}/R$
differences between the proton and neutron distributions in reasonable agreement 
with theoretical predictions~\cite{Krasznahorkay1991}.

The properties of IVGDR	have been extensively studied over the past decades. We refer to previous review articles for more details and references therein~\cite{berman1975,Goeke1982,Snover1986,Woude1987,Paar2007}, and the progress in more recent studies is addressed in this review.
%%%%%%%%%%%%%%%%%%%%%%%%%%%%%%%%

The experimental IVGDR properties and respective uncertainties 
have recently been updated in Ref.\cite{Plujko2011}, based on least-square fitting
of theoretical photo-absorption cross sections to experimental data. In the former
case, components corresponding to excitation of the IVGDR and quasi-deuteron
contribution to the experimental photo-absorption cross section have been taken
into account~\cite{Plujko2011}. A global analysis of the IVGDR for $A>80$ nuclei
allowed a study of the correlation of the resonance spreading with energy with 
high accuracy\cite{Junghans2008}. In heavy nuclei, departure from the shell closure
results in widening of the IVGDR spreading, and fits with a single Lorentzian to the
IVGDR data in deformed nuclei do not allow systematic  $A-$dependence  of 
the  widths. By considering the sensitivity to the details 
of the nuclear shape, a new parameterization of the energy dependence
of the electric dipole strength has been established for spherical, transitional, 
triaxial and well deformed nuclei~\cite{Junghans2008}. The properties of the IVGDR
have also been measured in the compound nuclei $^{80}$Zr and $^{81}$Rb, 
demonstrating the importance of these studies to obtain information
on the basic quantities in nuclear structure such as the shape evolution
and isospin symmetry~\cite{Ceruti2017}.

\begin{figure}[t!]
\centering
\includegraphics[width=0.5\linewidth,clip=true]{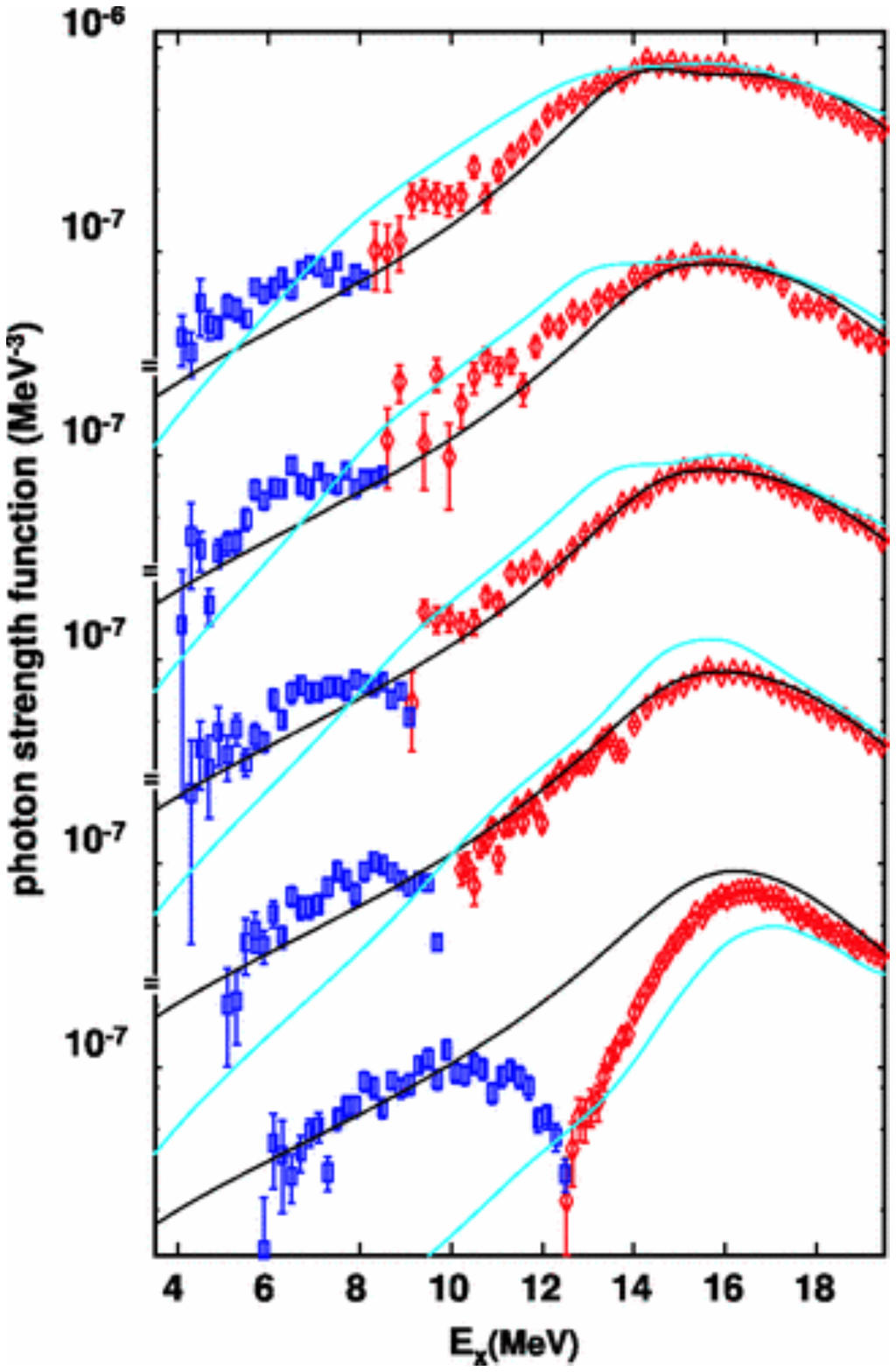}
\caption{\label{fig_dipole_erhard}Experimental photon strength functions
for $^{92,94,96,98,100}$Mo (from bottom to top) in comparison with the $E1$ 
strength parameterisation based on deformation (black solid line) and another
approach based on the RPA (cyan thin line). 
Figure taken from  Ref.~\cite{Erhard2010}.}
% corresponds to Fig.7. in Erhard2010
\end{figure}

Experimental study
of the electric dipole strength in Mo isotopes in Ref.~\cite{Erhard2010} includes 
two methods (i) photon scattering up to the neutron separation energies $S_n$
and (ii) photo-activation from $S_n$ toward the region of IVGDR resonance energies.
Fig.~\ref{fig_dipole_erhard} shows the experimental photon strength functions
for $^{92,94,96,98,100}$Mo in comparison with the $E1$
strength parameterisation based on deformation and simpler approach based on the RPA.
While previous parameterisations, based on a single Lorentzian, resulted in false extrapolation
to the low-energy tail, an updated parameterization~\cite{Junghans2008} provides improved description of the data below the neutron threshold due to the shift of one component of the IVGDR to lower excitation energy~\cite{Erhard2010}.
The results  provide the insight into the shape dependence of the dipole strength
in Mo isotopes, involving the effects of quadrupole deformation and triaxiality.
It has been shown that the parameterization for the dipole strength distribution
in $A>80$ deviates considerably from general prescription that has previously
been used, with possible implications to the nucleosynthesis network
calculations~\cite{Erhard2010}. In particular, dipole strength has important
impact on the neutron capture cross sections for the r-process simulations~\cite{Goriely2002,Goriely2004}. Since the experimental data are not available
for the whole nuclear chart,  large-scale QRPA calculations of the $E1$-strength have 
been performed to microscopically derive the radiative neutron capture 
cross sections. Because the QRPA cannot describe damping of the collective
motion, this effect has been taken into account through a folding procedure
of the dipole strength functions~\cite{Goriely2002,Goriely2004}.

From the theory side, isovector and isoscalar dipole excitations have been
investigated in light nuclei $^{9,10}$Be in the framework of antisymmetrized 
molecular dynamics with angular-momentum and parity projections~\cite{Enyo2016}.
The IVGDR shows two-peak structure, due to dipole excitation in the $2\alpha$ core
part with the prolate deformation. In addition, low-energy E1 resonance was predicted
due to oscillations of valence neutrons against the $2\alpha$ core.
For medium heavy to heavy spherical neutron-rich nuclei, investigation based on 
the collisional Landau kinetic theory, showed that the splitting of the giant multipole resonances in spherical neutron-rich nuclei originates from the interplay of the
 isovector and isoscalar sounds with different velocities~\cite{Kolomietz2006}. Within
 this approach, it has been shown that the calculated values of splitting energy
 the relative strength of the resonance peaks, appear in agreement with respective experimental data. It has been shown that the enhancement factor $\mathcal{K}$ in the sum of energy weighted strength for the IVGDR increases with $A$ due to boundary 
 condition on the moving nuclear surface~\cite{Kolomietz2009}.

\begin{figure}[t!]
\centering
\includegraphics[width=0.5\linewidth,clip=true]{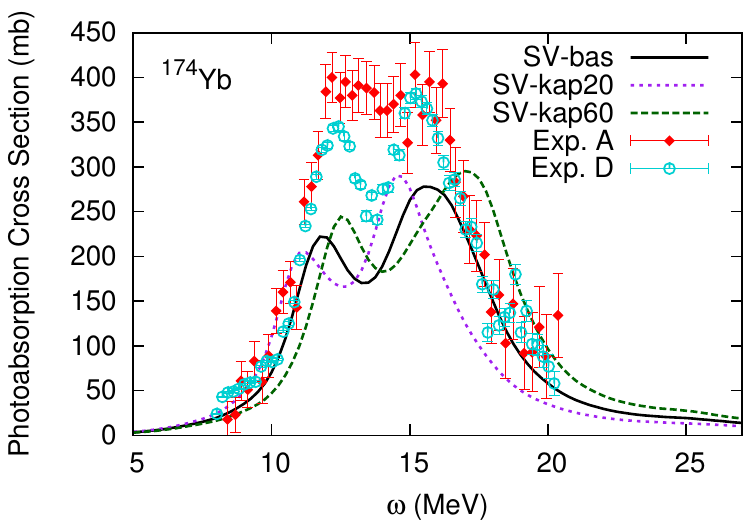}
\caption{\label{fig_dipole_oishi} The FAM-QRPA photo-absorption cross sections of $^{174}$Yb calculated with Skyrme parameterisations SV-bas, SV-kap20, and SV-kap60. For comparison, photo-absorption~\cite{Gurevich1981} and neutron yield~\cite{Goryachev1976} experimental data are shown. Figure taken from  Ref.~\cite{Oishi2016}.}
% corresponds to Fig.7. in Oishi2016
\end{figure}

%FAM etc.
The properties of IVGDR have been studied in heavy and superheavy nuclei
in the framework of separable RPA based on Skyrme parameterization SLy6~\cite{Kleinig2008}. The widths of the resonance appeared mainly determined
by the Landau fragmentation, that is strongly influenced by deformation effects.
The self-consistent RPA based on the FAM with Skyrme functionals in the three-dimensional
space has been applied in the description of electric dipole excitations~\cite{Inakura2009}, resulting with reasonable agreement for the IVGDR peak energies for heavy nuclei.
However, the energies of light nuclei have been systematically underestimated.
Further study of the IVGDR in heavy rare-earth isotopes based on FAM-QRPA,
showed that this highly efficient method mainly reproduces the resonance energy,
with some deficiencies in isotopes heavier than erbium. Fig.~\ref{fig_dipole_oishi}
shows the photo-absorption cross sections for $^{174}$Yb calculated using FAM-QRPA~\cite{Oishi2016} with Skyrme parameterisations SV-bas, SV-kap20, and SV-kap60, in comparison to the photo-absorption~\cite{Gurevich1981} and neutron yield~\cite{Goryachev1976} experimental data.
The model deficiencies could not be improved by adjusting the enhancement factor in the EWSR, through implementation of SV-kap20 and SV-kap60 parameterizations, suggesting that methods going beyond QRPA should be developed, or another 
optimisation of the EDF parameters should be introduced~\cite{Oishi2016}.

\begin{figure}[t!]
\centering
\includegraphics[width=0.5\linewidth,clip=true]{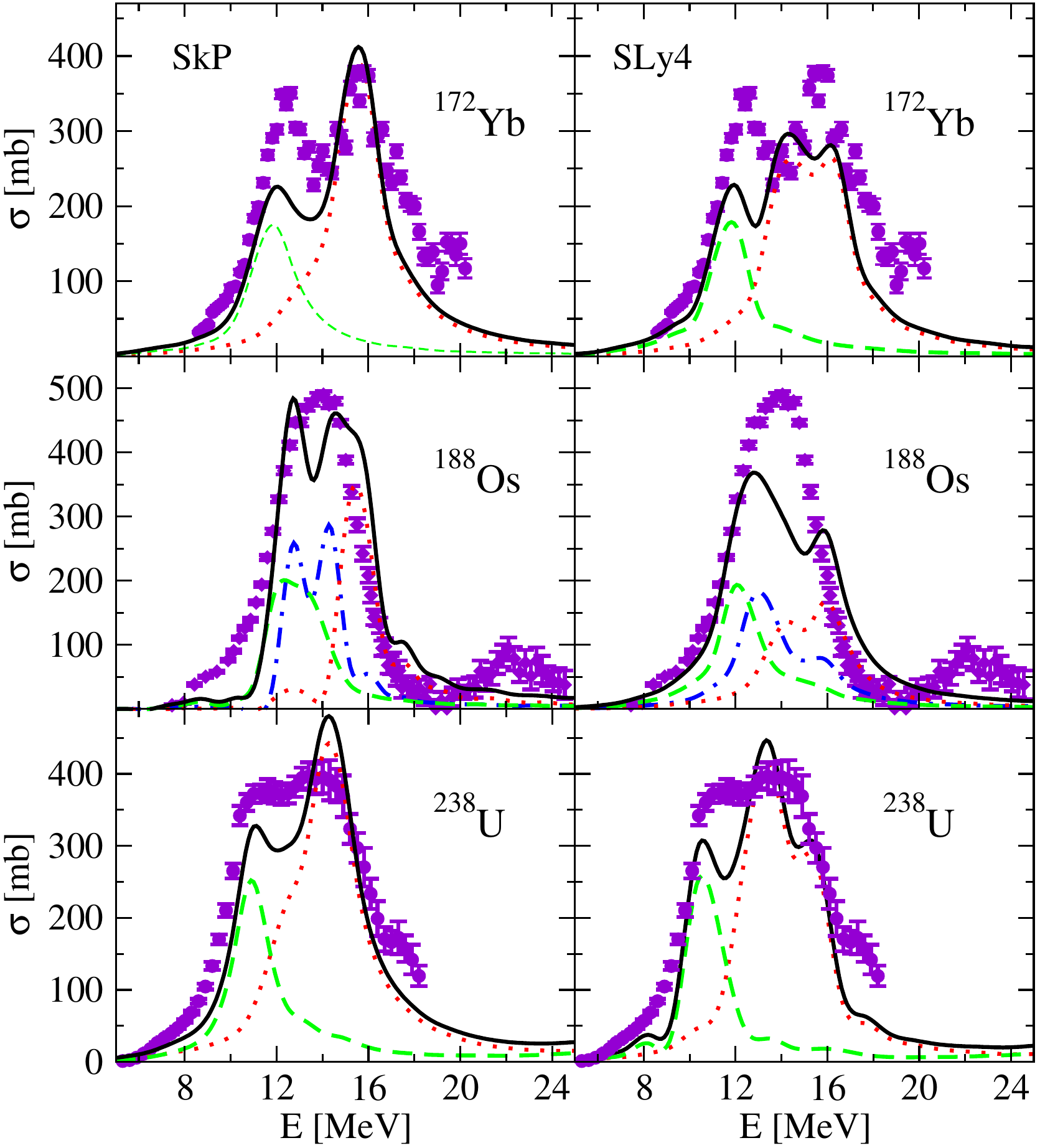}
\caption{\label{fig_dipole_stetcu} The photo-absorption cross sections in $^{172}$Yb,$^{188}$Os, and $^{238}$U calculated in the framework of 3D time-dependent DFT (solid line), in comparison to the experimental $(\gamma,n)$ cross sections (circles). Contributions to the cross sections are separately shown for exciting the nucleus along the long axis (dashed line), the short axis (dotted line; multiplied by 2 for $^{172}$Yb, $^{238}$U) and the third middle axis in the case of triaxial $^{188}$Os.
Figure taken from  Ref.~\cite{Stetcu2011}.}
% corresponds to Fig.2. in Stetcu2011
\end{figure}

In Ref.~\cite{Arteaga2009} the IVGDR strength has also been systematically 
studied in Sn isotopes with extreme neutron excess $(A\geq 132)$ in the framework of relativistic deformed QRPA~\cite{Arteaga2008}. Model calculations reproduce the main
properties of the IVGDR, reduction of the centroid energy and the splitting of the
response in two modes due to deformation. The GDR splitting depends linearly
on the deformation.
The IVGDR has been investigated in axially deformed $^{172}$Yb, $^{238}$U and triaxial nucleus $^{188}$Os in the three-dimensional time-dependent density functional theory,
based on superfluid local density approximation~\cite{Stetcu2011}. In the case of axially deformed nuclei $^{172}$Yb and $^{238}$U, calculations include two situations, perturbing the system along the longer and the shorter axis. For triaxial $^{188}$Os, three cases have been explored accordingly. The respective photo-absorption cross sections are shown in Fig.~\ref{fig_dipole_stetcu} for two Skyrme functionals, SkP and SLy4, resulting in reasonable agreement with the experimental $(\gamma,n)$ cross sections~\cite{Stetcu2011}.

Within a fully consistent HFB+QRPA with axial symmetry, based on the Gogny D1S effective interaction, isovector dipole response has been investigated in $^{26}$Ne, $^{28}$Si, and $^{24}$Mg, resulting in reasonable agreement with the experimental data~\cite{Peru2007}. The same framework, based on D1S and D1M Gogny functionals, has been employed in large-scale deformed QRPA calculations of the $\gamma$-ray strength function for astrophysical applications~\cite{Martini2016}. In order to account the effects going beyond the QRPA, phenomenological correlations have been included to reproduce quantitatively photoabsorption data~\cite{Martini2016}. Continuum effects on the IVGDR have been explored within continuum QRPA with Skyrme functional~\cite{Mizuyama2009} and continuum RPA for relativistic point coupling interactions~\cite{Daoutidis2009}. In Ref.~\cite{Mizuyama2009} a new formulation of the continuum QRPA has been established, in which the velocity-dependent terms of the Skyrme effective interaction are explicitly
treated, and their effects on the strength distribution and the transition density
of the low-lying surface modes and the giant resonances have been explored.
The decay properties of the IVGDR have been investigated in a semi-microscopic
approach based on the continuum RPA, phenomenological mean field and
Landau-Migdal $ph$ interaction~\cite{Urin2008}.

In Ref.~\cite{Simenel2009} couplings between dipole and quadrupole vibrations in tin isotopes have been investigated, using time-dependent HF based on Skyrme functional. Couplings of this kind provide a source of anharmonicity in the multiphonon spectrum, and also affects 
the dipole motion in a nucleus with a static or dynamical deformation induced by a 
quadrupole constraint or boost, respectively~\cite{Simenel2009}.

Fine structure of the IVGDR has been explored in $^{208}$Pb in the analysis of characteristic scales based on continuous wavelet transforms~\cite{Poltoratska2014}. From the comparison of the corresponding analyses of the $(p,p')$ experimental data and theoretical predictions, giant resonance decay mechanisms responsible for the fine structure have been identified. In model calculations major scales are present already at $ph$ level, and inclusion of complex configurations has limited impact on the wavelet power spectra and characteristic scales~\cite{Poltoratska2014}. Therefore, Landau damping has been suggested as the key mechanism responsible for the fine structure of the IVGDR~\cite{Poltoratska2014}.

\begin{figure}[t!]
\centering
\includegraphics[width=0.75\linewidth,clip=true]{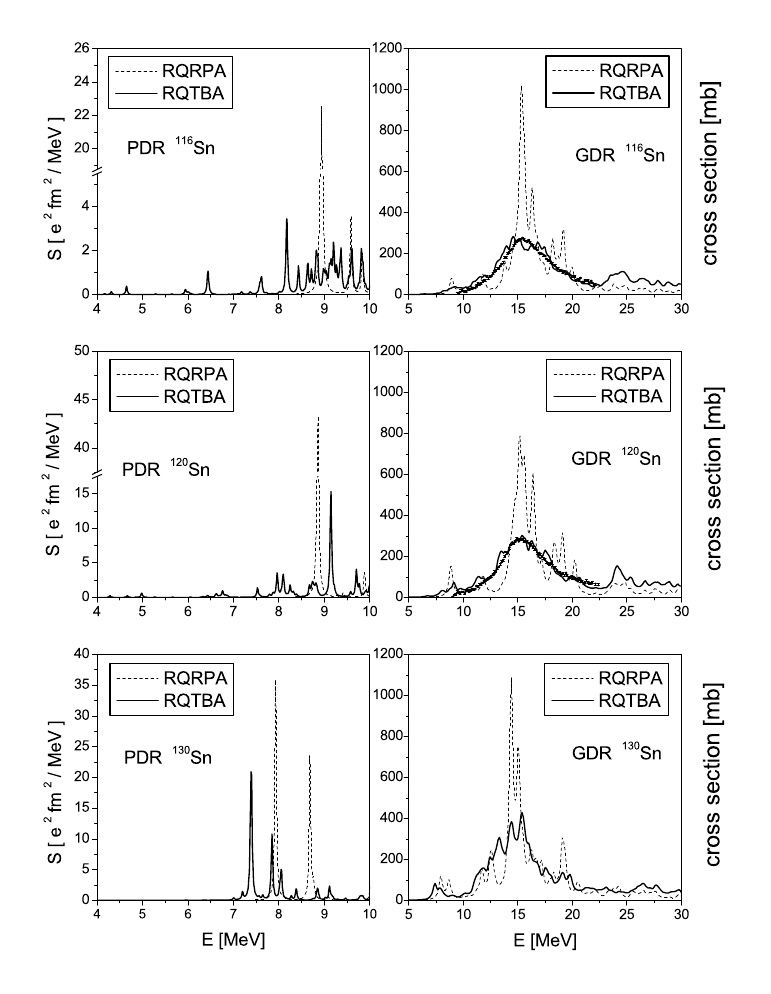}
\caption{\label{fig_dipole_litvinova} Isovector low-energy strength (left column)
and IVGDR spectra (right column) for $^{116,120,130}$Sn, calculated using 
relativistic QTBA and RQRPA. Experimental photoabsorption cross sections
are shown for $^{116,120}$Sn. Figure taken from  Ref.~\cite{Litvinova2008}.}
% corresponds to Fig.2. in Litvinova2008
\end{figure}

Electric dipole response has been investigated in open shell Sn isotopes and $(N=50)$ isotones 
in the relativistic QRPA extended by the quasi-particle-phonon coupling model using the
quasi-particle time blocking approximation (QTBA)~\cite{Litvinova2008} with NL3 parameterization~\cite{NL3}. Within this approach, by taking into account phonon-nucleon coupling vertices for $J\leq6$ and natural parity, the isovector dipole strength distributions have also been systematically 
investigated in $^{40-54}$Ca~\cite{Egorova2016}. Fig.~\ref{fig_dipole_litvinova} shows the isovector dipole strength distributions for $^{116,120,130}$Sn, based on relativistic QRPA and QTBA~\cite{Litvinova2008}. The envelopes of the calculated IVGDR for $^{116,120}$Sn successfully
describe measured photoabsorption cross sections, thus the key mechanism responsible for the
damping of the IVGDR has been taken into account in an appropriate approach.
Small deviations originate from
neglecting more complicated couplings beyond 2-quasi-particle - phonon, configurations by
the time blocking, discretized continuum, restriction of the phonon subspace only by
low-lying modes, and rather simple implementation of the pairing interaction. In the case 
of low-lying excitations (see Fig.~\ref{fig_dipole_litvinova}), more realistic description 
would necessitate including one-, two-, and possibly three-phonon configurations, as 
for example in the quasi-particle phonon model~\cite{Govaert1998}.

The properties of IVGDR in $^{40}$Ca and $^{208}$Pb have been addressed in the non-relativistic self-consistent framework based on the RPA extended by including the quasi-particle-phonon coupling at the level of the QTBA~\cite{Lyutorovich2015}. In addition to consistent calculation
of all matrix elements from the same Skyrme functional, single-particle continuum has been included to avoid discretization of the states usually employed in RPA and TBA. It has been shown that the inclusion of the phonon coupling in the TBA results in small shifts downwards of the resonance centroid energies~\cite{Lyutorovich2015}.

Isoscalar and isovector dipole strength distributions have also been investigated 
in relation to the Schiff moment~\cite{Auerbach2014}.

%%%%%%%%%%%%%%%%%%%%%%%%%%%%%%%%%%%%%%%%%%%%
\subsubsection{Dipole polarizability}
%%%%%%%%%%%%%%%%%%%%%%%%%%%%%%%%%%%%%%%%%%%%

In addition to the IVGDR excitation energy, another isospin sensitive observable can be
extracted from the isovector dipole strength distribution: the dipole polarizability $\alpha_D$ that 
corresponds to the sum of the inverse energy weighted transition strength (see also Sec.\ref{pheno}). The covariance
analysis within the energy density functional approach showed that the dipole polarizability
is strongly correlated with the neutron form factor, i.e., with the isovector properties of finite nuclei 
such as neutron skin thickness, as well as with the properties of the symmetry energy of
nuclear matter~\cite{reinhard10}. Therefore, in recent years the dipole polarizability attracted
considerable interest both in theoretical and experimental nuclear physics community.

The electric dipole polarizability has recently been measured in three nuclei using polarized
proton inelastic scattering at extreme forward angles~\cite{Tamii2009,Neveling2011}, a technique that allowed extraction of the $\alpha_D$ value over a wide energy range and with high resolution~\cite{Tamii2009,Neveling2011}. By implementing multipole decomposition of the angular distribution and by measuring all polarization transfer observables, high resolution electric dipole response of $^{208}$Pb has been obtained over a wide range of energies. By supplementing this measurement with
available data on electric dipole response in $^{208}$Pb toward higher energies, up to the
pion-production threshold~\cite{schelhaas1988,veyssiere1970}, the  value for
the electric dipole polarizability  $\alpha_D({}^{208}\textrm{Pb})=20.1 \pm 0.6$ fm$^3$ has 
been obtained~\cite{tamii11}. Within the same approach, the electric dipole strength of $^{120}$Sn 
has been measured in the interval between 5 and 22 MeV\,\cite{hashimoto15}, and supplemented
 with photo-absorption data up to 135 MeV\,\cite{lepretre1981}, resulting in $\alpha_D({}^{120}\textrm{Sn})=8.93 \pm 0.36$ fm$^3$~\cite{hashimoto15}. Finally, proton inelastic scattering 
has been used to measure electric dipole strength in $^{48}$Ca between 5 and 25 MeV\cite{Birkhan2017}. Additional contribution to E1 strength at low-energies has been extracted from $(\gamma,\gamma')$ reaction, while for energies from 25 to 60 MeV the photoabsorption data in $^{40}$Ca, with additional correction, have been used to determine the value $\alpha_D({}^{48}\textrm{Ca})=2.07 \pm 0.22$ fm$^3$~\cite{Birkhan2017}.

 The electric dipole polarizability has also been measured in unstable neutron-rich nucleus $^{68}$Ni,
in the approach based on Coulomb excitation in inverse kinematics and by measuring the invariant mass in the one- and two-neutron decay channels~\cite{Wieland:2009,rossi13}. The resulting
dipole polarizability, based on the measured energy range between 7.8 and 28.4 MeV, amounts
$\alpha_D({}^{68}\textrm{Ni})=3.40 \pm 0.23$ fm$^3$.
Measuring neutron rich nuclei opens the possibility of a precise determination of neutron skins
as a function of neutron excess~\cite{rossi13}. 

Results on $^{68}$Ni, $^{120}$Sn, and $^{208}$Pb provided important constraints on the isovector
properties of energy density functionals, i.e., the corresponding symmetry energy parameters~\cite{roca-maza2015b}. Correlations between the dipole polarizability, neutron skin thickness and the symmetry energy parameters have been studied within non-relativistic and relativistic
EDFs~\cite{roca-maza2015,roca-maza2015b}, and more recently also within {\it ab initio} 
models based on chiral effective field theory interactions~\cite{Birkhan2017}.

As pointed out in Ref.~\cite{roca-maza2015b}, experimental values on electric dipole polarizability for 
${}^{68}$Ni, ${}^{120}$Sn, and ${}^{208}$Pb reported in Refs.~\cite{rossi13,hashimoto15,tamii11}
cannot be directly compared with $\alpha_D$ values from the model calculations. Since for 
${}^{68}$Ni the dipole response has been measured in the energy interval between 7.8 MeV and 28.4 MeV,  the dipole response has to be extrapolated to lower and higher energy regions to cover 
complete range that would correspond to the model calculations. The strength below the neutron
threshold has been estimated from the tail of a Lorentzian-plus-Gaussian fit to the deconvoluted data,
where the Lorentzian extrapolates the giant dipole resonance to low energies and the Gaussian covers the low-energy pygmy dipole strength ~\cite{dominic,roca-maza2015b}. The strength above the upper experimental limit of 28.4 MeV \cite{rossi13} has been extrapolated from the same Lorentzian fit of the GDR strength~\cite{dominic,roca-maza2015b}. Finally, the corrected value of the dipole polarizability for  ${}^{68}$Ni, including contributions from extrapolated low-energy and high-energy regions, is $\alpha_D({}^{68}\textrm{Ni})=3.88 \pm 0.31$ fm$^3$  \cite{dominic}. 

In the case of reported $\alpha_D$ values for $^{120}$Sn~\cite{hashimoto15}, 
and $^{208}$Pb~\cite{tamii11}, for a direct comparison with model calculations, 
corrections are required in the energy region above 30 MeV due to contamination 
coming from non-resonant processes (the so-called quasi-deuteron effect \cite{schelhaas1988,lepretre1981}). After subtracting these contributions~\cite{atsushi,lepretre1981,schelhaas1988}, corrected values of the dipole
polarizability have been reported, $\alpha_D({}^{120}\textrm{Sn}) = 8.59 \pm 0.37$ fm$^3$ 
and $\alpha_D({}^{208}\textrm{Pb}) = 19.6 \pm 0.6$ fm$^3$~\cite{roca-maza2015b,atsushi}.
As pointed out in Ref.~\cite{roca-maza2015b}, quasi-deuteron excitations, if not properly subtracted, also lead to values of the experimental EWSR which are inaccurate, much more than for the dipole polarizability. 

In Ref.~\cite{piekarewicz12} the original experimental value on $\alpha_D$
in $^{208}$Pb~\cite{tamii11} has been used in theoretical analysis based on 
various non-relativistic and relativistic EDFs to constrain the neutron skin thickness $\Delta r_{\rm np}$.
It has been shown that precise measurements of $\Delta r_{\rm np}$ and $\alpha_D$ 
could significantly constrain the isovector sector of the EDFs~\cite{piekarewicz12,Piekarewicz2014}. 
The corrected experimental values for $\alpha_D$ in $^{68}$Ni, $^{120}$Sn
and $^{208}$Pb~\cite{roca-maza2015b}, as discussed above, have been implemented 
in the analysis of symmetry energy and its density dependence within the theory 
frameworks of non-relativistic and 
relativistic nuclear energy density functionals~\cite{roca-maza2015b}.
The calculated values of $\alpha_D$ are used to validate different correlations 
involving the symmetry energy at the saturation density $J$, the corresponding 
slope parameter $L$, and the neutron skin thickness $\Delta r_{\rm np}$.
In order to improve understanding of the correlations between $\alpha_D$ and
other isovector observables, in Ref.~\cite{roca-maza13a} droplet model has been
employed to establish analytic relations between the dipole polarizability, neutron 
skin thickness and the symmetry energy parameters. It has been shown that 
within the droplet model the product quantities $\alpha_D^{DM} J$ is linearly
correlated with neutron skin thickness $\Delta r_{\rm np}^{DM}$, in agreement with
Ref.~\cite{Satula2006}. 

\begin{figure}[t!]
\centering
\includegraphics[width=0.45\linewidth,clip=true]{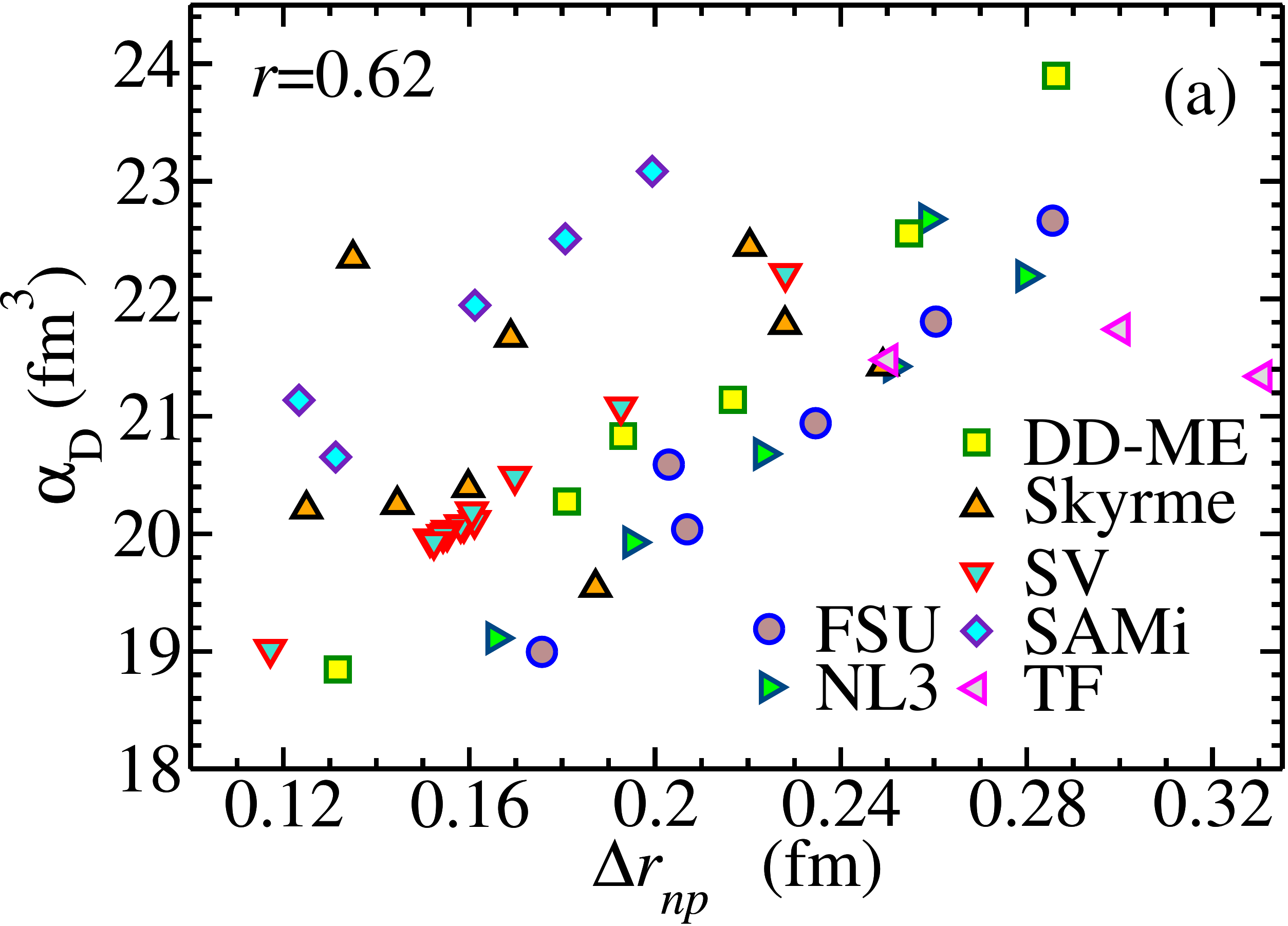}
\includegraphics[width=0.45\linewidth,clip=true]{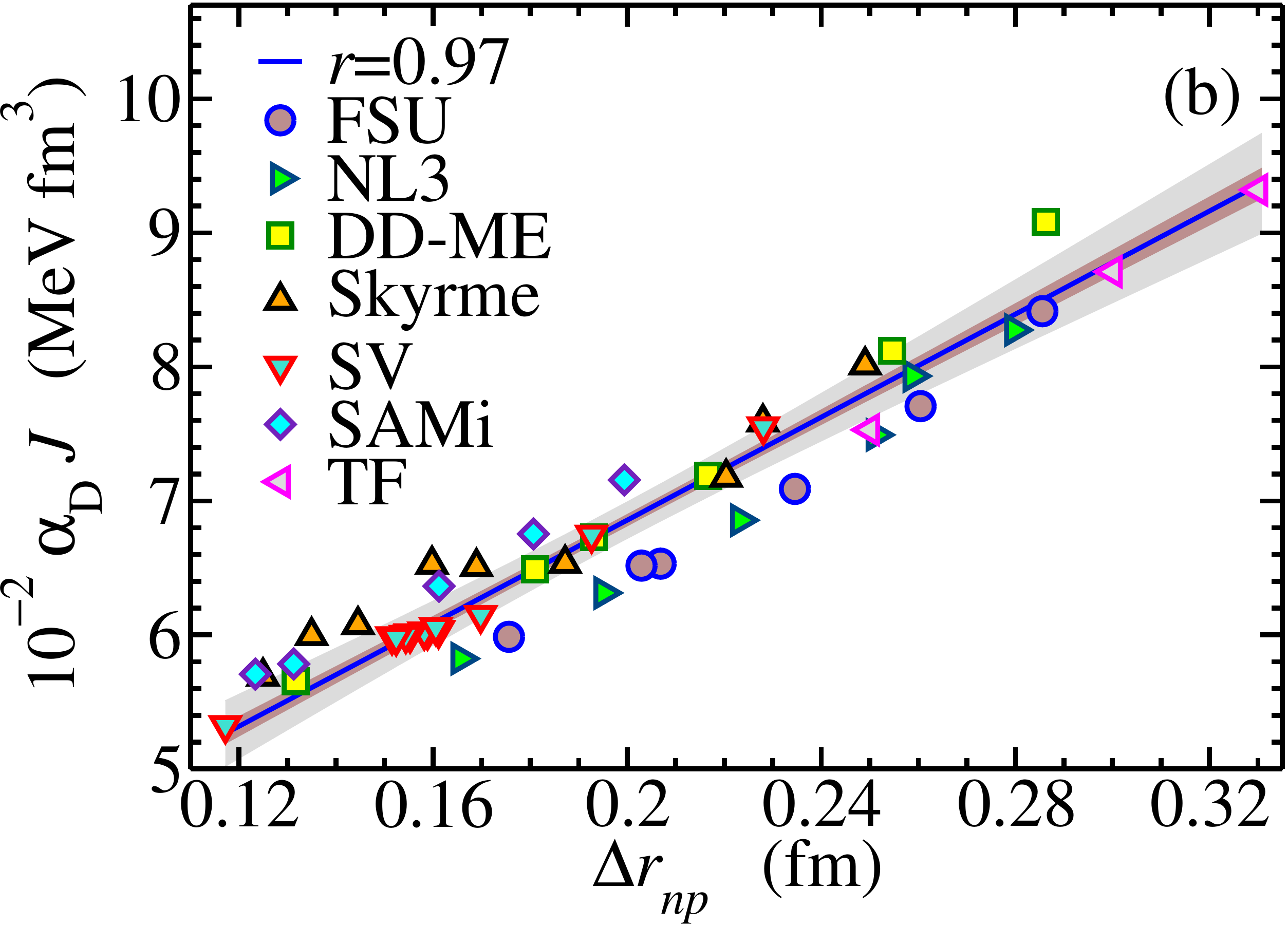}
\caption{\label{fig_polariz_roca_maza1} (a) Dipole polarizability $\alpha_D$ as a function of neutron skin thickness in $^{208}$Pb. (b) The same like at the left panel, 
but the dipole polarizability is multiplied by the symmetry energy at saturation $J$. See text for details on the interactions employed. Figure taken from  Ref.~\cite{roca-maza13}.}
% corresponds to Fig.1. in roca-maza13
\end{figure}

Following indications from the droplet model, the correlations between the
electric dipole polarizability and neutron skin thickness have been explored 
microscopically by implementing a large and representative set of energy density 
functionals that span a wide range of the values of isovector
quantities~\cite{roca-maza13a}. Model calculations include
families of systematically varied interactions obtained
by a variation of the parameters around an
optimal value, but without significantly compromising the
quality of the merit function. In Ref.~\cite{roca-maza13a} the following families have 
been used: non-relativistic Skyrme functionals SAMi~\cite{roca-maza12b}
and SV~\cite{sv}), meson-exchange covariant EDFs NL3/FSU~\cite{NL3,nl3fsu1,Todd2005,piekarewicz11} and TF~\cite{fattoyev13},
and covariant EDF with density-dependent meson-nucleon
couplings DD-ME~\cite{ddme}. Fig.~\ref{fig_polariz_roca_maza1} (left panel) shows the
calculated electric dipole polarizability as a function of the neutron skin thickness
in $^{208}$Pb, based on aforementioned functionals. The figure shows linear dependence within
each family of the functionals, however, a significant amount of scatter is obtained
between the results for different functionals, resulting in rather moderate correlation coefficient
amounting 0.62. However,  if $\alpha_D J$ is considered instead of $\alpha_D$, a large
spread in the model results is considerably reduced, resulting in the correlation 
coefficient 0.97 (Fig.~\ref{fig_polariz_roca_maza1} (right panel)). Clearly, the
correlation between $\alpha_D J$ and $\Delta r_{\rm np}$ suggested by the droplet
model approach has been confirmed by microscopic calculations~\cite{roca-maza13}.

By implementing strong correlations $\alpha_D J$ with $\Delta r_{\rm np}$ and 
slope of the symmetry energy $L$,  and using EDFs with the experimental data 
for $\alpha_D$ in $^{208}$Pb, the following relation between $J$ and $L$ has been obtained: $J=(24.5\pm 0.8)+(0.168\pm0.007)L$~\cite{roca-maza13}. 

\begin{figure}[t!]
\centering
\includegraphics[width=0.45\linewidth,clip=true]{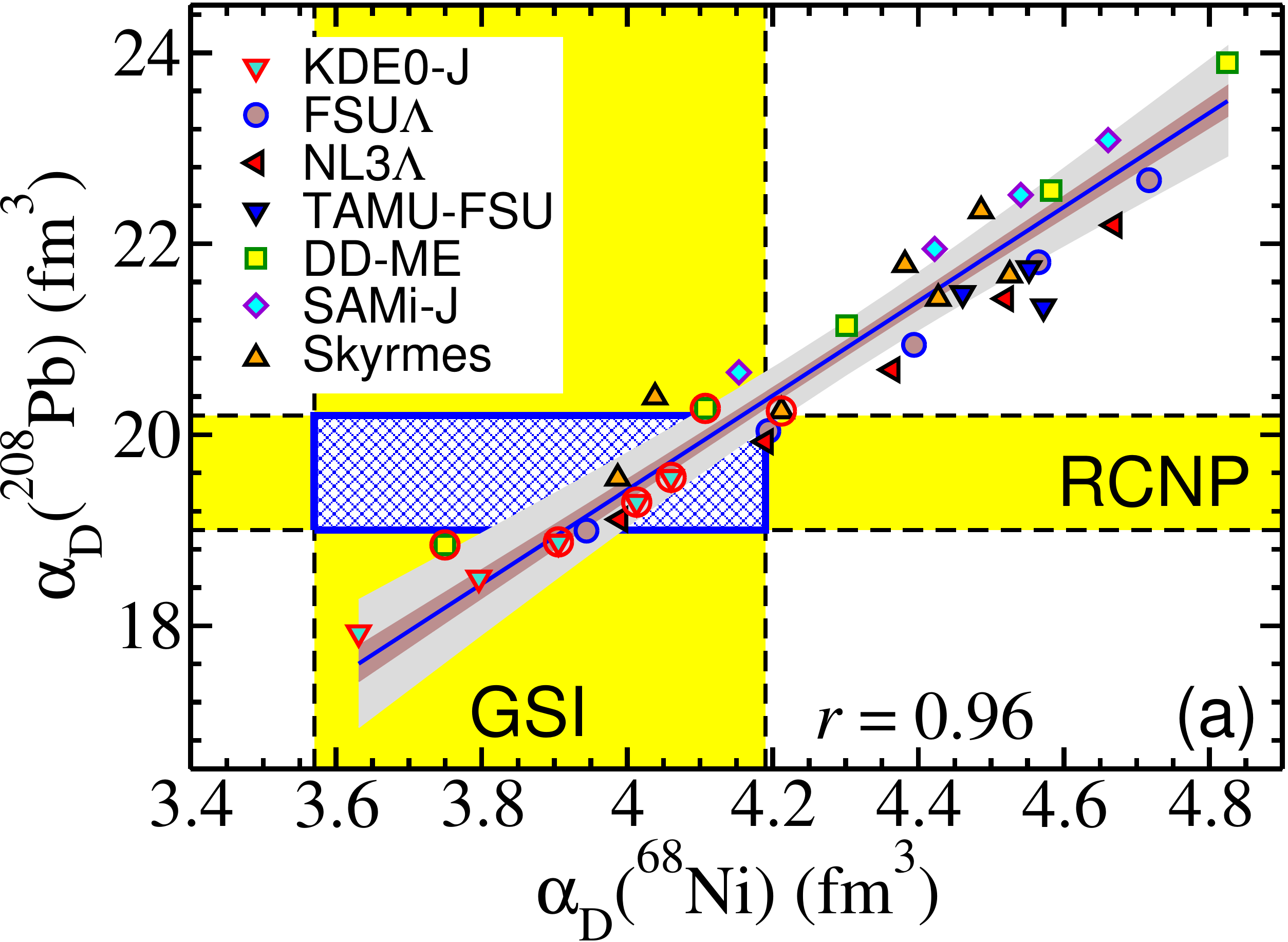}
\includegraphics[width=0.45\linewidth,clip=true]{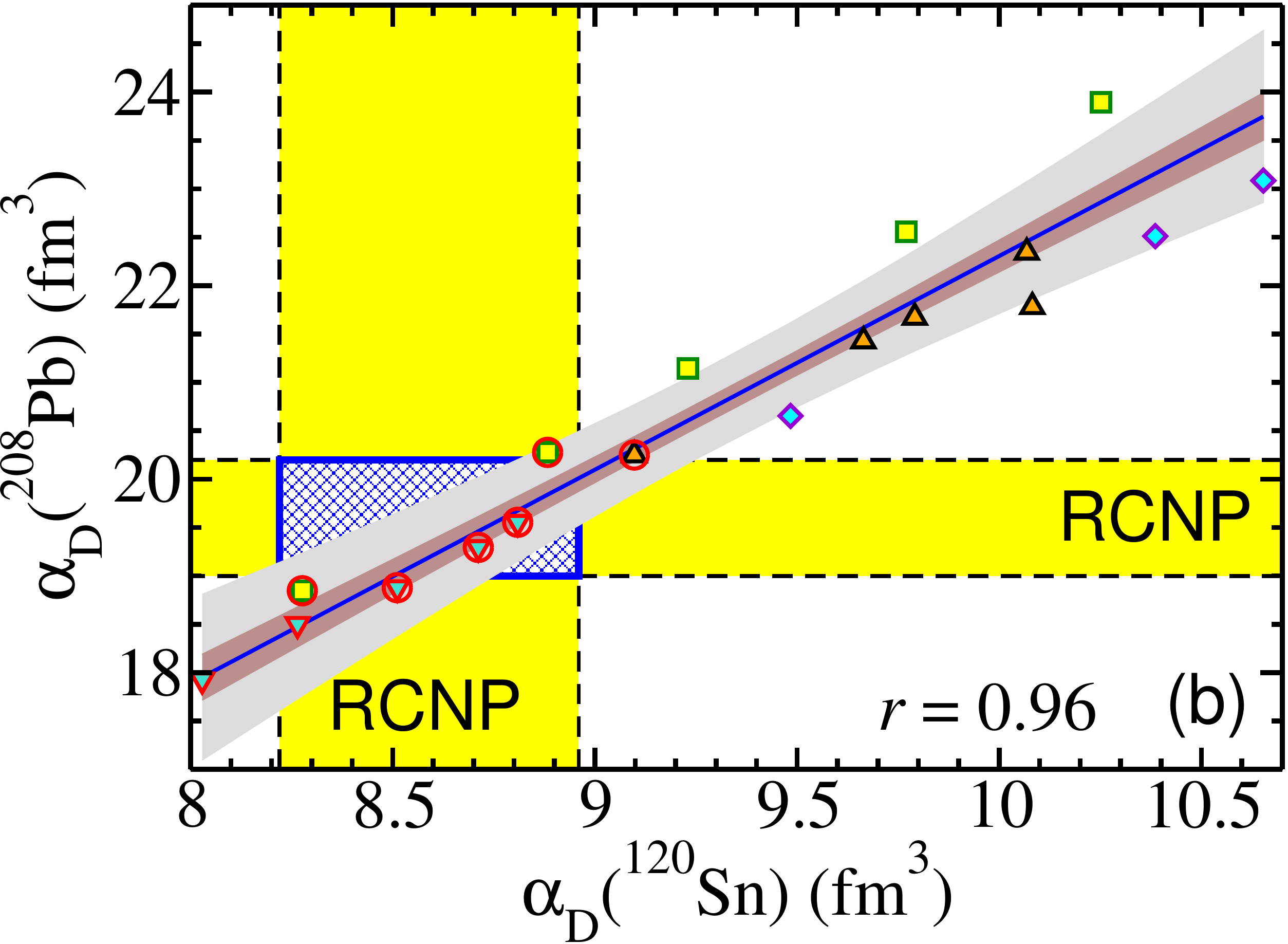}
\caption{\label{fig_polariz_roca-maza2} Calculated dipole polarizability in $^{68}$Ni, $^{120}$Sn, and $^{208}$Pb for various EDFs in comparison to experimental data (see text for details). Figure taken from  Ref.~\cite{roca-maza2015b}.}
% corresponds to Fig.4. in roca-maza2015b
\end{figure}

In the recent systematic study~\cite{roca-maza2015b}, corrected values of experimental data for $^{68}$Ni, $^{120}$Sn, and $^{208}$Pb (as discussed above), have been employed to constrain the symmetry energy parameters $J$, $L$,
and $\Delta r_{\rm np}$. Fig. ~\ref{fig_polariz_roca-maza2} shows the results
of model calculations for three respective nuclei, based on several families of
EDFs. These include non-relativistic Skyrme parameterizations SAMi-J\,\cite{roca-maza12b} and KDE0-J\,\cite{kde0}, and four relativistic families, NL3$\Lambda$, FSU$\Lambda$, TAMU-FSU\,\cite{NL3,nl3fsu1,piekarewicz11,fattoyev13}, and DDME~\cite{ddme2,ddme}. As shown in Fig.~\ref{fig_polariz_roca-maza2},
$\alpha_D$ in ${}^{208}$Pb appears strongly correlated to $\alpha_D$ in both ${}^{68}$Ni and  ${}^{120}$Sn. Horizontal and vertical bands in Fig.~\ref{fig_polariz_roca-maza2} denote the experimental values of the electric dipole polarizability corrected as discussed above, including error bars. A number of functionals (denoted by red
circles) reproduce the measured dipole polarizability in all three nuclei. By using this subset of EDFs,  the following values for the symmetry energy parameters have been obtained, $J\!=\!30 \text{-}35$ MeV, $L\!=\!20 \text{-} 66$ MeV; and the values for $\Delta r_{\rm np}$ in ${}^{68}$Ni, ${}^{120}$Sn, and ${}^{208}$Pb are in the ranges: 0.15\text{-}0.19 fm, 0.12\text{-}0.16 fm, and 0.13\text{-}0.19 fm, respectively~\cite{roca-maza2015b}.

In Ref.~\cite{Zhang2014} the dipole polarizability in $^{208}$Pb has been studied within the Skyrme HF+RPA using the MSL0-based family of parameterizations, as well as with an additional set of interactions obtained  with $\chi^2$ minimisation with systematically varied slope of the symmetry energy $L(\rho_c)$ at sub-saturation density $\rho_c=\text{0.11 fm}^{-3}$ ~\cite{Chen2010} (note that $L$ refers to $\rho_0$ if not specified as in the present case). From the comparison with the experimental
value of $\alpha_D$ in $^{208}$Pb, together with the symmetry energy at the sub-saturation density $\rho_c$=0.11 fm$^{-3}$ from the binding energy difference of heavy isotope pairs, the slope of the symmetry energy has been determined, $L(\rho_c)$=47.4$\pm$7.8 MeV. One should note that the analysis has been done by analyzing the predictions of only one family of interactions (MSL0) and that was based on the original 
experimental data~\cite{tamii11}, thus implementation of the necessary correction~\cite{roca-maza2015b} would somewhat lower the value for $L$ reported in ~\cite{Zhang2014}.

\begin{figure}[t!]
\centering
\includegraphics[width=0.45\linewidth,clip=true]{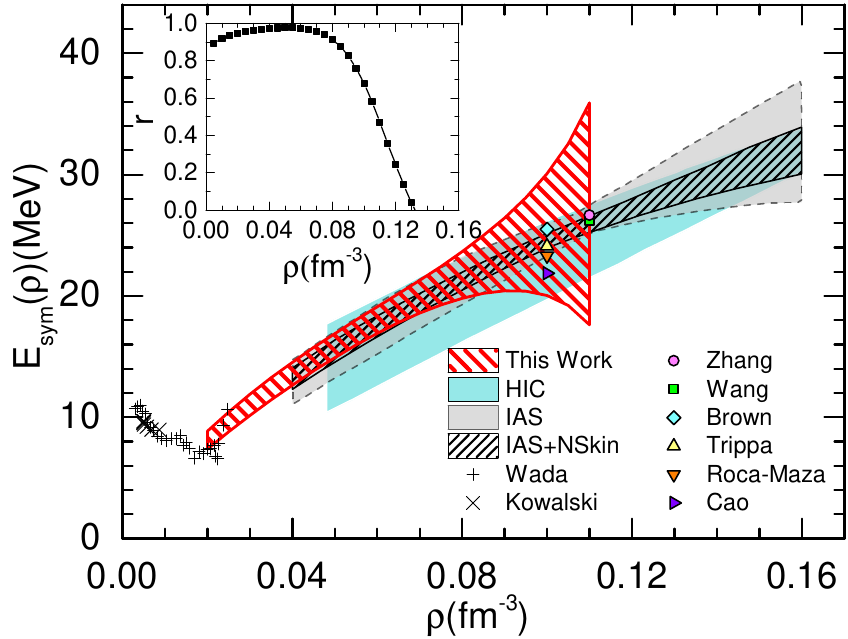}
\includegraphics[width=0.45\linewidth,clip=true]{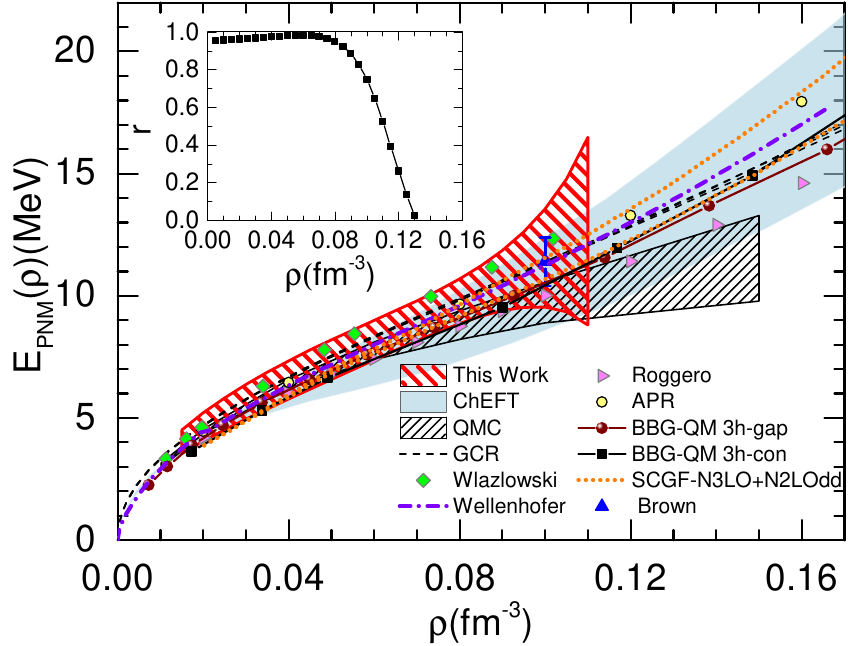}
\caption{\label{fig_polariz_zhang}Constraints on the symmetry energy (left panel) and the neutron matter EOS (right panel) as a function of density $\rho$.  The insets show the density dependence of the correlation $r$ between $\alpha_D$ and symmetry energy and neutron matter EOS, respectively. See text for details. Figure taken from  Ref.~\cite{Zhang2015}.}
% corresponds to Fig.2 & 3 in Zhang2015
\end{figure}

In another study, the electric dipole polarizability 
in $^{208}$Pb has been used as a probe of the symmetry energy and
neutron matter at sub-saturation densities, in particular around $\rho(r=0)/3$~\cite{Zhang2015}.
By exploiting the correlation between the value of $\alpha_D$ and the symmetry energy
at sub-saturation densities, the range of values for the symmetry energy below saturation 
has been obtained. Fig. ~\ref{fig_polariz_zhang} (left panel) shows a stringent constraint
obtained for the symmetry energy. Other constraints discussed in Ref.~\cite{Zhang2015}, shown in Fig.~\ref{fig_polariz_zhang}, include results from analyses of heavy ion collisions (HIC)~\cite{Tsang2009}, the Skyrme HF analyses of isobaric analog states (IAS) as well as combining additionally the neutron skin
thickness (IAS+NSkin) in Ref.~\cite{Danielewicz2014}, and six constraints on the value of $E_{\text{sym}}(\rho)$
around $2/3\rho_0$ from binding energy difference between heavy isotope pairs
(Zhang)~\cite{Zhang2013}, Fermi-energy difference in finite nuclei (Wang)~\cite{Wang2013}, properties of doubly magic nuclei (Brown)~\cite{Brown2013}, the giant dipole resonance in $^{208}$Pb (Trippa)~\cite{trippa08}, the giant quadrupole resonance in $^{208}$Pb (Roca-Maza)~\cite{roca-maza13} and the soft dipole excitation in $^{132}$Sn (Cao)~\cite{Cao2008}.
One should note that at low densities, below about $0.02$ fm$^{-3}$, clustering effects
are important, since they considerably increase the symmetry energy~\cite{Typel2010} .
The analysis in Ref.~\cite{Zhang2015} also indicates a strong correlation between $1/\alpha_{\mathrm{D}}$ in $^{208}$Pb and neutron matter energy 
$E_{\mathrm{PNM}}(\rho)$. Thus the experimental data on $\alpha_D$ in $^{208}$Pb 
has also been used to constrain the neutron matter EOS, as shown in the right panel in Fig.~\ref{fig_polariz_zhang} in the density interval $0.015$ fm$^{-3}<\rho<0.11$ $\mathrm{fm}^{-3}$~\cite{Zhang2015}. In Ref.~\cite{Zhang2016} the experimental value of dipole polarizability
in $^{208}$Pb, combined with the centroid energy of IVGDR has been used to constrain the
isovector effective mass of the Skyrme functionals, as well as to address the isospin splitting
of the effective mass.

The correlations between the slope parameter of the symmetry energy $L$ with $\alpha_D$ 
and $\alpha_D J$ have also been analyzed in Ref~\cite{Inakura2015} based on HF+RPA with Skyrme and Gogny functionals. The study confirmed  $\alpha_D J$ as a good quantity to constrain $L$, with $~\text{12 MeV}$ uncertainty. In Ref.~\cite{Knapp2015}, dipole transition strength in $^{208}$Pb has been studied in a self-consistent phonon coupling approach in a space spanned by one-phonon and two-phonon basis states using an optimized chiral two-body potential. In order to improve single-particle spectra, a phenomenological density-dependent term, derived from a contact three-body 
force, has been included in calculations~\cite{Knapp2015}. The results of model calculations demonstrate the role of the two-phonon states in enhancing the fragmentation of the strength.
Calculations of dipole polarizability showed that the phonon coupling does not influence
the value of $\alpha_D$~\cite{Knapp2015}. Another self-consistent approach, 
based on quasi-particle-phonon coupling at the level of time-blocking approximation,
has also been exploited in modeling dipole polarizability~\cite{Lyutorovich2015}.

\begin{figure}[t!]
\centering
\includegraphics[width=0.46\linewidth,clip=true]{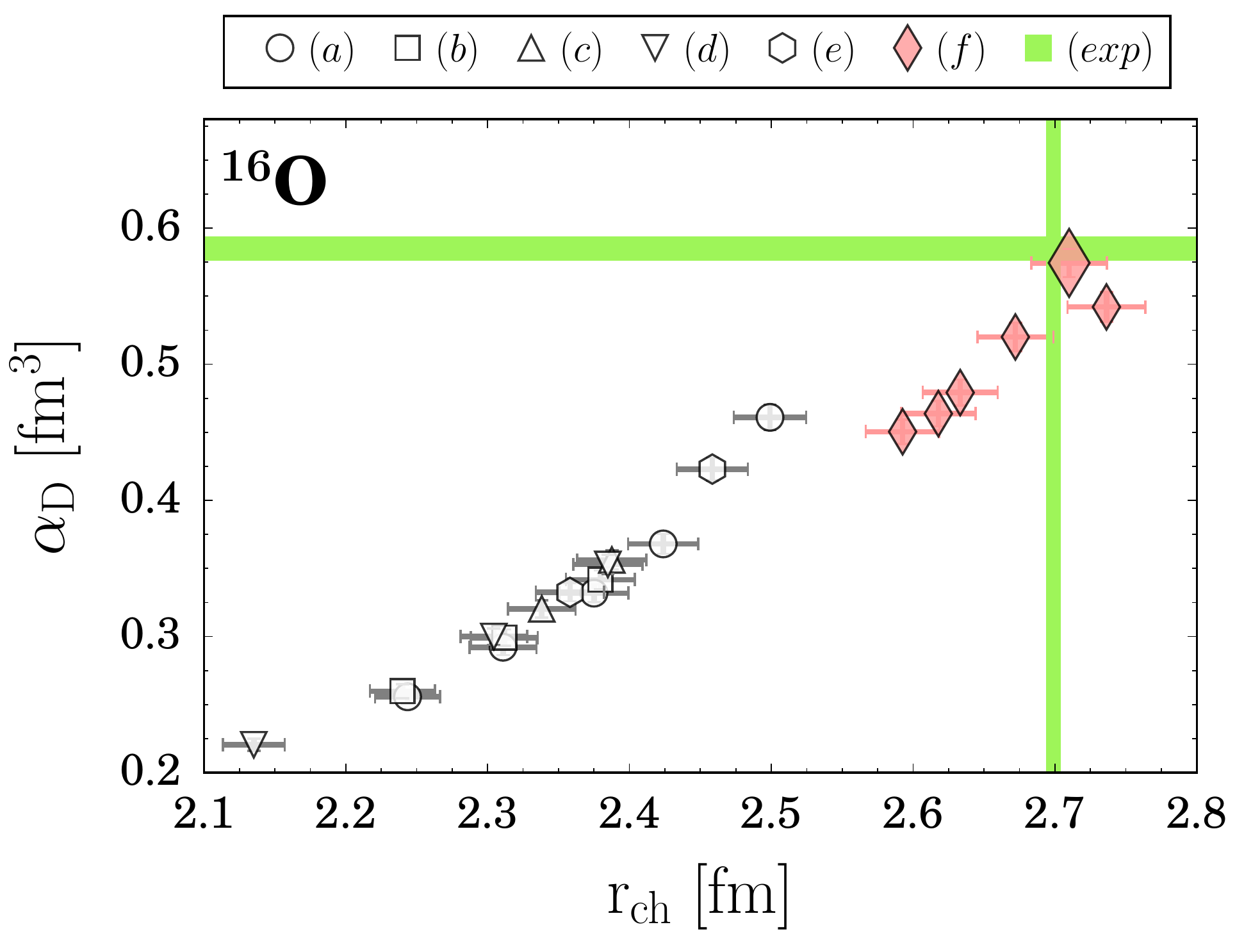}
\includegraphics[width=0.46\linewidth,clip=true]{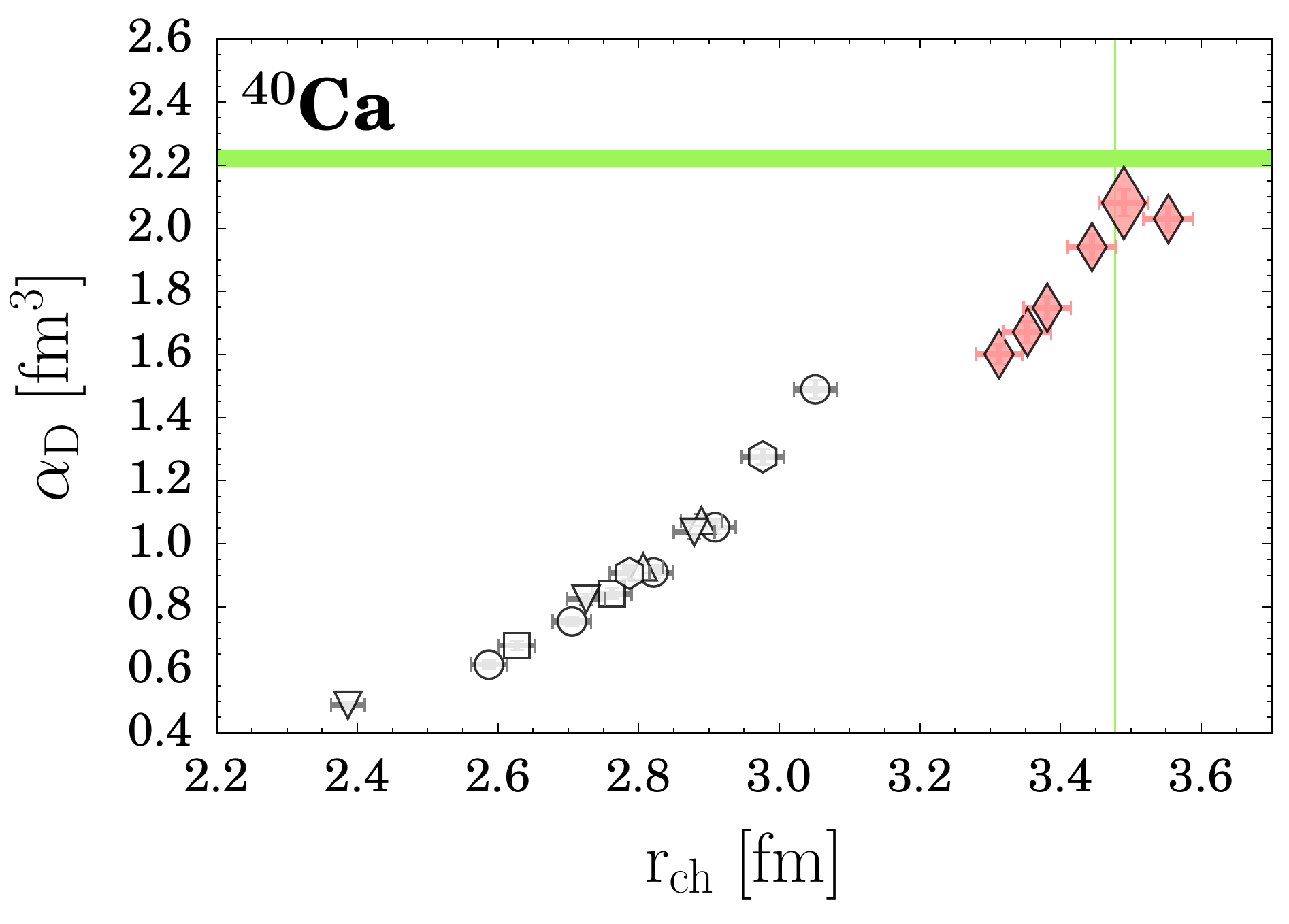}
\caption{\label{fig_polariz_Miorelli}Relationship $\alpha_D$ versus $r_{ch}$ in
  \textsuperscript{16}O and \textsuperscript{40}Ca. Results for
 $NN$ interactions include (empty symbols) $(a)$SRG evolved interaction~\cite{Entem2003}, $\Lambda=500$~MeV/c and $\lambda = \infty,3.5,3.0,2.5$ and $2.0 \ \rm{fm^{-1}}$, $(b)$ SRG evolved interaction~\cite{Entem2003} with $\Lambda=600$~MeV/c and $\lambda = 3.5,3.0$ and $2.5\ \rm{fm^{-1}}$, $(c)$ SRG evolved CD-BONN~\cite{Machleidt2001} with $\lambda = 4.0$ and $3.5\ \rm{fm^{-1}}$, $(d)$ V$_{low-k}$ evolved CD-BONN with $\lambda = 3.0,2.5$ and $2.0\ \rm{fm^{-1}}$ and $(e)$ V$_{low-k}$-evolved AV18~\cite{Wiringa1995} interaction and $\lambda = 3.0$ and $2.5\ \rm{fm^{-1}}$. The red diamonds $(f)$ denote
results including 3NF: the large one is from NNLO$_{\rm{sat}}$~\cite{Ekstrom2015} and the others from chiral interactions as in Ref.~\cite{Hebeler2011}.  Experimental data are denoted by green bands~\cite{Ahrens1975,angeli2013}. Figure taken from  Ref.~\cite{Miorelli2016}.}
% corresponds to Fig.11 in Miorelli2016
\end{figure}

%Electric dipole polarizability from first principles calculations~\cite{Miorelli2016}
Electric dipole polarizability has also been studied using the
interactions derived from chiral effective field theory.
In Ref.~\cite{Miorelli2016} the integral transforms have been combined with
the coupled-cluster method to compute the dipole polarizability in light
nuclei $^{4}$He, $^{16,22}$O, and $^{40}$Ca, using bound-state techniques. 
For the $\text{NNLO}_{sat}$ interaction, the model calculations reproduce 
experimental $\alpha_D$ values in $^{4}$He, $^{16}$O, however, in  $^{22}$O
the theoretical result exceeds the experimental value by a factor of two~\cite{Miorelli2016}.
Fig.~\ref{fig_polariz_Miorelli} shows the $\alpha_D-r_{\rm ch}$ dependence in $^{16}$O 
and $^{40}$Ca for various interactions, based on $\text{V}_{low-k}$~\cite{Bogner2003} and similarity renormalization group (SRG)~\cite{Bogner2007} to decouple high momentum modes 
and generate a set of phase-shift equivalent nucleon-nucleon($NN$) interactions. A range of values 
for $\lambda$, the parameter that measures spread of off-diagonal strength has been 
used in calculations~\cite{Bogner2007}. In addition, three-nucleon forces (3NF) have
been taken into account in model calculations~\cite{Miorelli2016}.
As shown in Fig.~\ref{fig_polariz_Miorelli}, a correlation between the electric dipole
polarizability and charge radii has been reproduced, as expected from nuclear 
droplet model in heavier systems. While $NN$ interaction alone systematically
underestimate measured values of $\alpha_D$ and $r_{\rm c}$, the inclusion of 3NFs
considerably improves agreement with data~\cite{Miorelli2016}.

In Ref.~\cite{Birkhan2017}, coupled-cluster method has been employed in {\it ab initio} 
calculations of dipole polarizability in $^{48}$Ca, by including two- and three-nucleon
interactions. Due to implementation of different chiral Hamiltonians, a range of values
of $\alpha_D$ is obtained, some within the range of experimental uncertainties. 
It is interesting to note that $\alpha_D$ value for NNLOsat interaction, that is additionally 
adjusted to improve the properties of finite nuclei, appears outside from the 
experimental range of values.  Coupled-cluster calculations are subject to the 
truncations in the model space, that could lead to a shift of $\approx$ 2 MeV~\cite{Birkhan2017}.
In addition, at low energies coupled-cluster triples corrections could also have important
role~\cite{Birkhan2017}, thus  {\it ab initio} description of dipole polarizability could be
further improved.

%Dipole polarizability has also been explored in weakly bound light nuclei, e.g. in borromean
%nucleus $^{6}$He~\cite{Parkar2011}
%
%%%%%%%%%%%%%%%%%%%%%%%%%%%%%%%%%%%%%%%%%%%%%%
\subsubsection{Pygmy dipole strength}
%%%%%%%%%%%%%%%%%%%%%%%%%%%%%%%%%%%%%%%%%%%%%%

The low-energy E1 transitions in nuclei, often termed as pygmy dipole strength (PDS) 
or pygmy dipole resonance (PDR), represent one of very active 
research topics in nuclear physics over the last decade~\cite{Paar2007, Savran2013}.
The PDS evolves as a characteristic feature in nuclei with neutron excess (but also in proton drip-line nuclei~\cite{Paar2005}) characterized by unique properties different than a
single-particle transition but also considerably different than highly collective
giant resonances. The general macroscopic picture of the PDS also seems to be rather
complex, e.g., ranging from the neutron skin oscillation against the core, toroidal motion, 
isoscalar and isovector mixing, etc.  The PDS is interesting not only as an unique excitation 
mode in nuclei, but also due to its astrophysical relevance, because in neutron-rich
nuclei it contributes significantly to the radiative neutron capture, that is essential in 
the r-process nucleosynthesis~\cite{Goriely2002,Goriely2004}. As pointed out in Ref.~\cite{Litvinova2009}, neutron capture rates are sensitive to the fine structure of the
 low-lying dipole strength, which emphasizes the importance of a reliable knowledge
on this mode. There is an extensive literature on the PDS, going 
beyond the scope of this review. An overview about recent developments in experimental 
approaches to study low-lying electric dipole strength related to the PDS in stable 
and radioactive atomic nuclei is given in Ref.~\cite{Savran2013}. The initial theoretical 
studies are reviewed in more details in Ref.~\cite{Paar2007}. Within this review,
we focus mainly on the implications of the PDS studies on the neutron skin
thickness and the symmetry energy of the EoS. Nevertheless, we also refer to some very
recent studies of relevance for the future investigations to constrain the neutron skin thickness and symmetry energy parameters from the PDS.

From the experimental side, the systematics of PDS measurements are given in
Ref.~\cite{Savran2013}, ranging from light toward heavy nuclei, various experimental methods and
energy ranges accessible in measurements.  In Ref.~\cite{Scheck2013}, the photo-response of $^{60}$Ni has been investigated in order to explore also the evolution of the PDS in a nucleus with rather small neutron excess. Recently the evidence of the PDS has also been identified in the $\gamma$-ray strength functions of $^{105-108}$Pd using charged particle
reactions $(^3\text{He},^3\text{He}'\gamma)$ and $(^3\text{He},\alpha\gamma)$~\cite{Eriksen2014}. By the same method, the evolution of the PDS in several Sn isotopes has been explored~\cite{Toft2011}. The PDS has also been studied in $^{86}$Kr, $^{136}$Ba, in 
N=50 isotones, etc. in photon-scattering experiments using bremsstrahlung produced with 
electron beams delivered by the linear accelerator ELBE~\cite{Massarczyk2012,Schwengner2013}.

As predicted by microscopic calculations based on the relativistic QRPA~\cite{Paar2007}, in the initial studies of the PDS using $(\gamma,\gamma')$ scattering, part of the transition strength was missing in the experimental result. 
One of the limitations of $(\gamma,\gamma')$ approach is the neutron threshold, thus 
measurements could only be performed  below its energy.
In more recent experiments of  $^{112,120}\text{Sn}(\gamma,\gamma')$ reactions, an additional method based on statistical model has been 
employed to estimate the missing strength due to unobserved decays to excited states, allowing
more reliable comparison with model calculations~\cite{Banu2014}. 
Another method to explore
the PDS properties, using $(p,p')$ scattering, allowed a consistent study of the E1 strength in $^{120}$Sn below and above threshold~\cite{Krumbholz2015}. The resulting B(E1) strength 
between 4 and 9 MeV appeared more than three times larger than in the case of $(\gamma,\gamma')$  experiment, despite including in the latter case corrections for unobserved
branching ratios and unresolved contributions in Ref.~\cite{Banu2014}.
As pointed out in Ref.~\cite{Krumbholz2015}, studying the PDS properties based on 
$(\gamma,\gamma')$ experiments should be carefully considered, as well as their
implementation in studies of the symmetry energy of the EoS.

Measurements based on $(\gamma,\gamma')$ reactions provide high resolution 
electromagnetic response below the neutron separation energy~\cite{tonchev2010,Tonchev2017}. Comparative study of experimental data with
theoretical predictions from the EDF plus 
quasi-particle phonon model allowed a separation of the PDS from both the tail of the giant dipole resonance and multi-phonon excitations~\cite{Tonchev2017}.
One of the essential properties of the PDS is its splitting into 
isoscalar and isovector component~\cite{Paar2009,Savran2013,Endres2012}.
Comparative studies of $(\gamma,\gamma')$ photon scattering experiments 
and measurements in $(\alpha,\alpha' \gamma)$ scattering allowed investigations of
the isospin character of the PDS~\cite{Endres2012}.  From the comparison of the
two respective experiments on $^{124}$Sn target, splitting of the low-energy excitation spectra
has been identified: lower-lying states of the PDS are observed in both kinds of
experimental approaches, while high-lying states of the PDS are excited only in $(\gamma,\gamma')$
scattering. Comparison with the relativistic quasi-particle time-blocking approximation(RQTBA) and
the quasi-particle phonon model showed that the low-lying isoscalar component  is dominated
by neutron skin oscillations, while the PDS states at higher energies have pronounced isovector component that corresponds to the low-energy tail of the IVGDR~\cite{Endres2012}.

In Ref.~\cite{Lanza2011} the perspectives of the heavy-ion inelastic scattering in neutron-rich
nuclei have been explored to resolve the properties of the PDS. The analysis based on
the HF+RPA with Skyrme functional showed how the combined information from 
reaction processes involving the Coulomb and different mixtures of isoscalar and
isovector nuclear reactions can provide the insight into the properties of these states.The isoscalar character of low-energy dipole excitations has also been explored using the transition densities from the RQTBA and semi-classical model to describe inelastic $\alpha - ^{124}$Sn scattering ~\cite{Lanza2014}.  In this way, the missing link has been established to directly compare the results from microscopic RQTBA calculations with the experimental data on $(\alpha,\alpha' \gamma)$ reaction.

\begin{figure}[t!]
\centering
\includegraphics[width=0.95\linewidth,clip=true]{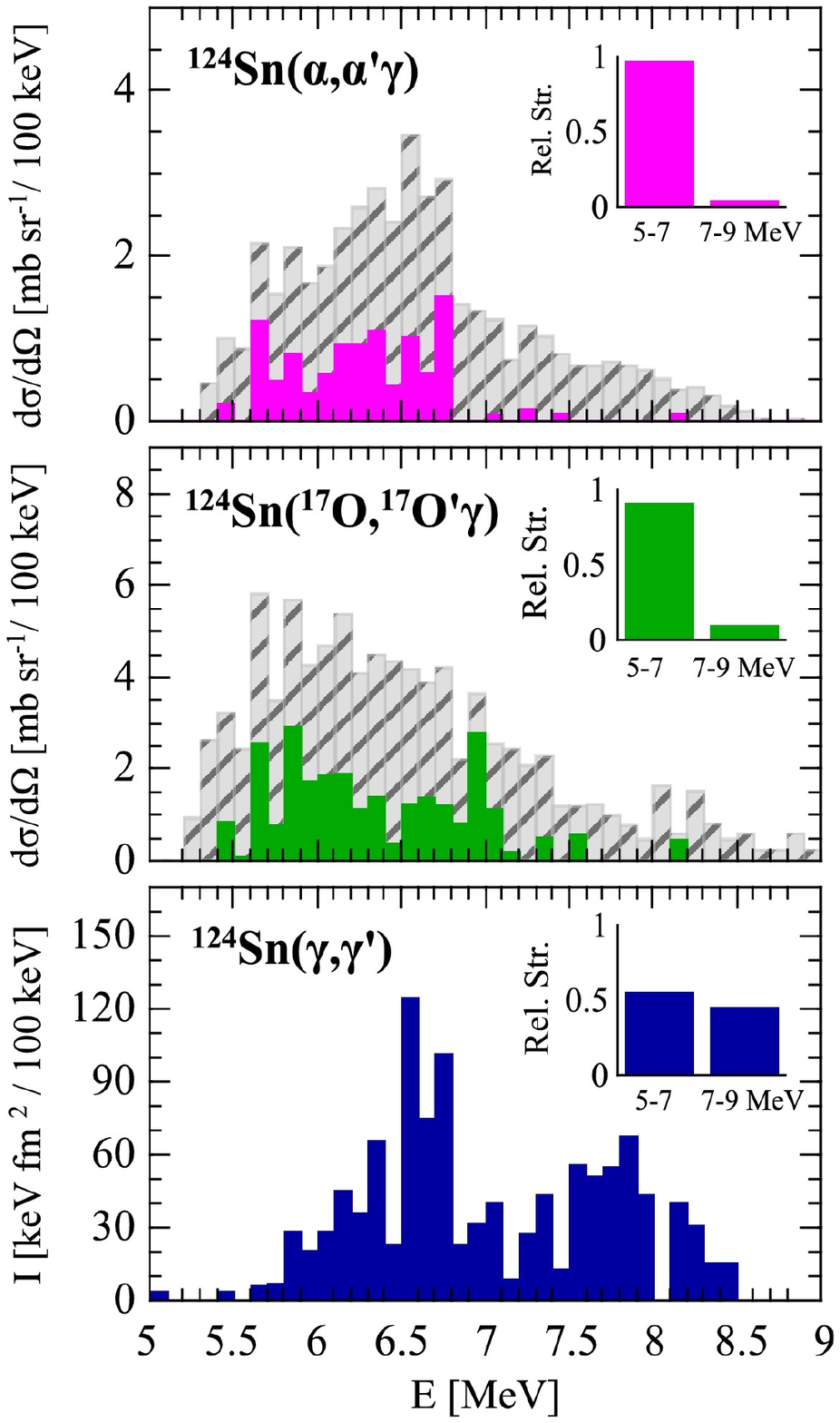}
\caption{\label{fig_pygmy_pellegri}. The reaction cross sections for $^{124}\text{Sn}(^{17}\text{O},^{17}\text{O'}\gamma)$ experiment in comparison with the results from $(\alpha,\alpha' \gamma)$~\cite{Endres2012} and $(\gamma,\gamma')$~\cite{Govaert1998} scattering.  See text for details. Figure taken from  Ref.~\cite{Pellegri2014}.}
%
% corresponds to Fig.3 in Pellegri2014
\end{figure}

In Ref.~\cite{Pellegri2014} (see also \cite{crespi2014}), the PDS has been investigated in $^{124}$Sn populated by
inelastic scattering of $^{17}$O, and $\gamma$ decays from excited states in $^{124}$Sn have
been measured together with the angular distribution of the scattered $^{17}$O ions. As 
a result, the isoscalar component of the $\text{1}^-$ excited states has been extracted.
Fig.~\ref{fig_pygmy_pellegri} shows the respective differential cross sections ~\cite{Pellegri2014}, measured in
$^{124}\text{Sn}(^{17}\text{O},^{17}\text{O'}\gamma)$ experiment in comparison with the
strengths measured in $(\alpha,\alpha' \gamma)$~\cite{Endres2012} and $(\gamma,\gamma')$~\cite{Govaert1998} scattering. The unresolved strength, corresponding to the total binned counts in the measured spectra are shown in grey. Clearly, one can observe in the figure the splitting of
the PDS states in two regions in all three cases. However, the strength is distributed 
differently due to the different nature of these states. In the case of the result for $(^{17}\text{O},^{17}\text{O'}\gamma)$, similar as for $(\alpha,\alpha' \gamma)$ reaction,  the low-lying part of the E1 strength appears to be characterised by isoscalar transition densities that are peaked on the surface which lead to an enhancement in the isoscalar E1 response, while the higher-lying states can be interpreted as transitions towards the GDR and, thus, are suppressed in the isoscalar channel. On the other side, for $(\gamma,\gamma')$ experiment the strengths
in two regions are similar. In another $(^{17}\text{O},^{17}\text{O'}\gamma)$ experiment, the
isoscalar component of the PDS has been extracted for $^{140}$Ce~\cite{Krzysiek2016}. In Ref.~\cite{Nakatsuka2017}, the isospin character of the PDS has been explored for the first time in unstable nucleus, in neutron-rich $^{20}$O~\cite{Nakatsuka2017}. By comparing two spectra,
obtained from a dominant isovector probe $(^{20}\text{O}+\text{Au})$ and a dominant isoscalar
probe $(^{20}\text{O}+\alpha)$, the evidence for different underlying structures has been obtained~\cite{Nakatsuka2017}.
Signatures of the collectivity in the PDS have been explored within the RPA~\cite{Co2009} and self-consistent RPA with a finite-range interaction~\cite{Co2013}, fully self-consistent Skyrme HF+RPA~\cite{roca-maza2012c}, fully self-consistent relativistic (Q)RPA~\cite{Litvinova2009b,Vretenar2012}, and the RQTBA~\cite{Litvinova2009b}. 
As shown in Refs.~\cite{roca-maza2012c,Vretenar2012}, for the isovector dipole operator, the reduced transition probability B(E1) of the PDS is rather small because of pronounced cancellation of neutron and proton partial contributions. On the other side, the isoscalar-reduced transition amplitude is mainly determined by neutron particle-hole configurations, most of which add coherently, and this results in a collective response of the PDSs to the isoscalar dipole operator~\cite{Vretenar2012}.
%
% Move to isoscalar GDR section
%Toroidal nature of the low-energy E1 mode~\cite{Repko2013} 

\begin{figure}[t!]
\centering
\includegraphics[width=0.46\linewidth,clip=true]{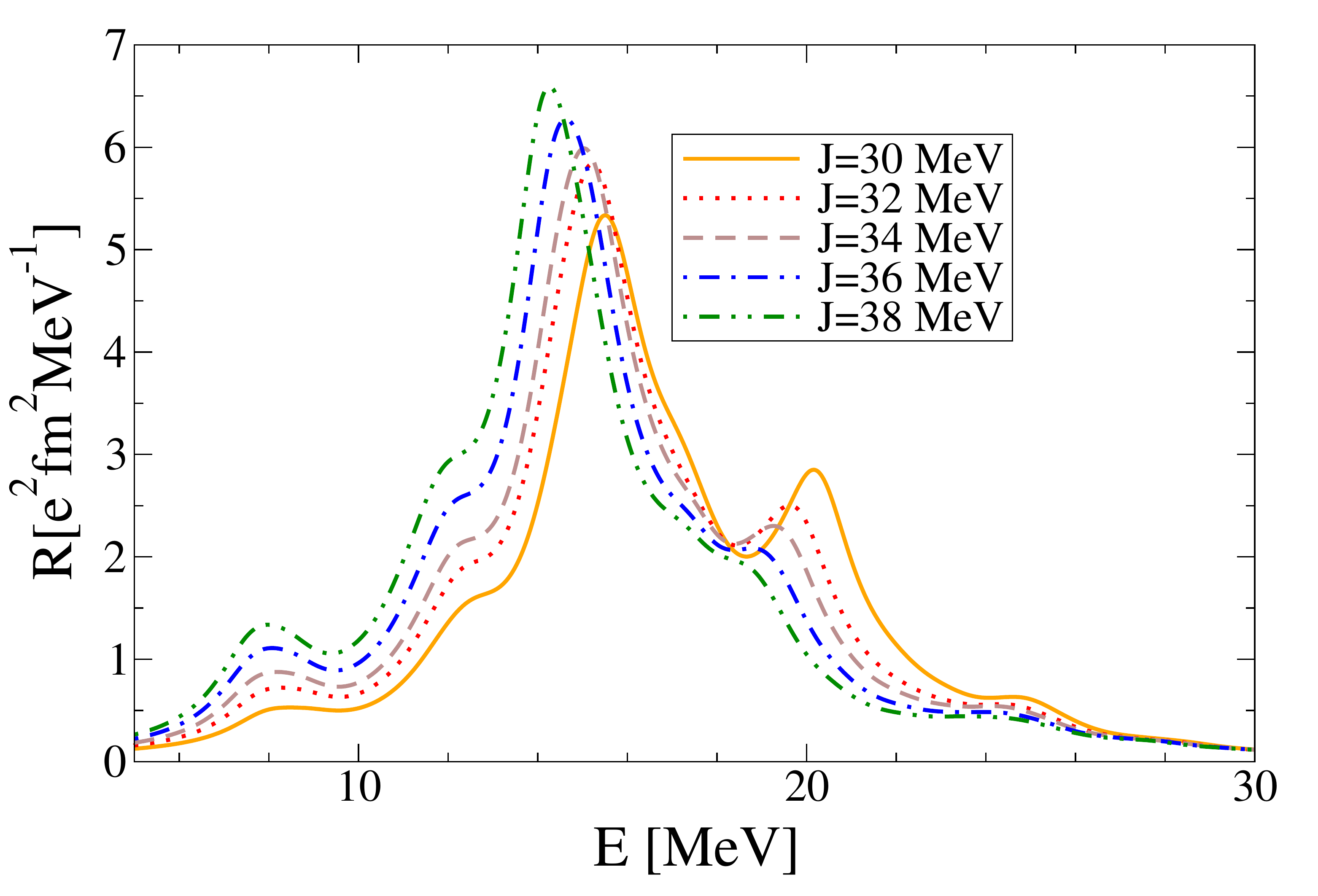}
\includegraphics[width=0.46\linewidth,clip=true]{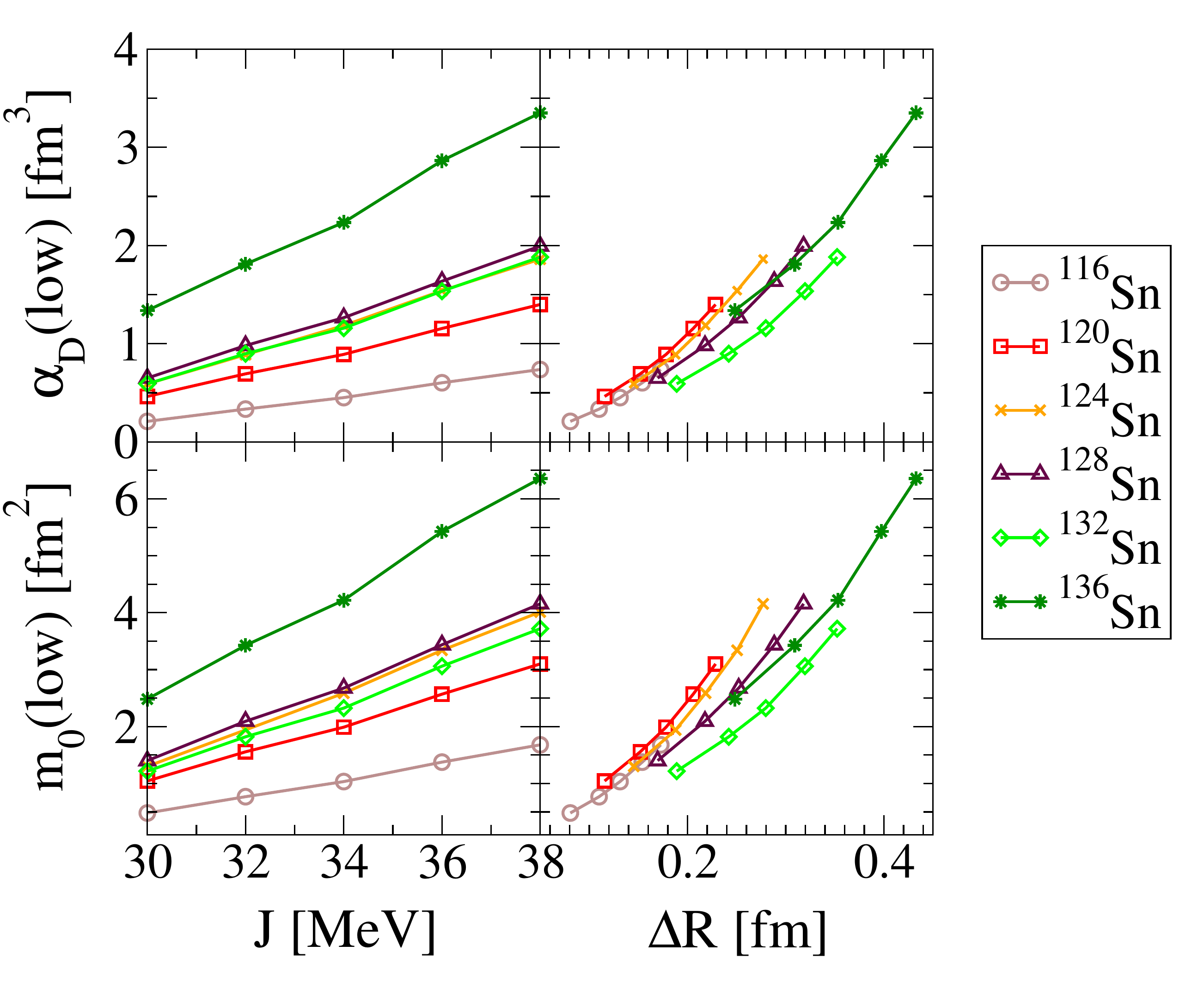}
\caption{\label{fig_pygmy_horvat}Left panel: The isovector dipole strength distribution for $^{132}$Sn based on relativistic RPA calculations with DD-ME~\cite{ddme} family of interactions. Right panel: the sum of  inverse energy weighted strength ($\alpha_D$) and  transition strength ($m_0$) below 10 MeV shown as functions of the symmetry energy at saturation $J$ and neutron skin thickness $\Delta r_{np}\equiv \Delta R$ for Sn isotopes. Figure taken from  Ref.~\cite{Horvat2015}.}
% corresponds to Fig.1 and Fig. 2 in Horvat2015
\end{figure}

As already discussed in previous sections, the properties of isovector dipole transitions (e.g., 
dipole polarizability, the IVGDR excitation energy, etc.) are correlated with the neutron-skin
thickness and symmetry energy parameters. Fig.~\ref{fig_pygmy_horvat} shows the
isovector dipole transition strength distribution for $^{132}$Sn~\cite{Horvat2015}, calculated with the relativistic RPA for the family of DDME interactions characterized by the symmetry energy at the saturation density, $J=$30,32,...,38 MeV~\cite{ddme}. In addition to the strong peak corresponding to the IVGDR, the pygmy strength appears in the low-energy region
at $E < \text{10}$ MeV. The transition strength of the main IVGDR peak is slightly increasing and shifting toward lower energies with increasing $J$. The PDS appears strongly dependent on $J$; its transition strength increases more than a factor of two within given range of $J$-values. 
 In the right panel in Fig.~\ref{fig_pygmy_horvat} the properties of the PDS are shown for even-even $^{116-136}$Sn  isotopes as a function of $J$ and neutron skin thickness $\Delta r_{np}\equiv \Delta R$. This includes the sum of inverse energy weighted strength ($\alpha_D\text{(low)}$) and  transition strength ($m_0$) at excitation energies below 10 MeV. Clearly, the PDS quantities $m_0$ and $\alpha_D\text{(low)}$ display a correlation with $J$ ($\Delta r_{np}$). Their slopes with respect to $J$ appear different, i.e., more neutron-rich (open shell) nuclei are more sensitive to the increase of the $J$ value. Therefore, by considering the respective experimental data on the PDS, the properties of the symmetry energy and neutron skin thickness could be estimated. In Ref.~\cite{piekarewicz11}, a strong correlation between the neutron skin thickness and the fraction of the
dipole polarizability exhausted by the PDS in $^{208}$Pb  has also been found in the analysis using FSUGold family of interactions.

In order to explore the relationships between various observables, properties of finite 
nuclei, and neutron matter properties, in Refs.~\cite{reinhard10,Reinhard2013} correlation analysis has been performed within self-consistent EDFs of the Skyrme type. It has been shown that while the correlation between the dipole polarizability and the neutron skin
thickness is strong, in the case of the low-energy dipole strength the respective
correlation is weak. This result is at variance with respect to the studies 
providing evidence that the correlation between the neutron skin thickness and PDS
is strong, with even larger sensitivity than the dipole polarizability~\cite{piekarewicz11,Horvat2015,Paar2014b}. Since the dipole polarizability is proportional to the inverse energy-weighted sum, the PDS exhausts its considerable fraction, thus similar correlation trend with respect to $\Delta r_{np}$ is expected for the pygmy strength and dipole polarizability, as shown in Fig.~6 in Ref.~\cite{piekarewicz11}. As pointed out in Ref.~\cite{Paar2014b}, both the PDS and dipole polarizability display strong sensitivity on the symmetry energy 
parameter $J$, that is more pronounced in neutron-rich isotopes. On the other side, the covariance analysis seems to be strongly cut-off dependent~\cite{Reinhard2013} and model dependent (see Fig.~\ref{fig-theo-9} above, and Fig. 5 in Ref.~\cite{Paar2014b}). 

The pioneering study in deriving  $J$ and $\Delta r_{np}$ from the PDS in unstable nuclei $^{130,132}$Sn has been performed in Ref.~\cite{klimkiewicz07}, based on the relativistic EDFs and experimental data on the PDS from Coulomb dissociation of high-energy radioactive beams of Sn isotopes. The following values have been obtained $J=\text{32.0}\pm\text{1.8}$MeV, $L\equiv 3p_0/\rho_0=43\pm 11$ MeV and $\Delta r_{np}=\text{0.24}\pm\text{0.04}$ fm for $^{132}$Sn, as well as $\Delta r_{np}=\text{0.18}\pm\text{0.035}$ fm for $^{208}$Pb.

\begin{figure}[t!]
\centering
\includegraphics[width=0.6\linewidth,clip=true,angle=90]{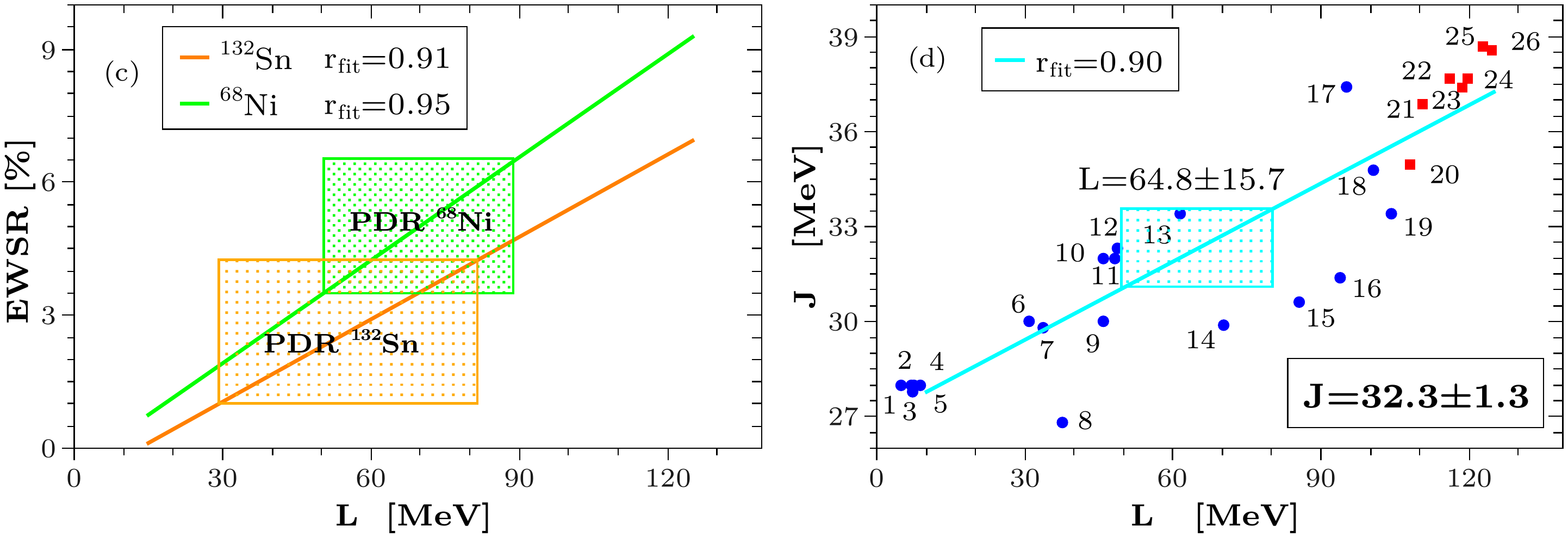}
\caption{\label{fig_pygmy_carbone1}Left panel: Correlation between the pygmy EWSR$[\%]$ for $^{68}$Ni and $^{132}$Sn (with respect to Thomas Reiche Kuhn sum rule) and $L$ obtained using a set of relativistic and non-relativistic interactions. Constraints from the experiments are shown. Right panel: Correlation between $L$ and $J$ is shown, with the values of $L$ and $J$ obtained from the weighted average of the results for $^{68}$Ni and $^{132}$Sn. For more details on the interactions employed see in ~\cite{carbone10}. Figure taken from Ref.~\cite{carbone10}.}
%
% corresponds to lower part of Fig.2 in carbone10
%
\end{figure}

In Ref.~\cite{Cao2008}, the symmetry energy has been explored in the study of IVGDR and
pygmy strength using different sets of relativistic EDFs. The comparison of the calculated PDS with the experimental data for $^{132}$Sn resulted in the symmetry energy at the density
$\rho=\text{0.1}$ fm$^{-3}$ within the range from 21.2 MeV to 22.5 MeV. By employing the experimental data on the PDS in $^{68}$Ni~\cite{Wieland:2009} and $^{132}$Sn~\cite{klimkiewicz07} the constraints on the slope of the symmetry energy $L$, and $\Delta r_{np}$ have been estimated in Ref.~\cite{carbone10}. The analysis is based on several Skyrme functionals, as well as on meson-exchange EDFs, in order to cover a broad range of $L$ values.
Fig.~\ref{fig_pygmy_carbone1} shows the linear fits of the theoretical results for the pygmy $EWSR$ vs. $L$ dependence, as well as for $J$ vs. $L$ relationship~\cite{carbone10}. By implementing the experimental values for the pygmy $EWSR$ for $^{68}$Ni and $^{132}$Sn, and evaluating the weighted average over the two nuclei, the following constraints have been obtained: $L=$64.8$\pm$15.7 MeV and $J=$ 32.3$\pm$1.3 MeV~\cite{carbone10}. Therefore, as discussed in Ref.~\cite{carbone10}, the PDS leads to estimate the parameters governing the density dependence of the symmetry energy. 

Within the harmonic oscillator shell model and a semi-classical Landau-Vlasov approach,
the PDS has been explored in neutron rich nuclei, in relation to the density dependence
of the symmetry energy~\cite{Baran2012}. Within the same approach, the correlation 
between the neutron skin thickness and the low-energy dipole response has been explored.
A linear correlation between the EWSR associated to the PDS and the $\Delta r_{np}$
has been confirmed~\cite{baran13}. In Ref.~\cite{Inakura2015}, a systematic analysis of the PDS relation to the symmetry energy parameters based on the HF+RPA using Skyrme functionals has been performed, indicating that the PDS makes the $\alpha_D J - L$ correlation strong and the slope of the linear function steep. A consistent study of various constraints on $J$ and $L$, based on relativistic family of DDME interactions~\cite{ddme}, including results based on the PDS in $^{68}$Ni, $^{130,132}$Sn, but also $\alpha_D$, IVGQR and AGDR excitation energies in $^{208}$Pb resulted in weighted average value
$J = \text{32.5} \pm \text{0.5}$ MeV and $L = \text{49.9} \pm \text{4.7}$ MeV. 

There are many very recent theoretical developments that could impact future studies 
of the EoS properties based on the PDS. One of the first {\it ab initio} studies of the PDS has only recently been reported, based on the approach that combines the Lorentz integral transform with the coupled cluster method and using the $N^3LO$ scheme~\cite{Bacca2014}.
There are, however, limitations of the model, which are consistent with the over-binding, too small charge radius and dipole polarizability for $^{40}$Ca. The role of the shell effects on the PDS has been explored within the self-consistent quasi-particle random-phase approximation with the D1S Gogny interaction and a continuum-RPA model with the SLy4 Skyrme force~\cite{Papakonstantinou2015}. It has been shown that due to the shell effects the isospin structure of the PDS may depend on the mass region. 

A number of studies includes developments of theory frameworks with explicit treatment of deformation effects, focused on the PDS. These include, e.g., axial-deformed QRPA with Gogny force~\cite{Peru2007,Peru2008,Martini2011} and deformed QRPA with particle-hole residual interaction derived from a Skyrme interaction through a
Landau-Migdal approximation~\cite{Yoshida2008}. In the framework of relativistic EDFs,
recent developments include deformed relativistic Hartree-Bogoliubov model and relativistic
RPA for axially symmetric nuclei~\cite{Arteaga2009}. Within this framework the dependence of the PDS on the difference of deformations for the ground state neutron and proton density distributions in $^{132-162}$Sn has been explored, showing their linear dependence.
In another approach, QRPA with the translational invariant Hamiltonian using 
deformed mean field potential has been employed in studies of electric dipole excitations, 
indicating that the strength below the threshold may be interpreted as the main fragments of the PDS~\cite{Guliyev2010}. Within a fully symmetry unrestricted (3D) time-dependent DFT approach based on Skyrme functional~\cite{Stetcu2011}, model calculations could not provide any evidence of a PDS in $^{172}$Yb using different Skyrme interactions, while there are experimental results indicating the PDS existence in this case~\cite{Voinov2001}.

\begin{figure}[t!]
\centering
\includegraphics[width=0.6\linewidth,clip=true]{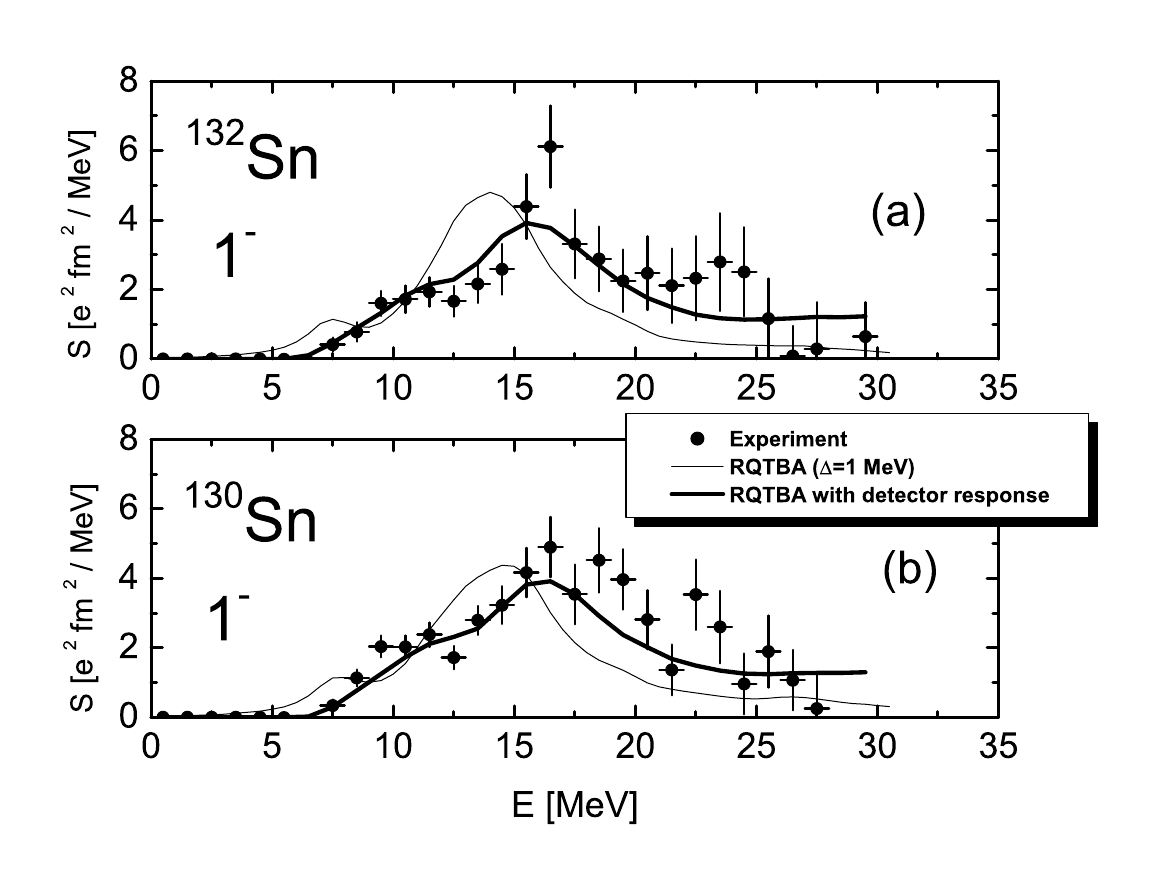}
\caption{\label{fig_pygmy_litvinova}The transition dipole strength distributions for $^{130}$Sn and $^{132}$Sn calculated within the RQTBA with 1 MeV smearing,  RQTBA
with the convolution using the detector response, in comparison to the experimental data~\cite{Adrich2005}.Figure taken from Ref.~\cite{Litvinova2009b}.}
%
% corresponds to lower part of Fig.1 in Litvinova2009b
%
\end{figure}

In order to establish a more direct link between theoretical descriptions on the PDS with 
the experimental data, the effects going beyond the QRPA need to be included, such as
coupling to collective vibrations that generates spectra with a multitude of two-quasi-particle plus phonon states, providing in this way a considerable fragmentation 
of the giant resonances as well as the PDS~\cite{Litvinova2008}. This has been achieved
within the relativistic quasi-particle time blocking approximation (RQTBA) based on the covariant EDF~\cite{Litvinova2009b,Egorova2016}. Fig.~\ref{fig_pygmy_litvinova} shows the
E1 transition spectra for $^{130,132}$Sn, obtained from the experiment and model calculations by the RQTBA~\cite{Litvinova2009b}. By convoluting the calculated spectra
with the detector response, realistic description of the experimental data~\cite{Adrich2005} has been achieved. Recent developments also include RQTBA upgrade with two-phonon model~\cite{Litvinova2013}.

The effects of the complex configurations on the properties of the PDS in neutron-rich Ca 
isotopes have been studied within the Skyrme QRPA calculations including the phonon-phonon coupling~\cite{Arsenyev2017}, as well as within the relativistic approach based on the RQTBA~\cite{Egorova2016}. In Ref.~\cite{Lyutorovich2015} the RPA has been extended by including the quasi-particle-phonon coupling at the level of the time-blocking approximation (TBA). All matrix elements were derived consistently from the given energy-density functional and calculated without any approximation. In addition, the single-particle continuum has been included, and in this way the artificial discretization usually implied in RPA and TBA
has been avoided~\cite{Lyutorovich2015}. Finally, in Ref.~\cite{Knapp2015} 
the equation of motion phonon method has been employed to perform a fully self-consistent calculation in a space spanned by one-phonon and two-phonon basis states using an optimized chiral two-body potential\cite{Knapp2015}. In this way, the role of 
the two-phonon states has been demonstrated in enhancing the fragmentation of the giant
resonances and low-energy excitations, thus reaching an improved consistency with the
experimental data. Clearly, significant effects that result in considerable fragmentation of
the PDS strength, and some redistribution of the strength in general, will have important implications for the future studies of the symmetry energy based on the PDS.

%%%%%%%%%%%%%%%%%%%%%%%%%%%%%%%%%%%%%%%%%%%%%%
\subsubsection{Isovector giant quadrupole resonance}

Another highly collective mode, the isovector giant quadrupole resonance (IVGQR), has been rather difficult to measure due to its high excitation energy, large width, relatively small excitation
cross section, and lack of a highly sensitive experimental probe~\cite{harakeh01,henshaw2011}.
Only recently a precise measurement of the IVGQR in $^{209}$Bi has been achieved, 
using polarized Compton scattering at the $\text{HI}\vec{\gamma}\text{S}$ facility and a 
novel technique that allowed precise extraction of the resonance parameters, in particular, the energy $E_x^{IV}=\text{23.0}\pm\text{0.13(stat.)}\pm\text{0.18(syst.)}$ MeV and the width $\Gamma = \text{3.9}\pm\text{0.7(stat.)}\pm\text{0.6(syst.)}$ MeV~\cite{henshaw2011}.
Since the restoring force of the IVGQR depends on the symmetry energy of the EOS, accurate measurement of the IVGQR excitation energy~\cite{henshaw2011} opened perspective 
for detailed studies of the symmetry energy.

In Ref.~\cite{roca-maza13} model calculations based on relativistic and non-relativistic EDFs
demonstrated a strong correlation between $(E_x^{IV})$ with the neutron skin thickness $\Delta r_{\rm np}$. Within a macroscopic model based on assumptions of the liquid drop model, an analytic expression has been derived~\cite{roca-maza13}, that expresses the symmetry 
energy $S(\rho_A)$ (where $\rho_A$ is the density at which the symmetry energy of the infinite system equals the symmetry energy of the finite nucleus with mass number $A$) in terms of the excitation energies of IVGQR ($E_x^{\rm IV}$), ISGQR ($E_x^{\rm IS}$), and the Fermi energy at nuclear saturation $\varepsilon_{{\rm F}_\infty}$, that is,
\begin{equation}\label{sa}
S(\rho_A) =\frac{\varepsilon_{{\rm F}_\infty}}{3}
\left\{\frac{A^{2/3}}{8\varepsilon_{{\rm F}_\infty}^2}
\left[\left(E_x^{\rm IV}\right)^2 - 2\left(E_x^{\rm IS}\right)^2\right] 
+ 1\right\} .
\end{equation}
By inserting the weighted averages of the experimental values for $E_x^{\rm IV}=22.7\pm0.2$ MeV and $E_x^{\rm IS}=10.9\pm0.1$ MeV (see Table I in Ref.~\cite{roca-maza13}), and by using $\rho_{A=208}=0.1$ fm${}^{-3}$, the following value for the symmetry energy has been obtained,
$S(0.1$ fm${}^{-3}) = 23.3\pm 0.6$ MeV~\cite{roca-maza13} (note that the reported error does not contain an estimation on the model error but just a propagation of the experimental errors), in agreement with the value from Ref.~\cite{trippa08}, $23.3$ MeV $\leq S(0.1$ fm${}^{-3})\leq 24.9$ MeV. 

Within  a droplet model approach, the neutron skin thickness $\Delta r_{\rm np}$ can also be 
explicitly related to the GQR excitation energies~\cite{roca-maza13},
\begin{equation}
\frac{\Delta r_{\rm np} - \Delta r_{\rm np}^{\mathrm{surf}}}{\langle r^2 \rangle^{1/2}} = \frac{2}{3}\left(I-I_C\right)
    \left\{ 1-\frac{\varepsilon_{{F}_\infty}}{3J}-\frac{3}{7}\frac{I_C}{I-I_C} - \frac{A^{2/3}}{24\varepsilon_{{F}_\infty}}\left[\frac{\left(E_x^{\rm IV}\right)^2 - 2\left(E_x^{\rm IS}\right)^2}{J}\right] \right\},
\label{dm-rnp-1}
\end{equation}
where $I=(N-Z)/A$ is the relative neutron excess, $I_C=e^2Z/(20JR)$ is a Coulomb correction to the neutron excess and $\Delta_{\rm np}^{surf}$ denotes the surface contribution to the neutron skin thickness~\cite{Centelles2010}. By employing this expression, the neutron skin thickness can be calculated from the experimental GQR energies.
One should note that the parameters $J$ and $\Delta r_{\rm np}^{\mathrm{surf}}$ contain a non-negligible uncertainty. At present, the value for $J$ could be adopted from systematic analyses~\cite{tsang12,lattimer2013}. In Ref.~\cite{roca-maza13} the value of somewhat larger uncertainty
has been employed,  $J$ = 32 $\pm$ 1 MeV, and $\Delta r_{\rm np}^{\mathrm{surf}}=\text{0.09}\pm\text{0.01}$ fm~\cite{Centelles2010} have been adopted.The resulting neutron skin thickness equals $\Delta r_{\rm np}=\text{0.22}\pm\text{0.02}$ fm, at the upper limit from systematic estimates~\cite{tsang12}.

\begin{figure}[t]
\centering
\includegraphics[width=0.45\linewidth,clip=true]{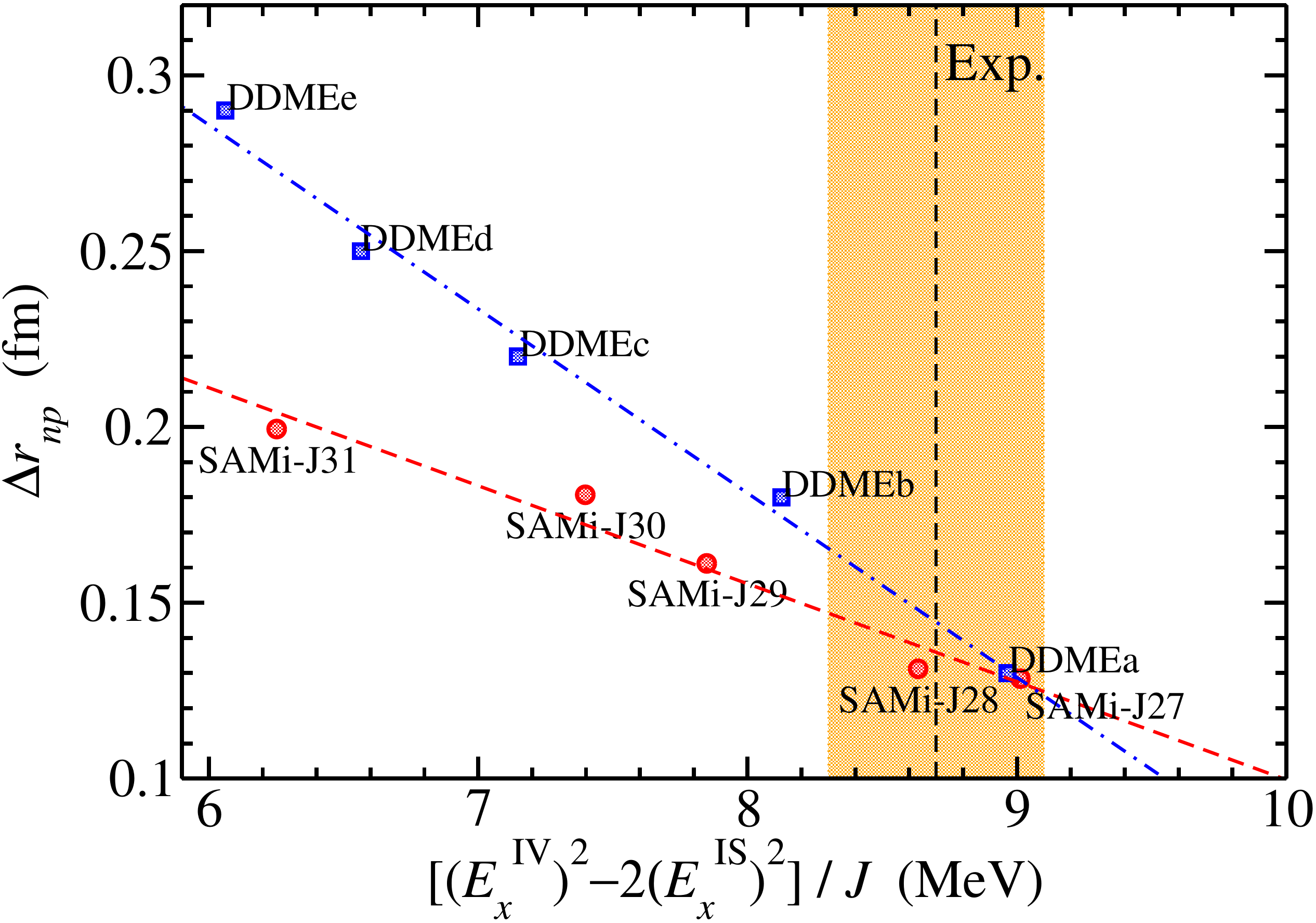}
\includegraphics[width=0.45\linewidth,clip=true]{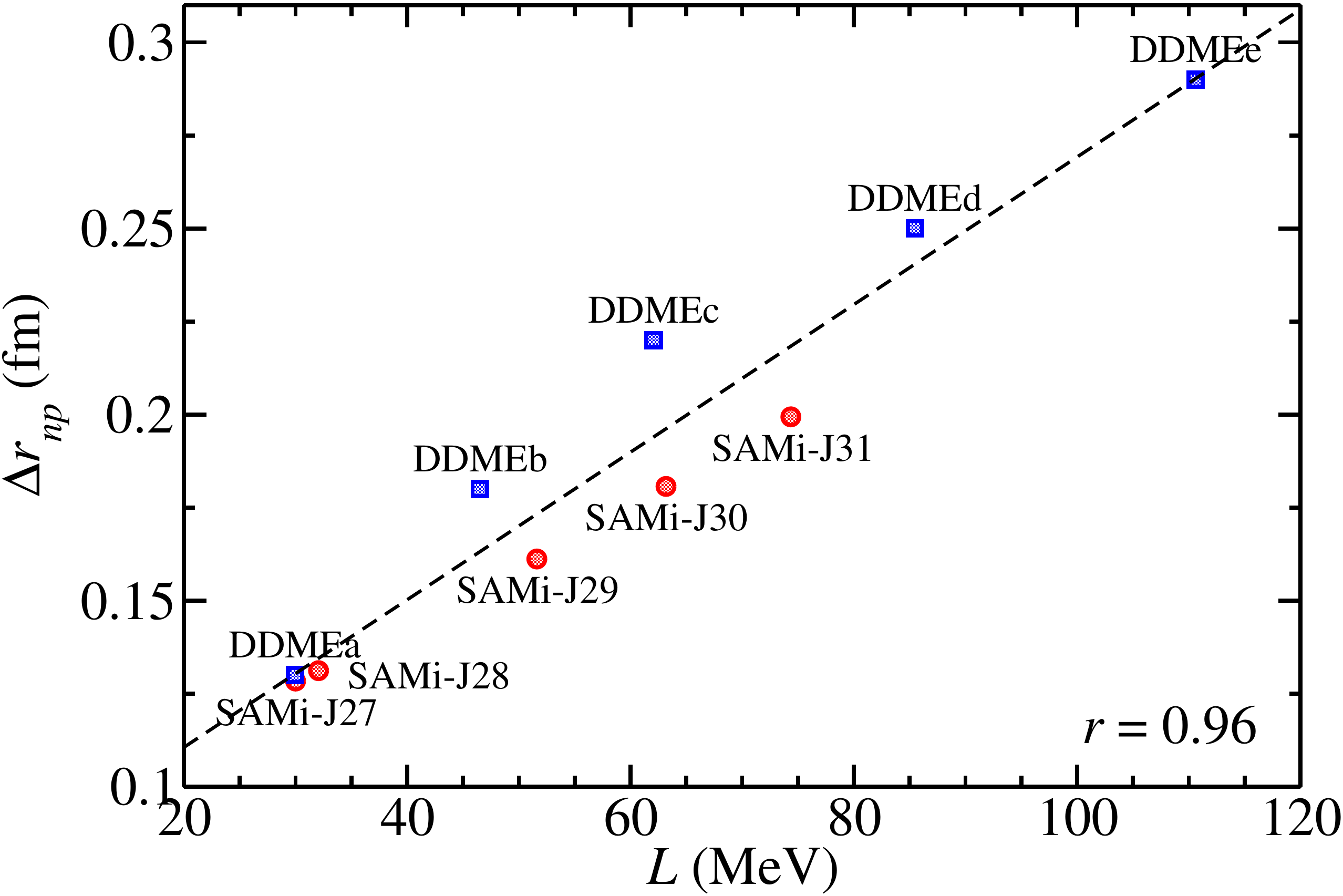}
\caption{\label{fig_gqr_skin_roca_maza} (Left panel) Neutron skin thicknesses $\Delta r_{\rm np}$ for ${}^{208}$Pb as functions of $[(E_x^{\rm IV})^2 - 2 (E_x^{\rm IS})^2] / J$, calculated with the SAMi-J~\cite{roca-maza12b} and DD-ME~\cite{ddme} functionals. The respective experimental
data and its uncertainty are shown as a vertical band.  (Right panel) Neutron skin thickness $\Delta r_{\rm np}$ for ${}^{208}$Pb
as a function of the slope parameter $L$ of the symmetry energy at saturation
density. Figure taken from  Ref.~\cite{roca-maza13}.}
%
% corresponds to Fig. 6  and Fig. 7 in roca-maza13
%
\end{figure}

The dependence of $\Delta r_{\rm np}$ on $\left[\left(E_x^{\rm IV}\right)^2 - 2\left(E_x^{\rm IS}\right)^2\right]/J$ established within the droplet model (Eq.~\ref{dm-rnp-1}), has also been tested in a microscopic analysis~\cite{roca-maza13} based on non-relativistic and relativistic EDFs, i.e., 
SAMi-J~\cite{roca-maza12b} and DD-ME~\cite{ddme} families of functionals. As shown 
in Fig.~\ref{fig_gqr_skin_roca_maza}, the two quantities are strongly correlated, however,
different slopes are obtained for SAMi-J and DD-ME functionals. The resulting range
of neutron skin thickness values is rather broad, $\Delta r_{\rm np} = 0.14 \pm 0.03$ fm,
however, consistent with previous studies: $\Delta r_{\rm np}=0.18\pm0.03$
fm \cite{tsang12}, and $\Delta r_{\rm np}=0.188\pm0.014$ fm \cite{Agrawal12}. In the next step,
slope parameter of the symmetry energy at saturation density $(L)$ has been determined 
from $\Delta r_{\rm np}$ , by exploiting a strong correlation between $\Delta r_{\rm np}$ for $^{208}$Pb and $L$ (see the right panel in Fig.~\ref{fig_gqr_skin_roca_maza} and Fig.\ref{fig-gs-18})~\cite{roca-maza13}.
The resulting value of slope parameter equals $L = 37 \pm18 $ MeV.

\begin{figure}[t]
\centering
\includegraphics[width=0.5\linewidth,clip=true]{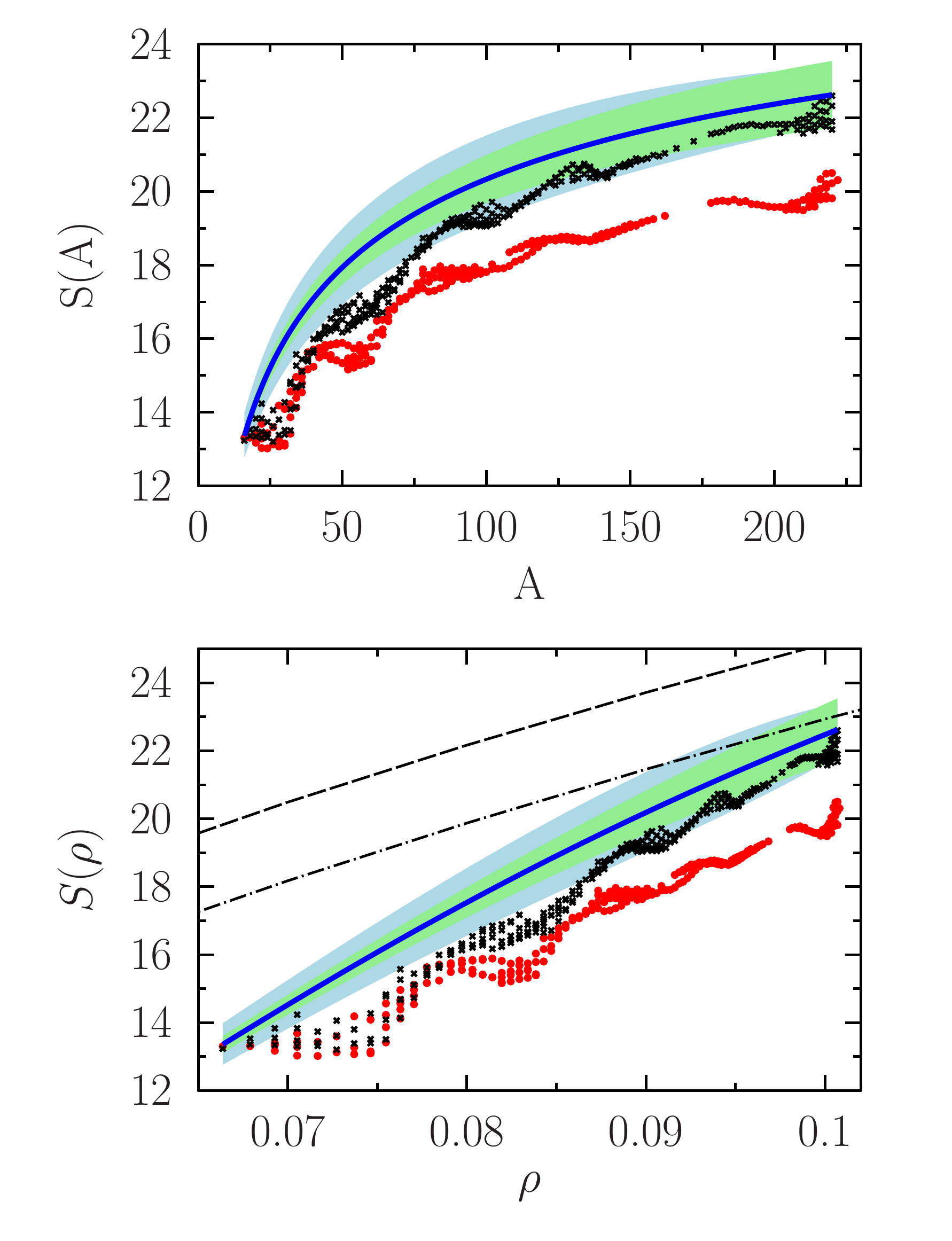}\\
\caption{\label{fig_ivgqr_scamps}(a) Symmetry energy as a function of the mass $A$ based on GQR excitation energies for the Sly4 (circles) and the SkM* (crosses) functionals. The blue solid line is based on the experimental data, the light blue area corresponds to the uncertainty
when accounting for the error bars on the IVGQR energy solely, while the light green area 
is the uncertainty when an error bars reduced to 500 
keV is assumed. (b) The symmetry energy of this analysis shown as a function of 
the density in comparison with the result from Eq. (4) of Ref. \cite{Chen2005} for the Sly4
(dashed line) and SkM* (dot-dashed line).Figure taken from  Ref.~\cite{Scamps2013}.}
%
% corresponds to Fig. 17 in Scamps2013
%
\end{figure}

The possibility to extract the density dependence of the symmetry energy from the IVGQR and ISGQR data has been explored in systematic calculations within a framework based on TDHF+BCS~\cite{Scamps2013} for two functionals, SLy4~\cite{Chabanat1998} and SkM*~\cite{Bartel1982}. As pointed out in Ref.~\cite{Chabanat1998} the TDHF+BCS approach is predictive for the energies of the low-lying state but not for $B(E2)$ values, while the QRPA, and thus TDHFB, improve the comparison but still miss part of the collectivity.
By implementing calculated values for $E_x^{\rm IV}$ and $E_x^{\rm IS}$ in
Eq.~\ref{sa} over a large number of nuclei, the symmetry energy has been determined as a
function of the mass number $A$ (see Fig.~\ref{fig_ivgqr_scamps}(a)). In the next step, mass dependence has been transformed into an effective density dependence following Ref.~\cite{centelles09},
\begin{eqnarray}
\rho &=& \rho_0 -\rho_0/(1 + cA^{1/3}), \label{eq:rhoa},
\end{eqnarray} 
where $c$ is set to impose $\rho=0.1$ fm$^{-3}$ for the $^{208}$Pb,  
and saturation density is set as $\rho_0 = 0.16$ fm$^{-3}$, leading to  $c= 0.28$~\cite{Chabanat1998}. The resulting symmetry energy as a function of density, extracted from the GQR excitation energies, is shown in Fig.~\ref{fig_ivgqr_scamps}(b), in comparison 
with the symmetry energy directly obtained for SLy4~\cite{Chabanat1998} and SkM*~\cite{Bartel1982} functionals. In addition, the symmetry energy has also been 
calculated directly from the experimental data on ISGQR and IVGQR~\cite{Bertrand1981,sims1997,Maeda2006}. Since data 
are rather limited, to smooth out the uncertainty on experimental collective
energies in order to cover a wider range of nuclei, a polynomial fit of
the experimental points has been used instead of the experimental
points themselves~\cite{Chabanat1998}. The resulting error bars appear 
rather large, and considerable differences could be observed between the
symmetry energy from the calculated and experimental GQR energies.
Due to differences in GQR excitation energies calculated using two 
different functionals, the resulting symmetry energy displays considerable
model dependence. In addition, the symmetry energies extracted
from the GQR are systematically lower than the ones obtained analytically in infinite
nuclear matter for the Sly4 and SkM* functionals. The slope
is also different between the infinite system and finite
system cases. As pointed out in Ref.~\cite{Chabanat1998},
considerable differences in the symmetry energy indicate the presence
of subtle finite size effects and model dependence in the description 
of the GQRs, that necessitate further studies.

In addition to giant quadrupole resonances, recent calculations with the HFB + quasi-particle phonon model (QPM) suggested the existence of low-energy quadrupole strength of a unique structure, denoted as pygmy quadrupole resonance~\cite{Tsoneva2011}. The analysis of spectral distributions, electric quadrupole response functions and transition densities of this low-energy strength showed that  it may have origin in the oscillations of neutron or proton skins against the nuclear core~\cite{Tsoneva2011}. In recent study a number of $\text{J}^{\pi} = \text{2}^+$ states in $^{124}$Sn
have been populated and identified with $(^{17}\text{O},^{17}\text{O'}\gamma)$ reaction, similar 
to the QPM prediction~\cite{Pellegri2015}. In recent study $(\alpha,\alpha'\gamma)$ and $(\gamma,\gamma')$ experiments were performed on $^{124}$Sn, thus providing complementary probes to populate $\text{J}^{\pi} = \text{2}^+$ states~\cite{Spieker2016}. The excitation of 
surface mode in  $(\alpha,\alpha'\gamma)$ experiment supports the quadrupole-type 
oscillation of the
neutron skin, as predicted by QPM calculations. Based on $\gamma$-decay branching 
ratios, non-statistical character of the E2 strength has been concluded, similar as in previous studies of the PDS~\cite{Spieker2016}. Further studies to resolve the properties of 
these low-energy quadrupole states are necessary, as well as to resolve its potential in
constraining the symmetry energy.

\subsection{Charge-exchange modes}

% Finite-amplitude method for charge-changing transitions in axially deformed nuclei  ~\cite{Mustonen2014}

\subsubsection{Isobaric analog resonance}
The basic charge-exchange mode of excitation is the isobaric analog state (IAS) or isobaric analog resonance (IAR). The theoretical operators that realize such a transition are the isospin Pauli matrix $\tau_-$ that exchanges neutrons into protons and $\tau_+$ that does the opposite. In both cases all the other quantum numbers are unchanged $\Delta J = \Delta L = \Delta S =\text{0}$. The IAR represents an important benchmark test for theoretical approaches to charge-exchange transitions. In addition, its energy represents a reference point in studies of the neutron skin thickness and symmetry energy when using  other charge-exchange modes,  e.g., Gamow-Teller, spin-dipole, and anti-analog giant dipole resonances, as will be discussed in the following subsections. We refer for more details on the IAR investigations in papers~\cite{Auerbach1972,Auerbach1983,Rumyantsev1994,Pham1995,Colo1998,Paar2004,roca-maza12b} and references therein.
\begin{figure}[t]
\centering
\includegraphics[width=0.5\linewidth,clip=true]{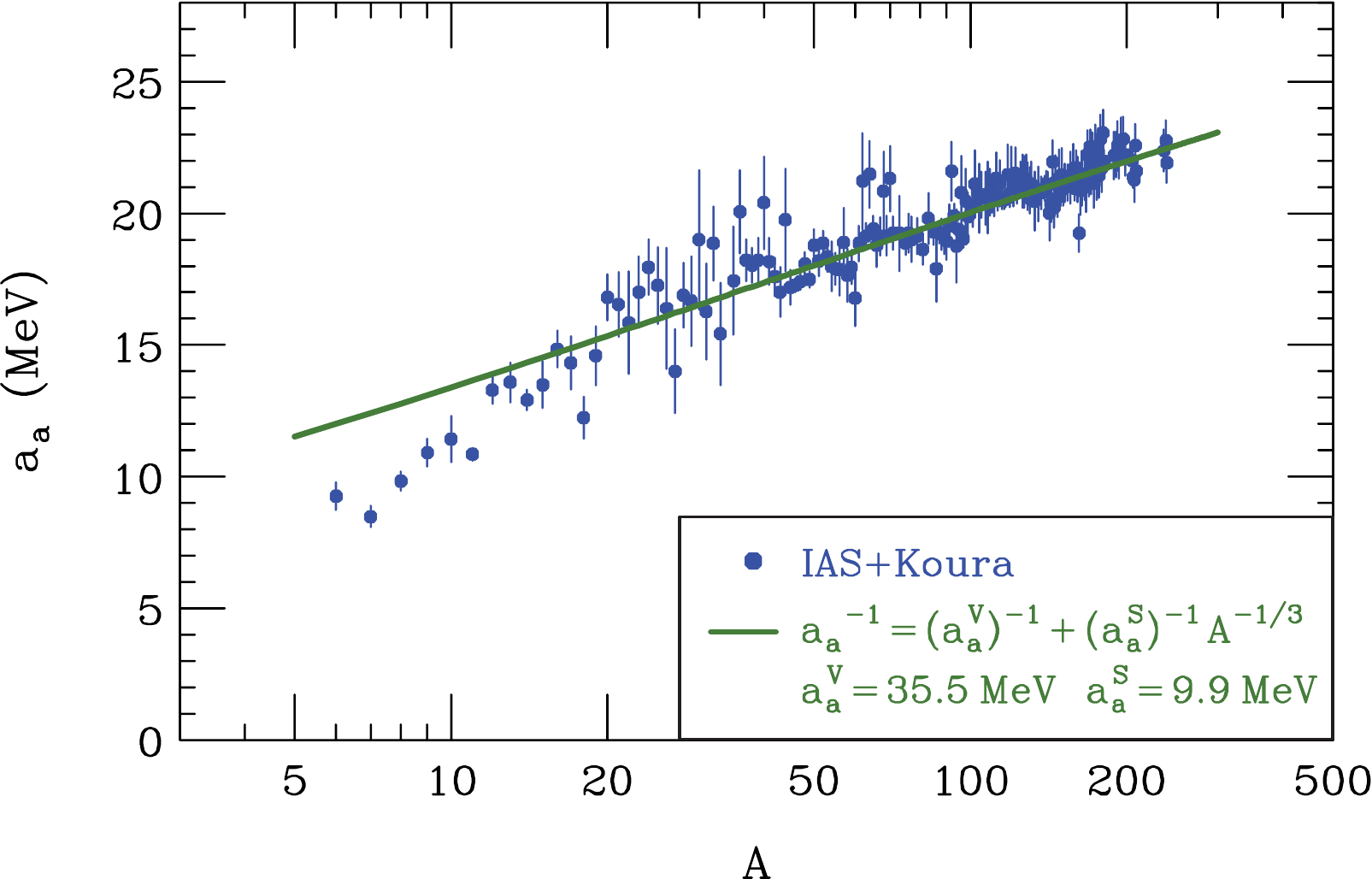}
\includegraphics[width=0.45\linewidth,clip=true]{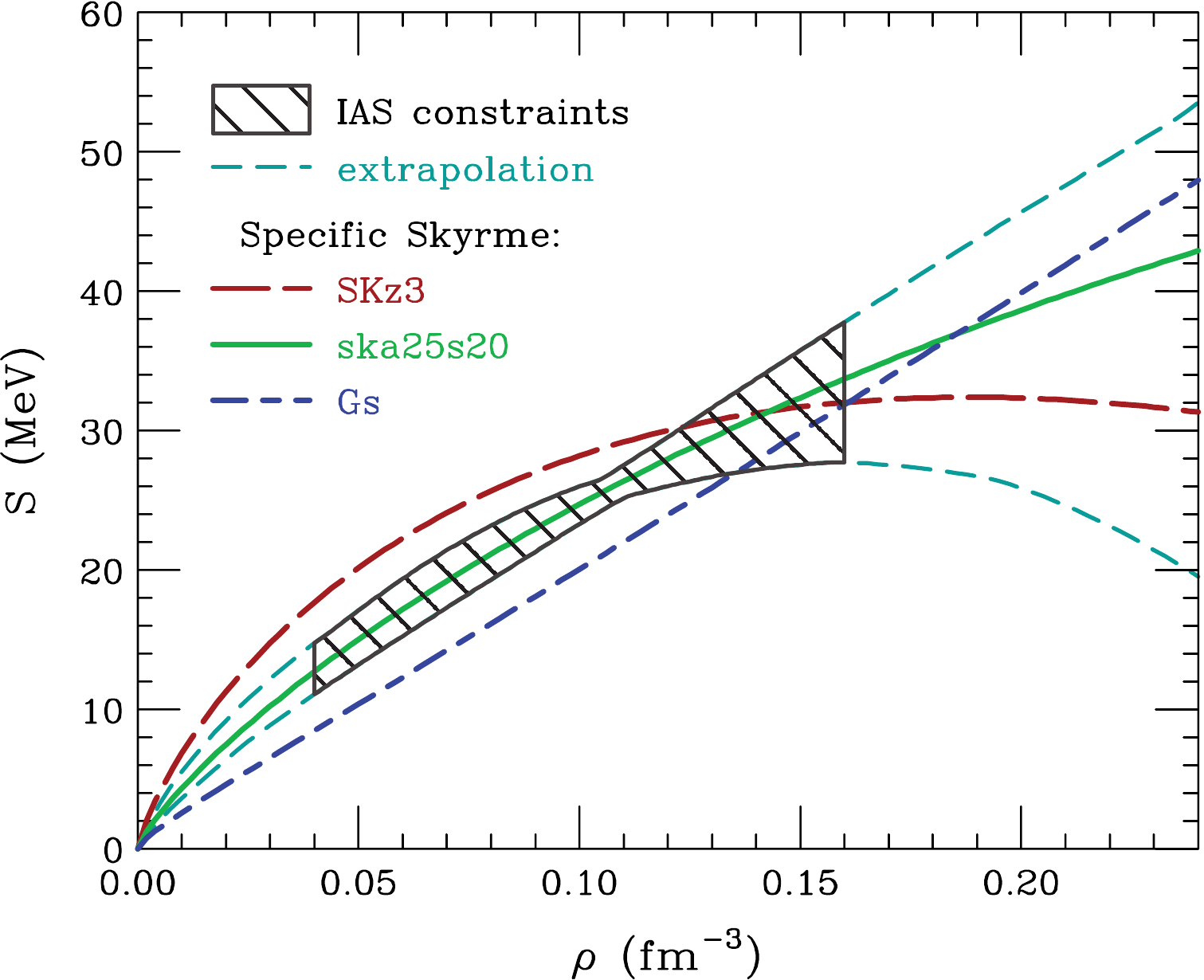}\\
\caption{\label{fig_ias_danielewicz}Left panel: The mass dependence of the asymmetry coefficient $a_A(A)$ with error estimates, extracted from excitation energies to ground-state IAS within individual isobaric chains with shell corrections applied. The line represents a fit at $A \geq \text{30}$, assuming a surface-volume competition in the asymmetry coefficient.
Right panel: Symmetry energy in uniform matter as  a function of density, showing the IAS constraints, their extrapolations to supra-normal $(\rho > \rho_0)$ and low $(\rho < \rho_0/4)$ densities and comparison with the results for three Skyrme parametrizations.
Figure taken from Ref.~\cite{Danielewicz2014}.}
%
% Figures 3 and 15 from Danielewicz2014
%
\end{figure}

The excitation energy of the IAS ($E_{\rm IAS}$) is defined as the energy difference between the analog state $\vert A\rangle$ and the parent state $\vert\pi\rangle$ (see Fig.\ref{fig_agdr_kras1}). The parent state is an eigenstate of the Hamiltonian $\mathcal{H}$ with $N$ neutrons and $Z$ protons and the analog state is defined \cite{Auerbach1972} as $\vert A\rangle\equiv T_-\vert\pi\rangle\langle\pi\vert T_+T_-\vert\pi\rangle^{-1/2}$ where $T_\pm\equiv \frac{1}{2}\sum_i^A\tau_\pm$. Hence,
\begin{equation}
  E_{\rm IAS} = \langle A\vert \mathcal{H}\vert A\rangle - \langle\pi\vert \mathcal{H}\vert\pi\rangle = \frac{\langle\pi\vert [T_+,[\mathcal{H},T_-]]\vert\pi\rangle}{\langle\pi\vert T_+T_-\vert\pi\rangle} \ .
\label{ed}    
\end{equation}  
Due to the structure of Eq.(\ref{ed}), $E_{\rm IAS}$ depends on isospin breaking parts of the $\mathcal{H}$ only. In nuclear physics, the main isospin breaking term is known to be due to the Coulomb interaction. Therefore, the bulk contribution to Eq.(\ref{ed}) will be due to the difference in the expectation value of the Coulomb matrix elements between proton and neutron distributions. Since we do not know the exact wave function for the parent state, one can approximately evaluate $E_{\rm IAS}$ by assuming the independent particle picture with single-particle states with good isospin quantum number in $\vert \pi\rangle$. This would lead to
\begin{equation}
  E_{\rm IAS}\approx\frac{1}{N-Z}\int \left[\rho_n(\vec{r})-\rho_p(\vec{r})\right]U_{\rm C}^{\rm direct}(\vec{r}) d\vec{r},
  \label{ed2}
\end{equation}
where $U^{\rm direct}_{\rm C}(\vec{r})$ is the direct part of the Coulomb energy potential
\begin{equation}
U_{\rm C}^{\rm direct}(\vec{r}) = \int\frac{e^2}{\vert\vec{r}^\prime-\vec{r}\vert}\rho_{\rm ch}(\vec{r}^\prime) d\vec{r}^\prime
\end{equation}

\begin{figure}[t]
\centering
\includegraphics[width=0.5\linewidth,clip=true]{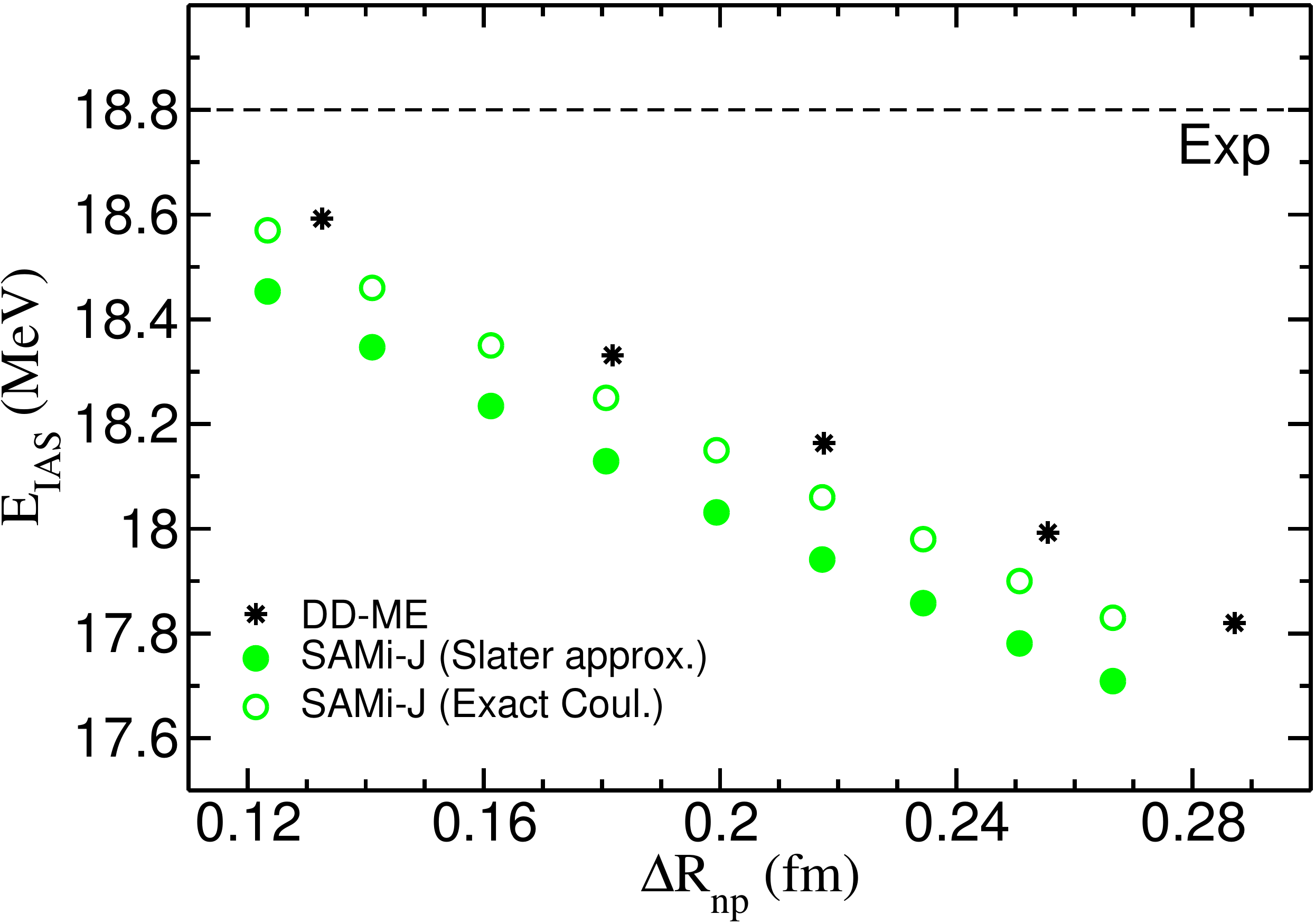}
\caption{\label{fig_ias_roca_maza} The relationship between the energy of the
IAS as a function of neutron skin thickness in $^{208}$Pb calculated using SAMI-J interaction 
with (i) the exact Coulomb exchange and (ii) the Slater approximation. The results are shown
in comparison with those of the DD-ME~\cite{ddme} family of interactions and experimental data.
Figure taken from Ref.~\cite{roca-maza2016}.}
%
% Figure 1 from roca-maza2016
%
\end{figure}

The exchange contribution can be evaluated in an analogous way. In order to gain some insights into the physics behind the IAS, we can resort on a simple model to evaluate Eq.(\ref{ed2}). Assuming a uniform neutron and proton distributions of radius $R_n$ and $R_p=R_{\rm ch}$ respectively, one can evaluate the Coulomb energy potential as
\[ U_{\rm C}^{\rm direct}(\vec{r}) =
  \begin{cases}
    \frac{Ze^2}{2R_p}\left(3-\frac{r^2}{R_p^2}\right)       & \quad \text{for } r< R_p\\
    \frac{Ze^2}{r} & \quad \text{for } r>R_p\\
  \end{cases}
\]
and, therefore, obtain a simple formula for estimating the main contribution to $E_{\rm IAS}$ in terms of physical quantities     
\begin{equation}
  E_{\rm IAS}\approx\frac{6}{5}\frac{Ze^2}{R_p}\left(1-\sqrt{\frac{5}{12}}\frac{N}{N-Z}\frac{\Delta R_{\rm np}}{R_p}\right)\equiv \Delta E_C(Z)\left(1-\sqrt{\frac{5}{12}}\frac{N}{N-Z}\frac{\Delta R_{\rm np}}{R_p}\right),
\label{toymodel}  
\end{equation}
where $\Delta E_C(Z)$ is the Coulomb energy difference between the parent $(N,Z)$ and daughter $(N-1,Z+1)$ nucleus within this simple approximation. This can be seen as follows
\begin{eqnarray}
  \Delta E_C(Z) &=& \frac{3}{5}\frac{(Z+1)Ze^2}{R_{\rm ch}^{(N-1,Z+1)}}-\frac{3}{5}\frac{Z(Z-1)e^2}{R_{\rm ch}^{(N,Z)}}\nonumber\\
  &=&\frac{3}{5}\frac{Z^2e^2\left(R_{\rm ch}^{(N,Z)} - R_{\rm ch}^{(N-1,Z+1)}\right)}{R_{\rm ch}^{(N,Z)}R_{\rm ch}^{(N-1,Z+1)}}+\frac{3}{5}\frac{Ze^2\left(R_{\rm ch}^{(N,Z)} + R_{\rm ch}^{(N-1,Z+1)}\right)}{R_{\rm ch}^{(N,Z)}R_{\rm ch}^{(N-1,Z+1)}},
\end{eqnarray}
where a competition between two terms exists. If we assume, as before, $R_p=R_{\rm ch}$ and $R_{\rm ch}^{(N,Z)} = R_{\rm ch}^{(N-1,Z+1)}$
\begin{equation}
  \Delta E_C(Z) = \frac{6}{5}\frac{Ze^2}{R_p} \ .
\end{equation}
According to Eq.(\ref{toymodel}), the $E_{\rm IAS}$ should decrease with increasing neutron skin thickness \cite{roca-maza2017b} as long as this simplified model keeps the relevant physics of the studied observables (see Fig.\ref{fig_ias_roca_maza}). 

In Ref.~\cite{tsang12}, it was pointed out that the energy differences between the ground state for a nucleus with $N>Z$ and the isobaric analog of the ground state of neighboring isobar are given by the symmetry energy as Eq.(\ref{toymodel}) confirms from a macroscopic perspective. In Ref.~\cite{Danielewicz2014} the IAS has been 
exploited to evaluate nuclear asymmetry coefficients $a_{A}(A)$\footnote{That we have approximately written as $a_A-a_{AS}A^{-1/3}$ in Sec.\ref{pheno}.} for the symmetry
energy term $(N-Z)^2/A$ in the Bethe-Weizs{\"a}cker mass formula~\cite{ringschuck},
on a nucleus-by-nucleus basis. Asymmetry coefficients obtained in this way 
resulted in a variation of values, from $a_A(A\sim\text{10})\approx \text{10 MeV}$
toward $a_A(A\sim\text{240})\approx \text{22 MeV}$. 
Fig.~\ref{fig_ias_danielewicz} (left panel) shows the mass dependence of the asymmetry coefficient $a_A(A)$, supplemented with error estimates, extracted from excitation energies to ground-state IAS within individual isobaric chains. The energies involved in the analysis are corrected for the shell effects with the estimate from Ref.~\cite{Koura2000,Koura2005}. Assuming a surface-volume competition in the asymmetry coefficient as the one used in Sec.\ref{pheno}, $a_A(A)$ has been
successfully fitted in the region $A \geq \text{30}$ (see figure). We note that the obtained coefficients (reported in the legend of the figure) are in reasonable agreement with the ones discussed in Sec.\ref{pheno}. 
In the consideration based on density functional theory, a relation has been established
between the densities of neutrons and protons in a nucleus and the asymmetry coefficient,
in the presence of Coulomb effects~\cite{Danielewicz2014}. By interpreting variation 
with mass in terms of dependence of the symmetry energy on density, constraints on 
the symmetry energy have been determined. Fig.~\ref{fig_ias_danielewicz} (right panel)
shows the symmetry energy in uniform matter as  a function of density, showing the IAS constraints from Ref.~\cite{Danielewicz2014}, their extrapolations to supra-normal $(\rho > \rho_0)$ and low $(\rho < \rho_0/4)$ densities. The results for the symmetry energy are compared with those
for the Skyrme parameterizations SKz3, ska25s20, and Gs~\cite{Margueron2002,Friedrich1986}. The ska25s20 
parameterization communicated for the work in Ref.~\cite{Danielewicz2014} appears
in agreement with the IAS constraint, while this is not the case for SKz3 and Gs interactions. 
By supplementing constraints based on the IAS with the information on neutron skin thicknesses, more tight limits for the symmetry energy have been obtained~\cite{Danielewicz2014}: $30.2$ MeV$<J<33.7$ MeV and $35$ MeV$<L<75$ MeV.

Over the past years, several self-consistent theory frameworks have been developed 
and benchmarked in studies of the IAS properties. Fully self consistent HF-BCS plus
charge-exchange QRPA has been developed using Skyrme functionals and 
density-dependent pairing force~\cite{Fracasso2005}. Calculated energies 
of the IAS along Sn isotope chain have been accurately reproduced.
In further development of the self-consistent HF+RPA\cite{roca-maza2016}, two points have been
improved, (i) the exchange term of the two-body Coulomb interaction is treated
exactly, and (ii) two parameters spin-orbit interaction is treated in a consistent
way within the EDF approach. Fig. ~\ref{fig_ias_roca_maza} shows the
dependence of the relationship between the energy of the IAS as a function of 
neutron skin thickness in $^{208}$Pb calculated using SAMi-J interaction 
with (i) the exact Coulomb exchange and (ii) the Slater approximation. In addition
to the results with the Skyrme functional, for comparison the results are shown
from the relativistic RPA calculations based on the DD-ME~\cite{ddme} set
of interactions, as well as the experimental data~\cite{Akimune1995}. Although
the results of model calculations appear rather close to the measured IAS energy, all the functionals underestimate experimental data. Although the
exact implementation of the exchange term of the two-body Coulomb 
interaction somewhat improved the situation with respect to the experimental
data, it did not resolve the systematic discrepancy \cite{roca-maza2017b}.

Fully self-consistent proton-neutron QRPA with axial symmetry has also been 
developed for the finite-range Gogny interaction and employed in description of
IAS and GTR~\cite{Martini2014}. The first relativistic RPA calculations of the
IAS properties have been reported in Ref.~\cite{Conti1998}, and further developments
for open shell nuclei include the relativistic Hartree-Bogoliubov model plus
relativistic QRPA~\cite{Paar2004}\footnote{Bogoliubov refers to a formalism to take into account pairing correlations in open shell nuclei \cite{ringschuck}.}. Since a complete QRPA formalism includes
T=1 and T=0 pairing channels, where T=1 channel includes the same pairing interaction
as in the ground state calculations, while T=0 channel remains open, and its strength
could be constrained on excitation properties such as low-lying GT strength 
or $\beta$-decay rates. The IAS has also been used as a benchmark test for
the recently established self-consistent relativistic QRPA based on the relativistic 
Hartree-Fock-Bogoliubov theory ~\cite{Liang2008}.

In order to provide microscopic description of the widths of IAS, a self-consistent
particle-phonon coupling model has been established and exploited in the case of 
$^{208}$Bi. It has been shown that quantitative agreement with experimental data
for the IAS energy and the width could only be reproduced if the effects of 
isospin-breaking nuclear forces are taken into account, in addition to Coulomb
force effects~\cite{Colo1998}.

\subsubsection{Gamow-Teller resonance}
The Gamow-Teller resonance is a collective excitation with $J^{\pi}={\text{1}}^+$ 
corresponding to coherent change of the nucleon spin and isospin orientations without
changing its orbital motion. The GT operator depends on the spin $\sigma$ and isospin operator $\tau_-$ that exchanges neutrons into protons (GT$^-$) or $\tau_+$ that exchanges protons into neutrons (GT$^+$). Hence, GT$^-$ transitions will dominate in nuclei with neutron excess. The GT$^-$ strength distribution is composed from 
three components~\cite{Paar2004}. Direct spin-flip transitions 
($\nu j = l + \frac{1}{2}$ $\rightarrow$ $\pi j = l - \frac{1}{2}$ where $\nu$ label refer to a neutron state and $\pi$ label to a proton state)
dominate the main resonance region. The low-energy tail
of the strength distribution corresponds to core-polarization 
($\nu j = l \pm \frac{1}{2}$ $\rightarrow$ $\pi j = l \pm \frac{1}{2}$), 
and back spin-flip 
($\nu j = l - \frac{1}{2}$ $\rightarrow$ $\pi j = l + \frac{1}{2}$)
transitions~\cite{Paar2004}.  Since in the GT$^-$ channel the mode is induced 
mainly by the excitations of excess neutrons, it can provide useful information about the role of neutrons in nuclear ground and excited states. 

The GTR has been predicted more than 50 years 
ago~\cite{Ikeda1963}, and in 1975 it has been experimentally confirmed in $(p,n)$ 
reactions~\cite{Doering1975}. Ever since, the GTR represents one of the most 
extensively investigated collective excitation in nuclear physics (for more details see e.g., 
~\cite{Osterfeld1992,Lutostansky2011}). 
More recent interest in the GTR is motivated by its
relevance for understanding nuclear structure and spin-isospin dependence of modern 
effective interactions~\cite{Bender2002,Marketin2012a,Litvinova2014,Ekstrom2014,Zelevinsky2017,Morita2017,Nabi2017,Ha2017}, nuclear beta decay~\cite{Madruga2016,Marketin2016,Moon2017} and beta delayed neutron emission~\cite{Folch2016}and double-beta decay~\cite{Faessler1998,Suhonen1998,Simkovic2011,Vergados2012,Menendez2014,Stefanik2015,Navas2015,Suhonen2017}. Detailed knowledge of GT$^{\pm}$ transitions, not only in stable nuclei but also away from the valley of stability, is of a particular 
importance for understanding weak interaction rates in stellar environment ~\cite{Langanke2000,Langanke2003,Janka2007,Noji2014,Paar2015}, r-process stellar nucleosynthesis~\cite{Arnould2007,Mori2016} and nuclear response to low-energy neutrinos of relevance for neutrino detectors and neutrino nucleosynthesis in stellar environment~\cite{Ejiri2000,Suzuki2003,Frekers2011,Cheoun2010,Paar2008,Paar2011,Karakoc2014}.

From the perspective of this review, the focus is on the role of the GTR, as 
a collective spin-isospin oscillation, in determining the neutron skin thickness 
and in this way constraining the neutron matter EOS. The spin-isospin 
characteristics of the GTR and the IAS are related through the Wigner
supermultiplet scheme. The Wigner SU(4) symmetry implies the degeneracy
of the GTR and IAS, that is, however, broken by the spin-orbit term of the
effective nuclear interaction~\cite{Gaponov1974,Gaponov1981}. As noted 
in Refs.~\cite{Gaponov1974,Gaponov1981}, the energy difference 
between the GTR and the IAS decreases with increasing asymmetry $(N-Z)/A$.
It is implicit, therefore, that the energy difference between the GTR and the IAS reflects the magnitude of the effective spin-orbit potential. With increasing the neutron 
number, the magnitude of the spin-orbit potential becomes 
reduced~\cite{Lalazissis1998}, and this is reflected in the larger spatial 
extension of the neutron density and increased neutron skin thickness. 

\begin{figure}[t!]
\centering
\includegraphics[width=0.47\linewidth,clip=true]{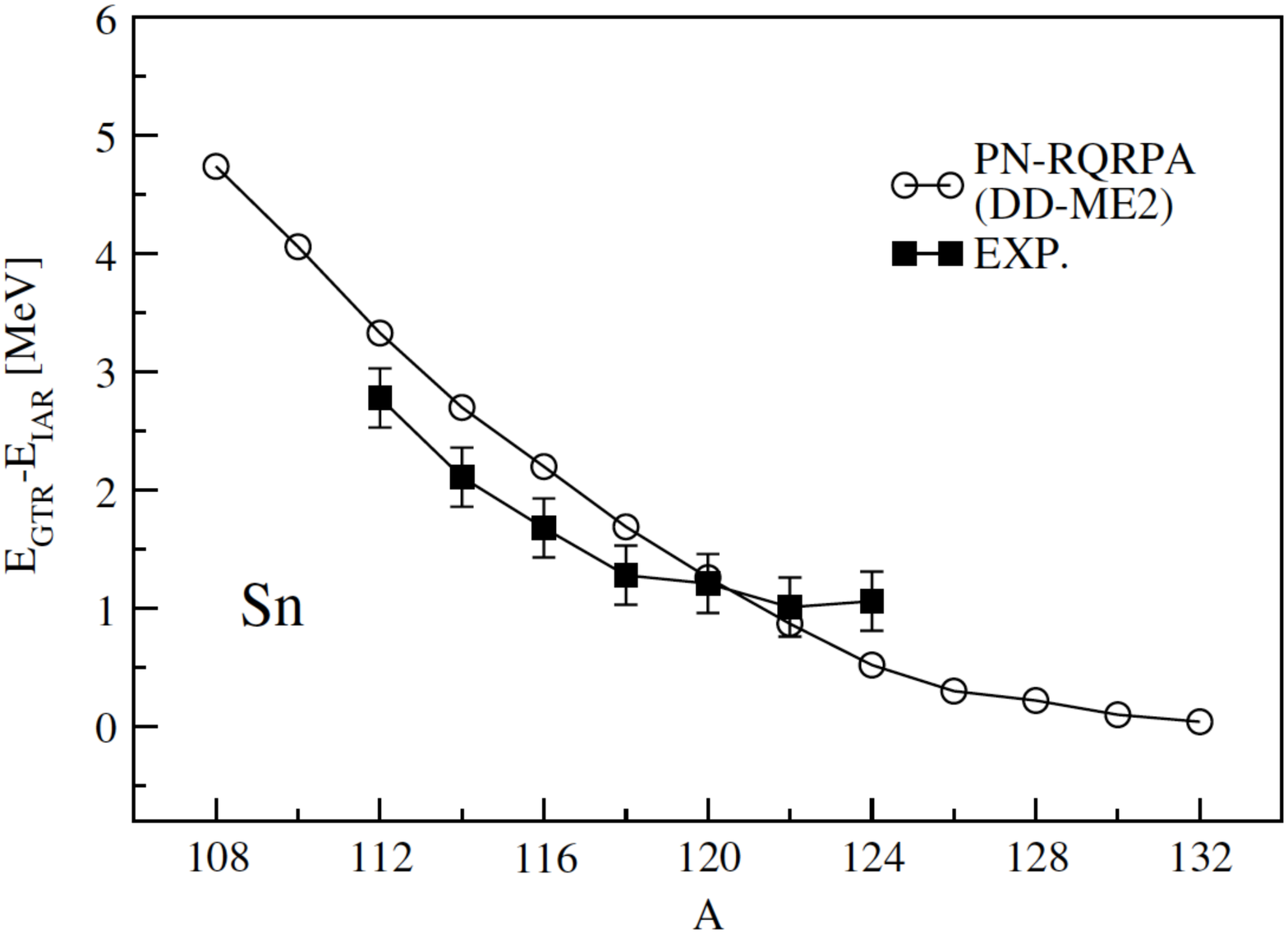}
\includegraphics[width=0.51\linewidth,clip=true]{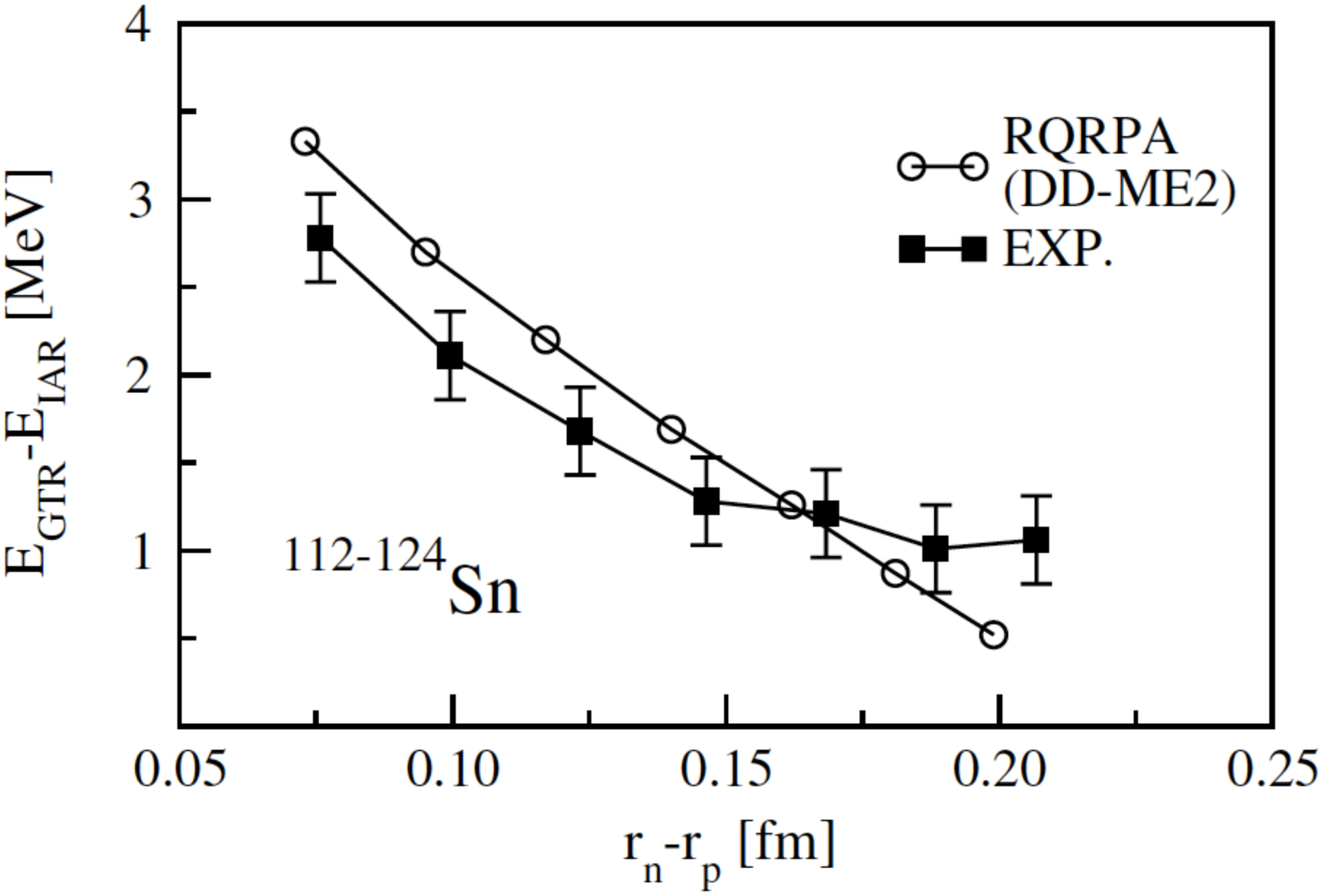}
\caption{\label{fig_gtr_paar}Differences between the GTR and IAR excitation energies for even-even isotopes $^{112-124}$Sn calculated with the RHB+RQRPA with DD-ME2~\cite{ddme2} interaction (left panel). The same excitation energy differences shown as a function of the neutron skin thicknesses calculated with the RHB model (right panel). Experimental data on the GTR and IAR energy differences are from Ref.~\cite{Pham1995} and the values for neutron skin thicknesses are from analysis in Ref.~\cite{Krasznahorkay1999}. Figure taken from  Ref.~\cite{Paar2007}.}
%
% corresponds to Fig.46 & 47 in Paar2007
%
\end{figure}

Accordingly, a method has been suggested to determine the difference
between the radii of the neutron and proton density distributions $\Delta r_{\rm np}$ 
along Sn isotope chain, based on measurements of GTR and IAS excitation
energies~\cite{Paar2003}. Fig.~\ref{fig_gtr_paar} (left panel) shows
the differences between the GTR and IAS excitation energies for 
even-even isotopes $^{112-124}$Sn, based
on relativistic Hartree-Bogoliubov (RHB) plus proton-neutron relativistic QRPA
calculations~\cite{Paar2007} with DDME2~\cite{ddme2} interaction.
The model calculations of the GTR-IAS energy spacings appear in
reasonable agreement with the experimental data from $(^3 He,t)$ 
reactions~\cite{Pham1995}. In the next step,
the differences between the GTR-IAS excitation energies are displayed
as a function of the neutron skin thickness of Sn isotopes under considerations,
calculated within the RHB model (right panel in Fig.~\ref{fig_gtr_paar}).
Model calculations established a strong correlation between the GTR-IAS
energy spacings and neutron skin thickness. Clearly, the implementation 
of experimental data on the GTR with respect to IAS allows 
constraining the value of $\Delta r_{\rm np}$. Fig.~\ref{fig_gtr_paar}
also shows the experimental data points for the excitation 
energies~\cite{Pham1995} and neutron skin thicknesses from the
analysis in Ref.~\cite{Krasznahorkay1999}. Some
differences exist between the theoretical and experimental results, especially
for more neutron-rich Sn isotopes. One should note that experimental 
values for $\Delta r_{\rm np}$ are model dependent and based on the analysis 
of isovector spin-dipole resonances. Clearly, more systematic studies, both theoretically
and experimentally are needed to exploit fully the GTR as a constraint for $\Delta r_{\rm np}$,
as well as the symmetry energy of the EOS.

From the experimental side, very recent progress on $\text{GT}^-$ excitation spectra includes high energy-resolution measurements based on $(^3 He,t)$ reactions for $^{82}$Se~\cite{Frekers2016}, $^{48}$Ti~\cite{Ganioglu2016}, $^{90}$Zr~\cite{Kalmykov2006} targets. In the case of high-resolution study of $^{42}\text{Ca}(^3He,t)^{42}\text{Sc}$ reaction, the GT transition spectra resulted in the appearance of the low-energy super-GT state at 0.611 MeV, while other states are suppressed~\cite{Fujita2015}. On the other side, measurements of $\text{GT}^+$ transitions also provided data for $^{45}\text{Sc}(t,^3He+\gamma)$~\cite{Noji2015} and $^{56}\text{Fe}(t,^3He)$\cite{Scott2014} reactions.  GT transitions have also been explored in $\beta$-decay of rare earth nuclei above $^{146}$Gd~\cite{Nacher2016}.

Microscopic modelling of the GTR has also progressed over the past few years.
Skyrme EDFs used in QRPA calculations have been improved, resulting in the SAMi functionals~\cite{roca-maza13b} with upgraded fitting protocol and improved spin-orbit potential needed for description of different proton high-angular momentum spin-orbit splittings. Furthermore, empirical hierarchy and positive values found in previous analysis of the spin $(G0)$ and spin-isospin $(G0')$ Landau-Migdal parameters, $0 < G0 < G0'$, have been taken care of. Another improvement of non-relativistic approaches is implementation of tensor correlation effects in QRPA description of the GTR, that result in redistribution of the main GTR strength to lower 
energy peaks, and also to high-energy tails~\cite{Severyukhin2013}. The analysis of GTR based on Skyrme EDF supplemented with the tensor terms, showed that the resonance centroids are sensitive to the adopted strengths of the triplet-even and triplet-odd tensor interactions~\cite{Bai2011}. In Ref.~\cite{Bai2009} the main GT peaks in $^{90}$Zr and $^{208}$Pb are
shifted downwards by about 2 MeV when tensor contributions are included.

The proton-neutron QRPA has also been established in the framework of 
D1M Gogny force, and employed in studies of even and odd $^{90-94}\text{Zr}$ isotopes~\cite{Deloncle2017}. The continuum RPA which treats the continuum part of the
single-particle spectrum without approximations has recently been extended for
studies of charge-exchange excitations, including Fermi, GTR and spin-dipole excitations,
indicating the importance of correct implementation of continuum configuration space and
tensor terms of the interaction~\cite{Donno2016}. Another relevant aspect is nuclear deformation effect on charge-exchange excited states, as shown in recent implementation of the
deformed quasi-particle random-phase approximation in the study of GT transitions for
$^{24,26}$Mg~\cite{Ha2016}.

In the relativistic framework, the GTR has also been described within the relativistic RHB + proton-neutron
RQRPA based on density dependent meson-nucleon vertex functions~\cite{Paar2004} and relativistic Hartree + RPA with non-linear $\sigma$-meson self-interaction terms~\cite{Ma2004}. 
In Ref.~\cite{Marketin2012a} the effects of finite momentum transfer, arising in the higher-order terms in the transition operator, have been explored in the case of GTR, indicating that the total strength is only slightly enhanced in nuclei with small neutron-to-proton ratios, while it is not affected with increasing neutron excess. Furthermore, fully self-consistent relativistic QRPA has been formulated in the canonical single-nucleon basis of the relativistic Hartree-Fock-Bogoliubov theory, and successfully applied in the description of charge-exchange excitations, including GTR~\cite{Liang2008,Niu2017}.

\begin{figure}[t!]
\centering
\includegraphics[width=0.45\linewidth,clip=true]{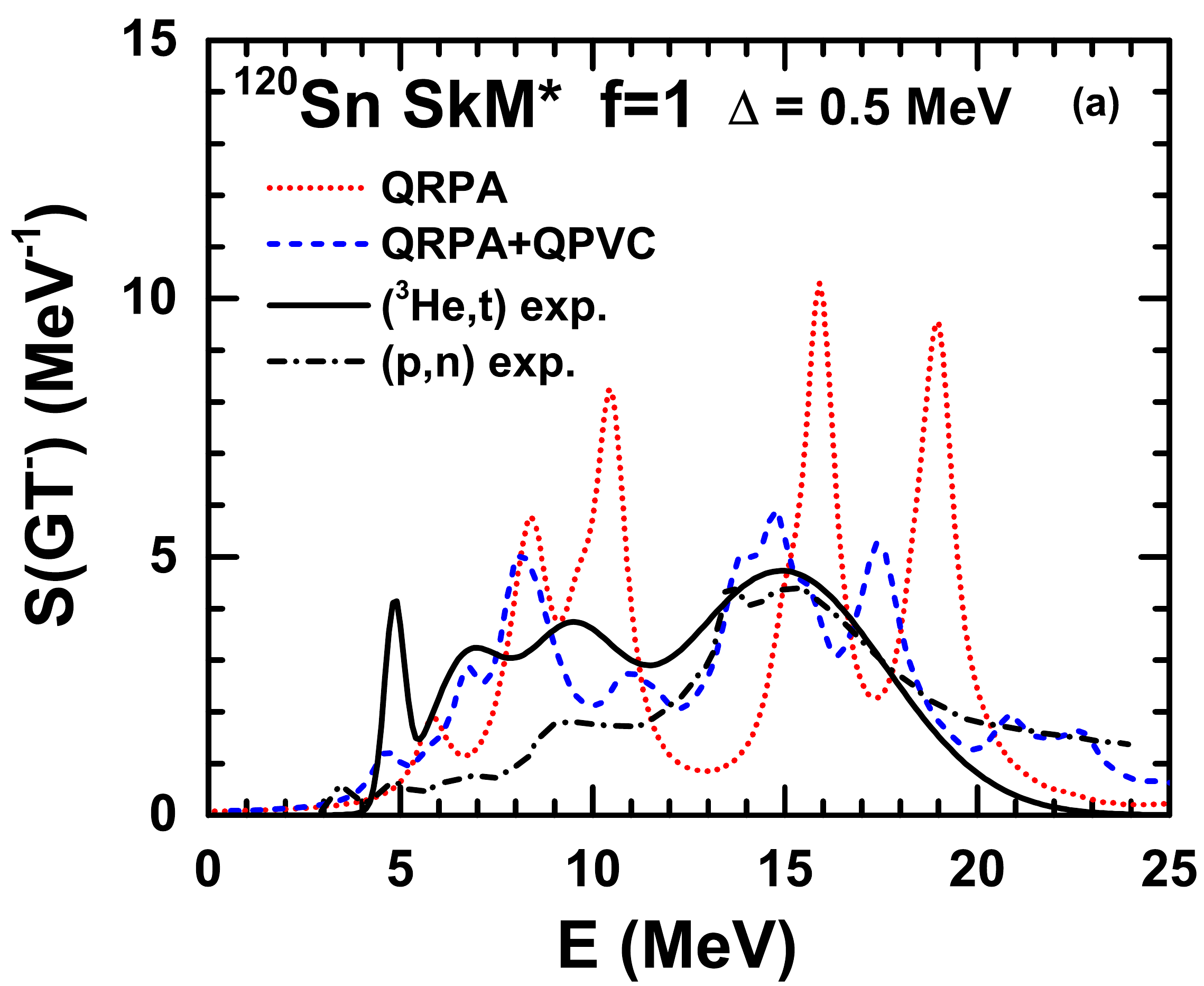}
\includegraphics[width=0.45\linewidth,clip=true]{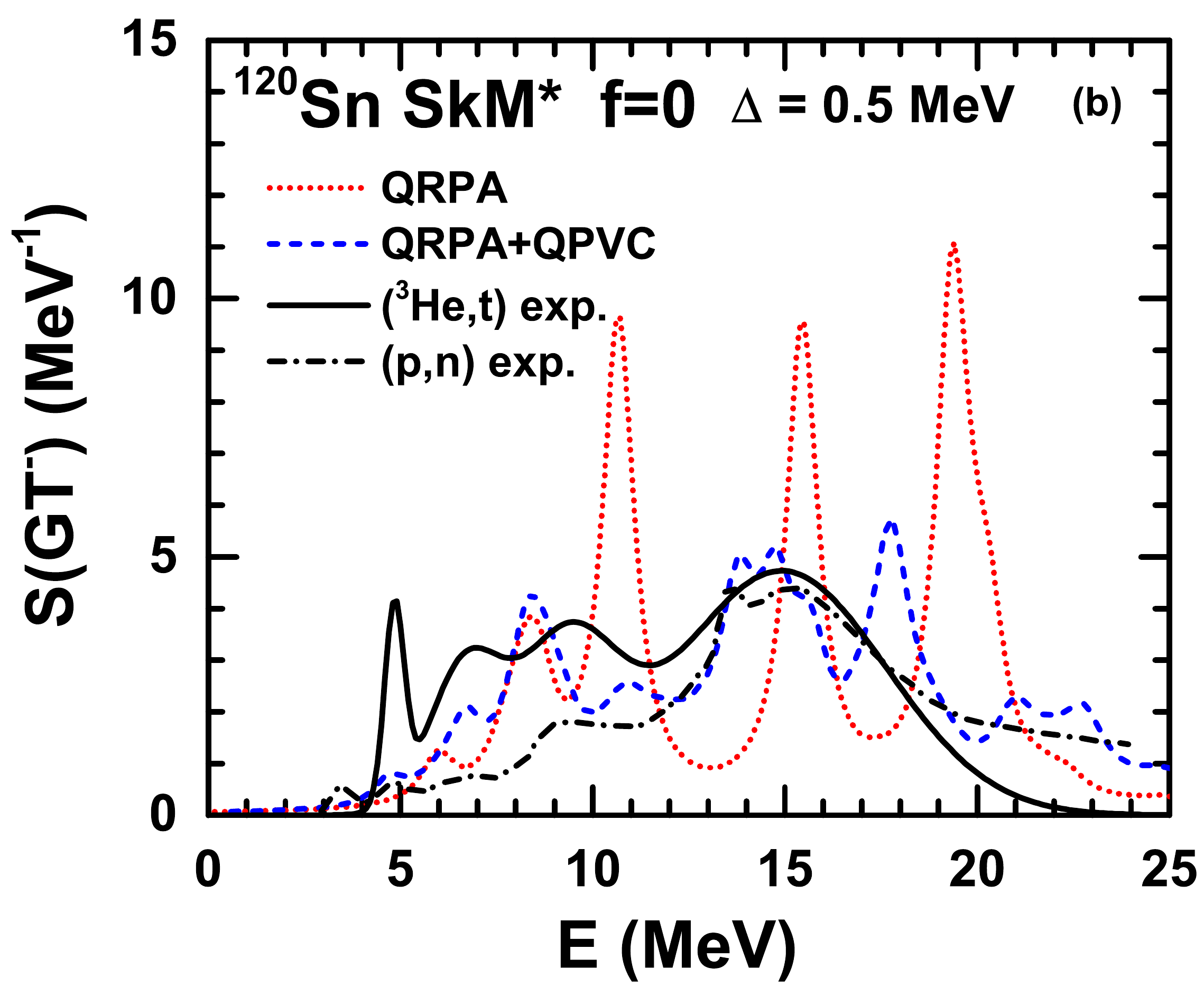}
\caption{\label{fig_gtr_niu}The GT strength distributions for $^{120}$Sn 
calculated by QRPA and QRPA+QPVC models with Skyrme interaction SkM*, 
with (a) and without (b) isoscalar pairing. Experimental data are from $(^3He,t)$ (scaled by a factor 1.6)~\cite{Pham1995} and $(p,n)$ reaction, normalized by the unit cross section~\cite{Sasano2009}. Figure taken from  Ref.~\cite{Niu2016}.}
\end{figure}

In order to account for the effects going beyond simple excitations at the $ph$ level, further improvements of theory frameworks are mandatory. In Ref.~\cite{Minato2016} the $2p2h$ effects on the GTR has been explored in neutron-rich nuclei $^{24}$O, $^{34}$Si, and $^{48}$Ca 
within the second Tamm-Dancoff approximation with Skyrme interaction. 
It has been shown that by including complex configurations considerable amount of
the total GT strength is shifted toward higher energy region above the GTR. However, for a complete understanding, the model should be further extended toward fully consistent
second RPA. Another model, based on self-consistent HF+RPA has been established in Ref.~\cite{niu15}, where the Skyrme interaction has been used to obtain the coupling between particles
and vibrations, in order to provide mixing of the GTR with a set of doorway states, leading to
the fragmentation of the overall transition strength. Further extension, based on self-consistent
QRPA plus quasi-particle-vibration coupling (QPVC) model with Skyrme interactions, successfully reproduced the GTR strength distribution in $^{120}$Sn (see Fig.~\ref{fig_gtr_niu}).
In the relativistic framework, the spreading effects in the GTR spectra have been accounted for within the proton-neutron relativistic quasi-particle time-blocking approximation, resulting in considerable fragmentation of the resonance along with quenching of the strength~\cite{Robin2016}.

Experimentally, the GTR exhausts only about 60$-$70\% of the well-known and model independent Ikeda sum rule (ISR), which corresponds to $m_0($GT$^{-})-m_0($GT$^{+})=3(N-Z)$. To explain this quenching problem, it has been proposed that the effects of the second-order configuration mixing (2 particle $-$ 2 hole correlations), or the coupling with the $\Delta-hole$ excitation had to be considered. The experimental analysis of ${}^{90}$Zr of Ref.\cite{Wakasa1997} indicates that most of the quenching has to be attributed to $2p-2h$ correlations, while the role played by the $\Delta$ isobar is much smaller. Recently, {\it {ab initio}} framework based on isospin-breaking coupled-cluster technique has also been tested in description of GT transitions in light systems $^{14}$C and $^{22,24}$O~\cite{Ekstrom2014}. Chiral interactions at next-to-next-to leading order have been optimized to observables in two- and three-nucleon systems and consistent two-body currents have been employed. As a result, the two-body currents reduce the Ikeda sum rule, corresponding to a quenching factor $\approx 0.84-0.92$ of the axial-vector coupling.

\subsubsection{Spin-dipole resonance}
%
%Remco Zeegers check
%

Excitation of spin-dipole resonance (SDR) represents another mode in 
the charge-exchange channel that allows to determine the neutron-skin thickness in finite nuclei and in this way to constrain the neutron matter EOS. The SDR corresponds to the spin-flip transition $\Delta S$ = 1, that is also a dipole mode with $\Delta L$ = 1, thus the SDR is composed from three components, $\Delta J^{\pi} = \text{0}^- , \text{1}^-$, and $\text{2}^-$. The SDR sum rule includes the difference of the respective $\beta^-$ and $\beta^+$
transitions,
\begin{equation}
m_0(SDR^-)-m_0(SDR^+)=\frac{9}{2\pi}(N\langle r_{\rm n}^2 \rangle-Z\langle r_{\rm p}^2 \rangle)\approx\frac{9}{2\pi}(N-Z)\langle r_{\rm p}^2 \rangle\left(1+\frac{2N}{N-Z}\frac{\Delta r_{\rm np}}{\langle r_{\rm p}^2 \rangle^{1/2}}+\frac{N}{N-Z}\frac{\Delta r_{\rm np}^2}{\langle r_{\rm p}^2 \rangle}\right),
\end{equation}
that involves dependence on the neutron and proton rms radii, the latter being commonly known from elastic electron scattering. The sum rule approach has been used in Ref.~\cite{Krasznahorkay1999} to extract the neutron skin thicknesses in several Sn isotopes from the experimental data on the SDR. Clearly, the SDR provides an useful excitation mode for the studies of isovector properties in finite nuclei and EOS.
In addition, the properties of SDR are of particular importance because of its contribution in the charged current neutrino-nucleus reactions~\cite{Suzuki2003,Cheoun2010,Paar2008,Paar2011,Paar2013}.

The first measurement of the SDR has been reported in the investigation of
$^{208}\text{Pb}(p,n)$ reaction~\cite{Horen1980}, and further properties have
been studied in Ref.~\cite{Gaarde1981}. The SDR has also been measured
in $(p,n)$ reaction with $^{90,92,94}$Zr isotopes, and its energy splitting with
respect to IVGDR has been explored~\cite{Austin2001}. 
Alternatively, the SDR has also been
studied in $(^3He,t)$ charge-exchange reaction, e.g. in Sn isotopes~\cite{Krasznahorkay1999}. In Ref.~\cite{Akimune2000}  
$^{208}\text{Pb}(^3He,t)^{208}\text{Bi}$ reaction at very forward scattering
angles has been employed to study the microscopic structure of the SDR 
in $^{208}\text{Bi}$.

\begin{figure}[t!]
\centering
\includegraphics[width=0.95\linewidth,clip=true]{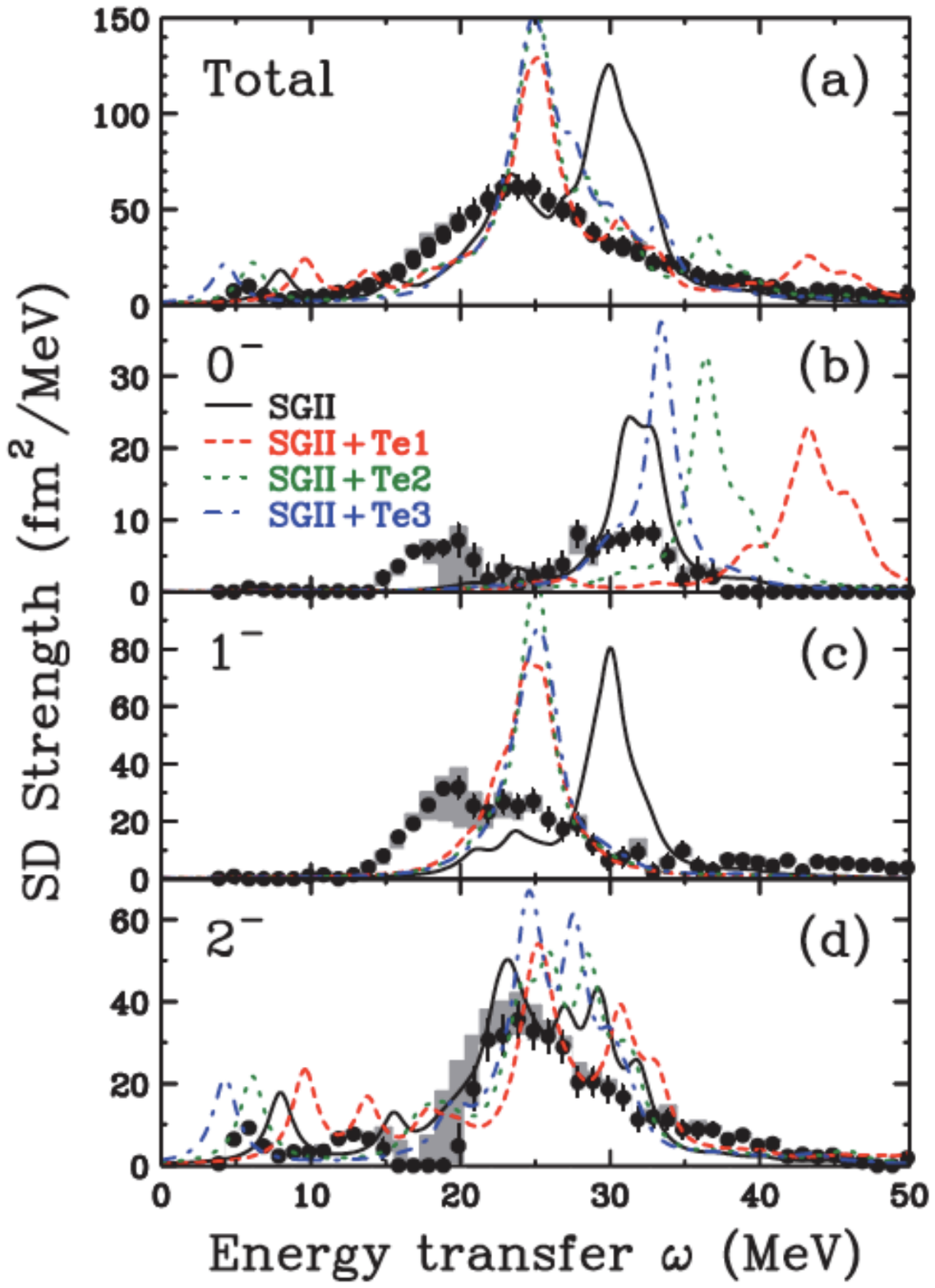}
\caption{\label{fig_sdr_wakasa}The SDR strength distributions from multipole 
decomposition analysis of the $^{208}\text{Pb}(p,n)$ reaction in comparison to
the HF+RPA calculations with SGII~\cite{sgII} interaction including tensor terms denoted 
as Te1, Te2, Te3~\cite{Bai2011}.  Figure taken from  Ref.~\cite{Wakasa2012}.}
%
% corresponds to Fig.19 in Wakasa2012
%
\end{figure}

In the early studies of the SDR~\cite{Horen1980,Gaarde1981,Austin2001}, measurements provided only a combined centroid energy of all three underlying components $\Delta J^{\pi} = \text{0}^- , \text{1}^-,$ and $\text{2}^-$. More recently some details about the underlying 
structure of the SDR became available.  In Ref.~\cite{Kawabata2002} cross sections 
and polarization transfer observables 
have been measured for the $^{16}\text{O}(p,p')$ reaction. In this way, the non-spin-flip
and spin-flip transition strengths were separated. It has been shown that part of the $1^-$
spin-flip strengths appears in the same energy region as the IVGDR states ($E_x=\text{20.9}, \text{22.1}, \text{24.0}$ MeV), while the resonances at $E_x=\text{19.0 and 20.4}$ MeV are observed only
in the spin-flip spectra, and are assigned as $2^-$ states~\cite{Kawabata2002}. The 
experimental results are consistent with the shell model calculations~\cite{Kawabata2002}.
In Ref.~\cite{Wakasa2012} a complete set of polarization transfer observables has been
measured for the $^{208}\text{Pb}(p,n)$ reaction. The polarisation observables have been, for the
first time, used to separate the SDR components into different spin-parity transfer contributions, $\Delta J^{\pi} = \text{0}^- , \text{1}^-,$ and $\text{2}^-$~\cite{Wakasa2012}. Fig.~\ref{fig_sdr_wakasa} shows the respective overall strength distribution for the SDR, as well as
separate distributions for the underlying three components. In addition to the strong SD resonance, there are also indications for possible appearance of soft spin-dipole modes in the low-energy region of excitation spectra~\cite{Stuhl2012}. The nature of this soft mode needs to be further investigated in more details.

In theoretical description of the SDR,  approaches based on the HF+RPA have recently
been extended to include the tensor terms of the Skyrme interactions~\cite{Bai2010,Bai2011}. 
The tensor force is known to have an important effect on the spin-orbit splitting
and properties of excited states~\cite{Brown2006,Colo2007,Lesinski2007,Zalewski2008,Yuksel2011,Yuksel2013,Sagawa2014}.
The tensor correlations result in multipole-dependent effects on the SDR transition
strengths, thus allowing improvements in the agreement of different SDR 
components with the experimental data~\cite{Bai2010,Bai2011}. The relevance of the 
tensor terms on the SDR is illustrated in Fig.~\ref{fig_sdr_wakasa}, where the 
results of HF+RPA calculations with SGII~\cite{sgII} interaction are shown in 
comparison to the experimental data. Tensor terms denoted as 
Te1, Te2, Te3~\cite{Bai2011} result in significant modifications
of the strength distributions for different components (especially $0^-$ and $1^-$)
as for the overall SDR strength distribution.

The properties of SDR have also been explored within covariant EDF approaches.
The first implementation of the fully self-consistent charge-exchange relativistic
RPA based on the relativistic Hartree-Fock model established a complete 
description of the SDR, without any additional adjustments of the model
parameters~\cite{Liang2008}. It has been shown that the isoscalar meson 
fields $(\sigma,\omega)$ play an essential role in spin-isospin resonances 
through the exchange terms~\cite{Liang2008}. Due to the Dirac sea effects
in the implementation of relativistic framework, the SDR sum rule reduction
has been found, $\text{6.4} \%$ for $^{90}$Zr and $\text{5.4}\%$ for 
$^{208}$Pb. The calculated SDR strength distributions for $^{90}$Zr 
reproduced the experimental data. 

\begin{figure}[t!]
\centering
\includegraphics[width=0.6\linewidth,clip=true]{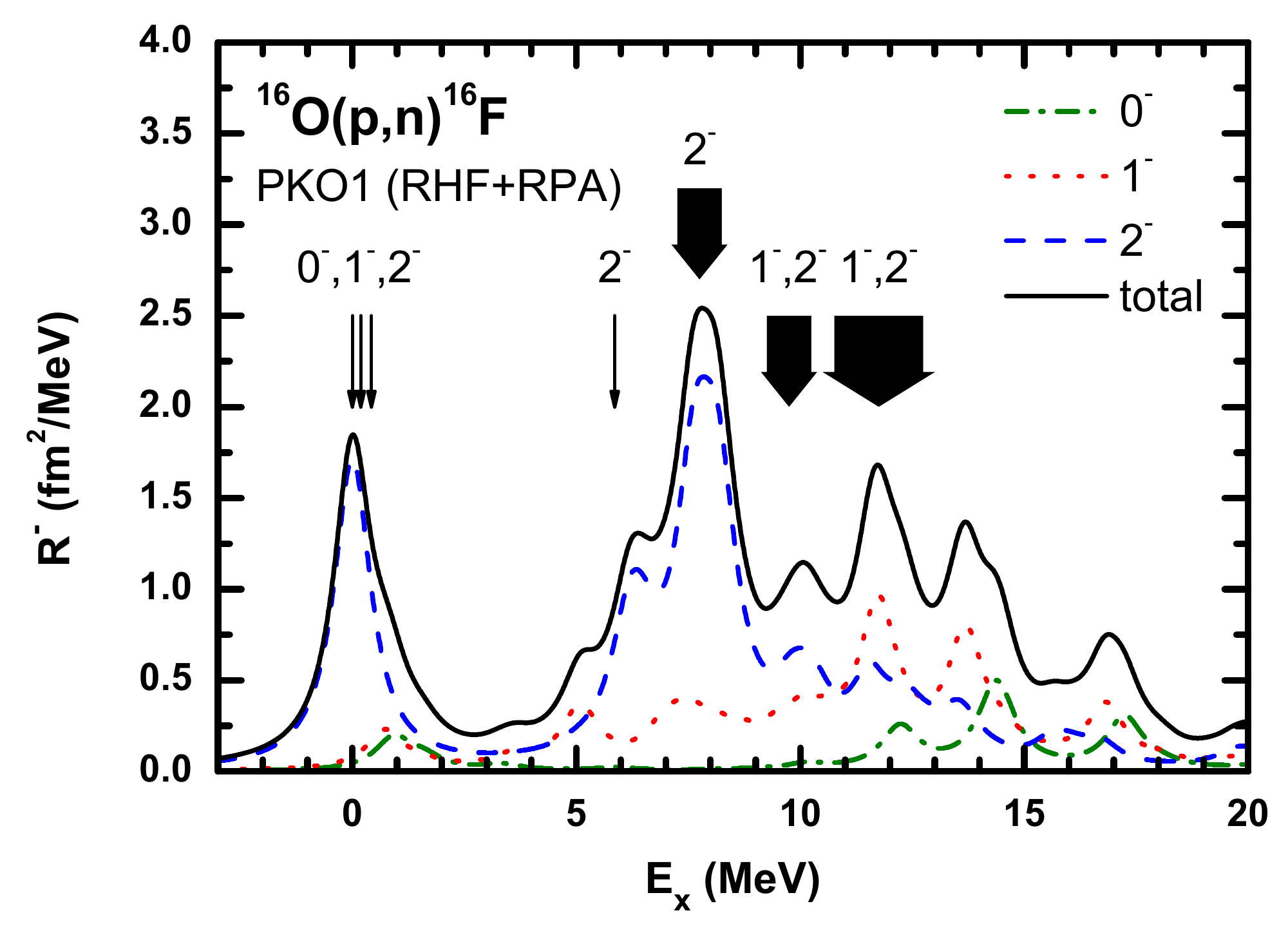}
\caption{\label{fig_sdr_liang}The SDR strength distributions for $^{16}$O calculated in the 
relativistic HF+RPA approach with PKO1~\cite{Long2006} interaction.  
Different components $\Delta J^{\pi} = \text{0}^- , \text{1}^-,$ and $\text{2}^-$,
are shown separately. Experimental data shown as arrows are 
adopted from Refs.~\cite{Wakasa2011,Tilley1993}.Figure taken from  Ref.~\cite{Liang2012a}.}
%
% corresponds to Fig.1 in Liang2012
%
\end{figure}

Further successful applications of the covariant DFT approach include 
analysis of spin-dipole excitations in $^{16}$O~\cite{Liang2012a}. 
Fig.~\ref{fig_sdr_liang} shows the respective strength distributions 
($T_-$ channel) based on relativistic HF+RPA with PKO1~\cite{Long2006}
interaction. From the comparison with the experimental data, one concludes
that the fine structure of excitation spectra has been successfully reproduced,
without any readjustments of the functional. The first implementations 
of this approach were non-local and require considerable numerical 
effort. In Ref.~\cite{Liang2012a} a new method has been introduced,
based on zero-range reduction and the Fierz transformation, 
in order to provide localized form of the Fock terms in the 
covariant DFT. This approach opened perspective for further
developments of local covariant density functionals with proper
isoscalar and isovector properties.

\begin{figure}[t!]
\centering
\includegraphics[width=0.6\linewidth,clip=true]{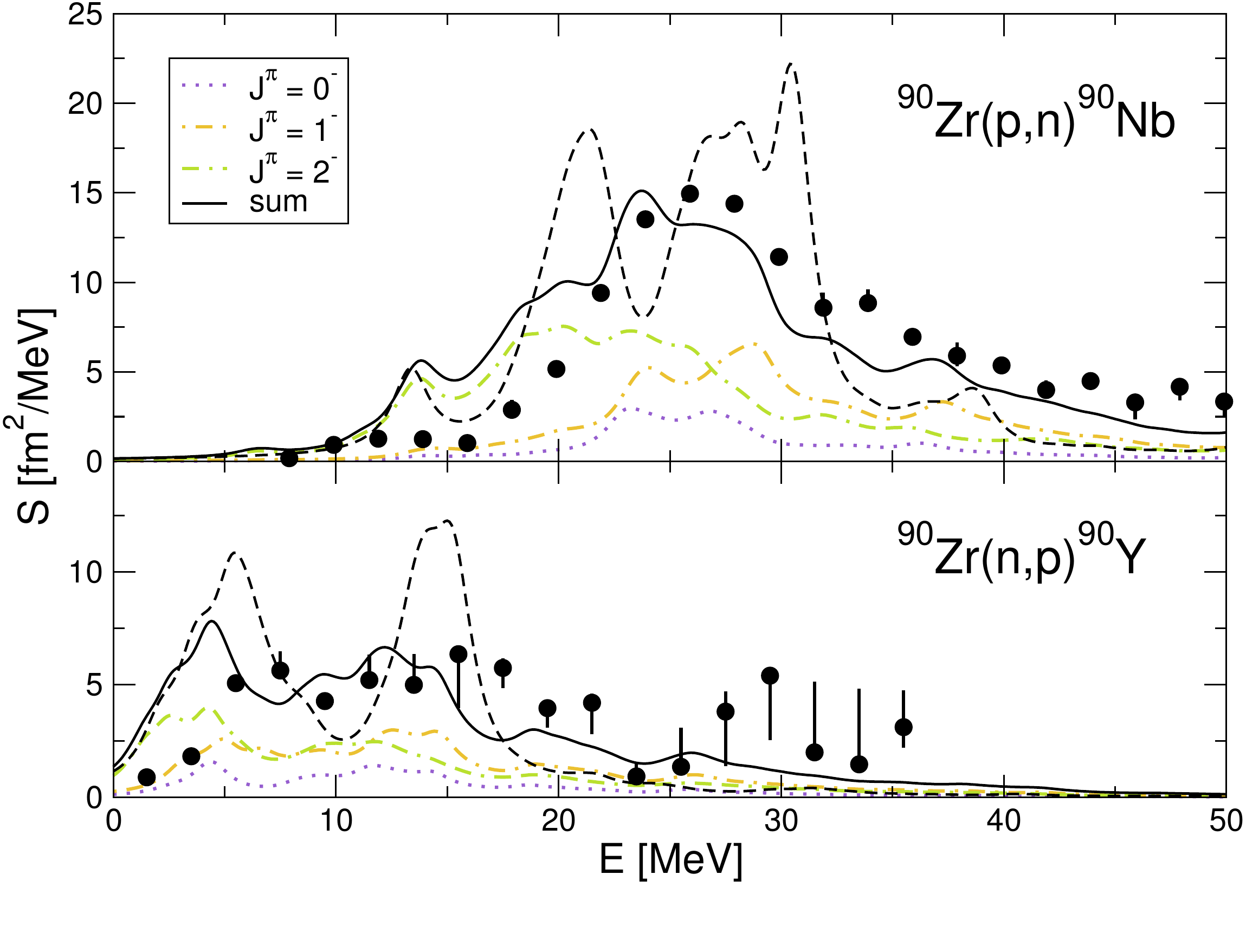}
\caption{\label{fig_sdr_marketin} The SDR transition strength distributions for
$T_-$ and $T_+$ channels (upper and lower panels, respectively) in $^{90}$Zr. 
The full line denotes complete RTBA results, while the dashed line corresponds
to the relativistic RPA strength distributions. Experimental data are 
adopted from Refs.~\cite{Wakasa1997,Yako2005}. Figure taken from  Ref.~\cite{Marketin2012}.}
%
% corresponds to Fig.1 in Marketin2012
%
\end{figure}

Further extensions of theory frameworks, going beyond the RPA, are necessary 
to describe the fine structure of the SDR excitation spectra. Already in the
early study in Ref.~\cite{Drozdz1987} the damping of the SDR has been 
explored in $^{90}$Zr within a microscopic framework which in addition to
$1p1h$ configurations includes coupling to $2p2h$ configurations. It resulted
in the energy dependent spreading width which shifts a large fraction of 
transition strength toward high excitation energies. More recently, within 
a consistent approach based on covariant DFT~\cite{Marketin2012},
charge-exchange version of the particle-vibration model has been established
and realized within relativistic time-blocking approximation (RTBA). 
Fig.~\ref{fig_sdr_marketin} shows the SDR dipole strength distributions
for  $^{90}$Zr, both for $T_-$ and $T_+$ channels, calculated with the RTBA
approach, in comparison to relativistic RPA results and experimental data~\cite{Wakasa1997,Yako2005}. Due to the particle-vibration coupling,
the SDR transition strength becomes fragmented and part of the strength is
shifted toward higher energies~\cite{Marketin2012}. Clearly, the implementation
of the effects going beyond the RPA are essential, and should be included in
the future approaches to constrain the symmetry energy from the SDR.

The implementation of the SDR in constraining the neutron-skin thickness 
and symmetry energy is rather limited. An overview of previous analyses is 
given in Ref.~\cite{colo15}. In addition to the first study for
Sn isotopes~\cite{Krasznahorkay1999}, neutron skin thickness has been
determined for $^{90}$Zr~\cite{Yako2006}. The analysis in the former case was limited
because only $T_-$ transition strength has been available from the experiment.
In the case of $^{90}$Zr, the SDR model-independent sum rule has been calculated explicitly using data from both $(p,n)$ and $(n,p)$ reactions, supplemented by the multipole decomposition analysis~\cite{Wakasa1997,Yako2005,Ichimura2006}. Clearly, the SDR approach to determine $\Delta r_{\rm np}$ and in this way to constrain the neutron matter EOS, requires experimental data, simultaneously from $(p,n)$ and $(n,p)$ or $(^3He,t)$ and $(t,^3He)$ reactions. In the case of $^{208}$Bi, multipole decomposition analysis has been performed for $(p,n)$ reaction~\cite{Wakasa2012}. Therefore, as pointed in Ref.~\cite{colo15}, it would be of paramount importance to perform the counter experiment
$^{208}\text{Pb}(n,p)$.

%
%This could be extended, based on ~\cite{colo15}
%
%

\subsubsection{Anti-analogue giant dipole resonance}

\begin{figure}[t!]
\centering
\includegraphics[width=0.6\linewidth,clip=true]{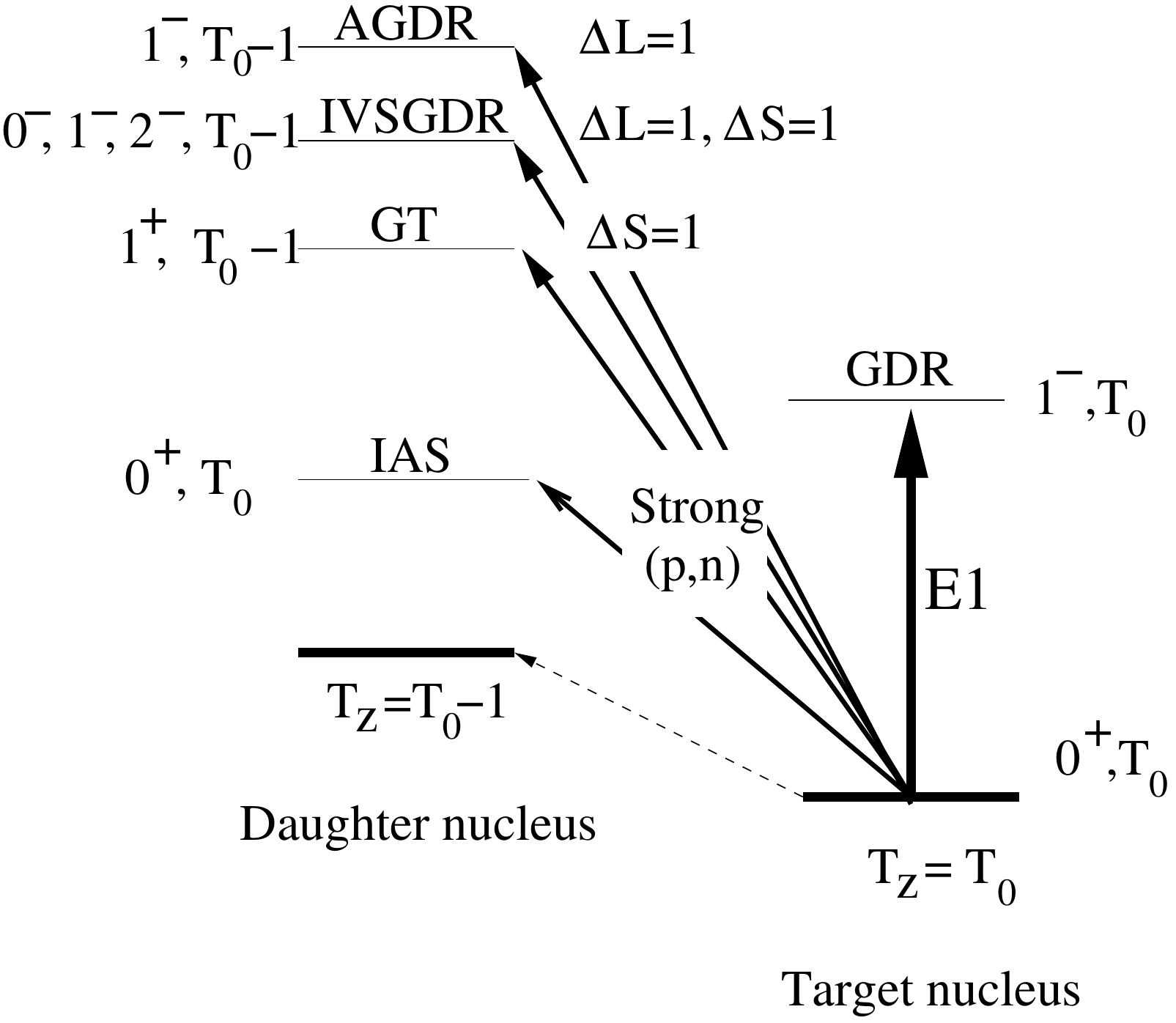}
\caption{\label{fig_agdr_kras1}The nuclear ground state ($T_z =
  T_0$) and excited states populated by strong $(p,n)$ reactions,
  corresponding to GTR, IVSGDR and AGDR in the 
  daughter nucleus ($T_z = T_0-1$). Figure taken from  Ref.~\cite{Krasznahorkay2013}.}
% corresponds to Fig. 1 in Krasznahorkay2013
\end{figure}

\begin{figure}[t!]
\centering
\includegraphics[width=0.6\linewidth,clip=true]{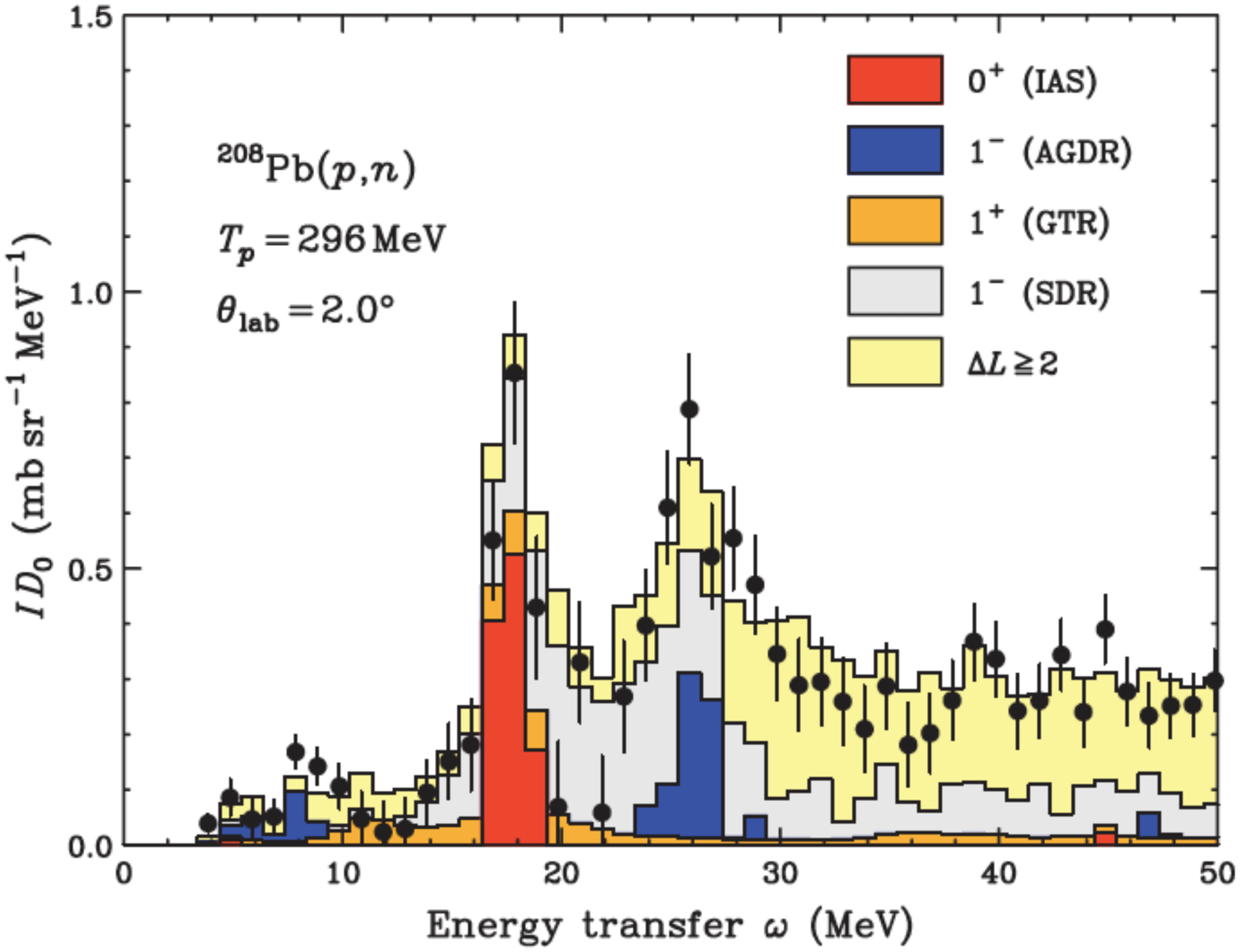}
\caption{\label{fig_agdr_yasuda}Result of multipole decomposition analysis for $^{208}$Pb reaction data. Relevant transitions of various charge-exchange excitation modes are separately denoted. Figure taken from  Ref.~\cite{Yasuda2013}.}
% corresponds to Fig. 2 in Yasuda2013
\end{figure}

Another charge-exchange excitation mode, observed in $(p,n)$ reactions,
is $\Delta J^{\pi}=1^-$, $\Delta L = 1$ resonant excitation, representing 
the anti-analog giant dipole resonance (AGDR), that is $T_0-1$ component
of the charge-exchange GDR, there $T_0$ denotes ground-state isospin 
of nuclear target~\cite{Auerbach1981,Osterfeld1981,Krasznahorkay2013}. In general, due to isovector
nature of the ($p$,$n$) reaction, the strength of
the E1 excitations is distributed into T$_0$-1, T$_0$ and T$_0$+1
components (see Fig.~\ref{fig_agdr_kras1}). However, from the 
consideration of relevant Clebsch-Gordan coefficients, 
the T$_0$-1 component (AGDR) is  favored compared to the T$_0$ and
T$_0$+1, by factors of $T_0$ and $2{T_0}^2$~\cite{Krasznahorkay2013}.  
Experimental data on the AGDR have been extracted from
$(p,n)$~\cite{Sterrenburg1980,Gaarde1981,Nishihara1985} and $(^3He,t)$ reactions~\cite{Pham1995}, because the observed $\Delta L =1$
resonance in general corresponds to the superposition of possible 
spin-flip dipole (IVSGDR) and non-spin-flip dipole (AGDR) modes.
The non-spin-flip transition is preferred at low bombarding energies 
below 50 MeV~\cite{Osterfeld1992}. In all experimental spectra,
a clear peak corresponding to $\Delta L = 1$ transfer has been observed
at an energy several MeV above the GTR.  Fig.~\ref{fig_agdr_yasuda}
shows a typical result of multipole decomposition analysis of the $(p,n)$ reaction 
on $^{208}$Pb target, where the contributions from various multipoles have been
separately shown, resulting in a clear evidence for the AGDR around 26 MeV.

\begin{figure}[t!]
\centering
\includegraphics[width=0.45\linewidth,clip=true]{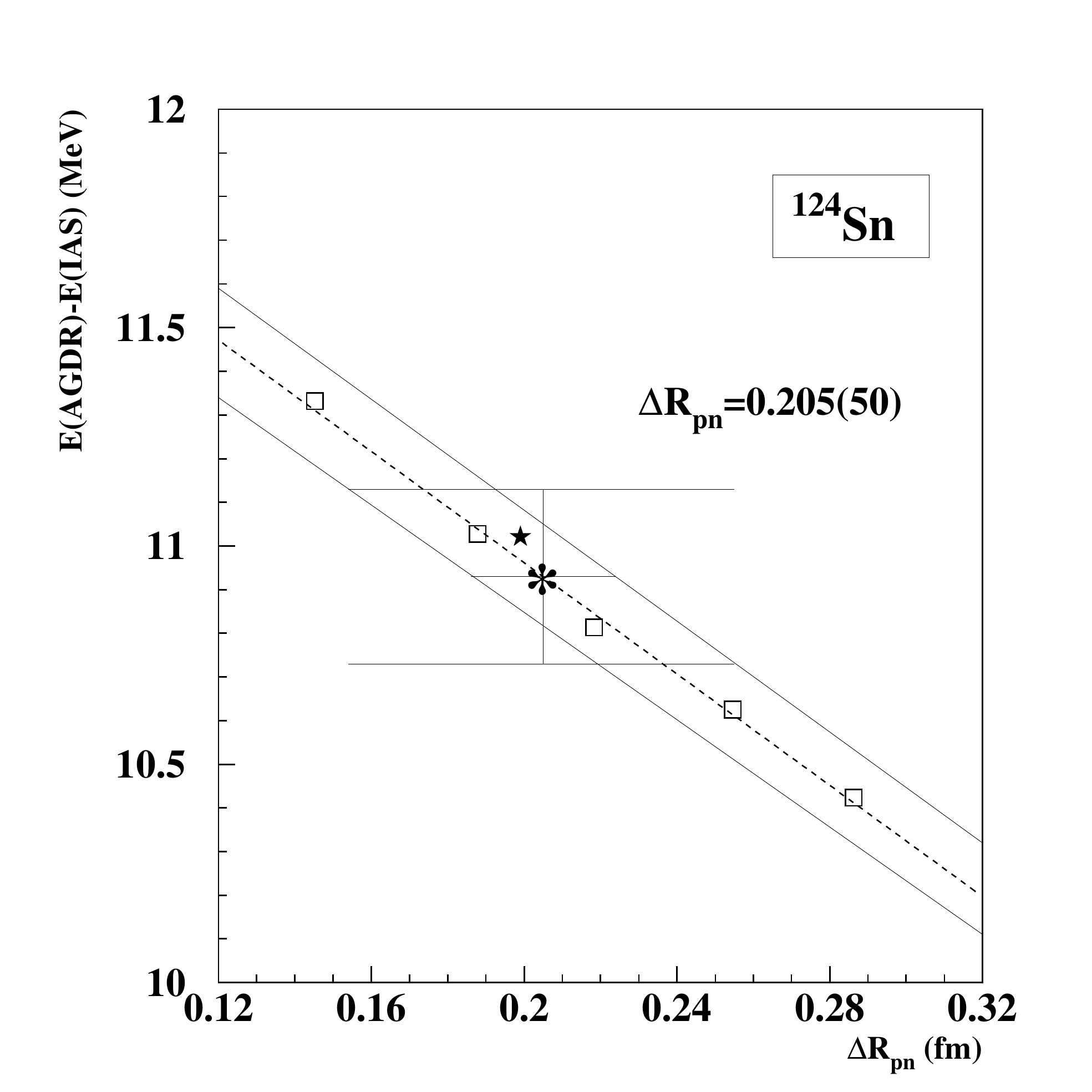}
\includegraphics[width=0.45\linewidth,clip=true]{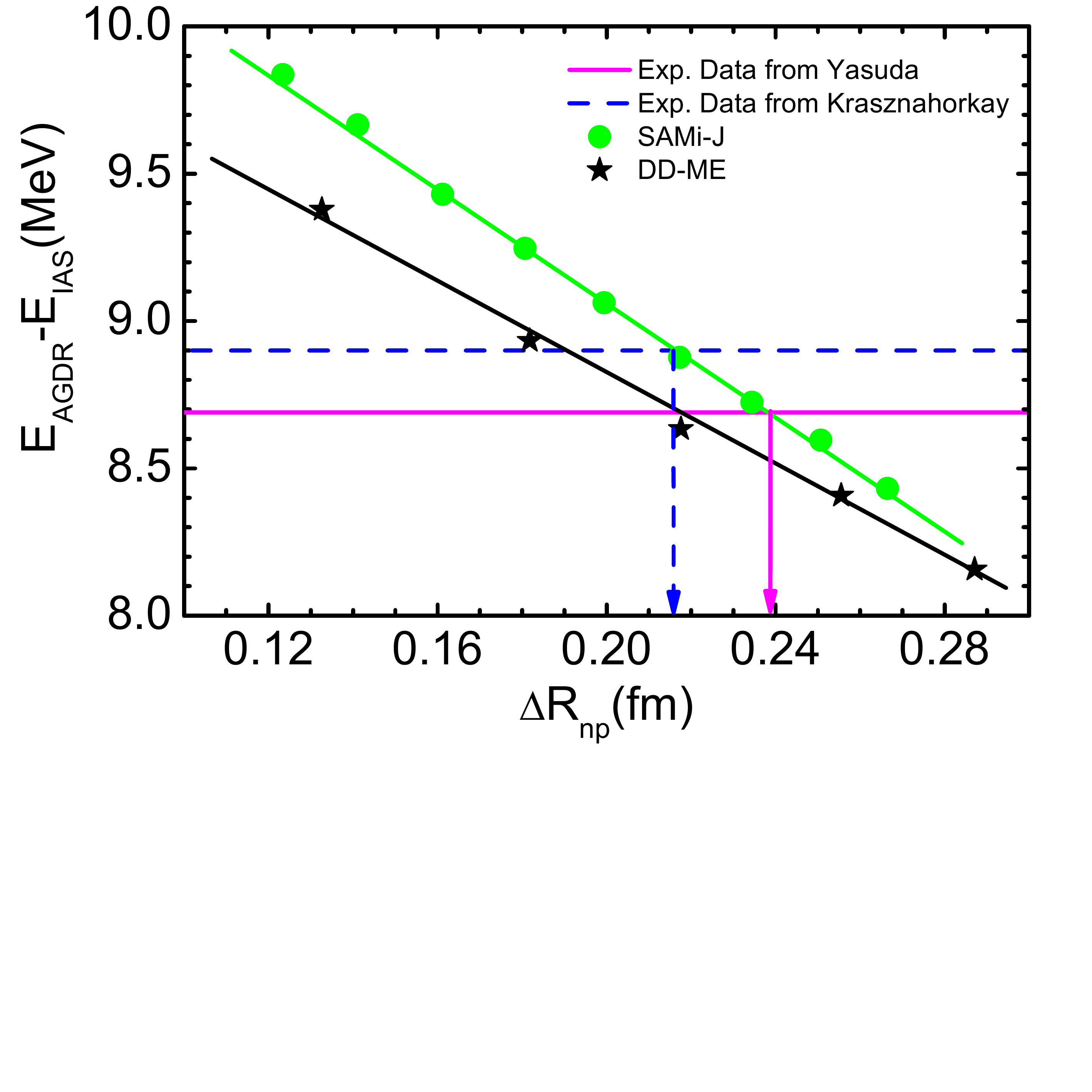}
\caption{\label{fig_agdr_kras2} Differences between the AGDR and IAS
excitation energies and neutron skin thicknesses for $^{124}$Sn 
calculated using DDME family of interactions~\cite{ddme,ddme2} (left panel).
The same differences are shown for $^{208}$Pb, with SAMi-J~\cite{roca-maza12b}
and DDME~\cite{ddme} families of interactions (right panel).
Experimental data on excitation energies for $^{124}$Sn~\cite{Sterrenburg1980}
and $^{208}$Pb ~\cite{Yasuda2013,Krasznahorkay} and extracted values of $\Delta r_{\rm np}$ are
denoted in the two panels. Figure taken from  Refs.~\cite{Krasznahorkay2013,roca-maza2016}.}
% corresponds to Fig. 4 in Krasznahorkay2013
%                 and   Fig. 2 in roca-maza2016
\end{figure}

As pointed out in Ref.~\cite{Krmpotic1983}, the excitation energy of the AGDR
depends sensitively on the neutron skin thickness $\Delta r_{\rm np}$. 
More recently it has been confirmed by microscopic calculations based on
the EDFs that the energy difference between the AGDR and isobaric analog 
state (IAS) is very sensitively related to the corresponding
neutron-skin thickness \cite{Krasznahorkay2013}. By comparing the
energy differences between  AGDR and IAS, calculated using various
EDFs with the experimental data, the values of $\Delta r_{\rm np}$, as well as
the symmetry energy parameters could be constrained. Linear dependence 
between the energy difference $E_{AGDR} - E_{IAS}$ has also been 
confirmed in the approach based on the nuclear droplet model~\cite{Cao2015}.
Clearly, measurements of the AGDR with rare isotope beams will open novel perspectives to determine the neutron skin thickness in exotic nuclei~\cite{Krasznahorkay2013b}.
In Ref.~\cite{Krasznahorkay2013}, a family of density dependent meson-exchange
covariant EDFs~\cite{ddme,ddme2} has been used within the relativistic QRPA to 
determine the properties of AGDR in Sn isotopes. 
Fig.~\ref{fig_agdr_kras2} shows the differences between  the AGDR and IAS
excitation energies for $^{124}$Sn calculated using proton-neutron relativistic
QRPA using five effective interactions spanning the range of values of the
symmetry energy at saturation, $J=$30,32,34,36, and 38 
MeV~\cite{Krasznahorkay2013}.The estimate of theoretical uncertainty of 10$\%$
is denoted by two parallel lines in the figure.

Comparison with the experimental data for $^{124}$Sn,
$E(AGDR) - E(IAS) = 10.93 \pm 0.20 $ MeV, obtained from the measured
value $E(AGDR) - E(IAS) = 10.60 \pm 0.20 $ MeV~\cite{Sterrenburg1980}
after correcting for the admixture of the spin dipole resonance, provided
a constraint on the neutron skin thickness, 
$\Delta r_{\rm np} = 0.205 \pm 0.050 $ fm~\cite{Krasznahorkay2013}.
In Ref.~\cite{Yasuda2013} the AGDR data from  $^{208}\text{Pb}(p,n)$ reaction have been separated from other excitations such as the spin-dipole resonance by
multipole decomposition. In this way, the energy difference between the AGDR and IAS has been found to be $\Delta E = 8.69 \pm 0.36$ MeV, with both statistical and systematic uncertainties included. The analysis based on the results of model calculations using the relativistic QRPA with DDME~\cite{ddme} interactions, 
resulted in the neutron skin thickness value for $^{208}$Pb, $\Delta r_{\rm np} = \text{0.216} \pm \text{0.046} \pm \text{0.015}$ fm~\cite{Yasuda2013}.

In Ref.~\cite{Cao2015} the energy differences $E_{AGDR} - E_{IAS}$ have been
explored in $^{208}$Pb using fully self-consistent HF plus charge-exchange RPA
based on Skyrme parameterizations SAMi-J~\cite{roca-maza12b} spanning
a range of values of symmetry energy at saturation density $J$. 
In a more recent study~\cite{roca-maza2016}, HF+ charge-exchange RPA calculations have been improved by (i) explicit treatment of the exchange term of the two-body Coulomb interaction, and (ii) consistent treatment of the spin-orbit interaction within the EDF theory. Model calculations for $^{208}$Pb based on the HF+RPA confirmed the linear correlation between the AGDR excitation energies and $\Delta r_{\rm np}$~\cite{Cao2015,roca-maza2016}. Fig. ~\ref{fig_agdr_kras2} (right panel) shows 
the differences $E_{AGDR} - E_{IAS}$ for $^{208}$Pb obtained within 
the Skyrme EDFs approach with SAMi-J interactions~\cite{roca-maza12b} 
and relativistic EDFs based on DDME~\cite{ddme} families of interactions.
The non-relativistic and relativistic model calculations indicate the same correlation, 
however, with a different slope. Constraints on the neutron skin thicknesses are
obtained from two different experimental data on the AGDR excitation energies~\cite{Yasuda2013,Krasznahorkay}

Since the measured data on the AGDR are available for $^{208}$Pb from 
two experiments~\cite{Yasuda2013,Krasznahorkay}, the average results obtained from the 
analysis based on SAMi interactions
give $\Delta r_{\rm np}=\text{0.236} \pm \text{0.018}$ fm, $J=\text{33.2} \pm \text{1.0}$ MeV,
and the slope parameter of the symmetry energy at saturation density 
$L=\text{97.3} \pm \text{11.2}$ MeV. The obtained values for $J$ and $L$ appear rather
large in comparison to the constraints obtained from other giant resonances.

\begin{figure}[t!]
\centering
\includegraphics[width=0.6\linewidth,clip=true]{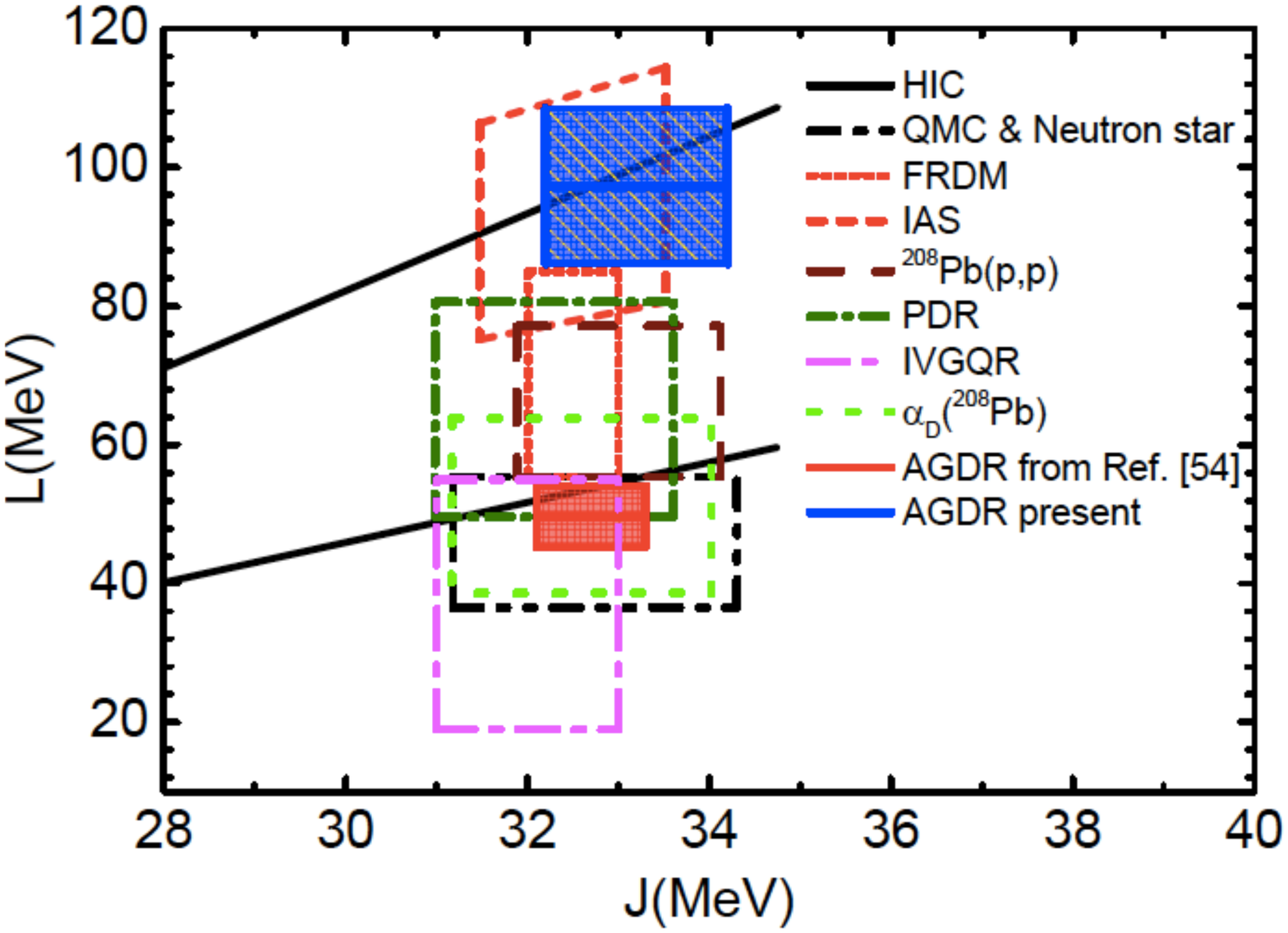}
\caption{\label{fig_agdr_cao} Constraints on the symmetry energy slope $L$ and symmetry energy $J$ at saturation density obtained from the analysis based on the AGDR, in comparison to the results from other approaches (see in ~\cite{Cao2015} for more details). Updated figure taken from private communcations~\cite{ligang2017}.}
%
%corresponds to Fig. 6 in Cao2015 
% included modified version of the figure by Cao, should we refer to it as Private communication?
%
\end{figure}

Fig.~\ref{fig_agdr_cao} shows the comparison of the constraints on the symmetry energy  $J$ and slope parameter $L$ at saturation density determined from the AGDR in $^{208}$Pb, in comparison to the ranges of values from other investigations based on modeling and experimental data on other nuclear phenomena (more details are given in Ref.~\cite{Cao2015}, figure is the update from ~\cite{ligang2017}). As shown in the figure, considerable differences are obtained for the AGDR constraint on $L$, that may have its origin in the model dependence in the EDFs employed in the analysis. Clearly, more theoretical and experimental investigations are necessary to establish more consistent role of the
AGDR in constraining the symmetry energy.

In a recent study~\cite{Yoshida2017}, non-spin-flip vector dipole response
in charge-exchange channel has been explored in neutron rich Ca, Ni, and Sn 
isotopes by using self-consistent 
proton-neutron QRPA based on Skyrme EDF. Model calculations 
show the emergence of low-lying resonance just below the AGDR,
that corresponds to the anti-analog of the pygmy dipole resonance (APDR).
The APDR mode is negatively correlated with the neutron skin thickness;
it shows strong enhancement when neutrons occupy low-$l$ orbital, but 
a moderate increase even for $l=\text{3}$~\cite{Yoshida2017}.
Due to its strong sensitivity with the shell structure, the APDR mode
may not be fully reliable to constrain the EOS properties.

%\section{Excitations in nuclei and EOS}
%do we need this section? Most of the material is already mentioned in subsections above.

%% file: conclusions.tex
\section{Conclusion}
\label{conclusions}

\begin{table}[t!]
\vspace{-0.5cm} 
\begin{center}
  \caption{Summary of selected constraints on the parameters of the EoS: recent results derived from the EDFs; recent experimental results for the $\rho_n$ in ${}^{208}$Pb; and the results from three recent and very exhaustive compilations \cite{lattimer2013,li2013,oertel2017}. Except for $\rho_0$ [fm$^{-3}$] all quantities are given in MeV. Note that instead of $K_{\rm sym}$ the related parameter $K_\tau$ is given. The relation between them is $K_\tau=K_{\rm sym} +L\left(\frac{K^\prime}{K_0}-6\right)$ or as it has been customary in different analyses $K_\tau=K_{\rm sym}-6L$, neglecting $K^\prime$. We also remind that around saturation $S(\rho=0.1$ fm${}^{-3})= J - L/8 + \mathcal{O}[\rho^3]$ and, therefore, $J$ and $L$ parameters are actually linearly correlated if $\mathcal{O}[\rho^3]$ are negligible.}
\begin{tabular}{lccl}
\hline\hline
EoS par. & Observable & Range & Comments\\ 
\hline
$\rho_0$ &$\langle r_{\rm ch}^2\rangle^{1/2}$ & 0.154-0.159 & Most accurate EDFs on $M(N,Z)$ and \\
         &            &       & $\langle r_{\rm ch}^2\rangle^{1/2}$ (see Sec.\ref{gs})\\
\hline
$e_0$        & $M(N,Z)$              & $-$16.2 - $-$15.6 & Most accurate EDFs on $M(N,Z)$ and \\
         &            &       & $\langle r_{\rm ch}^2\rangle^{1/2}$ (see Sec.\ref{gs})\\
\hline
$K_0$         & $M(N,Z)$ & 220-245& Most accurate EDFs on $M(N,Z)$ and \\
         &            &       & $\langle r_{\rm ch}^2\rangle^{1/2}$ (see Sec.\ref{gs})\\
              & ISGMR  & 220-260 & From EDFs in closed shell nuclei \cite{colo2008}\\
              & ISGMR  & 250-315 & Blaizot's formula [Eq.(\ref{eq-pheno-32}] \cite{stone2014}\\
              & ISGMR & $\sim 200$& EDF describing also open shell nuclei \cite{avogadro2013}\\
\hline
$J$           & $M(N,Z)$      & 29-35.6 & Most accurate EDFs on $M(N,Z)$ and \\
         &            &       & $\langle r_{\rm ch}^2\rangle^{1/2}$ (see Sec.\ref{gs})\\
         &IVGDR& $\sim 24.1(8) + L/8$& From EDF analysis \\
         &            &       &     [$S(\rho=0.1$ fm${}^{-3})=$ 24.1(8) MeV]\cite{trippa08}\\
         & PDS        & 30.2-33.8 & From EDF analysis \cite{klimkiewicz07}\\
         & PDS        & 31.0-33.6 & From EDF analysis \cite{carbone10}\\
              & $\alpha_D$       &$24.5(8)+0.168(7)L$& From EDF analysis ${}^{208}$Pb\cite{roca-maza13a}\\
              & $\alpha_D$       &30-35& From EDF analysis \cite{roca-maza2015b}\\ 
              & IAS and $\Delta r_{\rm np}$& 30.2-33.7& From EDF analysis \cite{Danielewicz2014}\\
              & AGDR &  31.2-35.4& From EDF analysis \cite{roca-maza2016}\\
              & PDS, $\alpha_D$, IVGQR, AGDR & 32-33 & From EDF analysis \cite{Paar2014}\\
              & compilation & 29.0-32.7 & \cite{lattimer2013}\\
              & compilation & 30.7-32.5 & \cite{li2013}\\
              & compilation & 28.5-34.9 & \cite{oertel2017}\\

\hline
$L$           & $M(N,Z)$      &  27-113   & Most accurate EDFs on $M(N,Z)$ \\
         &            &       & $\langle r_{\rm ch}^2\rangle^{1/2}$ (see Sec.\ref{gs})\\
              & $\rho_n$   & 40-110 & proton-${}^{208}$Pb scattering \cite{zenihiro2010}\\
              & $\rho_n$    & 0-60 & $\pi$ photoproduction (${}^{208}$Pb) \cite{tarbert2014}\\
              & $\rho_n$    &  30-80& antiprotonic at. (EDF analysis) \cite{centelles09,warda09}\\
              & $\rho_{weak}$&  > 20     & Parity violating scattering \cite{prex}\\
         & PDS        & 32-54& From EDF analysis \cite{klimkiewicz07}\\
         & PDS        &49.1-80.5 & From EDF analysis \cite{carbone10}\\
              & $\alpha_D$       &20-66& From EDF analysis \cite{roca-maza2015b}\\
              &IVGQR and ISGQR  & 19-55 & From EDF analysis \cite{roca-maza13}\\
              & IAS and $\Delta r_{\rm np}$& 35-75& From EDF analysis \cite{Danielewicz2014}\\
              & AGDR &  75.2-122.4& From EDF analysis \cite{roca-maza2016}\\
              & PDS, $\alpha_D$, IVGQR, AGDR & 45.2-54.6 & From EDF analysis \cite{Paar2014}\\
              & compilation & 40.5-61.9 & \cite{lattimer2013}\\
              & compilation & 42.4-75.4 & \cite{li2013}\\
              & compilation & 30.6-86.8 & \cite{oertel2017}\\
\hline
$K_{\tau}$     & $\rho_n$    & $-620$ - $-400$& antiprotonic at. (EDF analysis) \cite{centelles09}\\
              & ISGMR      &  $-650$ - $-450$ & $\alpha$-scattering Sn isotopes \cite{li2010}\\
              & ISGMR      &  $-630$ - $-480$ & $\alpha$-scattering Cd isotopes \cite{patel2012}\\
              & ISGMR      &  $-840$ - $-350$ & Blaizot's formula [Eq.(\ref{eq-pheno-32})] \cite{stone2014}\\
\hline\hline
\end{tabular}
\label{tab-conlusions}
\end{center}
\end{table}

The present status on the available constraints to the nuclear equation of state (EoS) around saturation density from ground and collective excited state properties of atomic nuclei has been reviewed. We have concentrated on the predictions based on self-consistent mean-field calculations, which can be considered as an approximate realization of an exact EDF. Those are derived from effective interactions commonly fitted to nuclear masses, charge radii and, in many cases, also to pseudo-data, often related to the nuclear matter properties. Nuclear EDFs constitute nowadays an unique tool to reliably and consistently access bulk ground state and collective excited state properties of atomic nuclei along the nuclear chart. Such properties are thought to be more clearly connected to the properties of the EoS than single-particle properties. Comparison with other approaches and constraints from observational data have been also also discussed. The main conclusion of this review is that the isospin dependence of the nuclear EoS around saturation density and, to a lesser extent, the nuclear matter incompressibility remain to be accurately determined. Experimental and theoretical efforts in finding and measuring observables specially sensitive to the EoS properties is of paramount importance, not only for low-energy nuclear physics but also for nuclear astrophysics applications. 

\subsection{Summary on the EOS parameters from different approaches}
In Table \ref{tab-conlusions} we summarize recent constraints from the EDFs on the parameters of the nuclear EoS at saturation density [see Eqs.(\ref{eq-pheno-4})-(\ref{eq-pheno-6})]. The saturation density ($\rho_0$) and energy ($e_0$) of the symmetric matter EoS have been shown to be reasonably well constrained in EDFs from the fit to nuclear masses and charge radii (see Table\ref{tab-gs-1}). For this reason, we give in the first two rows of 
Table~\ref{tab-conlusions} the predicted ranges for these two properties from some of the most accurate EDFs in the description of masses and charge radii (those in Table\ref{tab-gs-1}). In the third row of Table~\ref{tab-conlusions}, different estimates for the nuclear matter incompressibility ($K_0$) are given. This quantity is also sensitive to the fitting of nuclear masses although to a less extent than $\rho_0$ and $e_0$. In this case, three ranges are given for $K_0$ that come from an EDF analysis. First, by taking the range spanned by some of the  most accurate EDFs in the description of masses and, then, by reporting also ranges of different theoretical studies based on the excitation energy of the ISGMR as predicted by other EDFs than those detailed in Table\ref{tab-gs-1}. In the latter case, the study of the ISGMR in closed shell nuclei points towards very similar values of $K_0$ when compared to the range predicted by the EDFs of Table\ref{tab-gs-1} while the study of the ISGMR in open shell nuclei points towards lower values of $K_0$. In the fourth range shown for $K_0$ in Table\ref{tab-conlusions}, we give for comparison the latest analysis performed on the basis of the {\it macroscopic} Blaizot's formula where the predicted range is somewhat shifted to larger values. In the fourth and fifth rows, we give instead different estimates to the value of the symmetry energy at saturation ($J$) and the corresponding slope parameter ($L$). We remind here that $S(\rho=0.1$ fm${}^{-3})= J - L/8 + \mathcal{O}[(\rho-\rho_0)^3]$ and, therefore, $J$ and $L$ parameters are actually linearly correlated if $\mathcal{O}[(\rho-\rho)^3]$ are negligible. Most of these estimates come from different analyses made on the basis of EDF predictions for masses and collective excited state properties such as the IVGDR, $\alpha_D$, IAS, and the AGDR. We also give constraints derived from the PDS. We complement the constraints on $L$ with some recent experimental results on $\rho_n$ in ${}^{208}$Pb since the correlation $\Delta r_{\rm np}({}^{208}$Pb$)$ and $L$ is almost one to one for the EDFs. We finally complement the ranges given for $J$ and $L$ with the results from three recent and very exhaustive compilations \cite{lattimer2013,li2013,oertel2017}. In the last row, we complete Table\ref{tab-conlusions} by giving some ranges for the $K_\tau$ parameter instead of $K_{\rm sym}$. The relation between them is $K_\tau=K_{\rm sym} +L\left(\frac{K^\prime}{K_0}-6\right)$ or as it has been customary in different analyses $K_\tau=K_{\rm sym}-6L$, neglecting $K^\prime$. In this case, we only provide one estimate directly related to an analysis employing EDFs (antiprotonic atoms) while the other three are based on the Blaizot's formula and the experimental measurement of the ISGMR. The spread is large but different estimates seem to agree on the sign and magnitude.

For more details on analyses of the nuclear symmetry energy based on different theoretical frameworks, we refer the reader to Refs.\cite{tsang12,s-epja,Horowitz:2014bja}.